\documentclass[a4paper,10pt,twoside,notitlepage]{book}

 \usepackage{amsmath}
 \usepackage{amssymb}

 \usepackage{graphicx}

  \usepackage[cp1251]{inputenc}
  \usepackage[T2A]{fontenc}
  \usepackage[russian,english]{babel}

 \DeclareMathOperator{\sgn}{sgn}

 \renewcommand{\Re}{\mathop\mathrm{Re}\nolimits}
 \renewcommand{\Im}{\mathop\mathrm{Im}\nolimits}

\makeatletter

\renewenvironment{thebibliography}[1]
     {\clearpage
      \noindent\textbf{\large References}\medskip
      \list{\@biblabel{\@arabic\c@enumiv}}%
           {\settowidth\labelwidth{\@biblabel{#1}}%
            \leftmargin\labelwidth
            \advance\leftmargin\labelsep
            \@openbib@code
            \usecounter{enumiv}%
            \let\p@enumiv\@empty
            \renewcommand\theenumiv{\@arabic\c@enumiv}}%
      \sloppy
      \clubpenalty4000
      \@clubpenalty \clubpenalty
      \widowpenalty4000%
      \sfcode`\.\@m}

\begin{document}

\thispagestyle{empty}

\begin{center}

\vspace*{30mm}

\textbf{{\Large P}ROXIMITY AND {\Large J}OSEPHSON {\Large E}FFECTS\\[4mm]
IN {\Large S}UPERCONDUCTING {\Large H}YBRID {\Large S}TRUCTURES}

\end{center}

\newpage

\thispagestyle{empty}

\noindent Ph.D. committee:\\[5mm]
\hspace*{3mm} Prof. dr. J.~Mellema (University of Twente), Chairman and Secretary\\
\hspace*{3mm} Prof. dr. H.~Rogalla (University of Twente)\\
\hspace*{3mm} Dr. A.\,A.~Golubov (University of Twente)\\
\hspace*{3mm} Prof. dr. M.\,V.~Feigel'man (L.\,D.~Landau Institute for Theoretical Physics, Russia)\\
\hspace*{3mm} Dr. J.~Aarts (Leiden University)\\
\hspace*{3mm} Prof. dr. ing. B.~van~Eijk (University of Twente)\\
\hspace*{3mm} Prof. dr. D.\,J.~Van~Harlingen (University of Illinois at Urbana-Champaign, USA)\\
\hspace*{3mm} Prof. dr. P.\,J.~Kelly (University of Twente)\\
\hspace*{3mm} Prof. dr. Yu.\,V.~Nazarov (Delft University of Technology)

\vfill

\noindent The research described in this thesis was carried out as a collaboration between the L.\,D.~Landau Institute
for Theoretical Physics (Moscow, Russia) and the Low Temperature Division of the Department of Applied Physics at the
University of Twente (Enschede, The Netherlands).\\
The research has been supported by the D-Wave Systems Inc., the Russian Foundation for Basic Research, the Russian
Academy of Sciences, the Russian Ministry of Industry, Science and
Technology, the Research Centre J\"ulich (Landau Scholarship), and the Swiss National Foundation.\\[5mm]

\noindent Printed by PrintPartners Ipskamp, Enschede\\[5mm]

\noindent ISBN 90-365-1966-7

\newpage

\thispagestyle{empty}

\begin{center}

\vspace*{30mm}

\textbf{{\Large P}ROXIMITY AND {\Large J}OSEPHSON {\Large E}FFECTS\\[4mm]
IN {\Large S}UPERCONDUCTING {\Large H}YBRID {\Large S}TRUCTURES}

\vspace{30mm}

PROEFSCHRIFT

\vspace{25mm}

ter verkrijging van\\
de graad van doctor aan de Universiteit Twente,\\
op gezag van de rector magnificus,\\
prof. dr. F.\,A.~van Vught,\\
volgens besluit van het College voor Promoties\\
in het openbaar te verdedigen\\
op woensdag 12 november 2003 om 15.00 uur

\vspace{10mm}

door

\vspace{10mm}

\textbf{Yakov Fominov}

\vspace{10mm}

geboren op 6 mei 1976\\
~\\
~\\
te Moskou (Rusland)

\end{center}

\newpage

\thispagestyle{empty}

\noindent Dit proefschrift is goedgekeurd door:\\[5mm]
\hspace*{5mm} Prof. dr. H.~Rogalla (promotor)\\
\hspace*{5mm} Dr. A.\,A.~Golubov (assistent-promotor)

 \renewcommand{\@oddhead}{}
 \renewcommand{\@evenhead}{}
 \renewcommand{\@oddfoot}{\hfil\thepage\hfil}
 \renewcommand{\@evenfoot}{\hfil\thepage\hfil}

\tableofcontents \clearpage

\chapter*{\huge Introduction}
\addcontentsline{toc}{chapter}{Introduction}

\renewcommand{\@oddfoot}{}
\renewcommand{\@evenfoot}{}

\renewcommand{\@evenhead}
    {\raisebox{0pt}[\headheight][0pt]
     {\vbox{\hbox to\textwidth{\thepage \hfil \strut \textit{Introduction}}\hrule}}
    }

\renewcommand{\@oddhead}
    {\raisebox{0pt}[\headheight][0pt]
     {\vbox{\hbox to\textwidth{\textit{Introduction} \strut \hfil \thepage}\hrule}}
    }

\section*{Superconductivity}
\addcontentsline{toc}{section}{Superconductivity}

The mysterious effect of superconductivity (zero electric resistance) in a metal at low temperatures was discovered by
H.~Kamerlingh Onnes in 1911 \cite{Kamerlingh}. For a long time, the microscopic explanation of this phenomenon was
lacking, while a number of successful phenomenological approaches (the London theory \cite{Londons}, the
Ginzburg--Landau theory \cite{GL}) were employed to study various properties of superconductors.

During this time, the effect of superfluidity (flowing without dissipation) in the liquid helium $^4$He was
experimentally discovered by P.\,L.~Kapitza \cite{Kapitza} and theoretically explained by L.\,D.~Landau \cite{Landau}.
The basic mechanism behind the superfluidity is the Bose--Einstein condensation of helium atoms. Although at first
sight, liquid helium does not have much common with solid superconductor, actually the phenomena of superfluidity and
superconductivity are very close to each other. While the superfluidity is the flowing of helium atoms without
dissipation, the superconductivity is the flowing of electrons without dissipation. Thus superconductivity can be
imagined as the superfluidity of electronic liquid. At the same time, there is a crucial difference between the two
phenomena. While a macroscopic number of bosonic helium atoms undergo the Bose--Einstein condensation and occupy the
lowest energy level, the electrons, which are fermionic particles, cannot do this as they are not allowed to share the
same state.

A key step in unravelling the mystery of superconductivity was made by L.\,N.~Cooper in 1956 \cite{Cooper_pairing}, who
showed that if electrons attract then the electronic system becomes unstable against forming the so-called Cooper pairs
(the attraction that can overcome the Coulomb repulsion can be produced due to interaction of electrons with lattice
vibrations). The newly formed pairs of electrons are bosonic objects, hence they can undergo the Bose--Einstein
condensation and form a state necessary for the superfluidity! This idea became the keystone of the microscopic theory
of superconductivity which was finally built in the late fifties by J.~Bardeen, L.\,N.~Cooper, and J.\,R.~Schrieffer
(BCS) \cite{BCS} and N.\,N.~Bogolyubov \cite{Bogolyubov}. Shortly thereafter, L.\,P.~Gor'kov formulated the BCS theory
in the language of the Green functions \cite{Gor'kov}.

Later, the possibility of unconventional superconductivity, which is characterized by nonzero spin or orbital momentum
of the Cooper pairs, was discussed. The ordinary case corresponds to Cooper pairing between electrons with opposite
spins, the singlet pairing. The other, unconventional case, corresponds to pairing between electrons with the same spin
projection, the triplet pairing. Another nontrivial possibility is a nonzero orbital momentum. The superconducting state
is characterized by the order parameter $\Delta$, which describes the ``strength'' of the superconductivity and
determines the energy gap in the single-particle density of states. In the ordinary case, the energy gap $\Delta$ is the
same for all directions of electron's momentum. However, if the orbital momentum of the Cooper pairs is nonzero then the
superconductivity is anisotropic, which means that $\Delta$ depends on the direction of motion inside the superconductor
(an example of anisotropic superconductivity is established in the high-temperature superconductors
\cite{Van_Harlingen,Tsuei}).

Extensive investigations demonstrated that zero electric resistance is only one of the many manifestations of
superconductivity. Below we mention two of them which are particularly relevant for the subject of the thesis and appear
in its title.

When a superconductor contacts a normal metal, a number of phenomena known as the proximity effect takes place. The two
materials influence each other on a spatial scale of the order of the coherence length in the vicinity of the interface.
In particular, the superconducting correlations between quasiparticles are induced into the normal metal, because the
Cooper pairs penetrating into the normal metal have a finite lifetime there. Until they decay into two independent
electrons, they preserve the superconducting properties. Alternatively, the proximity effect can be viewed as resulting
from the fundamental process known as the Andreev reflection \cite{Andreev}. Imagine a low-energy electron impinging
from the normal metal onto the interface with the superconductor. A single electron can penetrate the superconductor
only if its energy is larger than the superconducting energy gap $\Delta$, while below it only Cooper pairs can exist.
Thus the low-energy electron cannot penetrate the superconductor. If there is no potential barrier at the interface, the
electron cannot be reflected back either, due to the momentum conservation law. The only way out of this contradiction
is the Andreev reflection: a Cooper pair goes into the superconductor, while a hole (``the antiparticle'' for the
electron) goes back into the normal metal in order to conserve charge.

Deep understanding of the nature of the superconducting state lead also to the discovery of one of the most spectacular
superconducting phenomena, the Josephson effect \cite{Josephson}. This occurs when two superconductors are connected via
a weak link, which can be either a nonsuperconducting material (insulator as in Ref.~\cite{Josephson} or metal) or a
geometrical constriction (the variety of Josephson junctions is reviewed in Ref.~\cite{Likharev}). The superconducting
condensate in each of the two weakly coupled superconductors is described by its wave function and the corresponding
phase. Josephson found that if the phases are different and the difference is $\varphi$, then the supercurrent
(nondissipative current) $I = I_c \sin\varphi$ arises across the junction in the absence of voltage. The quantity $I_c$
is called the critical current.

\section*{Motivation}
\addcontentsline{toc}{section}{Motivation}

From the most fundamental aspects of superconductivity let us now switch to the particular issues that became the
subject of the thesis.

Although the investigation of the proximity effect in SN systems\footnote{The notations: S
--- \textit{s}-wave superconductor, D
--- \textit{d}-wave superconductor, N --- normal metal, F --- ferromagnetic metal, I --- insulator, c
--- constriction.} was started about forty years ago \cite{Cooper,Werthamer,deGennes_review},
the technology allowing to produce and measure experimental samples of mesoscopic dimensions was achieved relatively
recently. In particular, it became possible to study SN structures consisting of thin layers (having thickness smaller
than the coherence length). Such structures behave as a single superconductor with nontrivial properties. The most basic
of them has already been studied mostly for the case of ideally transparent interface. At the same time, the
experimental progress requires the corresponding advances in theory, especially taking into account arbitrary interface
transparency. This crucial parameter determines the strength of the proximity effect and at the same time is not
directly measurable. From practical point of view, the SN proximity structures can be used as superconductors with
relatively easily adjustable parameters, in particular, the energy gap and the critical temperature. The parameters of
the proximity structures can be tuned, e.g., by varying the thicknesses of the layers. This method has already found its
application in superconducting transition edge bolometers and photon detectors for astrophysics (see, e.g.,
Refs.~\cite{Poelaert,Verhoeve}).

The physics of SF systems is even richer. In contrast to the SN case, the superconducting order parameter does not
simply decay into the nonsuperconducting metal but also oscillates. This behavior is due to the exchange field in the
ferromagnet that acts as a potential of different signs for two electrons in a Cooper pair and leads to a finite
momentum of the pair (similarly to the Larkin--Ovchinnikov--Fulde--Ferrell state in bulk materials \cite{LOFF}). This
oscillations reveal itself in nonmonotonic dependence of the critical temperature $T_c$ of SF systems as a function of
the F layers thickness, both in the cases of SF multilayers \cite{BK,Radovic} and bilayers \cite{Radovic,PKhI,Tagirov}.
At the same time, in most of the papers investigating this effect, the methods to calculate $T_c$ were approximate. An
exact method was proposed in Ref.~\cite{Radovic} for the limiting case of perfect interfaces and large exchange
energies. An exact method to calculate $T_c$ at arbitrary parameters of the system was lacking. The need for such a
method was also motivated by the experiment \cite{ROP} that did not correspond to the previously considered
approximations and limiting cases.

Another interesting effect in SF systems takes place if the magnetization of the ferromagnet is inhomogeneous. Then the
triplet superconducting component can arise in the system \cite{Bergeret,Kadigrobov}. Physically, this effect is similar
to generating the triplet correlations in magnetic superconductors \cite{JLTP} (with Cooper interaction only in the
singlet channel). Recently, it was demonstrated that the triplet component also arises in the case of several
homogeneous but noncollinearly oriented ferromagnets \cite{VBE}. However, the conditions at which the superconductivity
is not destroyed in this system were not found. The simplest system of the above type is an FSF trilayer. The answer to
the question about the conditions for the superconductivity to exist can be obtained when studying the critical
temperature $T_c$ of the system. At the same time, a method to calculate $T_c$ in a situation when the triplet component
is generated, was lacking. A possible practical application of FSF structures is a spin valve
\cite{Tagirov_PRL,Buzdin_EL}, a system that switches between superconducting and nonsuperconducting states when the
relative orientation of the magnetizations is varied. Although the superconductive spin valve is not yet experimentally
realized, the work in this direction has already started \cite{ANL}.

The Josephson effect in the systems containing ferromagnets (e.g., structures of the SFIFS, SIFIS or SFcFS type) also
has a number of peculiarities. Among them are the transition from the ordinary ($0$-state) to the so-called $\pi$-state
(in other words, the inversion of the critical current sign or the additional $\pi$ phase shift in the Josephson
relation
--- this effect was theoretically predicted in Refs.~\cite{Bulaevski,Buzdin_pi} and experimentally observed in
Ref.~\cite{Ryazanov}) when the two ferromagnets are aligned in parallel, the enhancement of the critical current by the
exchange field in the case of antiparallel orientation \cite{BVE}, and nonsinusoidal current--phase relation. Although
such effects have been studied before, it was often done in the simplest models and at the simplest assumptions about
system parameters. To achieve better understanding of this phenomena, one should study them at various conditions and
determine the physical mechanisms behind the effects. The interest to SFS junctions with nontrivial current--phase
relation is in particular due to their possible employment for engineering logic circuits of novel types, both classical
and quantum bits (see, e.g., Refs.~\cite{Beasley,Blatter}).

Another interesting type of nonuniform superconducting systems is a junction between superconductors of nontrivial
symmetry. The superconductors with the \textit{d}-wave symmetry of the order parameter are widely discussed because this
symmetry is realized in the high-temperature superconductors \cite{Van_Harlingen,Tsuei}. A possibility to implement a
so-called qubit (quantum bit) based on \textit{d}-wave junctions was proposed in Refs.~\cite{Ioffe,Zagoskin,Il'ichev}.
Quantum bit is, simply speaking, a quantum mechanical system with two states (which can be imagined as spin $1/2$).
While a classical bit can be either in one state or in the other, a qubit can also be in a superposition of the two
states. If a quantum computer is built of such qubits, it would have the advantage of natural computational parallelism
that can enormously speed up certain types of computational tasks. A possibility to implement a qubit based on DID
junction stems from the fact that the energy of such a junction as a function of the phase difference can have a
double-well form with two minima. This degeneracy of the ground state arises due to the nontrivial symmetry of the
superconductors. Due to tunneling between the wells, the ground state splits, and the two resulting levels effectively
form the quantum-mechanical two-state system. At the same time, the gapless nature of the \textit{d}-wave
superconductors leads to appearance of low-energy quasiparticles which can destroy the quantum coherence of the qubit
and hence hamper its successful functioning. Calculation of the corresponding decoherence time is necessary for
estimating the efficiency of the proposed qubits.

\section*{Outline of the thesis}
\addcontentsline{toc}{section}{Outline of the thesis}

In Chapter~1, the theory of superconductivity in thin SN sandwiches (bilayers) in the diffusive limit is developed, with
particular emphasis on the case of very thin superconductive layers, $d_S \ll d_N$. The proximity effect in the system
is governed by the interlayer interface resistance (per channel) $\rho_B$. The case of \textit{relatively} low
resistance (which can still have large absolute values) can be completely studied analytically. The theory describing
the bilayer in this limit is of BCS type but with the minigap (in the single-particle density of states) $E_g \ll
\Delta$ substituting the order parameter $\Delta$ in the standard BCS relations; the original relations are thus
severely violated. In the opposite limit of an opaque interface, the behavior of the system is in many respects close to
the BCS predictions. Over the entire range of $\rho_B$, the properties of the bilayer are found numerically. Finally, it
is shown that the results obtained for the bilayer also apply to more complicated structures such as SNS and NSN
trilayers, SNINS and NSISN systems, and SN superlattices.

In Chapter~2, we propose two exact methods to calculate the critical temperature $T_c$ of dirty SF bilayers at arbitrary
parameters of the system. The methods are applied to study the nonmonotonic behavior of the critical temperature versus
thickness of the F layer. Comparing our results with experimental data, we find good agreement. Then we study the
critical temperature of FSF trilayers, where the triplet superconducting component is generated at noncollinear
magnetizations of the F layers. We reduce the problem to the form that allows us to employ the exact numerical methods
developed earlier, and calculate $T_c$ as a function of the trilayer parameters, in particular, mutual orientation of
magnetizations. Analytically, we consider interesting limiting cases. Our results determine conditions which are
necessary for existence of the odd (in energy) triplet superconductivity in SF multilayers.

In Chapter~3, the quantitative theory of the Josephson effect in SFS junctions (with one or several F layers) is
presented in the dirty limit. Fully self-consistent numerical procedure is employed to solve the Usadel equations at
arbitrary values of the F-layers thicknesses, exchange energies, and interface parameters. In SFIFS junction, at
antiparallel ferromagnets' magnetizations the effect of the critical current enhancement by the exchange field is
observed, while in the case of parallel magnetizations the junction exhibits the transition to the $\pi$-state. In the
limit of thin F layers, we study these peculiarities of the critical current analytically and explain them
qualitatively; the scenario of the $0$--$\pi$ transition in our case differs from those studied before. The effect of
switching between $0$ and $\pi$ states by changing the mutual orientation of the F layers is demonstrated. Also, various
types of the current--phase relation $I(\varphi)$ in SFcFS point contacts and planar double-barrier SIFIS junctions are
studied in the limit of thin ferromagnetic interlayers. The physical mechanisms leading to highly nontrivial
$I(\varphi)$ dependence are identified by studying the spectral supercurrent density. In particular, these mechanisms
are responsible for the $0$--$\pi$ transition in SFS Josephson junctions.

In Chapter~4, we study the Josephson junction between two \textit{d}-wave superconductors, which is discussed as an
implementation of a qubit. We propose an approach that allows to calculate the decoherence time due to an intrinsic
dissipative process: quantum tunneling between the two minima of the double-well potential excites nodal quasiparticles
which lead to incoherent damping of quantum oscillations. In DID junctions of the mirror type, the contribution to the
dissipation from the nodal quasiparticles is superohmic and becomes small at small tunnel splitting of the energy level
in the double-well potential. For available experimental data, we estimate the quality factor.

\section*{Green functions in the theory of superconductivity}
\addcontentsline{toc}{section}{Green functions in the theory of superconductivity}

The quantum field theory methods turned out to be very powerful also in the theory of solid state systems \cite{AGD}.
The theory of superconductivity was formulated in this language, the language of the Green functions, by L.\,P.~Gor'kov
\cite{Gor'kov}. The Gor'kov equations were derived directly from the BCS Hamiltonian. They can be conveniently written
in the matrix form:
\begin{gather}
\begin{pmatrix}
\displaystyle (E+i0) +\frac 1{2m} \frac{\partial^2}{\partial \mathbf{r}_1^2} +E_\mathrm{F} - V(\mathbf{r}_1)
& i\Delta(\mathbf{r}_1) \\
i\Delta^* (\mathbf{r}_1) & \displaystyle - (E+i0) +\frac 1{2m} \frac{\partial^2}{\partial \mathbf{r}_1^2} +E_\mathrm{F}
- V(\mathbf{r}_1)
\end{pmatrix}
\times \notag \\
\times \Hat G (E, \mathbf{r}_1, \mathbf{r}_2) = \Hat 1 \, \delta(\mathbf{r}_1 - \mathbf{r}_2),
\end{gather}
where the retarded matrix Green function $\Hat G$ contains the standard Green function $G$ and the anomalous Green
functions $F$ and $\bar F$ that describe the superconducting correlations:
\begin{equation}
\Hat G =
\begin{pmatrix}
G & F \\
\bar F & \bar G
\end{pmatrix},
\end{equation}
$E_\mathrm{F}$ is the Fermi energy, $V(\mathbf{r}_1)$ is the impurity potential.

In real systems, solving the Gor'kov equations can be a formidable task. At the same time, the information contained in
those is often redundant. In particular, the Green functions entering the Gor'kov equation describe rapid oscillations
on the scale of the Fermi wave-length. Of course, there are situations when these oscillations play crucial role and
must be carefully described. At the same time, if characteristic scales in the problem are large compared to the
interatomic distance, then the rapid oscillations are averaged and only the slow part of the Green functions determines
the physical properties. For those cases, the Gor'kov equations can be simplified from the very beginning. The resulting
formalism is known as the quasiclassical method \cite{Eilenberger,LO_qc,Usadel}, and below we briefly outline the main
steps leading to it.

The Green functions entering the Gor'kov equations depend on two coordinates and energy. We can do the Wigner transform
over the coordinates, which means that we introduce the center of mass coordinate $\mathbf{r} =(\mathbf{r}_1 +
\mathbf{r}_2) /2$ and the relative coordinate $\boldsymbol\rho =\mathbf{r}_1 - \mathbf{r}_2$, and make the Fourier
transform over $\boldsymbol\rho$ resulting in the relative momentum $\mathbf{p}$. The above transformations are exact,
they are made simply for convenience. The next step is the quasiclassical approximation itself, the idea of which is
that the dependence of the Green function on the relative coordinate is fast, while the dependence on the center-of-mass
coordinate $\mathbf{r}$ is slow (the small parameter of this approximation is $T_c / E_\mathrm{F}$, with $T_c$ being the
critical temperature). The quasiclassical approximation implies that we average the Green functions over the rapid
oscillations described by the relative momentum $p$:
\begin{equation}
\Hat g(E,\mathbf{r},\mathbf{n}) = \frac i\pi \int d\xi\, \Hat G(E,\mathbf{r},\mathbf{p}).
\end{equation}
The integration over $p$, the absolute value of the relative momentum, is standardly rewritten as integration over $\xi
=p^2 /2m -E_\mathrm{F}$. The unit vector $\mathbf{n} = \mathbf{p} /p$ points in the direction of the relative momentum.
Finally, we should average the Green function over impurities.

The equation for the resulting quasiclassical function was derived by G.~Eilenberger \cite{Eilenberger} (see also
Ref.~\cite{LO_qc}); it reads
\begin{equation}
\mathbf{v}_\mathrm{F} \, \frac{\partial \Hat g}{\partial \mathbf{r}} + \left[ -i (E+i0)\, \Hat \sigma_3 + \Hat\Delta
+\frac{\left< \Hat g \right>}{2\tau},\, \Hat g \right] = 0, \qquad \Hat\Delta =
\begin{pmatrix}
0 & \Delta \\
\Delta^* & 0
\end{pmatrix},
\end{equation}
where the square brackets denote the commutator and the angular brackets denote averaging over directions of
$\mathbf{n}$ (angular averaging). $\mathbf{v}_\mathrm{F}$ is the Fermi velocity, $\sigma_3$ is the third Pauli matrix,
and $\tau$ is the time of the mean free path. The Green function entering this equation obeys the normalization
condition $\Hat g^2 =\Hat 1$.

The Eilenberger equation can be further simplified in the diffusive (dirty) limit. The physical reason is that due to
frequent scattering on impurities, the Green function becomes isotropic. The equation for the isotropic function
\begin{equation}
\Hat g(E,\mathbf{r}) = \left< \Hat g(E,\mathbf{r},\mathbf{n}) \right>
\end{equation}
was derived by K.\,D.~Usadel \cite{Usadel}:
\begin{equation}
D \frac\partial{\partial \mathbf{r}} \left( \Hat g \frac\partial{\partial \mathbf{r}} \Hat g \right)+ \left[ i (E+i0)\,
\Hat \sigma_3 - \Hat\Delta ,\, \Hat g \right] = 0, \qquad \Hat g^2 = \Hat 1.
\end{equation}

All forms of the equations discussed above must be complemented with the corresponding self-consistency equation for the
order parameter $\Delta$. For brevity, we write down the self-consistency condition only for the Usadel equation:
\begin{equation}
\Delta (\mathbf{r}) = -\frac{i\lambda} 2 \int_{-\omega_D}^{\omega_D} dE \, \tanh \left( \frac E{2T} \right)
f(E,\mathbf{r}),
\end{equation}
where $f$ is the anomalous component of the matrix Green function, $T$ is the temperature, $\lambda$ is the coupling
constant (effective parameter of electron-electron interaction), and the integration is cut off at the Debye energy
$\omega_D$. In particular, this equation yields the zero-temperature gap of the bulk superconductor $\Delta = 2\omega_D
\exp (-1/\lambda)$.

The quasiclassical approximation breaks down near interfaces between different materials, when the interfaces are sharp
on the scale of the Fermi wave-length (atomic scale). In order to use the quasiclassical equations for describing
nonuniform systems, one should complement them with the effective boundary conditions. The boundary conditions for the
Eilenberger equation were derived by A.\,V.~Zaitsev \cite{Zaitsev} and for the Usadel equation --- by M.\,Yu.~Kupriyanov
and V.\,F.~Lukichev \cite{KL}. The latter boundary conditions have the form
\begin{equation}
\sigma_l\, \Hat g_l \frac{\partial \Hat g_l}{\partial x} = \sigma_r\, \Hat g_r \frac{\partial \Hat g_r}{\partial x} =
\frac 1{2 R_B \mathcal{A}} \left[ \Hat g_l,\, \Hat g_r \right],
\end{equation}
where the indices $l$ and $r$ refer to the left- and right-hand side of the interface, respectively, $\sigma$ is the
conductivity of metals, $R_B$ is the total resistance of the interface, and $\mathcal{A}$ is its area. The $x$ axis is
normal to the interface.

\clearpage
\renewcommand{\@evenhead}{}

\chapter[Superconductivity in thin SN bilayers]{\huge Superconductivity in thin SN bilayers}

\renewcommand{\@evenhead}
    {\raisebox{0pt}[\headheight][0pt]
     {\vbox{\hbox to\textwidth{\thepage \hfil \strut \textit{Chapter \thechapter}}\hrule}}
    }

\renewcommand{\@oddhead}
    {\raisebox{0pt}[\headheight][0pt]
     {\vbox{\hbox to\textwidth{\textit{\leftmark} \strut \hfil \thepage}\hrule}}
    }

\section{Introduction}

It is well known that the majority of metallic superconductors is well described by the classical BCS theory of
superconductivity \cite{1:BCS}. One of the main qualitative features of the BCS theory is a simple relation between the
superconductive transition temperature $T_c$ and the low-temperature value of the energy gap for \textit{s}-wave
superconductors: $\Delta(0) = 1.76\, T_c$. Experimentally, violations of this simple relation are considered as a sign
of some unusual pairing symmetry or even of a non-BCS pairing mechanism. Recently, an evident example of such a
violation of the BCS theory predictions was found in experiments by Kasumov \textit{et al.} \cite{1:Kasumov}, who
studied the current-voltage characteristics of a carbon nanotube contact between two metallic bilayers (sandwiches) made
of ordinary metals, tantalum and gold. The observed value of the low-temperature Josephson critical current is 40 times
larger than the maximum expected (Ambegaokar-Baratoff) value \cite{1:AB} $I_c = \pi \Delta (0)/2 eR_\mathrm{tube}$,
where the energy gap of the bilayer $\Delta (0)$ is estimated from its transition temperature. The source of such
discrepancy is not clear at present. The most recent experiments \cite{1:Kasumov_new} demonstrate the existence of
intrinsic superconductivity in carbon nanotubes. However, the discrepancy could also be due to unusual superconductive
properties of the bilayers.

Although the proximity effect in SN bilayers is rather well studied, mostly the limit of the perfect interface (which is
called ``the Cooper limit'' in the thin bilayer case \cite{1:Cooper,1:deGennes_review}) or the opposite limit of opaque
interface has been explored. At the same time, as we demonstrate below, between the two limiting cases the
characteristics of the system and the relations between them, depending on the interface resistance, behave quite
nontrivially. The aim of the present chapter is to study the superconducting properties of a thin SN bilayer depending
on the interface resistance.

An essential feature of the experiment~\cite{1:Kasumov} was that the superconductive layer in the bilayer was very thin:
$d_S/d_N$ = 5~nm/100~nm = 1/20. In the present chapter, we investigate such a bilayer both analytically and numerically,
calculating quantities characterizing the superconductivity in this proximity system: the order parameter $\Delta$, the
density of the superconducting electrons $n$, the critical temperature $T_c$, the (mini)gap $E_g$ in the single-particle
density of states (DoS), the critical magnetic field $H_c$ parallel to the bilayer and the upper critical field $H_{c2}$
perpendicular to the bilayer. In our calculations, the parameter controlling the strength of the proximity effect is the
(dimensionless) resistance of the SN interface per channel $\rho_B$, which is related to the total interface resistance
$R_B$ by the Sharvin formula
\begin{equation} \label{rho_int}
R_B= \frac{R_q \rho_B}{N_\mathrm{ch}},
\end{equation}
where $R_q= \pi / e^2$ is the quantum resistance, and $N_\mathrm{ch} = \mathcal{A}/ (\lambda_F/2)^2$, with $\lambda_F$
being the Fermi wave-length, is the number of channels in the interface of area $\mathcal{A}$. We choose $\lambda_F$
referring to the S layer.

The values of the interface resistance can be divided into three ranges: (a) at large resistance, many characteristics
of the superconductor ($\Delta$, $n$, $H_c$, $H_{c2}$) are almost unaffected by the presence of the normal layer ---
this is the BCS limit (however, we note that $E_g$ does not coincide with the order parameter, and even vanishes as
$\rho_B$ increases); (b) at low resistance, the theory describing the bilayer is of BCS type but with the order
parameter $\Delta$ substituted by the minigap $E_g$ (for instance, $E_g = 1.76\, T_c$, whereas $E_g \ll \Delta$); the
original BCS relations are thus severely violated; (c) at intermediate resistance, the behavior of the system
interpolates between the above two regimes.

The violation of the BCS relations, corresponding to the parameters of the experiment~\cite{1:Kasumov}, appears to be
insufficient to explain the observed value of the critical current. Probably, the intrinsic superconductivity of the
carbon nanotubes \cite{1:Kasumov_new} plays an essential role. Therefore, the experiment~\cite{1:Kasumov} is mainly a
motivation for the theoretical study described below. The practical outcome is an estimate for the interface resistance,
obtained from the experimental values of the critical temperature and the parallel critical magnetic field (see
Sec.~\ref{sec:discussion} below).

\section{Method} \label{sec:usadel}

\subsection{Usadel equation}

Equilibrium properties of dirty systems are described \cite{1:LO} by the quasiclassical retarded Green function $\hat
g(\mathbf{r},E)$, which is a $2\times 2$ matrix in the Nambu space satisfying the normalization condition $\hat g^2
=\hat 1$. The retarded Green function obeys the Usadel equation \cite{1:Usadel}
\begin{equation} \label{usadel}
D\nabla (\hat g \nabla \hat g) + [i E \hat\sigma_3 - \hat\Delta,\,\hat g] =0.
\end{equation}
Here the square brackets denote the commutator, $D =\mathrm{v} l/3$ is the diffusion constant with $\mathrm{v}$ and $l$
being the Fermi velocity and the elastic mean free path, $E$ is the energy, whereas $\hat\sigma_3$ (the Pauli matrix)
and $\hat\Delta(\mathbf{r})$ are given by
\begin{equation}
\hat\sigma_3 = \begin{pmatrix} 1 & 0\\ 0 & -1 \end{pmatrix} ,\qquad \hat\Delta = \begin{pmatrix} 0 & \Delta\\ \Delta^* &
0 \end{pmatrix}.
\end{equation}
The order parameter $\Delta(\mathbf{r})$ must be determined self-consistently from the equation
\begin{equation}
\Delta (\mathbf{r}) = -\frac{i\lambda} 2 \int_{-\omega_D}^{\omega_D} dE \, \tanh \left( \frac E{2T} \right) f
(\mathbf{r},E),
\end{equation}
where $f$, the anomalous Green function, is the upper off-diagonal element of the $\hat g$ matrix, $\lambda$ is the
effective constant of electron-electron interaction in the S layer (whereas we assume $\lambda=0$ and hence $\Delta =0$
in the N layer), and integration is cut off at the Debye energy $\omega_D$ of the S material.

Equation (\ref{usadel}) should be supplemented with the appropriate boundary conditions at an interface, which read
\cite{1:KL}
\begin{equation} \label{interface}
\sigma_l \left( \hat g_l \nabla_\mathbf{n} \hat g_l \right) = \sigma_r \left( \hat g_r \nabla_\mathbf{n} \hat g_r
\right) = \frac{g_B}2 \left[ \hat g_l,\,\hat g_r \right],
\end{equation}
where the subscripts $l$ and $r$ designate the left and right electrode, respectively; $\sigma$ is the conductivity of a
metal in the normal state, and $g_B =G_B / \mathcal{A}$ (with $G_B = 1/ R_B$) is the conductance of the interface per
unit area when both left and right electrodes are in the normal state. $\nabla_\mathbf{n}$ denotes the projection of the
gradient upon the unit vector $\mathbf{n}$ normal to the interface.

We use the system of units in which the Planck constant and the speed of light equal unity: $\hbar=c=1$.

\subsection{Angular parameterization of the Green function}
\label{subsec:theta}

The normalization condition allows the angular parameterization of the retarded Green function:
\begin{equation}
\hat g = \begin{pmatrix} \cos\theta & e^{i\varphi} \sin\theta\\ e^{-i\varphi}\sin\theta & -\cos\theta
\end{pmatrix},
\end{equation}
where $\theta=\theta(\mathbf{r},E)$ is a complex angle which characterizes the pairing, and
$\varphi=\varphi(\mathbf{r})$ is the real superconducting phase. The off-diagonal elements of the matrix $\hat g$
describe \cite{1:LO} the superconductive correlations, vanishing in the bulk of a normal metal ($\theta =0$).

The Usadel equation takes the form
\begin{gather}
\frac D2 \nabla^2\theta + \left[ iE - \frac D2 \left( \nabla\varphi \right)^2 \cos\theta \right]
\sin\theta + \left| \Delta \right| \cos\theta = 0, \label{usadel_theta_1} \\
\nabla \left( \sin^2 \theta\, \nabla\varphi \right) = 0. \label{usadel_theta_2}
\end{gather}
The corresponding boundary conditions are
\begin{gather}
\sigma_l \nabla_\mathbf{n} \theta_l = g_B \left[ \cos (\varphi_r -\varphi_l ) \cos\theta_l
\sin\theta_r - \sin\theta_l \cos\theta_r \right], \label{b_1} \\
\sigma_r \nabla_\mathbf{n} \theta_r = g_B \left[ \cos\theta_l \sin\theta_r - \cos (\varphi_r
-\varphi_l )\sin\theta_l \cos\theta_r \right], \\
\sigma_l \sin^2 \theta_l\, \nabla_\mathbf{n} \varphi_l = \sigma_r \sin^2 \theta_r\, \nabla_\mathbf{n} \varphi_r = g_B
\sin (\varphi_r -\varphi_l) \sin\theta_l \sin\theta_r. \label{b_3}
\end{gather}
The self-consistency equation for the order parameter $\Delta(\mathbf{r})$ takes the form
\begin{equation} \label{Delta}
\left| \Delta \right| = \lambda \int_0^{\omega_D} dE\,\tanh\! \left( \frac E{2T} \right) \Im\left[\sin\theta\right] .
\end{equation}

The above equations are written in the absence of an external magnetic field. To take account of the magnetic field, it
is sufficient to substitute the superconducting phase gradient in the Usadel equations
(\ref{usadel_theta_1})--(\ref{usadel_theta_2}) by its gauge invariant form $2m \mathbf{v} =\nabla\varphi+2e\mathbf{A}$,
where $\mathbf{A}$ is the vector potential and $\mathbf{v}$ denotes the supercurrent velocity.

Physical properties of the system can be expressed in terms of the pairing angle $\theta(\mathbf{r},E)$. The
single-particle density of states $\nu(\mathbf{r},E)$ and the density of the superconducting electrons $n(\mathbf{r})$
are given by
\begin{align}
\label{nu} \nu &= \nu_0 \Re\,[\cos\theta],\\
\label{n} n &= \frac{2m\sigma}{e^2} \int_0^\infty dE\,\tanh\! \left( \frac E{2T} \right) \Im \left[\sin^2\theta\right],
\end{align}
where $m$ and $e$ are the electron's mass and the absolute value of its charge, and $\nu_0 = m^2 \mathrm{v} /\pi^2$ is
the normal-metal density of states at the Fermi level. The total number of single-particle states in a metal is the same
in the superconducting and normal states, which is expressed by the constraint
\begin{equation}
\int_0^\infty dE \left[ \nu(\mathbf{r},E) -\nu_0 \right] =0.
\end{equation}

Below, it will be sufficient to consider only positive energies, $E>0$.

\subsection{Simple example: the BCS case}

The simplest illustration for the above technique is the BCS case, when the order parameter
$\Delta(\mathbf{r})=\Delta_{BCS}$ is spatially constant. Its phase can be set equal to zero, $\varphi=0$. Then the
Usadel equations (\ref{usadel_theta_1})--(\ref{usadel_theta_2}) are trivially solved, and we can write the answer in
terms of the sine and the cosine of the pairing angle:
\begin{align}
\sin\theta_{BCS} (E) &= \frac{i\Delta_{BCS}}{\sqrt{E^2-\Delta_{BCS}^2}}, \label{sin_BCS}\\
\cos\theta_{BCS} (E) &= \frac E{\sqrt{E^2-\Delta_{BCS}^2}}. \label{cos_BCS}
\end{align}
An infinitesimal term $i0$ should be added to the energy $E$ to take the retarded nature of the Green function $\hat g$
into account, which yields
\begin{equation} \label{Im_sin_sq_BCS}
\Im \left[\sin^2 \theta_{BCS} (E) \right] =\frac \pi 2 \Delta_{BCS}\, \delta (E-\Delta_{BCS}).
\end{equation}

The usual BCS relations are straightforwardly obtained from Eqs. (\ref{Delta})--(\ref{n}) (for simplicity, we consider
the case of zero temperature):
\begin{align}
\Delta_{BCS} &= 2\omega_D \exp\left( -\frac 1\lambda \right),\\
\label{nu_BCS}
\nu_{BCS}(E) &= \left\{ \begin{array}{ll} 0, & \text{ at } E <\Delta_{BCS}\\
\nu_0 \frac E{\sqrt{E^2- \Delta_{BCS}^2}}, & \text{ at } E >\Delta_{BCS}
\end{array} \right. ,\\
\label{n_BCS} n_{BCS} &= \pi \frac{m\sigma}{e^2} \Delta_{BCS}.
\end{align}
The critical temperature must be determined from Eq. (\ref{Delta}) with vanishing $\Delta (T_c)$; the result is
\begin{equation}
\label{T_c_BCS} \Delta_{BCS} (0) = 1.76\; T_c^{BCS}.
\end{equation}

\section{Usadel equations for a thin bilayer} \label{sec:bilayer}

\begin{figure}
 \centerline{\includegraphics[width=40mm]{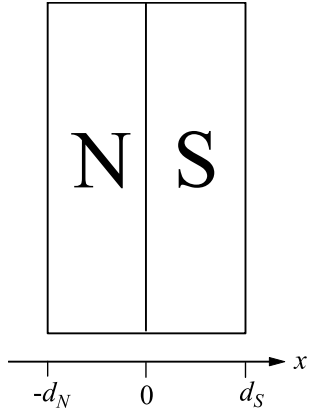}}
\caption{\label{fig:sketch_SN} SN bilayer. The N and S layers occupy the regions $-d_N<x<0$ and $0<x<d_S$,
respectively.}
\end{figure}

Let us consider a SN bilayer consisting of a normal metal ($-d_N<x<0$) in contact (at $x=0$) with a superconductor
($0<x<d_S$) --- see Fig.~\ref{fig:sketch_SN}. We choose the phase $\varphi$ of the order parameter $\Delta$ equal to
zero in the S layer. At the same time, we suppose that electron-electron interaction is absent in the normal layer:
$\lambda=0$, hence $\Delta=0$, although the superconductive correlations ($\theta \neq 0$) exist in the N layer due to
the proximity effect. The Usadel equations (\ref{usadel_theta_1})--(\ref{usadel_theta_2}) take the form
\begin{gather}
\frac{D_N}2 \frac{\partial^2\theta_N}{\partial x^2} +iE\sin\theta_N = 0, \label{usadel_bi_1} \\
\frac{D_S}2 \frac{\partial^2\theta_S}{\partial x^2} +iE\sin\theta_S+ \Delta\cos\theta_S = 0, \label{usadel_bi_2}
\end{gather}
where $\theta_N$ and $\theta_S$ denote the pairing angle $\theta$ at $x<0$ and $x>0$, respectively. The boundary
conditions~(\ref{b_1})--(\ref{b_3}) reduce to
\begin{equation} \label{gran_uslovija}
\sigma_N \frac{\partial\theta_N}{\partial x} = \sigma_S \frac{\partial\theta_S}{\partial x} = g_B
\sin(\theta_S-\theta_N).
\end{equation}

We assume that the layers are thin, hence the order parameter $\Delta$ can be regarded as constant in the
superconductive layer. The standard condition that the layers are thin is obtained if we compare their thicknesses to
the coherence lengths: $d_N \ll \sqrt{D_N/ \Delta_{BCS}}$ and $d_S \ll \sqrt{D_S/ \Delta_{BCS}}$. However, in the case
of low interface resistance, the condition can become stronger. The thin-layers assumption will be discussed in more
details in Sec.~\ref{sec:discussion}.

Equations (\ref{usadel_bi_1})--(\ref{usadel_bi_2}) can be integrated once, yielding
\begin{gather}
\frac{D_N}4 \left( \frac{\partial\theta_N}{\partial x} \right)^2 -iE\cos\theta_N = b_N, \label{integrated_1}\\
\frac{D_S}4 \left( \frac{\partial\theta_S}{\partial x} \right)^2 -iE\cos\theta_S + \Delta\sin\theta_S = b_S.
\label{integrated_2}
\end{gather}
The functions $b_N(E)$ and $b_S(E)$ are determined from the boundary condition $\partial\theta/\partial x=0$ at the
nontransparent outer surfaces of the bilayer, which give
\begin{align}
b_N(E) &= -iE\cos\theta_N(-d_N,E), \notag \\
b_S(E) &= -iE\cos\theta_S(d_S,E)+\Delta\sin\theta_S(d_S,E).
\end{align}
Let us denote $\theta_N(E)=\theta_N(-d_N,E)$, $\theta_S(E) =\theta_S(d_S,E)$. Due to small thickness of the layers, the
functions $\theta_N(x,E)$ and $\theta_S(x,E)$ are nearly spatially constant. However, in order to determine them, we
should take account of their weak spatial dependence and make use of the boundary conditions at the SN interface.
Substituting
\begin{align}
\theta_N(x,E) &= \theta_N(E)+\delta\theta_N(x,E), \notag \\
\theta_S(x,E) &= \theta_S(E)+\delta\theta_S(x,E)
\end{align}
into Eqs. (\ref{integrated_1})--(\ref{integrated_2}) and linearizing them with respect to $|\delta \theta_N(x,E)|$,
$|\delta\theta_S(x,E)| \ll 1$, we find the solution. Finally, the boundary conditions at the SN interface lead to
\begin{equation} \label{main}
-i\tau_N E\sin\theta_N(E) = i\tau_S E\sin\theta_S(E) + \tau_S\Delta\cos\theta_S(E) = \sin\left[ \theta_S(E)-\theta_N(E)
\right],
\end{equation}
where we have denoted
\begin{equation} \label{tau_original}
\tau_N = \frac{2\sigma_N d_N}{D_N\, g_B},\qquad \tau_S = \frac{2\sigma_S d_S}{D_S\, g_B}.
\end{equation}
Physically, $\tau_N$ and $\tau_S$ are the escape times from the respective layers (see Appendix~\ref{app:tau}). Using
the definition of the interface resistance per channel (\ref{rho_int}), we can represent these quantities as
\begin{equation}
\tau_N = 2\pi \frac{\mathrm{v}_N d_N}{\mathrm{v}_S^2} \rho_B,\qquad \tau_S = 2\pi \frac{d_S}{\mathrm{v}_S} \rho_B,
\end{equation}
with $\mathrm{v}_N$ and $\mathrm{v}_S$ being the Fermi velocities in the N and S layers. The ratio $\tau_N /\tau_S =
\mathrm{v}_N d_N /\mathrm{v}_S d_S$, which is independent of the interface properties, can also be interpreted as the
ratio of the \textit{total} densities of states (per energy interval) in the two layers:
\begin{equation}
\frac{\tau_N}{\tau_S} = \frac{\mathcal{A} d_N \nu_{0N}}{\mathcal{A} d_S \nu_{0S}}.
\end{equation}

Having solved the boundary conditions~(\ref{main}), we can determine all equilibrium properties of the system (because
knowledge of $\theta_S$, $\theta_N$ implies knowledge of the retarded Green function $\hat g$).

A useful representation of the boundary conditions~(\ref{main}) is obtained as follows. Excluding $\theta_N(E)$ from Eq.
(\ref{main}), we arrive at a single equation for the function $\theta_S(E)$, which can be written, in terms of
$Z=\exp(i\theta_S)$, as a polynomial equation
\begin{equation} \label{main_Z}
iC_6 Z^6 + C_5 Z^5 + iC_4 Z^4 + C_3 Z^3 + iC_2 Z^2 + C_1 Z +iC_0 = 0
\end{equation}
with real coefficients
\begin{align}
C_6 &= -\tau_N E \left(\frac{\tau_S}{\tau_N}\right)^2 \left[ 1+\frac\Delta E \right]^2, \notag \\
C_5 &= \left[ 1-\left( \tau_N E\right)^2 \right] \left(\frac{\tau_S}{\tau_N}\right)^2 \left[ 1+\frac\Delta
E \right]^2 -1, \notag \\
C_4 &= -\tau_N E \left(\frac{\tau_S}{\tau_N}\right)^2 \left[ 3\left( \frac\Delta E \right)^2+
2\frac{\Delta}{E} -1 \right], \notag \\
C_3 &= 2-2\left[ 1-\left( \tau_N E \right)^2 \right] \left(\frac{\tau_S}{\tau_N}\right)^2 \left[ 1-
\left( \frac{\Delta}E \right)^2 \right], \notag \\
C_2 &= -\tau_N E \left(\frac{\tau_S}{\tau_N}\right)^2 \left[ 3\left( \frac{\Delta}E \right)^2
-2\frac\Delta E -1 \right], \notag \\
C_1 &= \left[ 1-\left( \tau_N E\right)^2 \right] \left(\frac{\tau_S}{\tau_N}\right)^2 \left[ 1-\frac\Delta
E \right]^2 -1, \notag \\
C_0 &= -\tau_N E \left(\frac{\tau_S}{\tau_N}\right)^2 \left[ 1-\frac\Delta E \right]^2. \label{coeff}
\end{align}

During further analysis, the choice between the boundary conditions in the forms~(\ref{main}) and~(\ref{main_Z}) will be
a matter of convenience.

\subsection{Numerical results} \label{sec:numerical}

The solution of Eq. (\ref{main_Z}) can be found numerically. To this end, we solve the system of two nonlinear equations
for the functions $\Re Z(E)$ and $\Im Z(E)$, using the modified Newton method with normalization.

The solution depends on the bilayer's parameters: the thicknesses of the layers, characteristics of materials
constituting the bilayer, and the quality of the SN interface. This dependence enters Eqs. (\ref{main_Z}), (\ref{coeff})
via $\tau_N$ and $\tau_S$. For numerical calculations, we assume the characteristics of the bilayer to be the same as in
the experiment by Kasumov \textit{et al.} \cite{1:Kasumov} The superconductive layer is made of tantalum, $d_S=5$~nm,
and the normal layer is made of gold, $d_N=100$~nm. Approximate experimental values of the conductivities are
\cite{1:Kasumov_priv} $\sigma_S = 0.01\, \mu\Omega^{-1}\, \mathrm{cm}^{-1}$ and $\sigma_N = 1\, \mu\Omega^{-1}\,
\mathrm{cm}^{-1}$. In order to calculate the Fermi characteristics of tantalum and gold, we use the values of the Fermi
energy $E_\mathrm{F} (\mathrm{Ta}) = 11$~eV, $E_\mathrm{F} (\mathrm{Au}) = 5.5$~eV, and the free-electrons
model.\footnote{Certainly, the free-electron model does not describe the details of the electronic structure of
tantalum. However, this simplified model is used in the very derivation of the Usadel equation. Also, we do not expect
drastic dependence of our results on the Fermi characteristics. To check this, we have reproduced all the calculations
with slightly different (within 10-15\%) values of the Fermi energies, and found that the changes in the results amount
to, roughly speaking, a rescaling of the interface resistance $\rho_B$ within 10\%.}

\begin{figure} \vspace*{2mm}
 \centerline{\includegraphics[width=90mm]{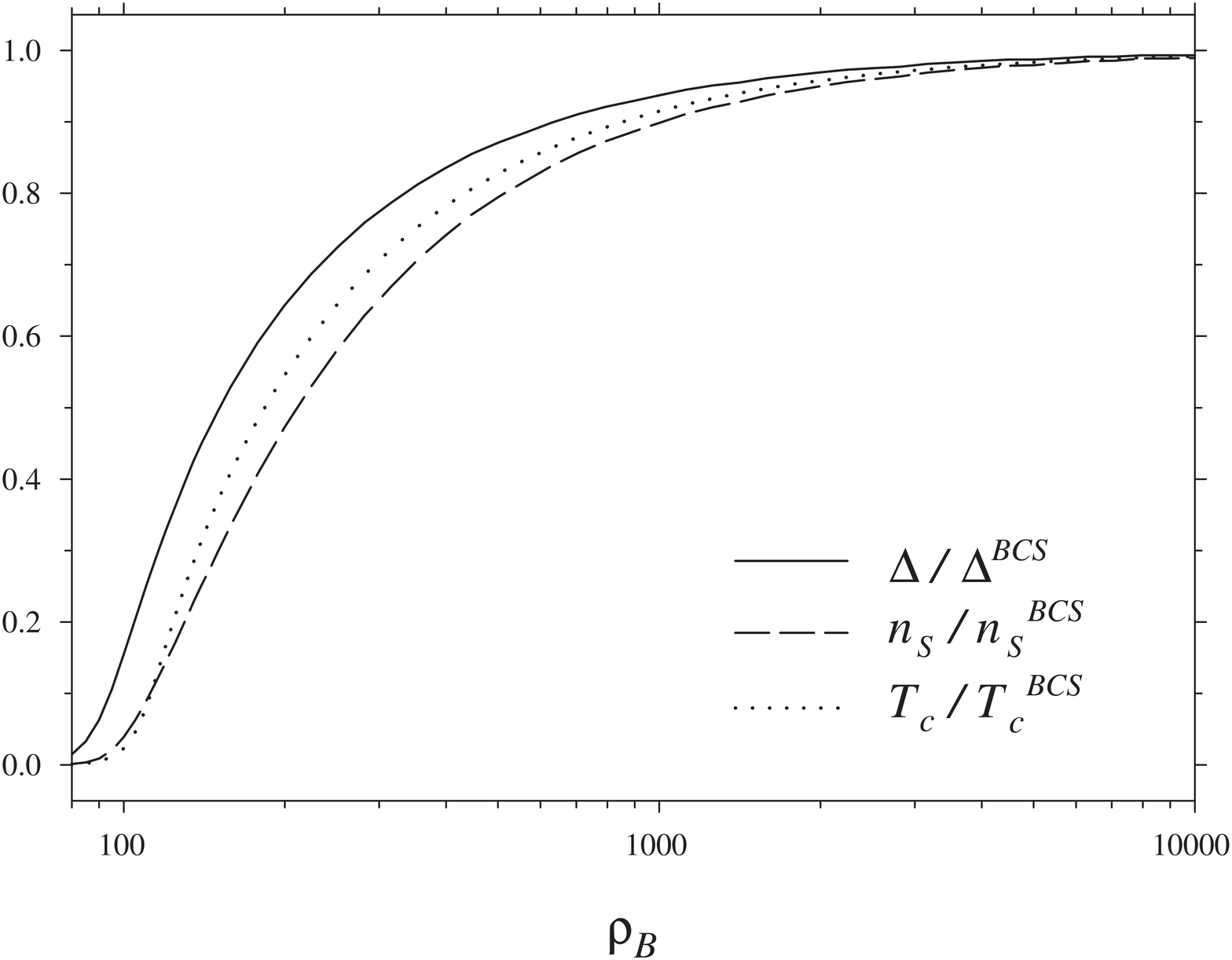}}
\caption{Dependence of the order parameter in the S layer $\Delta$, of the superconducting electrons' density in the S
layer $n_S$ at zero temperature, and of the bilayer's critical temperature $T_c$ on the interface resistance per channel
$\rho_B$. All the quantities are normalized by the corresponding BCS values. The discrepancy between the curves implies
a violation of the BCS relations between $\Delta$, $n_S$, and $T_c$. The choice of the bilayer's parameters correspond
to the experiment~\cite{1:Kasumov}, so that $\tau_S \Delta_{BCS} = 0.016\, \rho_B$, $\tau_N \Delta_{BCS} =
0.23\,\rho_B$.}
 \label{fig:wide_rho}
\end{figure}

\begin{figure} \vspace*{2mm}
 \centerline{\includegraphics[width=90mm]{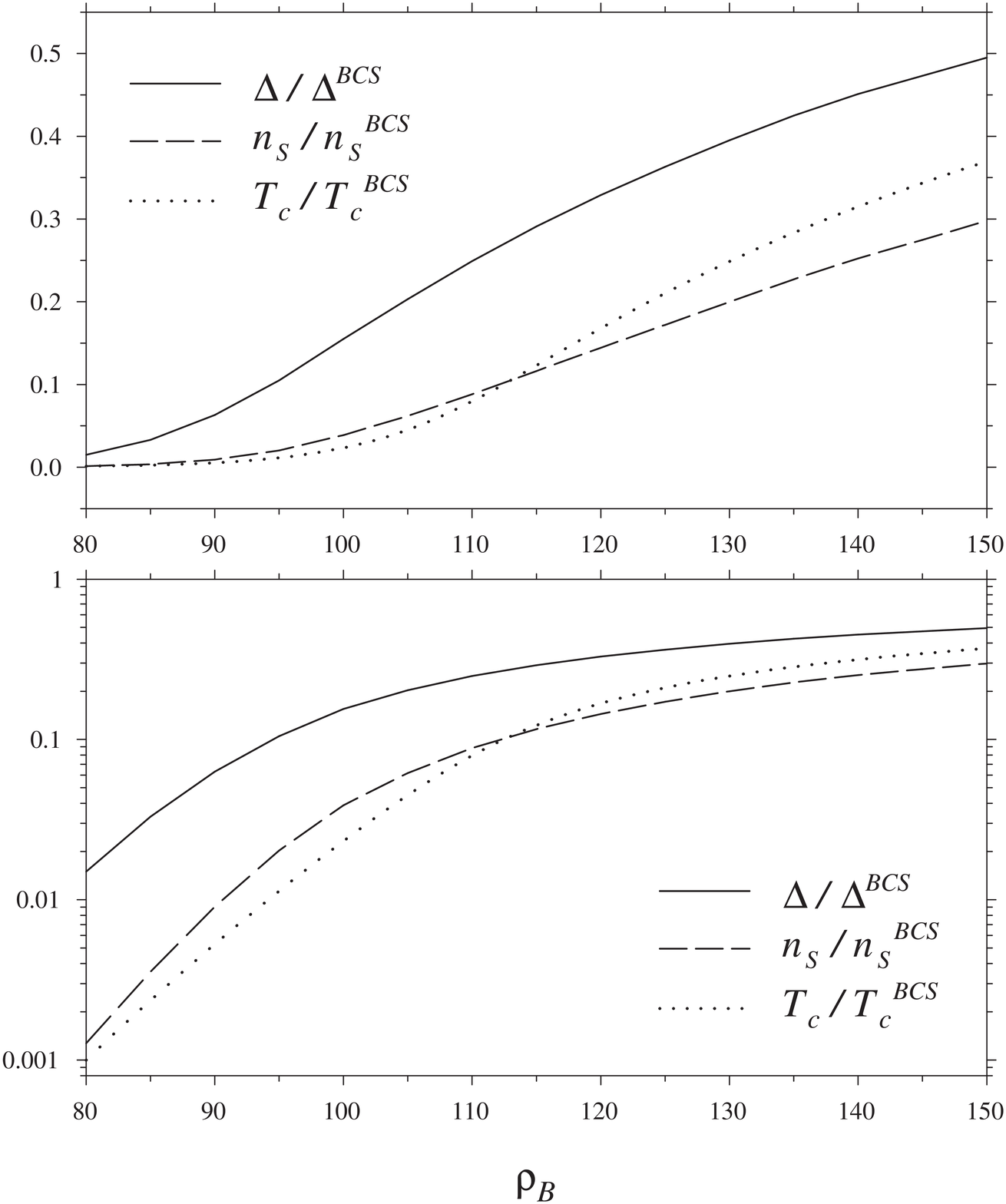}}
\caption{Zoomed part of Fig.~\protect\ref{fig:wide_rho}. In the shown range of relatively small resistance $\rho_B$, the
BCS relations between $\Delta$, $n_S$, and $T_c$ are severely violated. The upper and lower graphs differ only in the
scaling of the ordinate axis (normal and logarithmic, respectively).}
 \label{fig:narr_rho}
\end{figure}

\begin{figure} \vspace*{2mm}
 \centerline{\includegraphics[width=90mm]{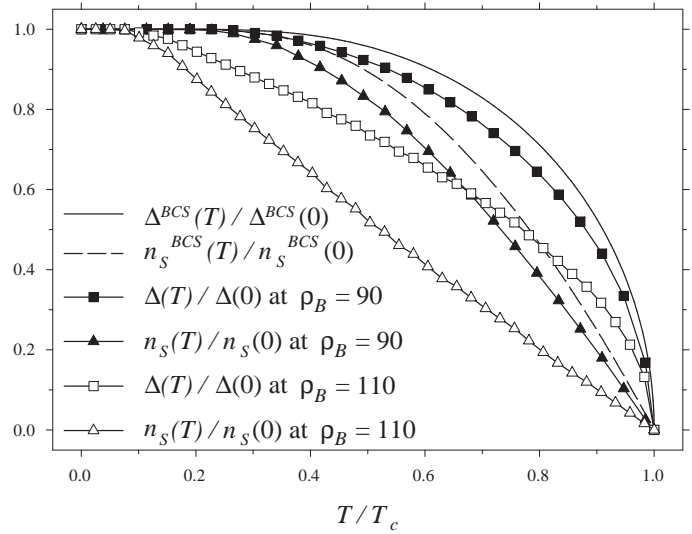}}
\caption{Temperature dependence of $\Delta$ and $n_S$ at $\rho_B =$ 90 and 110. The temperature is normalized by the
critical value $T_c$, which depends on $\rho_B$; $\Delta$ and $n_S$ are normalized by their zero-temperature values. The
choice of the bilayer's parameters implies the relations $\tau_S \Delta_{BCS} = 0.016\, \rho_B$, $\tau_N \Delta_{BCS} =
0.23\,\rho_B$. For comparison, the same dependence is also plotted for the BCS case.}
 \label{fig:t_depend}
\end{figure}

As a result, we obtain the relations $\tau_S \Delta_{BCS} = 0.016\, \rho_B$, $\tau_N \Delta_{BCS} = 0.23\,\rho_B$ (so
that $\tau_N /\tau_S \approx 14$), after which the solution of Eq. (\ref{main_Z}) depends only on the interface
resistance $\rho_B$. Having found the function $Z(E)$ [which is equivalent to finding $\theta_S(E)$], we start from the
case of zero temperature, $T=0$, and study the dependence of the order parameter $\Delta$ and of the superconducting
electrons' density in the S layer $n_S$ [Eqs. (\ref{Delta}), (\ref{n})] on $\rho_B$. The results are plotted in
Fig.~\ref{fig:wide_rho}, where we also show the $\rho_B$-dependence of the critical temperature $T_c$, determined by the
formula of Ref.~\cite{1:Khusainov1}.

The suppression of $\Delta$, $n_S$, and $T_c$, in comparison to their BCS values in the S layer, is a natural
consequence of proximity to the normal metal. At the same time, there is a possibility of BCS-like behavior, which
implies the BCS relations between the suppressed quantities and the coincidence of the three curves plotted in
Fig.~\ref{fig:wide_rho}. However, the curves split, and the difference between them is largest for relatively small
values of $\rho_B$. Figure~\ref{fig:narr_rho} presents the range $80 < \rho_B < 150$ on a larger scale.

Figure~\ref{fig:t_depend} shows the temperature dependence of the order parameter $\Delta$ and of the superconducting
electrons' density in the S layer $n_S$. Although the smaller $\rho_B$ the further it is from the BCS limit
(corresponding to $\rho_B\to\infty$), we observe that at $\rho_B =90$ the curves are closer to the BCS behavior than at
$\rho_B =110$. An explanation of this feature is given in Sec.~\ref{sec:anderson}.

\begin{figure}\vspace*{2mm}
 \centerline{\includegraphics[width=90mm]{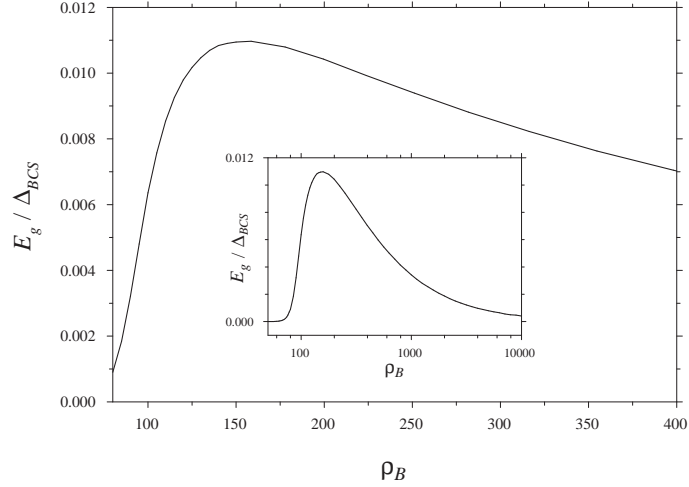}}
\caption{Minigap $E_g$ in the single-particle density of states versus $\rho_B$. The minigap is normalized by the BCS
gap value $\Delta_{BCS}$. $E_g$ is a nonmonotonic function of $\rho_B$, reaching its maximum at $\rho_B = 160$. The
inset shows $E_g (\rho_B)$ on a wider (logarithmic) scale over $\rho_B$. The choice of the bilayer's parameters implies
the relations $\tau_S \Delta_{BCS} = 0.016\, \rho_B$, $\tau_N \Delta_{BCS} = 0.23\,\rho_B$.}
 \label{fig:minigap}
\end{figure}

The DoS in the thin SN bilayer was studied by McMillan \cite{1:McMillan} and Golubov \cite{1:Golubov}. They demonstrated
that (in the case of thin S layer) there is a minigap $E_g$ that is much smaller than $\Delta_{BCS}$ (in the case of a
bulk superconductor and the perfect interface, the minigap was found in Ref.~\cite{1:Gol1}). The gap in the DoS is a
property of the bilayer as a whole and does not depend on the coordinate, while the energy dependence of the DoS is
different in the S and N layers. The results of the self-consistent calculation of the minigap versus the interface
resistance are shown in Fig.~\ref{fig:minigap}.

The minigap is nonmonotonic, and this fact is already contained in the results of McMillan \cite{1:McMillan}, who
obtained
\begin{align}
E_g &= \frac{\tau_S}{\tau_N+\tau_S} \Delta &&\text{at~~} \frac{\tau_S \tau_N \Delta}{\tau_S+\tau_N} \ll 1,
\label{small} \\
E_g &= \frac 1{\tau_N} &&\text{at~~} \frac{\tau_S \tau_N \Delta}{\tau_S+\tau_N} \gg 1. \label{large}
\end{align}
Formula (\ref{small}) does not determine the dependence of $E_g$ on the interface resistance self-consistently; however,
if we know that $\Delta$ is suppressed as the interface resistance is lowered, then we can conclude that the gap
decreases with decreasing $\rho_B$ in the corresponding range. At the same time, formula (\ref{large}) explicitly
contains (according to the definition of $\tau_N$) the inverse $\rho_B$-dependence. Therefore, $E_g$ reaches a maximum
at an intermediate $\rho_B$ corresponding to $\tau_S \tau_N \Delta / (\tau_S+\tau_N) \sim 1$, which yields $\rho_B\sim
140$. This estimate agrees with the numerical results (see Fig.~\ref{fig:minigap}).

\section{Anderson limit} \label{sec:anderson}

In the limit of relatively low interface resistance (following McMillan \cite{1:McMillan}, we call it the Anderson
limit), the theory describing the bilayer can be developed analytically. The condition defining this limit is $\tau_S
\Delta$, $\tau_N \Delta \ll 1$.

Previously, $T_c$ was calculated in this limit \cite{1:McMillan,1:Khusainov1,1:Golubov}, and the relation (\ref{small})
between the minigap and the order parameter in the S layer $\Delta$ was found, as well as the expression for the DoS
that has the standard BCS form with the gap $E_g$ was obtained \cite{1:McMillan}. At the same time, the relation between
$E_g$ (or $\Delta$) and the order parameter of the isolated S layer $\Delta_{BCS}$ was not found. Below we find the
minigap and other superconductive characteristics of the bilayer as functions of the interface resistance.

First of all, we need to determine $\theta (E)$ [or $Z(E)$] solving Eq. (\ref{main}) [or Eq. (\ref{main_Z})] over the
entire range of energies $E$.

In the region $E>\Delta$, the solution of Eq. (\ref{main_Z}) can be written as $Z=1+\delta Z$, with $|\delta Z| \ll 1$.
Keeping terms up to the second order in $\delta Z$, we obtain
\begin{equation} \label{more_1}
\delta Z = -\frac{\Delta \left( 1-i\tau_N E \right)}{E \left( \frac{\tau_S+\tau_N}{\tau_S} -i\tau_N E \right)}.
\end{equation}
This result is valid for arbitrary values of $\rho_B$.

At $E<\Delta$, the same calculation as for the minigap (\ref{small}) leads to the result
\begin{equation} \label{less_1}
\sin\theta_S = \sin\theta_N = \frac{iE_g}{\sqrt{E^2 -E_g^2}},
\end{equation}
with the minigap $E_g$ given by Eq. (\ref{small}). [To avoid confusion, we note that under the less strict limitations
for $\tau_S \Delta$, $\tau_N \Delta$ used in Eq. (\ref{small}), the BCS-like result (\ref{less_1}) is valid only up to
energies of the order of $E_g$.]

Now $\Im\,[\sin\theta_S]$ is readily calculated, and in the case of zero temperature, $\Delta$ can be found from the
self-consistency equation (\ref{Delta}), then the relation (\ref{small}) yields a formula for $E_g$. In the limit of the
perfect interface (the Cooper limit), which is determined by the condition $\tau_S\tau_N\omega_D/ (\tau_S+\tau_N) \ll
1$, we reproduce the classical result of Cooper and de~Gennes \cite{1:Cooper,1:deGennes_review}:
\begin{equation} \label{E_g_Cooper}
E_g (\rho_B\to 0)= 2\omega_D \exp \left( -\frac 1{\left< \lambda \right>} \right),
\end{equation}
with the effective pairing constant
\begin{equation}
\left< \lambda \right> =\frac{\tau_S}{\tau_S+\tau_N} \, \lambda.
\end{equation}
If the interface is imperfect, $\tau_S\tau_N\omega_D/ (\tau_S+\tau_N) \gg 1$, then the condition of the Anderson limit
nevertheless determines a wide range of the interface resistances, where we obtain
\begin{equation} \label{E_g}
\frac{E_g}{\Delta_{BCS}} = \left[ \frac{\tau_S \tau_N \Delta_{BCS}}{2 (\tau_S+\tau_N)} \right]^{\tau_N /\tau_S}.
\end{equation}
We emphasize that the Anderson limit does not reduce to the Cooper limit with small corrections. On the contrary, due to
the relation $\Delta \ll \omega_D$, the Cooper limit's condition is \textit{not} satisfied over the most part of the
Anderson limit's validity range; therefore, the minigap $E_g$ and the quantities calculated below differ drastically
from the Cooper limit expressions.

Now we proceed to calculate the density of the superconducting electrons in the S layer $n_S$. The main contribution to
the integral (\ref{n}) that determines $n_S$, comes from the vicinity of $E_g$. Inserting the infinitesimal imaginary
part of the energy into Eq. (\ref{less_1}), we obtain
\begin{equation}
\Im\,[\sin^2 \theta_S] = \frac\pi 2 E_g\; \delta(E-E_g),
\end{equation}
which immediately yields the density of the superconducting electrons $n_S$ at zero temperature:
\begin{equation} \label{n_S_anderson}
\frac{n_S}{n_S^{BCS}} = \frac{E_g}{\Delta_{BCS}} ,
\end{equation}
where $E_g$ is given by Eq. (\ref{E_g}).

The critical temperature $T_c$ of the bilayer in the Anderson limit [at $(\tau_S +\tau_N)/ \tau_S \tau_N \gg T_c$] was
found by Khusainov \cite{1:Khusainov1}. Employing our result (\ref{E_g}) for the minigap, we obtain the following
relation:
\begin{equation}
\frac{T_c}{T_c^{BCS}} = \frac{E_g}{\Delta_{BCS}}.
\end{equation}

Now we can discuss the general structure of the theory describing the bilayer in the Anderson limit. In the limit
$\rho_B\to 0$, our results for the pairing angle $\theta$ (which is constant over the entire bilayer, $\theta\equiv
\theta_S = \theta_N$) yield expressions which can be obtained from the BCS ones (\ref{sin_BCS})--(\ref{Im_sin_sq_BCS})
if we substitute the BCS order parameter $\Delta_{BCS}$ by the bilayer's minigap $E_g$. At $\rho_B
>0$, corrections to this simple result are small while the Anderson limit's conditions are satisfied. Therefore, we
obtain a BCS-type theory with $E_g$ substituting $\Delta_{BCS}$ in all formulas.

The results of this section immediately explain the numerical results in the limit of relatively small $\rho_B$, shown
in Fig.~\ref{fig:narr_rho}. As we have found, the Anderson limit implies the following relations between the quantities
under discussion:
\begin{align}
E_g &= 1.76\; T_c,\\
n_S &= \pi \frac{m\sigma_S}{e^2} E_g, \label{n_S_rel}
\end{align}
which substitute Eqs. (\ref{T_c_BCS}) and~(\ref{n_BCS}). For the Ta/Au bilayer to which the numerical results refer, the
Anderson limit is valid at $\rho_B <80$ (we see that the values of $\rho_B$ can be large although they are
\textit{relatively} small). Therefore, approaching $\rho_B =80$, the curves $n_S / n_S^{BCS}$ and $T_c / T_c^{BCS}$ tend
to coincide, while $\Delta / \Delta_{BCS}$ exceeds them by the large factor $(\tau_S+\tau_N) / \tau_S \approx 15$.

The temperature dependence of $\Delta$ and $n_S$, shown in Fig.~\ref{fig:t_depend}, is quite different at $\rho_B=90$
and $\rho_B=110$; at $\rho_B=90$, the curves are much closer to the BCS behavior. This is also explained by approaching
the Anderson limit, where the curves coincide with the BCS ones.

According to McMillan's results \cite{1:McMillan}, the DoS in the S and N layers coincide in the Anderson limit (this
also follows from Eq. (\ref{less_1})), having a BCS-like square-root singularity at $E=E_g$.

Over the whole range of the Anderson limit (including the region $\tau_S\tau_N\omega_D/ (\tau_S+\tau_N) \sim 1$, where
the crossover to the Cooper limit takes place), we obtain the result
\begin{equation} \label{E_g_Anderson}
\frac{E_g}{\Delta_{BCS}} = \left[ \frac{\Delta_{BCS}}{2\omega_D} \sqrt{1+ \left(
\frac{\tau_S\tau_N\omega_D}{\tau_S+\tau_N} \right)^2} \right]^{\tau_N /\tau_S},
\end{equation}
which reproduces Eq. (\ref{E_g}) at $\tau_S\tau_N\omega_D/ (\tau_S+\tau_N) \gg 1$ and the Cooper -- de~Gennes result
(\ref{E_g_Cooper}) at $\tau_S\tau_N\omega_D/ (\tau_S+\tau_N) \ll 1$. In the crossover region $\tau_S\tau_N\omega_D/
(\tau_S+\tau_N) \sim 1$ this formula should be considered as an interpolation, because the Debye energy $\omega_D$ only
determines the order of magnitude for the cutoff energy in the BCS theory.

A similar calculation for other superconductive characteristics shows that the relations
(\ref{n_S_anderson})--(\ref{n_S_rel}) between them and the minigap (\ref{E_g_Anderson}) are valid over the whole
Anderson limit.

\section{Parallel critical field} \label{sec:H_c}

We proceed to calculate the critical magnetic field $H_c$ directed along the plane of the bilayer. As it was mentioned
in Sec.~\ref{subsec:theta}, in the presence of an external magnetic field, the superconducting phase gradient in the
Usadel equations (\ref{usadel_theta_1})--(\ref{usadel_theta_2}) must be substituted by its gauge invariant form, which
can be expressed via the supercurrent velocity $\mathbf{v}$. The spatial distribution of $\mathbf{v}$ in the bilayer can
be found as follows.

Let us direct the $y$ axis along the magnetic field $\mathbf{H}$. The supercurrents $\mathbf{j}=-en\mathbf{v}$ are
directed along the bilayer and perpendicularly to $\mathbf{H}$, \textit{i.e.}, $\mathbf{j} =(0,0,j(x))$ and $\mathbf{v}
= (0,0,v(x))$. Near $H_c$, the magnetic field inside the bilayer is uniform, so the vector potential can be chosen as
$\mathbf{A}=(0,0,-xH)$. The supercurrent velocity distribution is determined by the equation $\nabla\times \mathbf{v}=
e\mathbf{H}/m$. Another essential point is the continuity of $\mathbf{v}$ at the SN interface, which follows from the
continuity of the superconducting phase $\varphi$ [see the boundary condition~(\ref{b_3})]. The result is
\begin{equation} \label{v}
v(x) = v_0-\frac{eH}m x,
\end{equation}
where $v_0$ is the supercurrent velocity at the interface, which must be determined from the condition that the total
charge transfer across the bilayer's cross-section is zero:
\begin{equation}
\int_{-d_N}^{d_S} j(x)\, dx=0,
\end{equation}
leading to
\begin{equation} \label{v_0}
v_0 =\left(\frac{eH}{2m}\right) \frac{n_S d_S^2 -n_N d_N^2}{n_S d_S + n_N d_N}.
\end{equation}
The density of the superconducting electrons is constant in each layer ($n_S$ and $n_N$).

Near $H_c$, the superconducting correlations are small, $|\theta| \ll 1$, and the Usadel equation~(\ref{usadel_theta_1})
for the paring angle $\theta(x,E)$ can be linearized:
\begin{gather}
\frac{D_N}2 \frac{\partial^2\theta_N}{\partial x^2} +\left( iE-2m^2 D_N\, \mathbf{v}^2 \right)\theta_N =0,
\label{usadel_H_c_1}\\
\frac{D_S}2 \frac{\partial^2\theta_S}{\partial x^2} +\left( iE -2m^2 D_S\, \mathbf{v}^2 \right)\theta_S+ \left| \Delta
\right| = 0. \label{usadel_H_c_2}
\end{gather}
At the same time, the second Usadel equation~(\ref{usadel_theta_2}) is trivial: its l.h.s. is proportional to
\begin{equation}
\nabla \left( \sin^2 \theta\,\, \mathbf{v} \right) =\sin 2\theta\,\, \nabla\theta\,\, \mathbf{v} + \sin^2 \theta\,\,
\nabla \mathbf{v},
\end{equation}
where both terms vanish due to the fact that $\nabla\theta$ is directed along the $x$ axis whereas $\mathbf{v}$ is
parallel to the $z$ axis.

The pairing angle $\theta$ is almost spatially constant in each layer; this allows us to average each of Eqs.
(\ref{usadel_H_c_1})--(\ref{usadel_H_c_2}) over the thickness of the corresponding layer, obtaining
\begin{align}
\left. \frac{\partial\theta_N}{\partial x} \right|_{x=0} &= \frac{2d_N}{D_N} \left( E_N -iE \right)
\theta_N, \label{after_avr_1} \\
\left. \frac{\partial\theta_S}{\partial x} \right|_{x=0} &= \frac{2d_S}{D_S} \left[ \left( iE -E_S \right)\theta_S+
\left| \Delta \right| \right], \label{after_avr_2}
\end{align}
where
\begin{equation}
E_N = 2m^2 D_N \left< \mathbf{v}^2 (x) \right>_N,\qquad E_S = 2m^2 D_S \left< \mathbf{v}^2 (x) \right>_S
\end{equation}
are $H$-dependent energies. Using Eqs. (\ref{v}), (\ref{v_0}), we express them via $H_c$ and the densities of the
superconducting electrons:
\begin{equation} \label{E_S}
E_S = \frac{D_S\, e^2 H_c^2}6 \left[ d_S^2+ 3 d_N^2 \frac{n_N^2 \left( d_S+d_N \right)^2} {\left( n_S d_S + n_N d_N
\right)^2} \right],
\end{equation}
and $E_N$ is obtained by the interchange of all the S and N indices.

Substituting (\ref{after_avr_1})--(\ref{after_avr_2}) into the boundary conditions~(\ref{gran_uslovija}) (which should
be linearized), we find
\begin{gather}
\theta_N = \frac{\tau_S \left| \Delta \right|}{\tau_S E_S+\tau_N E_N + \tau_S \tau_N E_S E_N -\tau_S
\tau_N E^2 -iE\left[ \tau_S+\tau_N + \tau_S\tau_N \left( E_S+E_N \right) \right]}, \notag \\
\theta_S = \left( 1+\tau_N E_N -i\tau_N E \right) \theta_N. \label{theta_for_H_c_2}
\end{gather}
The order parameter $\Delta$ cancels out from the self-consistency equation~(\ref{Delta}). However, the resulting
equation alone does not suffice for determining $H_c(T)$ because it contains $E_S$ and $E_N$, which are functions of
$n_N/n_S$. Therefore, to obtain a closed system, we must consider the self-consistency equation together with the
equation determining the ratio $n_N/n_S$; the latter equation is obtained from Eq. (\ref{n}). The resulting system of
two nonlinear equations for the quantities $H_c$ and $n_N/n_S$ is
\begin{align}
\ln\frac{2\omega_D}{\Delta_{BCS}} &= \int_0^{\omega_D} dE\, \tanh\!\left(\!
\frac E{2T} \!\right) \frac{\Im\theta_S}{\left| \Delta \right|}, \label{system_1} \\
\frac{n_N}{n_S} &= \frac{\sigma_N \int_0^\infty dE\, \tanh\!\left( \frac E{2T} \right) \Im\theta_N^2}{\sigma_S
\int_0^\infty dE\, \tanh\!\left( \frac E{2T} \right) \Im\theta_S^2}, \label{system_2}
\end{align}
with $\theta_N$ and $\theta_S$ given by Eqs. (\ref{theta_for_H_c_2}). The first equation of the system, Eq.
(\ref{system_1}), can be written via the digamma functions, thus taking exactly the same form as Eq. (\ref{digamma})
below (which determines the perpendicular upper critical field) if we denote $\mathcal{E}_S = E_S +1/\tau_S$,
$\mathcal{E}_N = E_N +1/\tau_N$.

In the limit $\rho_B \to\infty$, Eqs. (\ref{system_1})--(\ref{system_2}) lead to the BCS result. In this case, the
layers uncouple, the density of the superconducting electrons in the N layer vanishes, $n_N /n_S \to 0$, and Eq.
(\ref{system_1}) finally yields
\begin{equation} \label{like_MdG}
\ln\frac{T_c^{BCS}}T = \psi \left( \frac 12 +\frac{D_S \left[ eH_c^{BCS} d_S \right]^2}{12\pi T} \right) -\psi \left(
\frac 12 \right),
\end{equation}
which determines the parallel critical field $H_c^{BCS} (T)$ of a thin superconducting film.
\begin{figure}\vspace*{2mm}
 \centerline{\includegraphics[width=90mm]{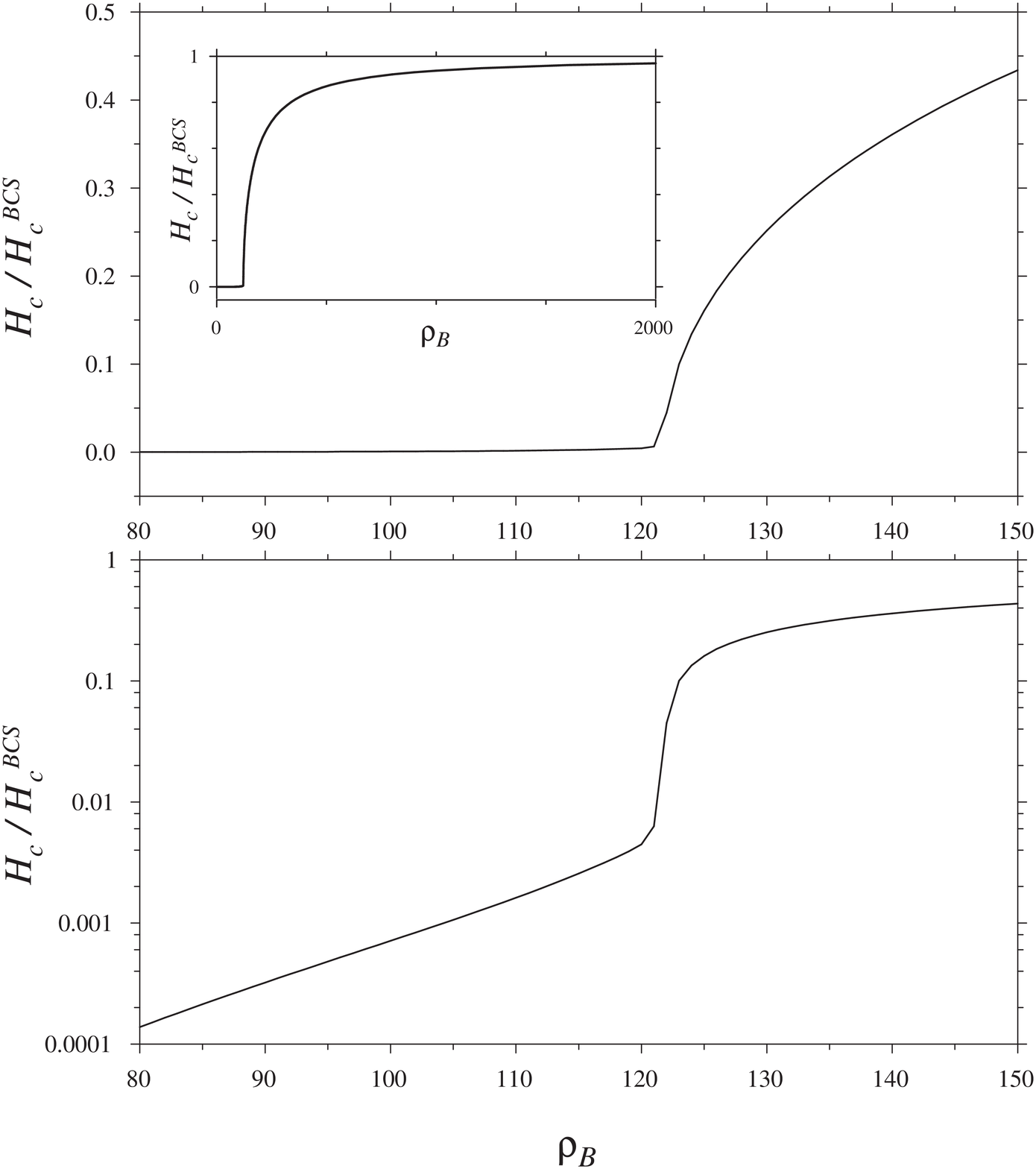}}
\caption{Parallel critical field $H_c$, normalized by the BCS value, versus $\rho_B$ at zero temperature. The upper and
lower graphs differ only in the scaling of the ordinate axis (normal and logarithmic, respectively). The nature of the
steep behavior of $H_c$ at $\rho_B =120$--$123$, which is best seen from the lower graph, is explained in the text. The
inset shows $H_c (\rho_B)$ on a wider scale over $\rho_B$. The choice of the bilayer's parameters implies the relations
$\tau_S \Delta_{BCS} = 0.016\, \rho_B$, $\tau_N \Delta_{BCS} = 0.23\,\rho_B$.}
 \label{fig:h_c}
\end{figure}
\begin{figure}\vspace*{2mm}
 \centerline{\includegraphics[width=90mm]{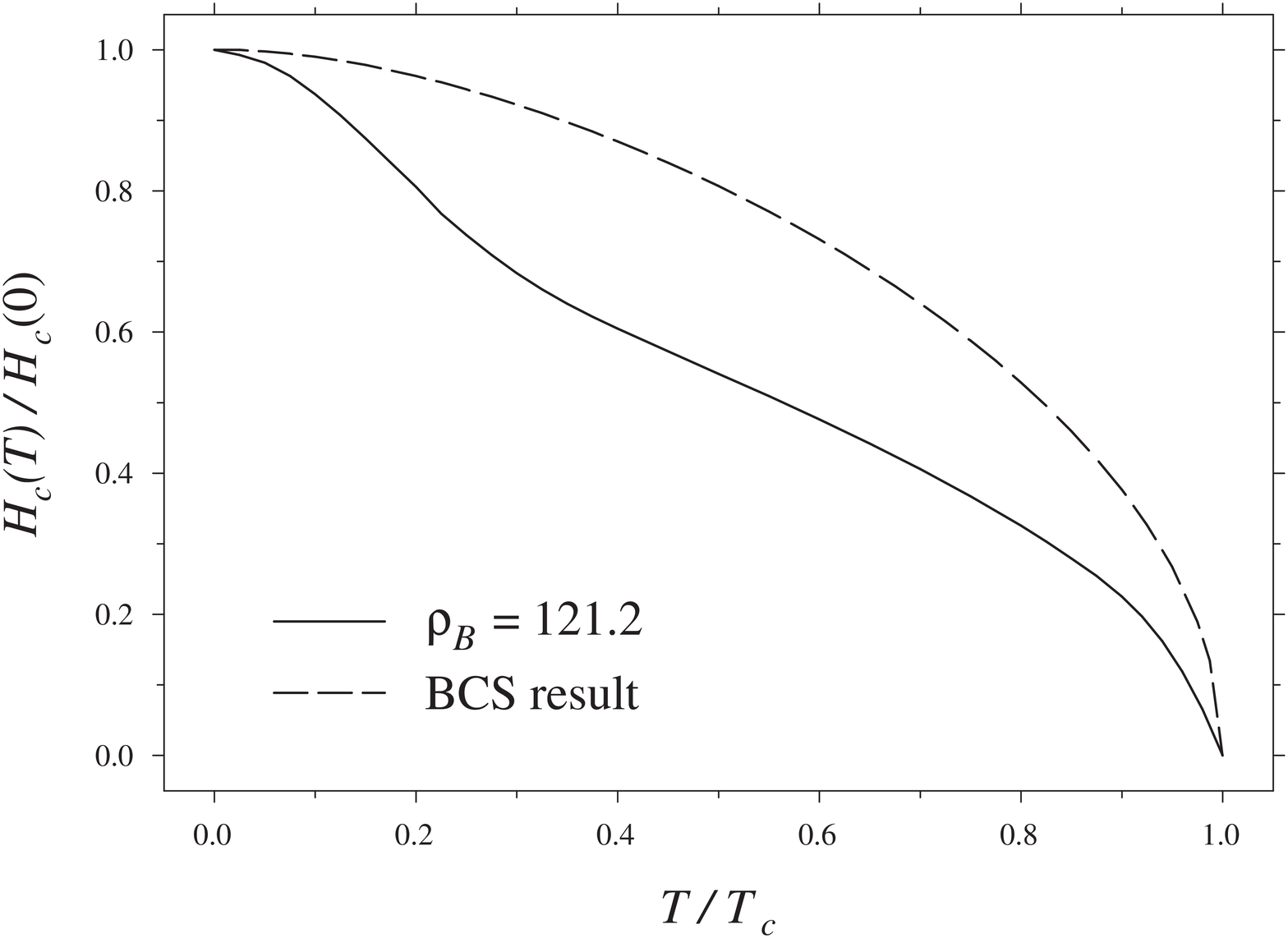}}
\caption{Temperature dependence of the parallel critical field $H_c$ at $\rho_B= 121.2$. The experimental value of
$H_c(0)$, analyzed with the use of the results shown in Fig.~\ref{fig:h_c}, suggests that this value of $\rho_B$
corresponds to the experiment by Kasumov \textit{et al.}~\protect\cite{1:Kasumov} The critical field is normalized by
its zero-temperature value, and the temperature is normalized by the corresponding $T_c$. The choice of the bilayer's
parameters implies the relations $\tau_S \Delta_{BCS} = 0.016\, \rho_B$, $\tau_N \Delta_{BCS} = 0.23\,\rho_B$. For
comparison, the same dependence is plotted for the BCS case.}
 \label{fig:h_c_vs_t}
\end{figure}
\begin{figure}\vspace*{2mm}
 \centerline{\includegraphics[width=90mm]{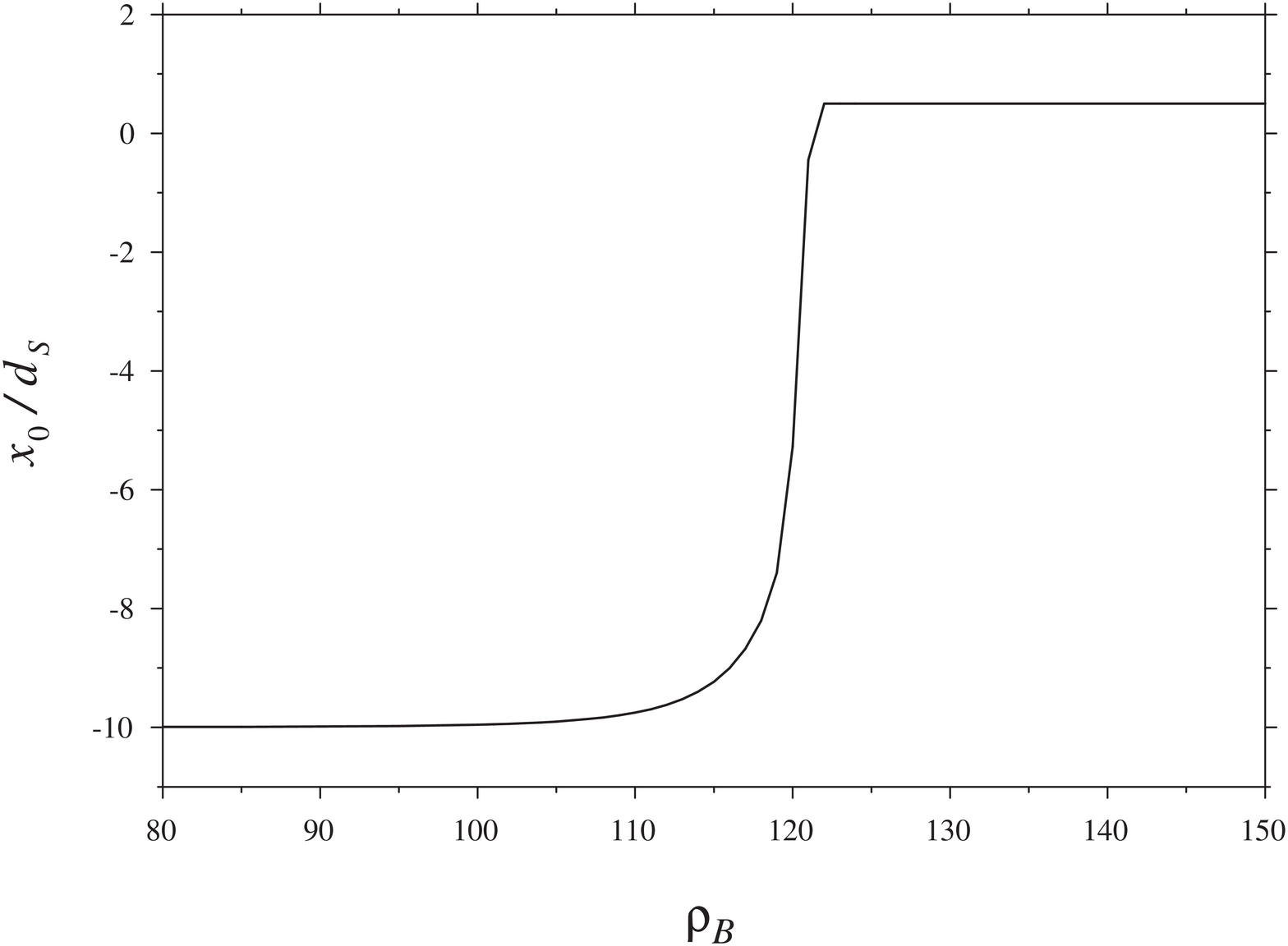}}
\caption{Position of the stationary point $x_0$ of the supercurrent distribution versus $\rho_B$ at zero temperature.
The coordinate $x_0$ is normalized by the S layer thickness $d_S$. The fast shift in $x_0$ from the center of the S
layer at large $\rho_B$ to (nearly) the center of the N layer at small $\rho_B$ corresponds to the steep drop in $H_c$,
shown in Fig.~\ref{fig:h_c}.}
 \label{fig:x_0}
\end{figure}

The system of equations~(\ref{system_1})--(\ref{system_2}) can be solved numerically at arbitrary values of the
temperature $T$ and the interface resistance $\rho_B$; the results for $H_c$ are presented in Figs.~\ref{fig:h_c},
\ref{fig:h_c_vs_t}.

A remarkable feature of the function $H_c (\rho_B)$ at zero temperature (Fig.~\ref{fig:h_c}) is the steep behavior of
$H_c$ at $\rho_B= 120$--$123$. This feature is due to rearrangement of the supercurrents inside the bilayer, which
occurs in the following way. The supercurrent velocity changes across the thickness of the bilayer according to the
simple linear law~(\ref{v}). This supercurrent distribution may be characterized by the position of the stationary point
$x_0$, where the supercurrent velocity is zero: $v(x_0) =0$, hence $x_0 = m v_0 /eH$. At large values of the interface
resistance $\rho_B$, the density of the superconducting electrons in the N layer is very small, $n_N /n_S \ll 1$, and
the supercurrents circulate only in the S part of the system; this case corresponds to
\begin{equation}
x_0 = \frac{d_S}2.
\end{equation}
Then, while decreasing $\rho_B$, a shift in $x_0$ occurs. Now the supercurrents in the S layer are not compensated (in
the sense of the charge transfer); therefore, they must be compensated by the supercurrents in the N layer, which are
enhanced due to significant increase in $n_N$. This situation corresponds to the beginning of the drop in $H_c$. The
ratio of the superconducting electrons' densities grows rapidly, approaching the Anderson limit value $n_N/n_S =
\sigma_N/\sigma_S$ (see Sec.~\ref{subsec:H_c_Anderson} below); simultaneously, $x_0$ tends to
\begin{equation} \label{x_0}
x_0 = \frac{\sigma_S d_S^2 -\sigma_N d_N^2}{2 \left( \sigma_S d_S + \sigma_N d_N \right)},
\end{equation}
and the steep drop in $H_c$ finishes. For the bilayer to which the numerical results refer, $d_S \ll d_N$ and $\sigma_S
\ll \sigma_N$, hence Eq. (\ref{x_0}) yields $x_0 \approx -d_N /2$.

This scenario is illustrated by Fig.~\ref{fig:x_0}, which has been obtained numerically.

The analytical solution of Eqs. (\ref{system_1})--(\ref{system_2}) at zero temperature in the Anderson limit is
presented below.

\subsection{$H_c$ at zero temperature in the Anderson limit}
\label{subsec:H_c_Anderson}

In the zero-temperature Anderson limit (defined by the conditions $\tau_S E_S$, $\tau_N E_N \ll 1$), the ratio of the
superconducting electrons' densities~(\ref{system_2}) becomes independent of the magnetic field, $n_N/n_S
=\sigma_N/\sigma_S$, and the self-consistency equation~(\ref{system_1}) yields
\begin{equation} \label{H_c}
\frac{\tau_S E_S +\tau_N E_N}{\tau_S +\tau_N} = \frac{\Delta_{BCS}}2 \left[ \frac{\Delta_{BCS}}{2\omega_D} \sqrt{1+
\left( \frac{\tau_S\tau_N\omega_D} {\tau_S+\tau_N} \right)^2} \right]^{\tau_N / \tau_S},
\end{equation}
which determines $H_c$ (at $\tau_S\tau_N\omega_D/ (\tau_S+\tau_N) \sim 1$ this formula should be considered as an
interpolation --- see the discussion in the end of Sec.~\ref{sec:anderson}). This result can be compared to the BCS
case, which corresponds to the limit $\rho_B\to\infty$. In this case, the density of the superconducting electrons in
the N layer vanishes, $n_N /n_S \to 0$, and the self-consistency equation yields
\begin{equation} \label{H_c_BCS}
E_S^{BCS} = \frac{\Delta_{BCS}}2,
\end{equation}
where $E_S^{BCS}$ is given by Eq. (\ref{E_S}) with $n_N=0$. Finally, we reproduce the result of Maki
\cite{1:Maki_parallel}:
\begin{equation} \label{Hc_BCS}
H_c^{BCS} = \frac{\sqrt 3 \, \Phi_0}{\pi\, \xi_{BCS}\, d_S},\qquad \xi_{BCS} = \sqrt{\frac{D_S}{\Delta_{BCS}}},
\end{equation}
where $\Phi_0 = \pi/e$ is the flux quantum, and $\xi_{BCS}$ is the correlation length in the dirty limit.

Remarking that the r.h.s. of Eq. (\ref{H_c}) is identical to $E_g/2$ with the minigap $E_g$ given by Eq.
(\ref{E_g_Anderson}), we see that equation~(\ref{H_c}), determining the parallel critical field of the bilayer in the
Anderson limit, is obtained from the BCS equation~(\ref{H_c_BCS}) if we substitute the order parameter $\Delta_{BCS}$ by
the minigap $E_g$ (in accordance with the results of Sec.~\ref{sec:anderson}) and the $H$-dependent energy $E_S^{BCS}$
by the corresponding averaged quantity $(\tau_S E_S +\tau_N E_N) / (\tau_S +\tau_N)$.

The explicit result for the parallel critical field of the bilayer, obtained from Eq. (\ref{H_c}), can be cast into a
BCS-like form:
\begin{equation} \label{H_c_BCS_like}
H_c = \frac{\sqrt 3 \Phi_0}{\pi \xi d_\mathrm{eff}}.
\end{equation}
The bilayer's correlation length $\xi$ is the characteristic space scale on which the order parameter (or the pairing
angle $\theta$, or the Green function) varies in the absence of the magnetic field. In the Anderson limit (under
discussion), the explicit formula for $\xi$ is a natural generalization of the BCS expression [see Eq. (\ref{Hc_BCS})]
which implies that $D_S$ must be substituted by the averaged diffusion constant $\left< D \right>$ and $\Delta_{BCS}$
must be substituted (in accordance with the results of Sec.~\ref{sec:anderson}) by the bilayer's characteristic energy
scale, the minigap $E_g$ [Eq. (\ref{E_g_Anderson})]:
\begin{equation} \label{xi}
\xi=\sqrt{\frac{\left< D \right>}{E_g}},\qquad \left< D \right> = \frac{\tau_S D_S+\tau_N D_N}{\tau_S +\tau_N}.
\end{equation}
The effective thickness of the bilayer in Eq. (\ref{H_c_BCS_like}) is
\begin{equation}
d_\mathrm{eff} = \frac{\sqrt{\left( \sigma_S d_S +\sigma_N d_N \right) \left( \sigma_S d_S^3 + \sigma_N d_N^3 \right) +
3\sigma_S \sigma_N d_S d_N \left( d_S+d_N \right)^2}}{\sigma_S d_S + \sigma_N d_N}.
\end{equation}
In the case of equal conductivities, $\sigma_S = \sigma_N$, the effective thickness is simply the geometrical one:
$d_\mathrm{eff} = d_S + d_N$. This case corresponds to a uniform density of the superconducting electrons, $n_S = n_N$,
which implies a continuous distribution of the supercurrents, centered at the middle of the bilayer [this can also be
seen from Eq. (\ref{x_0}) which yields $x_0 =(d_S-d_N)/2$ in the case $\sigma_S=\sigma_N$]. However, in a more subtle
situation when the conductivities are different, the density of the supercurrent experiences a jump at the SN interface;
this nontrivial supercurrent distribution results in the nonequivalence of $d_\mathrm{eff}$ to the geometrical thickness
of the bilayer.

\section{Perpendicular upper critical field} \label{sec:H_c2}

Now we turn to calculating the upper critical field $H_{c2}$ perpendicular to the plane of the bilayer.

As in the case of the parallel critical field, we start with discussing the supercurrent distribution, which is now a
function of the sample boundaries in the $yz$ plane, perpendicular to the magnetic field $\mathbf{H}$ (the magnetic
field is directed along the $x$ axis). The infinite bilayer under consideration can be thought of as a disk of a large
radius; let us assume $y=0$, $z=0$ at the axis of the disk. Then the supercurrent distribution is axially symmetric,
and, with the gauge chosen as $\mathbf{A}=\left[ \mathbf{Hr} \right]/2$, the superconducting phase must be constant,
$\varphi=0$, which yields a simple result for the supercurrent velocity: $\mathbf{v}=e\mathbf{A}/m$.

Near $H_{c2}$, the superconducting correlations are small, $|\theta| \ll 1$, and the Usadel equations can be linearized:
\begin{gather}
-\frac D2 \left( -i\nabla +2e \mathbf{A} \right)^2 \theta +iE\theta +\Delta = 0, \label{usadel_h_c2}\\
\mathbf{A} \nabla\theta = 0.
\end{gather}
The second of these equations is trivially satisfied because $\theta(\mathbf{r})$ is axially symmetric.

Thus, the Usadel equations reduce to the single Eq. (\ref{usadel_h_c2}) for the pairing angle $\theta(\mathbf{r},E)$.
Introducing the cylindrical coordinates $\mathbf{r} \leftrightarrow (x,\boldsymbol\rho)$ and denoting $\hat{\mathbf{P}}
=-i \nabla_{\boldsymbol\rho} +2e \mathbf{A} (\boldsymbol\rho)$, we rewrite this equation as
\begin{gather}
\frac{D_N}2 \frac{\partial^2 \theta_N}{\partial x^2} -\frac{D_N}2 \hat{\mathbf{P}}^2 \theta_N + iE\theta_N
= 0, \label{usadel_detailed_1}\\
\frac{D_S}2 \frac{\partial^2 \theta_S}{\partial x^2} -\frac{D_S}2 \hat{\mathbf{P}}^2 \theta_S + iE\theta_S +\Delta =0.
\label{usadel_detailed_2}
\end{gather}
We cannot solve these equations straightforwardly because near the upper critical field, the order parameter
$\Delta({\boldsymbol\rho})$ is a nontrivial unknown function of the in-plane coordinate $\boldsymbol\rho$ (while the
$x$-dependence is absent due to the small thickness of the bilayer). In this situation, we employ the following
approach.

Averaging each of Eqs. (\ref{usadel_detailed_1})--(\ref{usadel_detailed_2}) over the thickness of the corresponding
layer, we obtain
\begin{align}
\left. \frac{\partial\theta_N}{\partial x} \right|_{x=0} &=
\frac{2d_N}{D_N} \left( \frac{D_N}2 \hat{\mathbf{P}}^2 \theta_N -iE\theta_N \right), \label{after_avrg_1}\\
\left. \frac{\partial\theta_S}{\partial x} \right|_{x=0} &= \frac{2d_S}{D_S} \left( -\frac{D_S}2 \hat{\mathbf{P}}^2
\theta_S +iE\theta_S +\Delta \right) . \label{after_avrg_2}
\end{align}
The averaged pairing angles entering the r.h.s. of Eqs. (\ref{after_avrg_1})--(\ref{after_avrg_2}) are
\begin{align}
\theta_N (\boldsymbol\rho,E) &= \frac 1{d_N} \int_{-d_N}^0 dx\, \theta_N (x,\boldsymbol\rho,E),\\
\theta_S (\boldsymbol\rho,E) &= \frac 1{d_S} \int_0^{d_S} dx\, \theta_S (x,\boldsymbol\rho,E).
\end{align}
Substituting Eqs. (\ref{after_avrg_1})--(\ref{after_avrg_2}) into the boundary conditions~(\ref{gran_uslovija}) (which
should be linearized), we obtain a system of two differential equations for the function $\theta(\boldsymbol\rho,E)$:
\begin{equation}
\tau_N \left( \frac{D_N}2 \hat{\mathbf{P}}^2 \theta_N -iE\theta_N \right) = \tau_S \left( -\frac{D_S}2
\hat{\mathbf{P}}^2 \theta_S +iE\theta_S +\Delta \right) = \theta_S-\theta_N. \label{gr_us}
\end{equation}

From the vicinity of the superconductive transition it follows that the pairing angle $\theta$ depends on the order
parameter $\Delta$ linearly:
\begin{align}
\theta_N (\boldsymbol\rho,E) &= \frac{\Delta \left(\boldsymbol\rho\right)}{\alpha_N\left( E\right)},\\
\theta_S (\boldsymbol\rho,E) &= \frac{\Delta \left(\boldsymbol\rho\right)}{\alpha_S \left( E\right)},
\end{align}
where the functions $\alpha_N (E)$ and $\alpha_S (E)$ are spatially independent. Then Eqs. (\ref{gr_us}) can be
rewritten as
\begin{gather}
\frac{D_N}2 \hat{\mathbf{P}}^2 \Delta (\boldsymbol\rho) = \left[ iE + \frac1{\tau_N} \left( \frac{\alpha_N
\left( E \right)}{\alpha_S \left( E \right)} -1 \right) \right] \Delta (\boldsymbol\rho), \\
\frac{D_S}2 \hat{\mathbf{P}}^2 \Delta (\boldsymbol\rho) = \left[ iE + \frac 1{\tau_S} \left( \frac{\alpha_S \left( E
\right)}{\alpha_N \left( E \right)} -1 \right) + \alpha_S (E) \right] \Delta (\boldsymbol\rho).
\end{gather}

We see that the order parameter must be an eigenfunction of the differential operator $\hat{\mathbf{P}}^2$. Moreover, in
order to obtain the largest value of $H_{c2}$, we should choose the eigenfunction corresponding to the lowest eigenvalue
(in complete analogy with Refs.~\cite{1:Abrikosov,1:HW}). The solution of the emerging eigenvalue problem is readily
found thanks to its formal equivalence to the problem of determining the Landau levels of a two-dimensional particle
with the ``mass'' $1/D$ and the charge $-2e$ in the uniform magnetic field $\mathbf{H}$ directed along the third
dimension. The lowest Landau level is $DeH$; the function $\alpha_S (E)$ is straightforwardly determined,
\begin{equation}
\alpha_S (E) = D_S\, eH -iE +\frac{\tau_N \left( D_N\, eH -iE \right)}{\tau_S \left[ 1+\tau_N \left( D_N\, eH -iE
\right) \right]},
\end{equation}
and we substitute $\theta_S (\boldsymbol\rho,E)$ into the self-consistency equation~(\ref{Delta}). The order parameter
$\Delta (\boldsymbol\rho)$ cancels out, and the resulting equation, which determines $H_{c2}(T)$, can be cast into the
form
\begin{gather}
\ln\frac{T_c^{BCS}}T = -\frac{\tau_N}{\tau_S+\tau_N} \ln\sqrt{1+\left(
\frac{\tau_S+\tau_N}{\tau_S\tau_N\omega_D} \right)^2} -\psi \left( \frac 12 \right) \notag \\
\textstyle +\frac 12 \left[ 1+\frac{\mathcal{E}_S-\mathcal{E}_N}{\sqrt{\left( \mathcal{E}_S-\mathcal{E}_N \right)^2 +
4/\tau_S\tau_N}} \right] \psi \left( \frac 12 +\frac 1{4\pi T} \left[ \mathcal{E}_S+\mathcal{E}_N +\sqrt{\left(
\mathcal{E}_S-\mathcal{E}_N \right)^2 + \frac 4{\tau_S\tau_N}} \right] \right) \notag \\
\textstyle +\frac 12 \left[ 1-\frac{\mathcal{E}_S-\mathcal{E}_N}{\sqrt{\left( \mathcal{E}_S-\mathcal{E}_N \right)^2 +
4/\tau_S\tau_N}} \right] \psi \left( \frac 12 +\frac 1{4\pi T} \left[ \mathcal{E}_S+\mathcal{E}_N -\sqrt{\left(
\mathcal{E}_S-\mathcal{E}_N \right)^2 + \frac 4{\tau_S\tau_N}} \right] \right) , \label{digamma}
\end{gather}
where
\begin{equation}
\mathcal{E}_S = D_S\, eH_{c2}+ \frac 1{\tau_S} ,\qquad \mathcal{E}_N = D_N\, eH_{c2}+ \frac 1{\tau_N}
\end{equation}
are $H$-dependent energies. The logarithmic term in the r.h.s. of Eq. (\ref{digamma}) takes account of the finiteness of
the Debye energy $\omega_D$; it becomes important only in the limit of the perfect interface (the Cooper limit),
\textit{i.e.}, when $\tau_S\tau_N\omega_D /(\tau_S+\tau_N) \ll 1$. At $\tau_S\tau_N\omega_D/ (\tau_S+\tau_N) \sim 1$ the
logarithmic term should be considered as an interpolation --- see the discussion in the end of Sec.~\ref{sec:anderson}.

In the limit $\rho_B\to\infty$, Eq. (\ref{digamma}) reproduces the classical result of Maki \cite{1:Maki} and de~Gennes
\cite{1:deGennes} for the BCS case (see also the book~\cite{1:deGennes_book}):
\begin{equation} \label{MdG}
\ln\frac{T_c^{BCS}}T = \psi \left( \frac 12 +\frac{D_S\, eH_{c2}^{BCS}}{2\pi T} \right) -\psi \left( \frac 12 \right),
\end{equation}
which is valid for bulk superconductors and superconductive layers of arbitrary thickness (when the magnetic field is
directed perpendicularly to them).

\begin{figure}\vspace*{2mm}
 \centerline{\includegraphics[width=90mm]{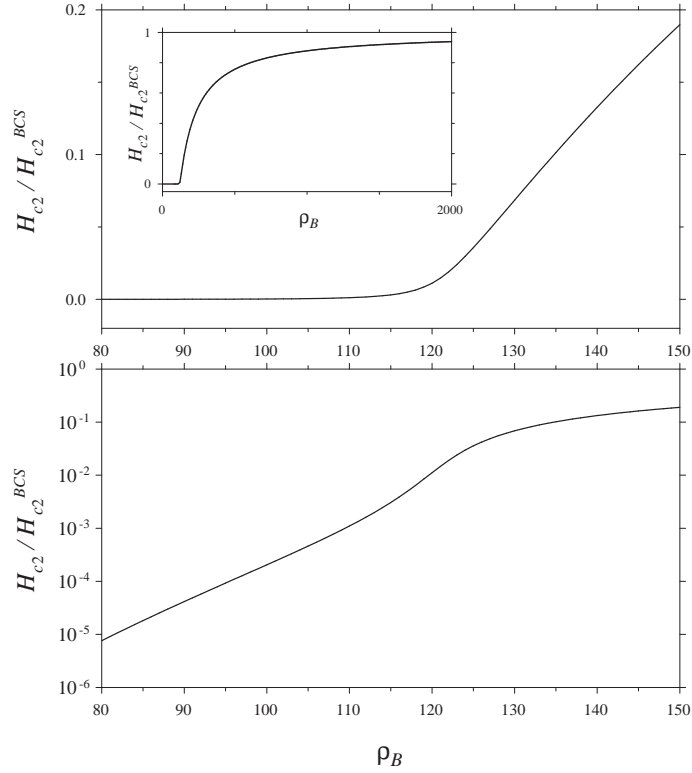}}
\caption{Perpendicular upper critical field $H_{c2}$, normalized by its BCS value, versus $\rho_B$ at zero temperature.
The upper and lower graphs differ only in the scaling of the ordinate axis (normal and logarithmic, respectively). The
inset shows $H_{c2} (\rho_B)$ on a wider scale over $\rho_B$. The choice of the bilayer's parameters implies the
relations $\tau_S \Delta_{BCS} = 0.016\, \rho_B$, $\tau_N \Delta_{BCS} = 0.23\,\rho_B$.}
 \label{fig:h_c2}
\end{figure}
\begin{figure}\vspace*{2mm}
 \centerline{\includegraphics[width=90mm]{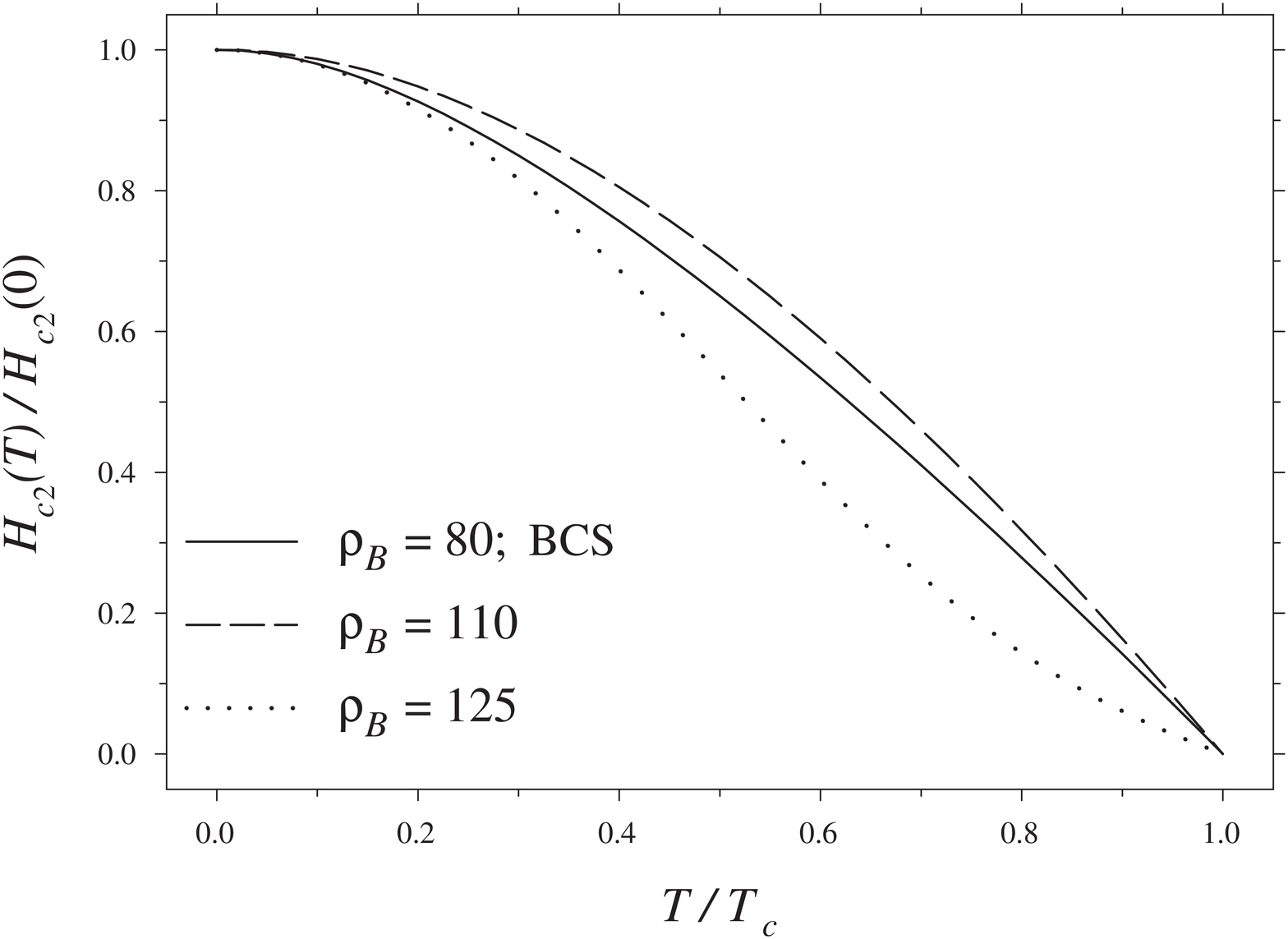}}
\caption{Temperature dependence of the perpendicular upper critical field $H_{c2}$ at $\rho_B=80$, 110, 125, and in the
BCS case. In each case, the critical field is normalized by its zero-temperature value, and the temperature is
normalized by the corresponding $T_c$. According to the results of Sec.~\ref{sec:anderson}, the curves in the BCS and
Anderson ($\rho_B=80$) limits coincide. At intermediate values of $\rho_B$, the curves can lie both above ($\rho_B=110$)
and below ($\rho_B=125$) the BCS curve. The choice of the bilayer's parameters implies the relations $\tau_S
\Delta_{BCS} = 0.016\, \rho_B$, $\tau_N \Delta_{BCS} = 0.23\,\rho_B$.}
 \label{fig:h_c2_vst}
\end{figure}

Equation~(\ref{digamma}) can be solved numerically at arbitrary values of the temperature $T$ and the interface
resistance $\rho_B$; the results for $H_{c2}$ are presented in Figs.~\ref{fig:h_c2}, \ref{fig:h_c2_vst}.

The analytical solution of Eq. (\ref{digamma}) at zero temperature in the Anderson limit is presented below.

\subsection{$H_{c2}$ at zero temperature in the Anderson limit}

In the zero-temperature Anderson limit (defined by the conditions $D_S\, eH_{c2} \ll 1/\tau_S$, $D_N\, eH_{c2} \ll
1/\tau_N$), Eq. (\ref{digamma}) yields
\begin{equation} \label{H_c2}
\frac{H_{c2}}{H_{c2}^{BCS}} = \frac{\left( \tau_S+\tau_N \right) D_S}{\tau_S D_S + \tau_N D_N} \left[
\frac{\Delta_{BCS}}{2\omega_D} \sqrt{1+ \left( \frac{\tau_S\tau_N\omega_D}{\tau_S+\tau_N} \right)^2}
\right]^{\tau_N/\tau_S} .
\end{equation}
(at $\tau_S\tau_N\omega_D/ (\tau_S+\tau_N) \sim 1$ this formula should be considered as an interpolation --- see the
discussion in the end of Sec.~\ref{sec:anderson}). Here the zero-temperature BCS value of the upper critical field, as
follows from Eq. (\ref{MdG}), is
\begin{equation}
H_{c2}^{BCS} =\frac{\Delta_{BCS}}{2eD_S} = \frac{\Phi_0}{2\pi\xi_{BCS}^2} .
\end{equation}

It is instructive to rewrite the perpendicular upper critical field of the bilayer~(\ref{H_c2}) in the standard BCS-like
form
\begin{equation} \label{H_c2_BCSlike}
H_{c2} =\frac{\Phi_0}{2\pi\xi^2},
\end{equation}
where $\xi$ is the bilayer's correlation length given by Eq. (\ref{xi}) [the physical interpretation of this result for
$\xi$ precedes Eq. (\ref{xi})].

In the Cooper limit (\textit{i.e.}, at $\tau_S\tau_N\omega_D/ (\tau_S+\tau_N) \ll 1$) Eqs. (\ref{H_c2}),
(\ref{H_c2_BCSlike}) reproduce the result of Refs.~\cite{1:Radovic_Hc2,1:Luders}.

\section{SNS, NSN, SNINS, NSISN, and superlattices}
\label{sec:SNS}

Our results for $\Delta$, $n_S$, $T_c$, $E_g$, and $H_{c2}$ (\textit{i.e.}, all the results except $H_c$) can be
directly applied to more complicated structures such as SNS and NSN trilayers, SNINS and NSISN systems, and SN
superlattices.

Let us consider, for example, a symmetric SNS trilayer consisting of two identical S layers of thickness $d_S$ separated
by a N layer of thickness $2 d_N$. The SN interfaces can have arbitrary (but equal) resistances. As before, the $x$ axis
is perpendicular to the plane of the structure. This trilayer can be imagined as composed of two identical bilayers
perfectly joined together along the N sides. Indeed, the pairing angle $\theta$ has zero $x$-derivative on the outer
surfaces of the bilayers, thus producing the correct (symmetric in the $x$-direction) solution for $\theta$ in the
resulting trilayer. Consequently, the symmetric SNS trilayer has exactly the same physical properties [$\Delta$, $n_S$,
$T_c$, $E_g$, $H_{c2}$] as the SN bilayer considered in the present chapter. The only point where the above reasoning
fails is the calculation of the parallel critical field $H_c$. In this case, the combination of the supercurrent
distributions in the two bilayers does not yield the correct distribution in the resulting SNS trilayer, which implies
that the Usadel equations for the two systems are different.

Evidently, the above reasoning, based on the formal equivalence of the outer-surface boundary condition for the bilayer
to the symmetry-caused condition in the middle of the SNS trilayer, also holds for symmetric NSN trilayers (N layers of
thickness $d_N$, S layer of thickness $2 d_S$, identical SN interfaces) and SN superlattices (N layers of thickness $2
d_N$, S layers of thickness $2 d_S$, identical SN interfaces). Moreover, the same applies to systems composed of two
bilayers in \textit{nonideal} contact with each other: SNINS and NSISN (where I stands for an arbitrary potential
barrier), because the presence of a potential barrier does not violate the applicability of the symmetry argument. Thus,
all the results obtained for the bilayer (except $H_c$) are also valid for these structures.

\section{Discussion} \label{sec:discussion}

Now we turn to a possible experimental application of our results. Our results provide a method for determining
$\rho_B$, a very important parameter of the bilayer which is not directly measurable. By analyzing the experimental
\cite{1:Kasumov,1:Kasumov_priv} values $T_c =0.4$~K and $H_c =0.1$~T, we get $\rho_B\approx 111$ and $\rho_B\approx
121$, respectively. Within the experimental accuracy of the bilayer's parameters, the two estimates for $\rho_B$ should
be considered close. Interestingly, the value $\rho_B\approx 121$ extracted from the measured value of $H_c$ corresponds
to the extremely narrow region of the steep drop in $H_c (\rho_B)$ (see Fig.~\ref{fig:h_c}).

An essential property of the bilayer used throughout the chapter is its small thickness. Now we shall argue that the
bilayer employed in the experiment by Kasumov \textit{et al.} \cite{1:Kasumov} (and to which our numerical results
refer) can be considered thin. The Usadel equations~(\ref{usadel_bi_1})--(\ref{usadel_bi_2}) imply that the
characteristic spatial scale of the bilayer's properties variation is $\sqrt{D_{N,S}/E_0}$ for the N and S layers,
respectively. Here $E_0$ is the characteristic energy scale for the self-consistency equation. As the interface
resistance decreases from infinity to zero, $E_0$ first increases from $\Delta_{BCS}$ to $(\tau_S^{-1}+\tau_N^{-1})$,
and then --- to $\omega_D$ (the crossovers between these regimes occur at such $\rho_B$ that the corresponding
expressions are of the same order). The above estimates show that the experiment \cite{1:Kasumov} corresponds to the
case $E_0\sim \Delta_{BCS}$. Therefore, the thicknesses of the layers ($d_N =$ 100~nm and $d_S =$ 5~nm) must be small
compared to $\sqrt{D_{N,S}/ \Delta_{BCS}}$, which equals 194~nm and 16~nm for the N and S layer, respectively. We thus
conclude that the condition of thin bilayer is approximately satisfied.

Finally, we wish to remark on a peculiarity of real systems which can be relevant when one compares our findings with an
experiment. The point is that during the fabrication of a bilayer, the interface between S and N materials cannot be
made ideally uniform. In other words, the local interface resistance possesses spatial fluctuations. At the same time,
as we have shown, the bilayer's properties are highly sensitive to the interface quality, which could lead to
complicated behavior not reducing to the simple averaging of the interface resistance embodied in $\rho_B$. One
possibility could be a percolation-like proximity effect. We leave the study of inhomogeneity effects for further
investigation.

\section{Conclusions} \label{sec:conclusion}

In this chapter, we have studied, both analytically and numerically, the proximity effect in a thin SN bilayer in the
dirty limit. The strength of the proximity effect is governed by $\rho_B$, the resistance of the SN interface per
channel.

The quantities calculated were $\Delta$, the order parameter; $n_S$, the density of the superconducting electrons in the
S layer; $T_c$, the critical temperature; $E_g$, the minigap in the density of states; $H_c$ and $H_{c2}$, the critical
magnetic field parallel to the bilayer and the upper critical field perpendicular to the bilayer.

These quantities were calculated numerically over the entire range of $\rho_B$. For this purpose, the characteristics of
the bilayer were assumed to be the same as in the experiment by Kasumov \textit{et al.} \cite{1:Kasumov} (Ta/Au bilayer,
$d_S/d_N =1/20$). In the limit of an opaque interface, $\Delta$, $n_S$, $T_c$, $H_c$, and $H_{c2}$ approach their BCS
values. At the same time, $E_g$ does not coincide with the order parameter $\Delta$, and $E_g\to 0$ when
$\rho_B\to\infty$, although in general, the energy dependence of the DoS in the S and N layers approaches the BCS and
normal-metal results, respectively.

The minigap $E_g$ demonstrates nonmonotonic behavior as a function of $\rho_B$. Analytical results for the two limiting
cases of small and large $\rho_B$ show that in the Anderson limit, $E_g$ increases with increasing $\rho_B$, whereas in
the limit of an opaque interface, $E_g$ tends to zero. Thus, $E_g$ reaches its maximum in the region of intermediate
$\rho_B$.

Also in the region of moderate resistances, a jump of the parallel critical field (due to a redistribution of
supercurrents in the bilayer) is discovered.

The most interesting case of relatively low interface resistance (the Anderson limit) has been considered analytically.
The simple BCS relations between $\Delta$, $n_S$, $T_c$, $H_c$, $H_{c2}$ are substituted by similar ones with $E_g$
standing instead of $\Delta$. The relation between the minigap $E_g$ and the order parameter $\Delta$ in this limit is
expressed by Eq. (\ref{small}), implying that in the case where $\tau_S < \tau_N$, the BCS relations are strongly
violated (by more than the order of magnitude for the above-mentioned Ta/Au bilayer). The DoS in the S and N layers
coincide, showing BCS-like behavior with the standard peculiarity at $E=E_g$. It should be emphasized that absolute
values of $\rho_B$ corresponding to the Anderson limit can be large; for the experiment \cite{1:Kasumov} this limit is
already valid at $\rho_B <80$.

All the results (except $H_c$) obtained for the bilayer also apply to more complicated structures such as SNS and NSN
trilayers, SNINS and NSISN systems, and SN superlattices.

\clearpage
\renewcommand{\@evenhead}{}

\chapter[Proximity effect in SF systems]{\huge Proximity effect in SF systems}

\section{Nonmonotonic critical temperature in SF bilayers} \label{sec:tcSF}

\renewcommand{\@evenhead}
    {\raisebox{0pt}[\headheight][0pt]
     {\vbox{\hbox to\textwidth{\thepage \hfil \strut \textit{Chapter \thechapter}}\hrule}}
    }

\renewcommand{\@oddhead}
    {\raisebox{0pt}[\headheight][0pt]
     {\vbox{\hbox to\textwidth{\textit{\leftmark} \strut \hfil \thepage}\hrule}}
    }

\subsection{Introduction}

Superconductivity and ferromagnetism are two competing orders: while the former ``prefers'' an antiparallel spin
orientation of electrons in Cooper pairs, the latter forces the spins to align in parallel. Therefore, their coexistence
in one and the same material is possible only in a narrow interval of parameters; hence the interplay between
superconductivity and ferromagnetism is most conveniently studied when the two interactions are spatially separated. In
this case the coexistence of the two orders is due to the proximity effect. Recently, much attention has been paid to
properties of hybrid proximity systems containing superconductors (S) and ferromagnets (F); new physical phenomena were
observed and predicted in these systems \cite{2:Ryazanov,2:Kontos,2:Radovic,2:Tagirov_PRL,2:Buzdin_DOS,2:Nazarov_DOS}.
One of the most striking effects in SF layered structures is highly nonmonotonic dependence of their critical
temperature $T_c$ on the thickness $d_F$ of the ferromagnetic layers. Experiments exploring this nonmonotonic behavior
were performed previously on SF multilayers such as Nb/Gd \cite{2:Jiang}, Nb/Fe \cite{2:Muhge}, V/V-Fe \cite{2:Aarts},
and Pb/Fe \cite{2:Lazar}, but the results (and, in particular, the comparison between the experiments and theories) were
not conclusive.

To perform reliable experimental measurements of $T_c(d_F)$, it is essential to have $d_F$ large compared to the
interatomic distance; this situation can be achieved only in the limit of weak ferromagnets. Active experimental
investigations of SF bilayers and multilayers based on Cu-Ni dilute ferromagnetic alloys are carried out by several
groups \cite{2:Ryazanov_JETPL,2:Aarts_recent}. In SF bilayers, they observed nonmonotonic dependence $T_c(d_F)$. While
the reason for this effect in multilayers can be the $0$--$\pi$ transition \cite{2:Radovic}, in a bilayer system with a
single superconductor this mechanism is irrelevant, and the cause of the effect is interference of quasiparticle,
specific to SF structures.

In the present chapter, motivated by the experiments of Refs.~\cite{2:Ryazanov_JETPL,2:Aarts_recent} we theoretically
study the critical temperature of SF bilayers. Previous theoretical investigations of $T_c$ in SF structures were
concentrated on systems with thin or thick layers (compared to the corresponding coherence lengths); with SF boundaries
having very low or very high transparencies; the exchange energy was often assumed to be much larger than the critical
temperature; in addition, the methods for solving the problem were usually approximate
\cite{2:Radovic,2:Aarts,2:Lazar,2:BVK,2:Demler,2:PKhI,2:Tagirov,2:Tagirov_PRL}. The parameters of the experiments of
Refs.~\cite{2:Ryazanov_JETPL,2:Aarts_recent} do not correspond to any of the above limiting cases. In the present
chapter we develop two approaches giving the opportunity to investigate not only the limiting cases of parameters but
also the intermediate region. Using our methods, we also confirm different types of nonmonotonic $T_c(d_F)$ behavior,
found previously \cite{2:Radovic,2:PKhI,2:Tagirov}: the minimum of $T_c$ and the reentrant superconductivity. Comparison
of our theoretical predictions with the experimental data shows good agreement.

A number of methods can be used for calculating $T_c$. When the critical temperature of the structure is close to the
critical temperature $T_{cS}$ of the superconductor without the ferromagnetic layer, the Ginzburg--Landau (GL) theory
applies. However, $T_c$ of SF bilayers may significantly deviate from $T_{cS}$, therefore we choose a more general
theory valid at arbitrary temperature --- the quasiclassical approach \cite{2:Usadel,2:LO,2:RS}. Near $T_c$ the
quasiclassical equations become linear. In the literature the emerging problem is often treated with the help of the
so-called ``single-mode'' approximation \cite{2:Demler,2:PKhI,2:Tagirov,2:Tagirov_PRL}, which is argued to be
qualitatively reasonable in a wide region of parameters. However, this method is justified only in a specific region of
parameters which we find below. Moreover, below we show examples when this method fails even qualitatively. Thus there
is need for an exact solution of the linearized quasiclassical equations. The limiting case of perfect boundaries and
large exchange energy was treated by Radovi\'c \textit{et al.} \cite{2:Radovic}.

Based on the progress achieved for calculation of $T_c$ in SN systems (where N denotes a nonmagnetic normal material)
\cite{2:Golubov1}, we develop a generalization of the single-mode approximation --- the multimode method. Although this
method seems to be exact, it is subtle to justify it rigorously. Therefore we develop yet another approach (this time
mathematically rigorous), which we call ``the method of fundamental solution''. The models considered previously
\cite{2:Radovic,2:Aarts,2:Lazar,2:BVK,2:Demler,2:PKhI,2:Tagirov,2:Tagirov_PRL} correspond to limiting cases of our
theory.


\subsection{Model} \label{sec:model}

\begin{figure}
 \centerline{\includegraphics[width=40mm]{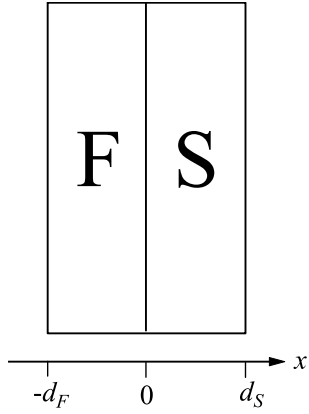}}
\caption{\label{fig:fig1_SF} SF bilayer. The F and S layers occupy the regions $-d_F<x<0$ and $0<x<d_S$, respectively.}
\end{figure}
We assume that the dirty-limit conditions are fulfilled, and calculate the critical temperature of the bilayer within
the framework of the linearized Usadel equations for the S and F layers (the domain $0<x<d_S$ is occupied by the S
metal, $-d_F<x<0$ --- by the F metal, see Fig.~\ref{fig:fig1_SF}). Near $T_c$ the normal Green function is
$G=\sgn\omega_n$, and the Usadel equations for the anomalous function $F$ take the form
\begin{gather}
\label{U_1} \xi_S^2 \pi T_{cS} \frac{d^2 F_S}{dx^2}- |\omega_n| F_S +\Delta=0,\qquad 0<x<d_S,\\
\label{U_2} \xi_F^2 \pi T_{cS} \frac{d^2 F_F}{dx^2} -(|\omega_n|+ih \sgn\omega_n) F_F=0, \qquad -d_F<x<0, \\
\label{U_3} \Delta\ln\frac{T_{cS}}{T} = \pi T \sum_{\omega_n} \left( \frac\Delta{|\omega_n|} -F_S \right)
\end{gather}
(the order parameter $\Delta$ is nonzero only in the S part). Here $\xi_S=\sqrt{D_S/2\pi T_{cS}}$, $\xi_F=\sqrt{D_F/2\pi
T_{cS}}$ are the coherence lengths, while the diffusion constants can be expressed via the Fermi velocity and the mean
free path: $D=vl/3$; $\omega_n=\pi T(2n+1)$ with $n=0,\pm 1, \pm 2,\ldots$ are the Matsubara frequencies; $h$ is the
exchange energy; and $T_{cS}$ is the critical temperature of the S material. $F_{S(F)}$ denotes the function $F$ in the
S(F) region. We use the system of units in which Planck's and Boltzmann's constants equal unity, $\hbar=k_B=1$.

Equations (\ref{U_1})--(\ref{U_3}) must be supplemented with the boundary conditions at the outer surfaces of the
bilayer:
\begin{equation}
\frac{dF_S(d_S)}{dx}=\frac{dF_F(-d_F)}{dx}=0,
\end{equation}
as well as at the SF boundary \cite{2:KL}:
\begin{align}\label{bound_1_}
& \xi_S\frac{dF_S(0)}{dx} =\gamma\xi_F \frac{dF_F(0)}{dx}, && \gamma =\frac{\rho_S\xi_S}{\rho_F\xi_F}, \\
& \xi_F\gamma_B \frac{dF_F(0)}{dx} = F_S(0)-F_F(0), && \gamma_B =\frac{R_B \mathcal{A}}{\rho_F\xi_F}. \label{bound_2_}
\end{align}
Here $\rho_S$, $\rho_F$ are the normal-state resistivities of the S and F metals, $R_B$ is the resistance of the SF
boundary, and ${\cal A}$ is its area. The above boundary conditions were derived for SN interfaces \cite{2:KL} (N is a
normal metal); their use in the SF case is justified by the small parameter $h/E_\mathrm{F} \ll 1$ ($E_\mathrm{F}$ is
the Fermi energy). Indeed, the interface has atomic (by the order of magnitude) thickness. While the exchange energy is
small compared to the Fermi energy, the characteristic length for magnetic properties is much larger than the atomic
scale. Therefore the boundary conditions are determined by the properties of the interface itself but not by the
properties of the contacting metals. In the limit of strong ferromagnets, this condition fails and the
Kupriyanov--Lukichev boundary conditions lose its validity. However, in the thesis we consider only weak ferromagnets.

The Usadel equation in the F layer is readily solved:
\begin{gather}
F_F=C(\omega_n) \cosh\left( \widetilde k_h [x+d_F]\right), \label{F_F} \\
\widetilde k_h=\frac 1{\xi_F} \sqrt{\frac{|\omega_n|+ih \sgn\omega_n}{\pi T_{cS}}},\notag
\end{gather}
and the boundary condition at $x=0$ can be written in closed form with respect to $F_S$:
\begin{gather}
\xi_S \frac{dF_S(0)}{dx} =\frac\gamma{\gamma_B +B_h(\omega_n)} F_S(0), \label{bound_Phi} \\
B_h = \left[ \widetilde k_h \xi_F \tanh (\widetilde k_h d_F) \right]^{-1}.\notag
\end{gather}

This boundary condition is complex. In order to rewrite it in a real form, we do the usual trick and go over to the
functions
\begin{equation}
F^\pm =F(\omega_n)\pm F(-\omega_n).
\end{equation}
According to the Usadel equations (\ref{U_1})--(\ref{U_3}), there is the symmetry $F(-\omega_n)=F^*(\omega_n)$ which
implies that $F^+$ is real while $F^-$ is a purely imaginary function.

The symmetric properties of $F^+$ and $F^-$ with respect to $\omega_n$ are trivial, so we shall treat only positive
$\omega_n$. The self-consistency equation is expressed only via the symmetric function $F_S^+$:
\begin{equation} \label{self_cons}
\Delta\ln\frac{T_{cS}}T = \pi T \sum_{\omega_n>0} \left(\frac{2\Delta}{\omega_n}-F_S^+ \right),
\end{equation}
and the problem of determining $T_c$ can be formulated in a closed form with respect to $F_S^+$ as follows. The Usadel
equation for the antisymmetric function $F_S^-$ does not contain $\Delta$, hence it can be solved analytically. After
that we exclude $F_S^-$ from boundary condition (\ref{bound_Phi}) and arrive at the effective boundary conditions for
$F_S^+$:
\begin{equation} \label{bound_p}
\xi_S \frac{dF_S^+ (0)}{dx}= W(\omega_n) F_S^+ (0),\qquad\frac{dF_S^+ (d_S)}{dx}=0,
\end{equation}
where
\begin{gather}
W(\omega_n) =\gamma \frac{A_S (\gamma_B+\Re B_h)+ \gamma}{A_S |\gamma_B+B_h|^2 +\gamma (\gamma_B+\Re B_h)},
\label{W_def} \\
A_S= k_S \xi_S \tanh (k_S d_S),\quad k_S=\frac 1{\xi_S} \sqrt{\frac{\omega_n}{\pi T_{cS}}} \notag.
\end{gather}
The self-consistency equation (\ref{self_cons}) and boundary conditions (\ref{bound_p})--(\ref{W_def}), together with
the Usadel equation for $F_S^+$:
\begin{equation}
\label{usadel_b} \xi_S^2 \pi T_{cS} \frac{d^2 F_S^+}{dx^2}- \omega_n F_S^+ +2\Delta=0
\end{equation}
will be used below for finding the critical temperature of the bilayer.

The problem can be solved analytically only in limiting cases (see Appendix~\ref{ap:sec:analytics}). In the general
case, one should use a numerical method, and below we propose two methods for solving the problem exactly.

\subsection{Multimode method} \label{sec:multi-mode}

\subsubsection{Starting point: the single-mode approximation and its applicability}

In the single-mode approximation (SMA) one seeks the solution of the problem (\ref{self_cons})--(\ref{usadel_b}) in the
form
\begin{gather}
F_S^+(x,\omega_n)= \mathrm{f} (\omega_n)\cos\left(\Omega\frac{x-d_S}{\xi_S}\right), \label{Phi_sm} \\
\Delta(x)=\delta \cos\left(\Omega\frac{x-d_S}{\xi_S}\right). \label{Delta_sm}
\end{gather}
This anzatz automatically satisfies boundary condition (\ref{bound_p}) at $x=d_S$.

The Usadel equation (\ref{usadel_b}) yields
\begin{equation}
\mathrm{f}(\omega_n)=\frac{2\delta}{\omega_n+ \Omega^2 \pi T_{cS}},
\end{equation}
then the self-consistency Eq. (\ref{self_cons}) takes the form ($\delta$ and $\Omega$ do not depend on $\omega_n$)
\begin{equation} \label{Omega}
\ln\frac{T_{cS}}{T_c}=\psi\left(\frac 12+ \frac{\Omega^2}2 \frac{T_{cS}}{T_c}\right)- \psi\left(\frac 12\right),
\end{equation}
where $\psi$ is the digamma function.

Boundary condition (\ref{bound_p}) at $x=0$ yields
\begin{equation} \label{bound_s_mode}
\Omega\tan\left(\Omega \frac{d_S}{\xi_S} \right) = W(\omega_n).
\end{equation}
The critical temperature $T_c$ is determined by Eqs. (\ref{Omega}), (\ref{bound_s_mode}).

Although this method is popular, it is often used without pointing out the limits of its applicability. The explicit
formulation of the corresponding condition is as follows \cite{2:BVK}: the single-mode method is correct only if the
parameters of the system are such that $W$ can be considered $\omega_n$-independent [because the left-hand side of Eq.
(\ref{bound_s_mode}) must be $\omega_n$-independent].

Appendix~\ref{ap:sma} demonstrates examples of the SMA validity and corresponding analytical results.

In one of experimentally relevant cases, $h/\pi T_{cS} > 1$, $d_F \sim \xi_F$, the SMA is applicable if $\sqrt{h/\pi
T_{cS}}\gg 1 /\gamma_B$ (see Appendix~\ref{ap:sma} for details).

\subsubsection{Inclusion of other modes}

\begin{figure}
 \centerline{\includegraphics[width=70mm]{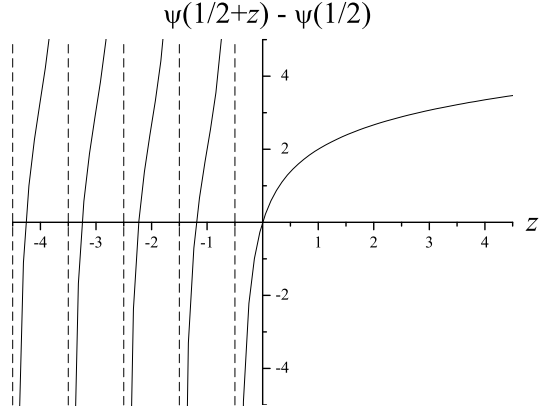}}
\caption{\label{fig:digamma} Plot of the $\psi(1/2+z)-\psi(1/2)$ function. At any positive $y$, the equation
$y=\psi(1/2+z)-\psi(1/2)$ has one positive and infinite number of negative solutions. The positive solution is employed
in the single-mode approximation [see Eq. (\ref{Omega})], while taking into account the negative solutions [which
implies imaginary $\Omega$ in Eq. (\ref{Omega})] we obtain the exact multimode method.}
\end{figure}

The single-mode approximation implies that one takes the (only) real root $\Omega$ of Eq. (\ref{Omega}). An exact
(multimode) method for solving problem (\ref{self_cons})--(\ref{usadel_b}) is obtained if we also take imaginary roots
into account --- there is infinite number of these \cite{2:Golubov1}, see Fig.~\ref{fig:digamma}.

Thus we seek the solution in the form
\begin{gather}
F_S^+(x,\omega_n) = \mathrm{f}_0(\omega_n)\cos\left(\Omega_0 \frac{x-d_S}{\xi_S} \right) + \sum_{m=1}^{\infty}
\mathrm{f}_m (\omega_n)\frac{\cosh\left(\Omega_m \frac{x-d_S}{\xi_S} \right)} {\cosh\left(\Omega_m
\frac{d_S}{\xi_S}\right)}, \label{Phi} \\
\Delta(x) = \delta_0 \cos\left(\Omega_0\frac{x-d_S}{\xi_S}\right) + \sum_{m=1}^{\infty} \delta_m
\frac{\cosh\left(\Omega_m\frac{x-d_S}{\xi_S}\right)} {\cosh\left(\Omega_m\frac{d_S}{\xi_S}\right)}. \label{Delta_}
\end{gather}
(The normalizing denominators in the $\cosh$-terms have been introduced in order to increase accuracy of numerical
calculations.) This anzatz automatically satisfies boundary condition (\ref{bound_p}) at $x=d_S$.

Substituting the anzatz [Eqs. (\ref{Phi})--(\ref{Delta_})] into the Usadel equation (\ref{usadel_b}), we obtain
\begin{align}
\mathrm{f}_0(\omega_n) & =\frac{2\delta_0}{\omega_n+\Omega_0^2 \pi T_{cS}},\\
\mathrm{f}_m(\omega_n) & =\frac{2\delta_m}{\omega_n-\Omega_m^2 \pi T_{cS}},\quad m=1,2,\ldots, \notag
\end{align}
then the parameters $\Omega$ are determined by the self-consistency equation (\ref{self_cons}) ($\delta$ and $\Omega$ do
not depend on $\omega_n$):
\begin{align}
\ln\frac{T_{cS}}{T_c} & =\psi\left(\frac 12+ \frac{\Omega_0^2}2 \frac{T_{cS}}{T_c}\right)-
\psi\left(\frac 12\right), \label{Om} \\
\ln\frac{T_{cS}}{T_c} & =\psi\left(\frac 12- \frac{\Omega_m^2}2 \frac{T_{cS}}{T_c}\right)- \psi\left(\frac
12\right),\quad m=1,2,\ldots \notag
\end{align}
From Eqs. (\ref{Om}) and properties of the digamma function \cite{2:digamma} it follows that the parameters $\Omega$
belong to the following intervals:
\begin{gather}
0< \Omega_0^2 < \frac 1{2\exp(C)},\\
\frac{T_c}{T_{cS}} (2m-1) < \Omega_m^2 < \frac{T_c}{T_{cS}} (2m+1),\quad m=1,2,\ldots,\notag
\end{gather}
where $C \approx 0.577$ is Euler's constant.

Boundary condition (\ref{bound_p}) at $x=0$ yields the following equation for the amplitudes $\delta$:
\begin{eqnarray}
&& \delta_0 \frac{W(\omega_n) \cos\left(\Omega_0 d_S/\xi_S\right) -\Omega_0\sin\left( \Omega_0
d_S/\xi_S\right)}{\omega_n+\Omega_0^2 \pi T_{cS}} \notag \\
&& + \sum_{m=1}^\infty \delta_m \frac{W(\omega_n) + \Omega_m \tanh\left(\Omega_m d_S/\xi_S\right)}{\omega_n-\Omega_m^2
\pi T_{cS}} =0. \label{bound_m_mode}
\end{eqnarray}
The critical temperature $T_c$ is determined by Eqs. (\ref{Om}) and the condition that Eq. (\ref{bound_m_mode}) has a
nontrivial ($\omega_n$-independent) solution with respect to $\delta$.

Numerically, we take a finite number of modes: $m=0,1,\ldots,M$. To take account of $\omega_n$-independence of the
solution, we write down Eq. (\ref{bound_m_mode}) at the Matsubara frequencies up to the $N$th frequency:
$n=0,1,\ldots,N$. Thus we arrive at the matrix equation $K_{nm}\delta_m=0$ with the following matrix $\hat K$:
\begin{gather}
K_{n0} = \frac{W(\omega_n)\cos\left(\Omega_0 d_S/\xi_S \right) -\Omega_0\sin\left( \Omega_0
d_S/\xi_S\right)} {\omega_n/\pi T_{cS} +\Omega_0^2}, \notag \\
\label{K_def} K_{nm} = \frac{W(\omega_n) +\Omega_m \tanh\left(\Omega_m d_S/\xi_S\right)}{\omega_n/\pi
T_{cS} -\Omega_m^2}, \\
n=0,1,\ldots,N,\quad m=1,2,\ldots,M. \notag
\end{gather}
We take $M=N$, then the condition that Eq. (\ref{bound_m_mode}) has a nontrivial solution takes the form
\begin{equation} \label{K}
\det\hat K=0.
\end{equation}

Thus the critical temperature $T_c$ is determined as the largest solution of Eqs. (\ref{Om}), (\ref{K}).

\subsection{Method of fundamental solution} \label{sec:fund_sol}

By definition, the fundamental solution $\mathcal{G}(x,y;\omega_n)$ (which is also called the Green function) of problem
(\ref{bound_p})--(\ref{usadel_b}) satisfies the same equations, but with the delta-functional ``source'' \cite{2:Morse}:
\begin{gather}
\xi_S^2 \pi T_{cS} \frac{d^2 \mathcal{G}(x,y)}{dx^2}- \omega_n \mathcal{G}(x,y) = - \delta(x-y),
\label{usadel_G} \\
\xi_S \frac{d \mathcal{G}(0,y)}{dx}= W(\omega_n) \mathcal{G}(0,y),\qquad\frac{d \mathcal{G}(d_S,y)}{dx}=0.
\end{gather}
The fundamental solution can be expressed via solutions $v_1$, $v_2$ of Eq. (\ref{usadel_G}) without the delta-function,
satisfying the boundary conditions at $x=0$ and $x=d_S$, respectively:
\begin{equation}
\mathcal{G}(x,y;\omega_n) = \frac{k_S / \omega_n}{\sinh( k_S d_S) +(W/ k_S \xi_S) \cosh\left( k_S d_S\right)}
\times\left\{
\begin{aligned} v_1(x) v_2(y),\quad x\leqslant y\\ v_2(x) v_1(y),\quad y\leqslant x
\end{aligned} \right. , \label{G}
\end{equation}
where
\begin{subequations}
\begin{align}
v_1(x) & =\cosh( k_S x) +(W/ k_S \xi_S) \sinh( k_S x),\\
v_2(x) & =\cosh\left( k_S [x-d_S] \right). \label{v_2}
\end{align}
\end{subequations}

Having found $\mathcal{G}(x,y;\omega_n)$, we can write the solution of Eqs. (\ref{bound_p})--(\ref{usadel_b}) as
\begin{equation} \label{fsplus}
F_S^+ (x;\omega_n)=2\int_0^{d_S} \mathcal{G}(x,y;\omega_n) \Delta(y) dy.
\end{equation}
Substituting this into the self-consistency equation (\ref{self_cons}), we obtain
\begin{equation}
\Delta(x) \ln\frac{T_{cS}}{T_c} = 2\pi T_c \sum_{\omega_n>0} \left[ \frac{\Delta(x)}{\omega_n}-\int_0^{d_S}
\mathcal{G}(x,y;\omega_n) \Delta(y) dy \right]. \label{sc_Green}
\end{equation}
This equation can be expressed in an operator form: $\Delta\ln(T_{cS}/T_c)=\hat L\Delta$. Then the condition that Eq.
(\ref{sc_Green}) has a nontrivial solution with respect to $\Delta$ is expressed by the equation
\begin{equation} \label{det}
\det\left(\hat L- \hat 1 \ln\frac{T_{cS}}{T_c} \right)=0.
\end{equation}
The critical temperature $T_c$ is determined as the largest solution of this equation.

Numerically, we put problem (\ref{sc_Green}), (\ref{det}) on a spatial grid, so that the linear operator $\hat L$
becomes a finite matrix.

\subsection{Numerical results} \label{sec:num_res}

In Secs. \ref{sec:multi-mode}, \ref{sec:fund_sol} we developed two methods for calculating the critical temperature of a
SF bilayer. Specifying parameters of the bilayer we can find the critical temperature numerically. It can be checked
that the multimode method and the method of fundamental solution yield equivalent results. However, at small
temperatures $T_c \ll T_{cS}$, the calculation time for the multimode method increases. Indeed, the size of the matrix
$\hat K$ [Eq. (\ref{K_def})] is determined by the number $N$ of the maximum Matsubara frequency $\omega_N$, which must
be much larger than the characteristic energy $\pi T_{cS}$; hence $N\gg T_{cS}/T_c$. Therefore, at low temperatures we
use the method of fundamental solution.

\subsubsection{Comparison with experiment}

\begin{figure}
 \centerline{\includegraphics[width=90mm]{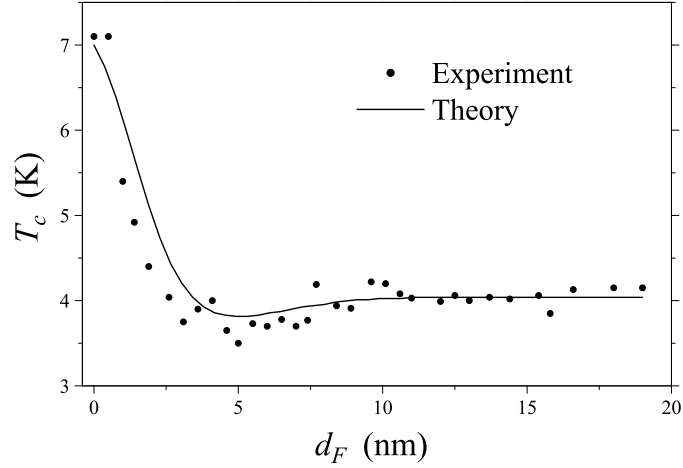}}
\caption{\label{fig:ryazanov} Theoretical fit to the experimental data of Ref.~\cite{2:Ryazanov_JETPL}. In the
experiment, Nb was the superconductor (with $d_S=11$\,nm, $T_{cS}=7$\,K) and Cu$_{0.43}$Ni$_{0.57}$ was the weak
ferromagnet. From our fit we estimate $h\approx 130$\,K and $\gamma_B\approx 0.3$.}
\end{figure}
Using our methods we fit the experimental data of Ref.~\cite{2:Ryazanov_JETPL}; the result is presented in
Fig.~\ref{fig:ryazanov}. Estimating the parameters $d_S=11$\,nm, $T_{cS}=7$\,K, $\rho_S=7.5$\,$\mu\Omega$\,cm,
$\xi_S=8.9$\,nm, $\rho_F=60$\,$\mu\Omega$\,cm, $\xi_F=7.6$\,nm, $\gamma=0.15$ from the experiment \cite{2:private_com},
and fitting only $h$ and $\gamma_B$, we find good agreement between our theoretical predictions and the experimental
data.

The fitting procedure was the following: first, we determine $h\approx 130$\,K from the position of the minimum of
$T_c(d_F)$; second, we find $\gamma_B\approx 0.3$ from fitting the vertical position of the curve.

The deviation of our curve from the experimental points is small; it is most pronounced in the region of small $d_F$
corresponding to the initial decrease of $T_c$. This is not unexpected because, when $d_F$ is of the order of a few
nanometers, the thickness of the F film may vary significantly along the film (which is not taken into account in our
theory), and the thinnest films can even be formed by an array of islands rather than by continuous material. At the
same time, we emphasize that the minimum of $T_c$ takes place at $d_F\approx 5$\,nm, when with good accuracy the F layer
has uniform thickness.

\subsubsection{Various types of $T_c(d_F)$ behavior}

The experimental results discussed above represent only one possible type of $T_c(d_F)$ behavior. Now we address the
general case; we obtain different kinds of $T_c(d_F)$ curves depending on parameters of the bilayer.

\begin{figure}
 \centerline{\includegraphics[width=90mm]{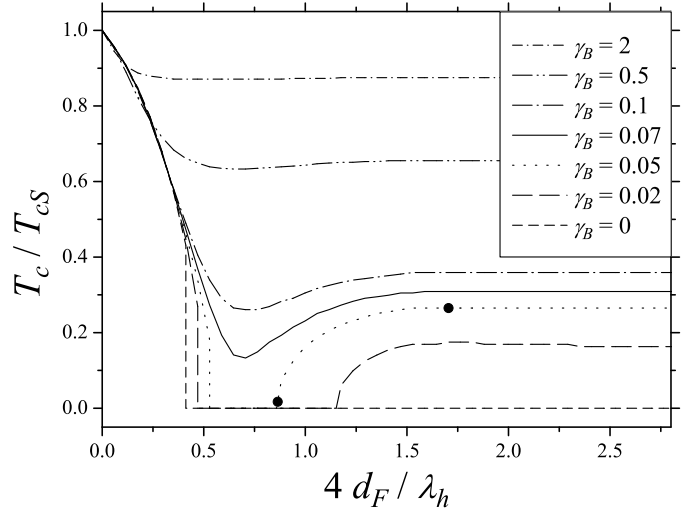}}
\caption{\label{fig:tc_sf} Characteristic types of $T_c(d_F)$ behavior. The thickness of the F layer is measured in
units of the wavelength $\lambda_h$ defined in Eq. (\ref{lambda_ex}). The curves correspond to different values of
$\gamma_B$. The exchange energy is $h=150$\,K; the other parameters are the same as in Fig.~\ref{fig:ryazanov}. One can
distinguish three characteristic types of $T_c(d_F)$ behavior \cite{2:Radovic,2:PKhI,2:Tagirov}: 1)~nonmonotonic decay
to a finite $T_c$ with a minimum at particular $d_F$ ($\gamma_{b}=2$; $0.5$; $0.1$; $0.07$), 2)~reentrant behavior
($\gamma_B=0.05$; $0.02$), 3)~monotonic decay to $T_c=0$ at finite $d_F$ ($\gamma_B=0$). The bold points indicate the
choice of parameter corresponding to Fig.~\ref{fig:Delta}.}
\end{figure}

To illustrate, in Fig.~\ref{fig:tc_sf} we plot several curves for various values of $\gamma_B$ [we recall that
$\gamma_B\propto R_B$, where $R_B$ is the resistance of the SF interface in the normal state
--- see Eq. (\ref{bound_2_})]. The exchange energy is $h=150$\,K; the other parameters are the
same as in Fig.~\ref{fig:ryazanov}.

We confirm three characteristic types of $T_c(d_F)$ behavior, found previously in limiting cases or by approximate
methods in Refs.~\cite{2:Radovic,2:PKhI,2:Tagirov}: 1)~at large enough interface resistance, $T_c$ decays
nonmonotonically to a finite value exhibiting a minimum at a particular $d_F$, 2)~at moderate interface resistance,
$T_c$ demonstrates the reentrant behavior: it vanishes in a certain interval of $d_F$, and is finite otherwise, 3)~at
low enough interface resistance, $T_c$ decays monotonically vanishing at finite $d_F$. A similar succession of
$T_c(d_F)$ curves as in Fig.~\ref{fig:tc_sf} can be obtained by tuning other parameters, e.g., the exchange energy $h$
or the normal resistances of the layers (the parameter $\gamma$).

A common feature seen from Fig.~\ref{fig:tc_sf} is saturation of $T_c$ at large $d_F \gtrsim \lambda_h$. This fact has a
simple physical explanation: the suppression of superconductivity by a dirty ferromagnet is only due to the effective F
layer with thickness on the order of $\lambda_h$, adjacent to the interface (this is the layer explored and ``felt'' by
quasiparticles entering from the S side due to the proximity effect).

\begin{figure}
 \centerline{\includegraphics[width=90mm]{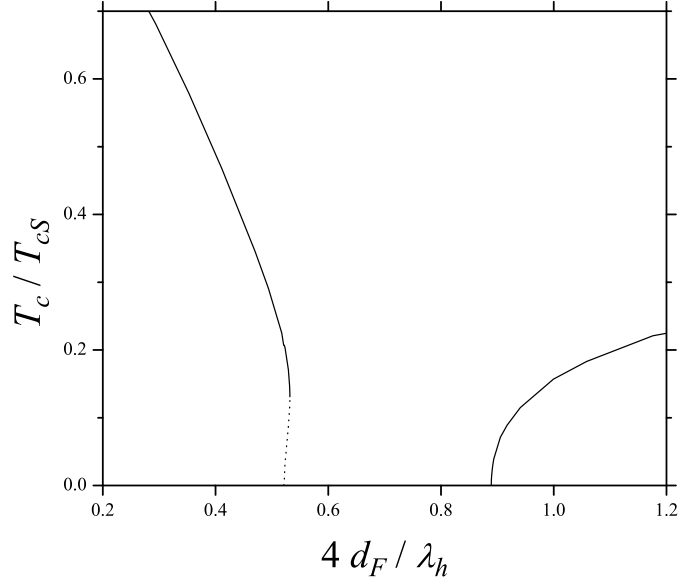}}
\caption{\label{fig:zagib} Change of the phase transition's order. This phenomenon manifests itself as discontinuity of
$T_c(d_F)$: the critical temperature jumps to zero abruptly without taking intermediate values. Formally, $T_c$ becomes
a double-valued function, but the smaller solution is physically unstable (dotted curve). For illustration we have
chosen the curve from Fig.~\ref{fig:tc_sf} corresponding to $\gamma_B=0.05$.}
\end{figure}

It was shown by Radovi\'{c} \textit{et al.} \cite{2:Radovic} that the order of the phase transition may change in
short-periodic SF superlattices, becoming the first order. We also observe this feature in the curves of types 2) and 3)
mentioned above. This phenomenon manifests itself as discontinuity of $T_c(d_F)$: the critical temperature jumps to zero
abruptly without taking intermediate values (see Figs.~\ref{fig:tc_sf}, \ref{fig:zagib}). Formally, $T_c$ becomes a
double-valued function, but the smaller solution is physically unstable (dotted curve in Fig.~\ref{fig:zagib}).

An interesting problem is determination of the tricritical point where the order of the phase transition changes. The
corresponding result for homogeneous bulk superconductors with internal exchange field was obtained a long time ago in
the framework of the Ginzburg--Landau theory \cite{2:Sarma}. However, the generalization to the case when the GL theory
is not valid has not yet been done. We note that the equations used in Refs.~\cite{2:Radovic,2:PKhI} were applied beyond
their applicability range because they are GL results valid only when $T_c$ is close to $T_{cS}$.

\subsubsection{Comparison between single- and multimode methods}

\begin{figure}
 \centerline{\includegraphics[width=90mm]{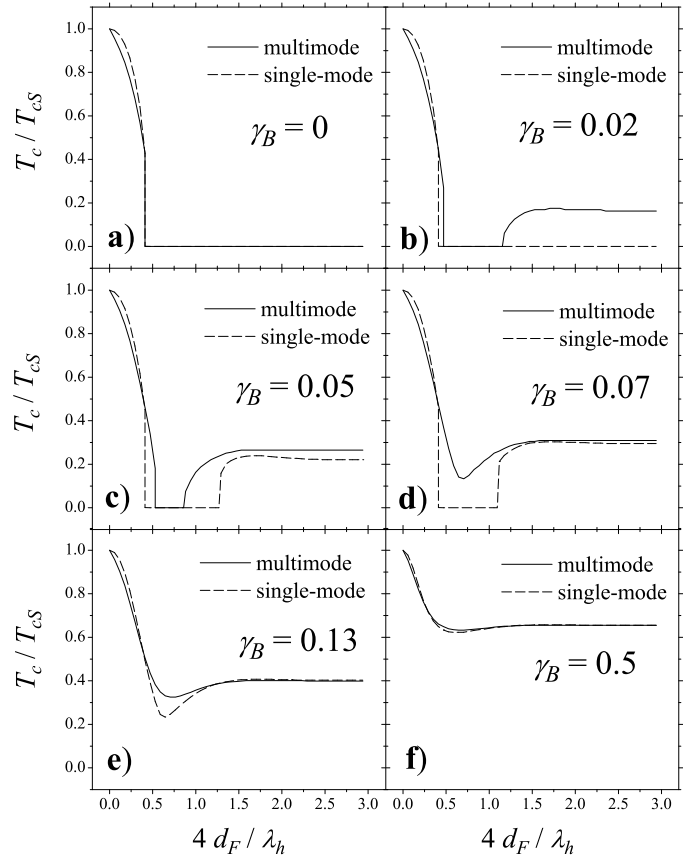}}
\caption{\label{fig:compare} Comparison between single- and multimode methods. The parameters are the same as in
Fig.~\ref{fig:tc_sf}. Generally speaking, the results of the single-mode and multimode (exact) methods are
quantitatively and even qualitatively different: b), c), d), and e). However, sometimes the results are close: a) and
f). Thus the single-mode approximation can be used for quick estimates, but reliable results should be obtained by one
of the exact (multimode or fundamental-solution) techniques.}
\end{figure}

A popular method widely used in the literature for calculating the critical temperature of SF bi- and multi-layers is
the single-mode approximation. The condition of its validity was formulated in Sec.~\ref{sec:multi-mode}. However, this
approximation is often used for arbitrary system's parameters. Using the methods developed in
Secs.~\ref{sec:multi-mode}, \ref{sec:fund_sol}, we can check the actual accuracy of the single-mode approximation. The
results are presented in Fig.~\ref{fig:compare}.

We conclude that although at some parameters the results of the single-mode and multimode (exact) methods are close
(Figs.~\ref{fig:compare} a,f), in the general case they are quantitatively and even qualitatively different
[Figs.~\ref{fig:compare} b,c,d,e --- these cases correspond to the most nontrivial $T_c(d_F)$ behavior]. Thus to obtain
reliable results one should use one of the exact (multimode or fundamental-solution) techniques.

\subsubsection{Spatial dependence of the order parameter}

According to the general theory of second-order phase transitions, the nonzero order parameter characterizes the
``ordered'' phase, in particular, the superconducting state. In the BCS theory, the role of the order parameter is
usually played by $\Delta$. However, this choice of the order parameter is not a good one when we consider proximity
systems, in which a superconductor contacts a normal metal or a ferromagnet. The order parameter $\Delta$, defined as
\begin{equation}
\Delta(x) = \lambda\, \pi T \sum_{\omega_n} F(x,\omega_n)
\end{equation}
is zero in the nonsuperconducting part of the structure, if the pairing constant $\lambda$ is zero. Nevertheless, the
superconducting correlations are induced into the nonsuperconducting metal, and these correlations are adequately
described not by $\Delta$ but by the anomalous Green function $F$.

Thus it is reasonable to define also a ``proximity order parameter'':
\begin{equation} \label{order_parameter}
F(x,\tau=0) = T \sum_{\omega_n} F(x,\omega_n),
\end{equation}
where $\tau$ denotes the imaginary time [in the S metal $F(x,\tau=0)\propto \Delta(x)$]. This function is real due to
the symmetry relation $F(-\omega_n)=F^*(\omega_n)$. The proximity effect in the SF bilayer is characterized by the
spatial behavior of this proximity order parameter.

\begin{figure}
 \centerline{\includegraphics[width=90mm]{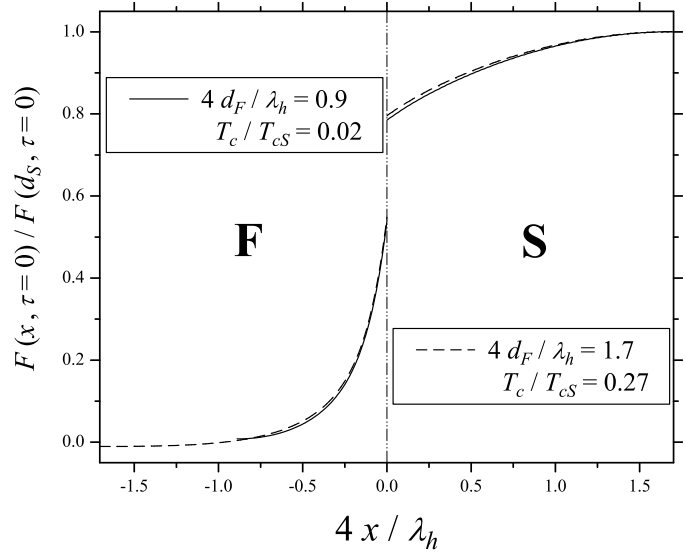}}
\caption{\label{fig:Delta} Spatial dependence of the proximity order parameter normalized by its value at the outer
surface of the S layer. Two cases are shown differing by the thickness of the F layer $d_F$ (and by the corresponding
$T_c$) at $\gamma_B=0.05$. The other parameters are the same as in Fig.~\ref{fig:tc_sf}, where the chosen cases are
indicated by the bold points. Although the critical temperatures differ by more than the order of magnitude, the
normalized proximity order parameters are very close to each other, which means that the value of $T_c$ has almost no
effect on the shape of $F(x,\tau=0)$. The jump at the SF interface is due to its finite resistance. With an increase of
$d_F$ the proximity order parameter starts to oscillate, changing its sign (this can be seen for the dotted curve,
although negative values of the proximity order parameter have very small amplitudes).}
\end{figure}
We illustrate this dependence in Fig.~\ref{fig:Delta}, which shows two cases differing by the thickness of the F layer
$d_F$ (and by the corresponding $T_c$). Although the critical temperatures differ by more than the order of magnitude,
the normalized proximity order parameters are very close to each other, which means that the value of $T_c$ has almost
no effect on the shape of $F(x,\tau=0)$. Details of the calculation are presented in Appendix~\ref{ap:sec:space dep
expl}.

Another feature seen from Fig.~\ref{fig:Delta} is that the proximity order parameter in the F layer changes its sign
when the thickness of the F layer increases (this feature can be seen for the dotted curve, although negative values of
the order parameter have very small amplitudes). We discuss this oscillating behavior in the next section.

\subsection{Discussion} \label{sec:discussion_}

\subsubsection{Qualitative explanation of the nonmonotonic $T_c(d_F)$ behavior}

A qualitative explanation of the nonmonotonic $T_c(d_F)$ behavior in SF bilayers has been presented in several papers
(see, e.g., Ref.~\cite{2:PKhI}). Below we present another interpretation.

\begin{figure}
 \centerline{\includegraphics[width=70mm]{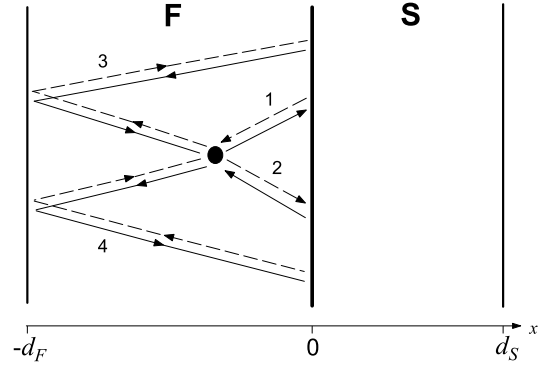}}
\caption{\label{fig:intrfrnc} Four types of trajectories contributing (in the sense of Feynman's path integral) to the
anomalous wave function of correlated quasiparticles in the ferromagnetic region. The solid lines correspond to
electrons, the dashed lines --- to holes; the arrows indicate the direction of the velocity.}
\end{figure}

The thickness of the F layer at which the minimum of $T_c(d_F)$ occurs, can be estimated from qualitative arguments
based on the interference of quasiparticles in the ferromagnet. Let us consider a point $x$ inside the F layer.
According to Feynman's interpretation of quantum mechanics \cite{2:Feynman}, the quasiparticle wave function may be
represented as a sum of wave amplitudes over all classical trajectories; the wave amplitude for a given trajectory is
equal to $\exp(iS)$, where $S$ is the classical action along this trajectory. We are interested in an \textit{anomalous}
wave function of correlated quasiparticles, which characterizes superconductivity; this function is equivalent to the
anomalous Green function $F(x)$. To obtain this wave function we must sum over trajectories that (i) start and end at
the point $x$, (ii) change the type of the quasiparticle (i.e., convert an electron into a hole, or vice versa). There
are four kinds of trajectories that should be taken into account (see Fig.~\ref{fig:intrfrnc}).
Two of them (denoted 1 and 2) start in the direction toward the SF interface (as an electron and as a hole), experience
the Andreev reflection, and return to the point $x$. The other two trajectories (denoted 3 and 4) start in the direction
away from the interface, experience normal reflection at the outer surface of the F layer, move toward the SF interface,
experience the Andreev reflection there, and finally return to the point $x$. The main contribution is given by the
trajectories normal to the interface. The corresponding actions are
\begin{gather}
S_1=-Qx-\alpha,\\
S_2=Qx-\alpha, \\
S_3=-Q(2d_F+x)-\alpha,\\
S_4=Q(2d_F+x)-\alpha
\end{gather}
(note that $x<0$), where $Q$ is the difference between the wave numbers of the electron and the hole, and
$\alpha=\arccos(E/\Delta)$ is the phase of the Andreev reflection. To make our arguments more clear, we assume that the
ferromagnet is strong, the SF interface is ideal, and consider the clean limit first: then
$Q=k_e-k_h=\sqrt{2m(E+h+E_\mathrm{F})}- \sqrt{2m(-E-h+E_\mathrm{F})}\approx 2h/v$, where $E$ is the quasiparticle
energy, $E_\mathrm{F}$ is the Fermi energy, and $v$ is the Fermi velocity. Thus the anomalous wave function of the
quasiparticles is
\begin{equation}
F(x)\propto \sum_{n=1}^4 \exp(iS_n)\propto \cos(Qd_F) \cos\left(Q[d_F+x] \right).
\end{equation}

The suppression of $T_c$ by the ferromagnet is determined by the value of the wave function at the SF interface:
$F(0)\propto \cos^2 (Qd_F)$. The minimum of $T_c$ corresponds to the minimum value of $F(0)$ which is achieved at
$d_F=\pi/2Q$. In the dirty limit the above expression for $Q$ is replaced by
\begin{equation} \label{lambda_ex}
Q = \sqrt{\frac{h}{D_F}}\equiv \frac{2\pi}{\lambda_h}
\end{equation}
(here we have defined the wavelength of the oscillations $\lambda_h$); hence the minimum of $T_c(d_F)$ takes place at
\begin{equation}
d_F^\mathrm{(min)} = \frac\pi 2 \sqrt{\frac{D_F}{h}} =\frac{\lambda_h}4.
\end{equation}
For the bilayer of Ref.~\cite{2:Ryazanov_JETPL} we obtain $d_F^\mathrm{(min)}\approx 7$\,nm, whereas the experimental
value is $5$\,nm (Fig.~\ref{fig:ryazanov}); thus our qualitative estimate is reasonable.

The arguments given above seem to yield not only the minimum but rather a succession of minima and maxima. However,
numerically we obtain either a single minimum or a minimum followed by a weak maximum (Fig.~\ref{fig:tc_sf}). The reason
for this is that actually the anomalous wave function not only oscillates in the ferromagnetic layer but also decays
exponentially, which makes the amplitude of the subsequent oscillations almost invisible.

Finally, we note that our arguments concerning oscillations of $F(x)$ also apply to a half-infinite ferromagnet, where
we should take into account only the trajectories 1 and 2 (see Fig.~\ref{fig:intrfrnc}). This yields
$F(x)\propto\cos(Qx)$ (another qualitative explanation of this result can be found, for example, in
Ref.~\cite{2:Demler}).

\subsubsection{Multilayered structures}

The methods developed and the results obtained in this chapter apply directly to more complicated symmetric multilayered
structures in the $0$-state such as SFS and FSF trilayers, SFIFS and FSISF systems (I denotes an arbitrary potential
barrier), and SF superlattices. In such systems an SF bilayer can be considered as a unit cell, and joining together the
solutions of the Usadel equations in each bilayer we obtain the solution for the whole system (for more details see
Sec.~\ref{sec:SNS}).

Our methods can be generalized to take account of possible superconductive and/or magnetic $\pi$-states (when $\Delta$
and/or $h$ may change their signs from layer to layer). In this case the system cannot be equivalently separated into a
set of bilayers. Mathematically, this means that the solutions of the Usadel equations lose their purely cosine form
[see Eqs. (\ref{F_F}), (\ref{Phi_sm}), (\ref{Delta_sm}), (\ref{Phi}), (\ref{Delta_}), (\ref{v_2})] acquiring a sine part
as well.

\subsection{Conclusions} \label{sec:conclusions}

In the present section, we have developed two methods for calculating the critical temperature of a SF bilayer as a
function of its parameters (the thicknesses and material parameters of the layers, the quality of the interface). The
multimode method is a generalization of the corresponding approach developed in Ref.~\cite{2:Golubov1} for SN systems.
However, the rigorous justification of this method is not clear. Therefore, we propose yet another approach --- the
method of fundamental solution, which is mathematically rigorous. The results demonstrate that the two methods are
equivalent; however, at low temperatures (compared to $T_{cS}$) the accuracy requirements are stricter for the multimode
method, and the method of fundamental solution is preferable. Comparing our method with experiment we obtain good
agreement.

In the general case, we confirm three characteristic types of the $T_c(d_F)$ behavior, found previously in limiting
cases or by approximate methods: 1)~nonmonotonic decay of $T_c$ to a finite value exhibiting a minimum at particular
$d_F$, 2)~reentrant behavior, characterized by vanishing of $T_c$ in a certain interval of $d_F$ and finite values
otherwise, 3)~monotonic decay of $T_c$ and vanishing at finite $d_F$. Qualitatively, the nonmonotonic behavior of
$T_c(d_F)$ is explained by interference of quasiparticles in the F layer, which can be either constructive or
destructive depending on the value of $d_F$.

Using the developed methods we have checked the accuracy of the widely used single-mode approximation. We conclude that
although at some parameters the results of the single-mode and exact methods are close, in the general case they are
quantitatively and even qualitatively different. Thus, to obtain reliable results one should use one of the exact
(multimode or fundamental-solution) techniques.

The spatial dependence of the order parameter (at the transition point) is shown to be almost insensitive to the value
of $T_c$.

The methods developed and the results obtained in this section, apply directly to more complicated symmetric
multilayered structures in the $0$-state such as SFS and FSF trilayers, SFIFS and FSISF systems, and SF superlattices.
Our methods can be generalized to take account of possible superconductive and/or magnetic $\pi$-states (when $\Delta$
and/or $h$ may change their signs from layer to layer).

In several limiting cases, $T_c$ is considered analytically.

\clearpage

\section{Triplet proximity effect in FSF trilayers}

\subsection{Introduction}

A striking feature of the proximity effect between singlet superconductors and inhomogeneous ferromagnets is the
possibility of generating the triplet superconducting component \cite{2:Bergeret,2:Kadigrobov}. Recently, it was shown
that the triplet component also arises in the case of several homogeneous but differently oriented ferromagnets
\cite{2:VBE}. Physically, the generating of the triplet component in SF systems \cite{2:Bergeret,2:Kadigrobov,2:VBE} is
similar to the case of magnetic superconductors \cite{2:JLTP}.

In Ref.~\cite{2:VBE}, the Josephson effect was studied having in mind that the superconductivity in the system is not
destroyed by the ferromagnets. However, this issue requires separate study.

Although the SF proximity effect is rather well studied, the influence of the \textit{mutual orientation} of F layers
magnetizations (exchange fields) on $T_c$ of layered SF structures has been mostly considered basing on the cases of
parallel (P) and antiparallel (AP) alignment \cite{2:Tagirov_PRL,2:Buzdin_EL,2:Khusainov,2:Buzdin_cm,2:Deutscher,2:ANL}.
At the same time, those are the only cases when the triplet component is absent.

A FSF trilayer with homogeneous but noncollinear magnetizations of the F layers is the simplest example of a layered
structure in which the triplet component is generated. The triplet component (superconducting correlations between
quasiparticles with parallel spins) arises due to a mechanism that is similar to the one described in
Ref.~\cite{2:Kadigrobov}, with the difference that instead of local magnetic inhomogeneity we deal with magnetic
inhomogeneity of the structure as a whole. This mechanism can be described in terms of the Andreev reflection at FS
interfaces. In the case of a single FS interface, the Andreev reflection of a spin-polarized electron impinging on the
FS interface from the F side, generates the singlet superconducting correlations in the ferromagnet. At the same time,
in the FSF trilayer, we must take into account that the Andreev reflection from the FS interface is nonlocal in space:
it takes place on a scale of the order of the coherence length in the vicinity of the interface. If the S layer is thin,
this process ``touches'' the second F layer. If its magnetization is noncollinear with the first one, it acts as a spin
splitter, inducing the opposite spin component and hence the triplet superconducting correlations between the input and
the output of the Andreev reflection.

The critical temperature of the noncollinear FSF system was studied in Ref.~\cite{2:Buzdin}. However, in that work the
triplet component was not taken into account. Thus calculation of $T_c$ in the noncollinear FSF trilayer is still an
open question.

\begin{figure}
 \centerline{\includegraphics[width=90mm]{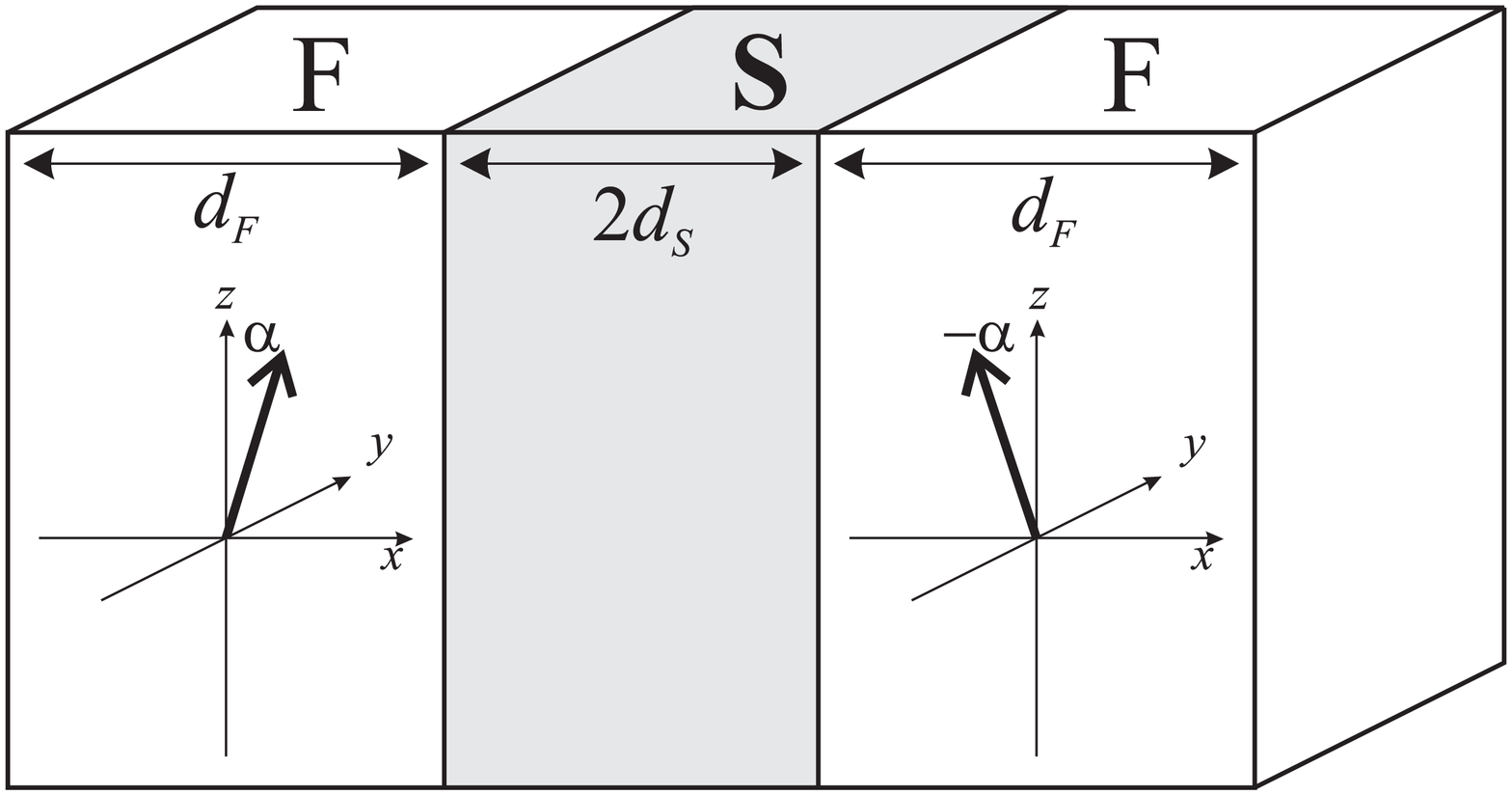}}
\caption{FSF trilayer. The system is the same as in Ref.~\cite{2:Buzdin}. The thickness of the S layer is $2d_S$, of
each F layer --- $d_F$. The center of the S layer corresponds to $x=0$. The thick arrows in the F layers denote the
exchange fields $\mathbf{h}$ lying in the $(y,z)$ plane. The angle between the in-plane exchange fields is $2\alpha$.}
 \label{fig:FSF}
\end{figure}

In this chapter we study the critical temperature of a FSF trilayer at arbitrary angle between the in-plane
magnetizations (see Fig.~\ref{fig:FSF}), which makes it necessary to take the triplet component into account.

\subsection{General description}

We consider the dirty limit, which is described by the Usadel equations. Near $T_c$, the Usadel equations are linearized
and contain only the anomalous Green function $\widehat F$ \cite{2:Bergeret}:
\begin{gather}
\frac D2 \frac{d^2 \widehat F}{dx^2} - |\omega_n| \widehat F +\Delta \widehat \sigma_3 -\frac i2
\sgn\omega_n \left( \widehat F \widehat H^* +\widehat H \widehat F \right) =0, \label{Usadel} \\
\widehat F = \begin{pmatrix} f_{\uparrow\downarrow} & f_{\uparrow\uparrow} \\ f_{\downarrow\downarrow} &
f_{\downarrow\uparrow} \end{pmatrix} . \notag
\end{gather}
Here $D$ is the diffusion constant ($D_S$ and $D_F$ for the S and F layers), $\omega_n=\pi T(2n+1)$ are the Matsubara
frequencies, and $\widehat \sigma_3$ is the third Pauli matrix. The function $\widehat F$ is a matrix in the spin space.
The $f_{\uparrow\uparrow}$ and $f_{\downarrow\downarrow}$ components describe the triplet superconducting correlations.
In the P and AP cases it is sufficient to consider only the scalar equation for the singlet component
$f_{\uparrow\downarrow}$.

Equation (\ref{Usadel}) is written in the general case when both order parameter and exchange field are present. In our
system, in the F layers the order parameter is absent, $\Delta=0$, while
\begin{equation}
\widehat H = h \left( \widehat \sigma_2 \sin\alpha + \widehat \sigma_3 \cos\alpha \right)
\end{equation}
at the exchange field $\mathbf{h} = h(0,\sin\alpha,\cos\alpha)$. $h$ is the exchange energy, and $\alpha$ describes the
direction of the in-plane magnetization.

In the S layer, the exchange energy is zero, while the superconducting order parameter obeys the self-consistency
equation
\begin{equation}
\Delta\ln\frac{T_{cS}}{T} = \pi T \sum_{\omega_n} \left( \frac\Delta{|\omega_n|} -f_{\uparrow\downarrow} \right),
\end{equation}
where $T_{cS}$ is the critical temperature of the S material. In the case of a single S layer, $\Delta$ can be chosen
real.

The boundary conditions at the outer surfaces of the trilayer are
\begin{equation} \label{vacuum}
\frac{d \widehat F_F}{dx} = 0,
\end{equation}
while at the SF interfaces
\begin{align}
& \xi_S \frac{d \widehat F_S}{dx} = \gamma \xi_F \frac{d \widehat F_F}{dx}, &&
\gamma = \frac{\rho_S \xi_S}{\rho_F \xi_F}, \label{bound_1} \\
& \pm \xi_F \gamma_B \frac{d \widehat F_F}{dx} = \widehat F_S -\widehat F_F, && \gamma_B = \frac{R_B \mathcal{A}}{\rho_F
\xi_F}. \label{bound_2}
\end{align}
Here $\xi_{S(F)}=\sqrt{D_{S(F)}/2\pi T_{cS}}$ and $\rho_{S(F)}$ are the coherence lengths and the normal state
resistivities of the S and F metals, $R_B$ is the total resistance of the SF boundary, and $\mathcal A$ is its area. The
$\pm$ sign in the l.h.s. of Eq. (\ref{bound_2}) refers to the left and right SF interface, respectively. The above
boundary conditions were derived for SN interfaces \cite{2:KL} (N is a normal metal); their use in the SF case is
justified by the small parameter $h/E_\mathrm{F} \ll 1$ ($E_\mathrm{F}$ is the Fermi energy).

Our strategy is to reduce the problem to the S layer only, with effective boundary conditions.

We expand the Green function $\widehat F$ in the basis of the Pauli matrices $\widehat \sigma_1$, $\widehat \sigma_2$,
$\widehat \sigma_3$, and the unit matrix $\widehat\sigma_0$. It can be shown that the solution has the form
\begin{equation}
\widehat F = f_0 \widehat\sigma_0 + f_1 \widehat\sigma_1 + f_3 \widehat\sigma_3.
\end{equation}
The $f_0$ component is imaginary, while $f_1$ and $f_3$ are real. The relations $f_0(-\omega_n)=-f_0(\omega_n)$,
$f_1(-\omega_n)=-f_1(\omega_n)$, $f_3(-\omega_n)=f_3(\omega_n)$ make it sufficient to consider only positive Matsubara
frequencies.

The $f_1$ component describes a special type of triplet condensate \cite{2:Bergeret,2:VBE}, odd in frequency
[$f_1(-\omega_n)=-f_1(\omega_n)$] and even in the relative momentum of electrons in the Cooper pair,\footnote{The
relative momentum of electrons in the Cooper pair should not be confused with the momentum corresponding to the $x$
coordinate. The latter describes the center of mass of the Cooper pair, while the relative momentum is not an argument
of the quasiclassical Green functions which are integrated over it.} which is similar to the one proposed by Berezinskii
\cite{2:Berezinskii}. It is independence on the momentum direction that allows the triplet condensate to survive in the
diffusive limit, in contrast to the standard (odd in momentum) case \cite{2:Larkin}. This Berezinskii-type triplet phase
is characterized by zero orbital momentum of the triplet Cooper pairs. The odd dependence on frequency implies
(spontaneous) breaking of the time-reversal symmetry.

Equation (\ref{Usadel}) yields three coupled scalar equations (we consider $\omega_n>0$):
\begin{gather}
\frac D2 \frac{d^2 f_0}{dx^2} - \omega_n f_0 -i h f_3 \cos\alpha =0, \notag \\
\frac D2 \frac{d^2 f_1}{dx^2} - \omega_n f_1 + h f_3 \sin\alpha =0, \label{sys_gen} \\
\frac D2 \frac{d^2 f_3}{dx^2} - \omega_n f_3 -i h f_0 \cos\alpha -h f_1 \sin\alpha +\Delta =0. \notag
\end{gather}
Analyzing symmetries implied by Eqs. (\ref{sys_gen}) and geometry of the system, we conclude that $f_0(x)=f_0(-x)$,
$f_1(x)=-f_1(-x)$, $f_3(x)=f_3(-x)$. Thus we can consider only one half of the system, say $x<0$, while the boundary
conditions at $x=0$ are
\begin{equation} \label{bound_components}
\frac{d f_0}{dx} = 0,\qquad f_1 =0,\qquad \frac{d f_3}{dx} =0.
\end{equation}

Below we shall use the following wave vectors:
\begin{gather}
k_F = \sqrt{2\omega_n / D_F},\qquad k_h = \sqrt{h / D_F},\qquad \widetilde k_h = \sqrt{k_F^2 + 2i k_h^2}, \notag \\
k_S = \sqrt{2\omega_n / D_S}. \label{k}
\end{gather}
The solution in the left F layer, satisfying the boundary condition (\ref{vacuum}), has the form
\begin{gather}
\widehat F_F = C_1 \left( i \widehat\sigma_0 \sin\alpha +\widehat\sigma_1 \cos\alpha
\right) \cosh\left[ k_F \left( x+d_S +d_F \right) \right] + \notag \\
+ C_2 \left( \widehat\sigma_0 \cos\alpha + i\widehat\sigma_1 \sin\alpha +\widehat\sigma_3
\right) \cosh\left[ \widetilde k_h \left( x+d_S +d_F \right) \right] + \notag \\
+ C_3 \left( \widehat\sigma_0\cos\alpha  + i\widehat\sigma_1 \sin\alpha -\widehat\sigma_3 \right) \cosh\left[ \widetilde
k_h^* \left( x+d_S +d_F \right) \right] .
\end{gather}
The matrix boundary condition (\ref{bound_2}) yields three scalar equations, which allow to express the coefficients
$C_1$, $C_2$, $C_3$ in terms of the components $f_0$, $f_1$, $f_3$ of the Green function on the S side of the FS
interface:
\begin{align}
C_1 &= \frac{ -i f_0 \sin\alpha+ f_1 \cos\alpha } { 1+\gamma_B A_F } , \notag \\
C_2 &= \frac{ f_0 \cos\alpha -i f_1 \sin\alpha +f_3 } { 2 \left( 1+\gamma_B A_h \right) }, \\
C_3 &= \frac{ f_0 \cos\alpha -i f_1 \sin\alpha -f_3 } { 2 \left( 1+\gamma_B A_h^* \right) }, \notag
\end{align}
where we have introduced the following notations:
\begin{gather}
A_F = k_F \xi_F \tanh (k_F d_F), \qquad
A_h = \widetilde k_h \xi_F \tanh (\widetilde k_h d_F), \notag \\
V_F = \frac{\gamma A_F}{1+\gamma_B A_F}, \qquad V_h = \frac{\gamma A_h}{1+\gamma_B A_h}. \label{AO}
\end{gather}
Then the boundary condition (\ref{bound_1}) yields
\begin{align}
\xi_S \frac{d f_0}{dx} &= f_0 \left( V_F \sin^2 \alpha + \Re V_h   \cos^2 \alpha \right) - \notag \\
&\phantom{=} -i f_1 \left( V -\Re V_h \right) \sin\alpha \cos\alpha + i f_3 \Im V_h \cos\alpha, \label{f_0} \\
\xi_S \frac{d f_1}{dx} &= i f_0 \left( V_F - \Re V_h \right) \sin\alpha \cos\alpha + \notag \\
&\phantom{=} +f_1 \left( V_F \cos^2 \alpha + \Re V_h \sin^2 \alpha \right) + f_3 \Im V_h \sin\alpha, \\
\xi_S \frac{d f_3}{dx} &= if_0\Im V_h \cos\alpha -f_1 \Im V_h \sin\alpha + f_3 \Re V_h. \label{f_3}
\end{align}
Thus the Green function of the F layer is eliminated, and we obtain equations for the S layer only. Moreover, we can
proceed further, because in the S layer the unknown function $\Delta(x)$ (which must be determined self-consistently)
only enters the equation for the $f_3$ component [the last of Eqs. (\ref{sys_gen})]. At the same time, taking boundary
conditions (\ref{bound_components}) into account, we can write $f_0 = B_0 \cosh (k_S x)$, $f_1 = B_1 \sinh (k_S x)$.
Excluding $B_0$ and $B_1$ from the boundary conditions (\ref{f_0})--(\ref{f_3}), we arrive at the effective boundary
condition for $f_3$:
\begin{equation}
\xi_S \frac{d f_3}{dx} = \mathcal{W} f_3,
\end{equation}
where
\begin{equation} \label{W1}
\mathcal{W} = \Re V_h + \frac{\left( \Im V_h \right)^2} {k_S \xi_S A(\alpha) +\Re V_h},
\end{equation}
and the angular dependence is determined by
\begin{equation}
\label{A} A = \frac{k_S \xi_S \tanh (k_S d_S) +V_F \left[ \sin^2 \alpha +\tanh^2 (k_S d_S) \cos^2 \alpha \right]} {k_S
\xi_S \left[ \cos^2 \alpha +\tanh^2 (k_S d_S) \sin^2 \alpha \right] +V_F \tanh (k_S d_S)}.
\end{equation}

Effectively, we obtain the following problem:
\begin{gather}
\Delta \ln \frac{T_{cS}}T = 2\pi T \sum_{\omega_n>0} \left( \frac\Delta{\omega_n} - f_3 \right), \label{1} \\
\frac{D_S}2 \frac{d^2 f_3}{dx^2} - \omega_n f_3 +\Delta =0, \label{2} \\
\xi_S \frac{d f_3(-d_S)}{dx} = \mathcal{W}(\omega_n) f_3(-d_S),\qquad \frac{d f_3(0)}{dx} =0 \label{3}
\end{gather}
All information about the F layers is contained in a single function $\mathcal{W}$, all information about the
misorientation angle
--- in its part $A(\alpha)$. Knowledge of $\mathcal{W}$ is already sufficient to draw several general conclusions about
the behavior of $T_c$. First, if the S layer is thick, i.e. $d_S \gg \xi_S$, then $\tanh (k_S d_S) \approx 1$ at
characteristic frequencies, and $T_c$ does not depend on $\alpha$. Qualitatively, this happens because the effect of
mutual orientation of the F layers is due to ``interaction'' between the two SF interfaces, which is efficient only in
the case of thin S layer. Second, $T_c$ does not depend on $d_F$ if $d_F \gg \xi_F$. Qualitatively, this is due to the
fact that the superconducting correlations penetrate from the S to F layer only on the scale $\xi_F$, and if $d_F$ is
even larger, it does not affect $T_c$.

The triplet component is ``nonmonotonic'' as a function of $\alpha$: it vanishes at $\alpha=0$ and $\alpha=\pi/2$ (P and
AP case, respectively), and arises only between the two boundary values. However, the $T_c(\alpha)$ dependence is always
monotonic ($T_c$ monotonically grows with increasing $\alpha$). It can be directly proven from the monotonic behavior of
$A(\alpha)$, and, hence, $\mathcal{W}$. This rigorously derived conclusion disproves the result obtained by the
approximate single-mode method in Ref.~\cite{2:Khusainov}, where it was claimed that $T_c$ in the AP configuration can
be smaller than in the P case.

\subsection{Numerical results in the general case}

At arbitrary parameters of the system, the effective problem (\ref{1})--(\ref{3}) can be solved numerically by the
methods developed in Secs. \ref{sec:multi-mode}, \ref{sec:fund_sol}.

\begin{figure}
 \centerline{\includegraphics[width=90mm]{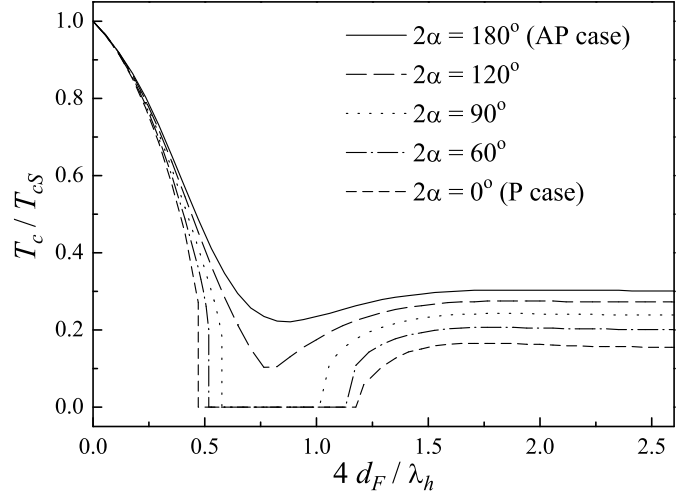}}
\caption{Critical temperature $T_c$ vs. thickness of the F layers $d_F$, which is normalized on the wavelength of the
singlet component oscillations $\lambda_h = 2\pi/k_h$. Parameters $d_S /\xi_S =1.2$, $h/\pi T_{cS} =6.8$, $\gamma=0.15$,
$\gamma_B=0.02$ are close to the experiment~\cite{2:Ryazanov_JETPL}. The curves are calculated at different angles
$2\alpha$ between the in-plane exchange fields in the F layers. A minimum of the $T_c(d_F)$ dependence and reentrant
behavior were obtained for collinear orientations in Refs.~\cite{2:Radovic,2:PKhI,2:Tagirov}.}
 \label{fig:Tc_dF}
\end{figure}

\begin{figure}
 \centerline{\includegraphics[width=90mm]{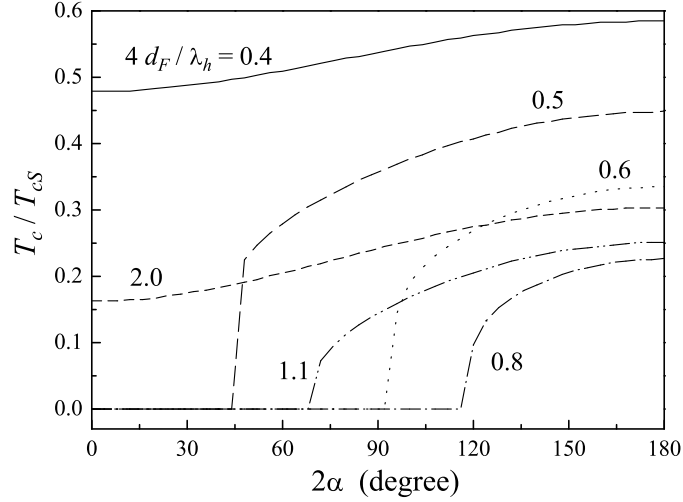}}
\caption{$T_c$ vs. misorientation angle $2\alpha$. The curves correspond to different thicknesses of the F layers $d_F$.
The parameters are the same as in Fig.~\ref{fig:Tc_dF}.}
 \label{fig:Tc_alpha}
\end{figure}

Numerical results obtained by the exact methods are shown in Figs.~\ref{fig:Tc_dF}, \ref{fig:Tc_alpha}. A question
arises: why is there pronounced angular dependence in the case $d_S> \xi_S$, when the S layer is not thin? The answer is
that the condition $d_S \ll \xi_S = \sqrt{D_S /2\pi T_{cS}}$ is a \textit{sufficient} condition of thin S layer, whereas
the \textit{necessary} condition is weaker: $d_S \ll \xi=\sqrt{D_S /2\pi T_c}$, since the characteristic energy for a
particular system is $\pi T_c$ with its own value of $T_c$. The two conditions become essentially different if $T_c$ is
notably suppressed, and in this case $T_c$ can exhibit pronounced angular dependence at $d_S \ll \xi$, while it is
possible to have $d_S > \xi_S$.

Experimentally, the conditions for observing the angular dependence of $T_c$ are more easily met when $T_c$ is
essentially (but not completely) suppressed. Accordingly, the effect of $\alpha$ on $T_c(d_F)$ dependence is most
pronounced near the reentrant behavior. Experimental detection of the reentrant behavior was reported in
Ref.~\cite{2:reentrant}.

\subsection{Analytical results for the case of thin S layer}

If $d_S \ll \xi_S$, then $\Delta$ is constant. The Usadel equation (\ref{2}) can be solved, and the equation determining
$T_c$ takes the form
\begin{equation} \label{ln_thin}
\ln \frac{T_{cS}}{T_c} = 2\pi T_c \sum_{\omega_n>0} \left( \frac 1{\omega_n} - \frac 1{\omega_n +\mathcal{W} \pi T_{cS}
\xi_S/d_S} \right),
\end{equation}
where $\mathcal{W}$ is given by Eq. (\ref{W1}) with simplified function $A(\alpha)$:
\begin{equation} \label{A_simple}
A = \frac{k_S^2 \xi_S d_S +V_F \left[ \sin^2 \alpha +(k_S d_S)^2 \cos^2 \alpha \right]} {k_S \xi_S \left[ \cos^2 \alpha
+(k_S d_S)^2 \sin^2 \alpha \right] +V_F k_S d_S}.
\end{equation}

For the P and AP alignments, under additional assumption of strong ferromagnetism ($h\gg \pi T_{cS}$), we obtain:
\begin{gather}
\ln\frac{T_{cS}}{T_c^P} = \Re \psi \left( \frac 12 +\frac{V_h}2 \frac{\xi_S}{d_S}
\frac{T_{cS}}{T_c^P} \right) - \psi \left( \frac 12 \right), \label{eq_parall} \\
\ln\frac{T_{cS}}{T_c^{AP}} = \psi \left( \frac 12 +\frac{\mathcal{W}}2 \frac{\xi_S}{d_S} \frac{T_{cS}}{T_c^{AP}} \right)
- \psi \left( \frac 12 \right), \label{eq_anti}
\end{gather}
where $\psi$ is the digamma function, $V_h$ is determined by Eqs. (\ref{AO}) with $\widetilde k_h =(1+i) k_h$, and in
the region of parameters, where $T_c \ne 0$ [the corresponding conditions can be extracted from the results for the
critical thickness --- see Eqs. (\ref{dscr}), (\ref{dscr_cond}) below], we may write
\begin{equation} \label{Wb}
\mathcal{W} = \Re V_h + \frac{d_S}{\xi_S} \left( \Im V_h \right)^2.
\end{equation}
Due to symmetry, the result for the P case (\ref{eq_parall}) reproduces that for the SF bilayer with S layer of
thickness $d_S$ [Eq. (\ref{thin_S})]. In the AP case, if the second terms in the r.h.s. of Eq. (\ref{Wb}) can be
neglected (e.g., at $k_h d_F \gg 1$ in the region of parameters where $T_c \ne 0$), then $\mathcal{W} = \Re V_h$ and we
reproduce the result of Ref.~\cite{2:Buzdin_cm}. However, the second term becomes essential in the Cooper limit, defined
by conditions $d_S \ll \sqrt{D_S / 2 \omega_D}$, $d_F \ll \min( \sqrt{D_F /2 \omega_D}, k_h^{-1} )$, $\gamma_B =0$, with
$\omega_D$ the Debye energy of the S material. In this case $\Re V_h=0$ and Eqs. (\ref{eq_anti}), (\ref{Wb}) reproduce
the result of Tagirov \cite{2:Tagirov_PRL}.

The critical thickness $d_{cS}$ of the S layer, below which the superconductivity vanishes, immediately follows from
Eqs. (\ref{eq_parall}), (\ref{eq_anti}) for the P and AP cases:\footnote{When calculating $d_{cS}$, we assume that the
phase transition is of the second order. However, this issue may require a separate study. In principle, under some
circumstances, the order of the phase transition in SF systems can change from second to first one
--- see Sec.~\ref{sec:num_res}.}
\begin{equation} \label{dscr}
\frac{d_{cS}^P}{\xi_S} = 2 \exp(C)\, \left| V_h \right|,\qquad \frac{d_{cS}^{AP}}{\xi_S} = 2 \exp(C)\, \mathcal{W}
\end{equation}
at
\begin{equation} \label{dscr_cond}
\frac{d_{cS}}{\xi_S} \ll 1.
\end{equation}
Here $C \approx 0.577$ is Euler's constant. Condition (\ref{dscr_cond}) is necessary for applicability of Eqs.
(\ref{dscr}). If this condition is not satisfied, then Eqs. (\ref{dscr}) only tell us that at $d_S /\xi_S \ll 1$ the
superconductivity is certainly absent, i.e., $T_c=0$. According to the monotonic growth of $T_c(\alpha)$, the function
$d_{cS} (\alpha)$ decreases monotonically, hence $d_{cS}^P > d_{cS}^{AP}$. At $\gamma_B=0$, $k_h d_F \gg 1$, Eqs.
(\ref{dscr}) reproduce the results of Ref.~\cite{2:Buzdin} for the P and AP cases.

The $T_c(\alpha)$ dependence can be most easily studied in the Cooper limit. In this case a simple analysis can be done
already on the level of the Usadel equations, and the system is described as a uniform layer with the effective exchange
energy\footnote{Since $\omega_n$ was neglected in comparison with $h$ in the Usadel equation, the result of the Cooper
limit is valid only at $\tau_S \gg \tau_F$.}
\begin{equation} \label{h_eff}
h_\mathrm{eff} = \frac{\tau_F}{\tau_S}\, h \cos\alpha ,
\end{equation}
where $\tau_{S(F)} = 2 d_{S(F)} R_B \mathcal{A} / \rho_{S(F)} D_{S(F)}$ (similarly to Eqs. (\ref{tau_original})). The
accuracy of this result is limited to the first order over $h$, which becomes insufficient in the vicinity of
$\alpha=\pi/2$. At $\alpha =\pi/2$, the first-order effect of $h$ vanishes, while a more accurate analysis
(Ref.~\cite{2:Tagirov_PRL} and Eqs. (\ref{eq_anti}), (\ref{Wb})) reveals the second-order effect of $h$ on $T_c$.

Let us now consider the same limit as in  Ref.~\cite{2:Buzdin}:
\begin{gather}
d_S \ll \xi_S,\quad k_h d_F \gg 1,\quad h \gg \pi T_{cS},\quad \gamma_B =0, \\
\gamma k_h \xi_F \frac{d_S}{\xi_S} \ll 1. \label{conds}
\end{gather}
The condition to have superconductivity at least at some orientations has the form $d_{cS}^{AP} < d_S \ll \xi_S$, and in
the case under discussion, Eqs. (\ref{dscr}), (\ref{dscr_cond}) yield:
\begin{equation} \label{cond_S}
2 \exp(C)\, \gamma k_h \xi_F < \frac{d_S}{\xi_S} \ll 1,
\end{equation}
hence condition (\ref{conds}) becomes redundant.

Starting from Eqs. (\ref{ln_thin}), (\ref{W1}), (\ref{A_simple}), we finally obtain the following equation for $T_c$:
\begin{equation} \label{ln_final}
\ln \frac{T_{cS}}{T_c} = Q\, \psi \left( \frac 12 +\frac{\Omega_1}{2\pi T_c} \right) +R\, \psi \left( \frac 12
+\frac{\Omega_2}{2\pi T_c} \right) -\psi \left( \frac 12 \right),
\end{equation}
where
\begin{gather}
Q = \frac 12 +\frac{\sin^2 \alpha}{2\sqrt{\sin^4 \alpha -4 \cos^2 \alpha}},\qquad R = 1-Q, \notag \\
\Omega_{1,2} = \frac{d_0}{d_S} \pi T_{cS} \left( 1+\cos^2 \alpha \pm \sqrt{\sin^4 \alpha -4 \cos^2 \alpha} \right),
\label{legend1} \\
d_0 = \gamma k_h \xi_F \xi_S /2. \notag
\end{gather}
In the P and AP cases, where the triplet component is absent, Eqs. (\ref{ln_final}), (\ref{legend1}) reproduce the
results of Refs.~\cite{2:Buzdin_EL,2:Buzdin}. At the same time, at a noncollinear alignment the results are clearly
different.

\begin{figure}
 \centerline{\includegraphics[width=90mm]{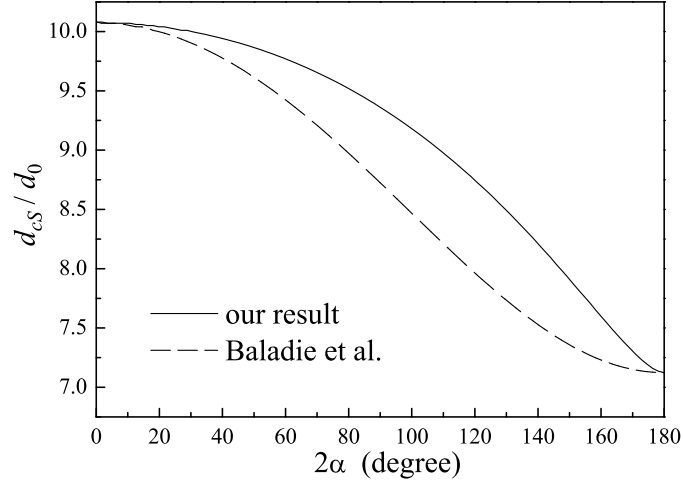}}
\caption{Critical thickness of the S layer $d_{cS}$ vs. misorientation angle $2\alpha$. Dashed line is the result of
Baladi\'{e} et al.~\cite{2:Buzdin}, obtained without account of the triplet component.}
 \label{fig:d_sc}
\end{figure}

The critical thickness is found from Eqs. (\ref{ln_final}), (\ref{legend1}):
\begin{equation} \label{dscr_alpha}
\frac{d_{cS} (\alpha)}{d_0} = 4\sqrt{2} \exp(C)\, \cos\alpha \left( \frac{1+\cos^2 \alpha + \sqrt{\sin^4 \alpha -4
\cos^2 \alpha}} {1+\cos^2 \alpha - \sqrt{\sin^4 \alpha -4 \cos^2 \alpha}} \right)^ {\textstyle\frac{\sin^2 \alpha} {2
\sqrt{\sin^4 \alpha -4 \cos^2 \alpha}}}.
\end{equation}
Although the square root in this expression can become imaginary, the whole expression remains real ($z^i$ is real if
$|z|=1$). Figure \ref{fig:d_sc} illustrates the result (\ref{dscr_alpha}).

\subsection{Conditions of existence of the odd triplet superconductivity in SF multilayers}

Now we turn to analyze the conditions of applicability for the results reported in Ref.~\cite{2:VBE}. A noncollinear FSF
trilayer is a unit cell of the multilayered structure studied in that work. The main result of Ref.~\cite{2:VBE}, the
Josephson current due to the long-range triplet component, requires that the S layer is thin $d_S \ll \xi_S$, while the
F layers are thick for the singlet component and moderate for the triplet one: $k_h^{-1} \ll \xi_F < d_F$ \cite{2:VBE}.
In this case the condition that superconductivity is not completely suppressed at least in the vicinity of the AP
alignment [Eqs. (\ref{dscr}), (\ref{dscr_cond})] takes the form
\begin{equation} \label{cond_S1}
4 \exp(C)\, \gamma k_h \xi_F \frac{1+2 \gamma_B k_h \xi_F}{(1+2 \gamma_B k_h \xi_F)^2 +1} < \frac{d_S}{\xi_S} \ll 1.
\end{equation}
At $\gamma_B=0$ (as it was assumed in Ref.~\cite{2:VBE}), this yields
\begin{equation}
2 \exp(C)\, \gamma k_h \xi_F < \frac{d_S}{\xi_S} \ll 1,
\end{equation}
which is a rather strong condition for $\gamma$, since $k_h \xi_F \gg 1$. Finite interface resistance relaxes this
condition: already at $\gamma_B \gtrsim 1$, Eq. (\ref{cond_S1}) yields
\begin{equation}
2 \exp(C)\, \frac\gamma{\gamma_B} < \frac{d_S}{\xi_S} \ll 1.
\end{equation}

The condition that superconductivity exists at all orientations has the form similar to Eq. (\ref{cond_S1}) but with the
corresponding expression for $d_{cS}^P$ instead of $d_{cS}^{AP}$ in the l.h.s. This only leads to a minor difference,
since the two critical thicknesses are of the same order: $d_{cS}^P = \sqrt{2}\, d_{cS}^{AP}$ at $\gamma_B =0$, while
$d_{cS}^P = d_{cS}^{AP}$ at $\gamma_B > 1$.

\subsection{Conclusions}

In this section, we have studied $T_c$ of a FSF trilayer as a function of its parameters, in particular, the angle
$2\alpha$ between magnetizations of the F layers. At noncollinear orientation of magnetizations, we take into account
the triplet superconducting component generated in the system due to nonlocality of the Andreev reflection.

The $T_c(\alpha)$ dependence becomes pronounced when the S layer is thin, and can lead to switching between
superconducting and non-superconducting states as the angle is varied. In the general case, we reduce the problem to the
form that allows us to employ the exact numerical methods of Sec.~\ref{sec:tcSF}. In the most interesting limiting
cases, we analytically analyze $T_c$ and the critical thickness of the S layer.

Our results directly apply to multilayered SF structures, where a FSF trilayer is a unit cell. We have formulated the
conditions which are necessary for existence of recently investigated odd triplet superconductivity in SF multilayers
\cite{2:VBE}. The conditions are severe in the case of transparent interfaces, but become realizable at moderate
interface transparency.

\clearpage
\renewcommand{\@evenhead}{}

\chapter[Josephson effect in SFS junctions]{\huge Josephson effect in SFS junctions}

\renewcommand{\@evenhead}
    {\raisebox{0pt}[\headheight][0pt]
     {\vbox{\hbox to\textwidth{\thepage \hfil \strut \textit{Chapter \thechapter}}\hrule}}
    }

\renewcommand{\@oddhead}
    {\raisebox{0pt}[\headheight][0pt]
     {\vbox{\hbox to\textwidth{\textit{\leftmark} \strut \hfil \thepage}\hrule}}
    }

\begin{figure}
 \centerline{\includegraphics[width=70mm]{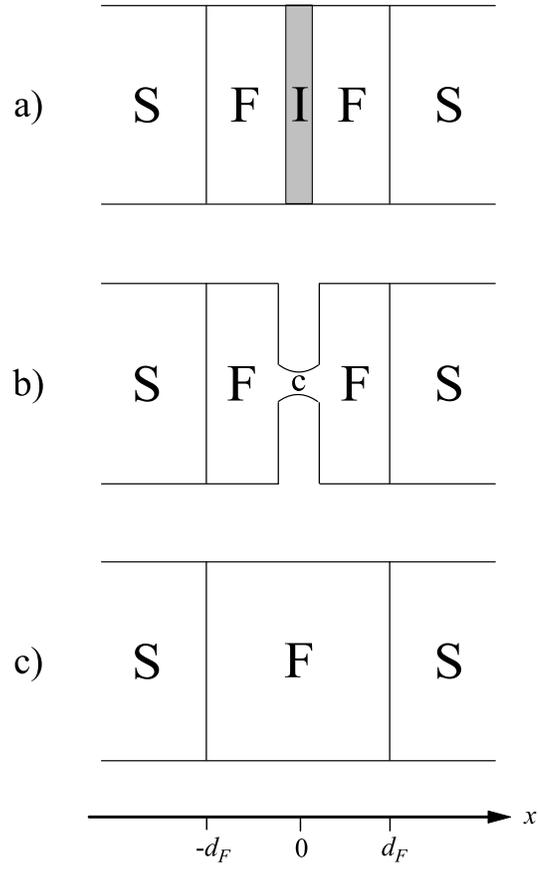}}
\caption{\label{fig:sketch} Types of SFS systems considered in the present chapter.}
\end{figure}

\section{Critical current in SFIFS junctions}
\label{sec:SFIFS}

\subsection{Introduction}

Josephson structures involving ferromagnets as weak link material are presently a subject of intensive study. The
possibility of the so-called ``$\pi$-state'', characterized by the negative sign of the critical current $I_c$, was
theoretically predicted in SFS Josephson junctions
\cite{3:Bulaevski,3:Buzdin_pi,3:Buzdin1,3:Buzdin2,3:Tanaka,3:KK_p,3:Dobro,3:Yip,3:Wilhelm,3:Fogelstrom,3:Radovic01,
3:Chtch,3:Barash_Bobkova}. The first experimental observation of the crossover from $0$- to $\pi$-state was reported by
Ryazanov \textit{et al.} \cite{3:Ryazanov} and explained in terms of temperature-dependent spatial oscillations of
induced superconducting ordering in the diffusive F layer.

More recently a number of new phenomena was predicted in junctions with more than one magnetically ordered layer. First,
the crossover to the $\pi$-state was predicted in Ref.~\cite{3:KK_p} for the case of parallel magnetizations in SFIFS
Josephson junctions even in the absence of the order parameter oscillations in thin F layers. Second, the possibility of
the critical current enhancement by the exchange field in SFIFS junctions with thin F layers and antiparallel
magnetization directions was discussed in the regimes of small S layer thicknesses \cite{3:BVE} and bulk S electrodes
\cite{3:KK_a}. However, in the case of thin S layers, Ref.~\cite{3:BVE} dealt with an idealized model in the tunneling
limit, which lead to a divergency of the critical current at zero temperature; moreover, the physical mechanisms of the
discovered enhancement effect were unclear. At the same time, for the bulk S case an approximate method was used in
Refs.~\cite{3:KK_p,3:KK_a} and a part of the results was obtained beyond its applicability range (this will be shown
below
--- see discussion of Fig.~\ref{fig:fig1}).

The above intriguing scenario motivated us to attack the problem of the Josephson effect in SFIFS junctions by
self-consistent solution of the Usadel equations for arbitrary thicknesses of the F layers and arbitrary barrier
transparencies. Below we show that the $0$--$\pi$ transition in the case of parallel orientation of the exchange fields
$h$ or enhancement of $I_c$ by $h$ in the antiparallel case with thin F layers occur when the effective energy shift in
the ferromagnets (due to the exchange field) becomes equal to a local value of the effective energy gap induced into the
F layers. Under this condition a peak in the local density of states (DoS) near the SF interfaces is shifted to zero
energy. In the models with DoS of the BCS type this leads to a logarithmic divergency of $I_c$ in the antiparallel case
at zero temperature, similarly to the well known Riedel singularity of the \textit{ac} supercurrent in SIS tunnel
junctions at voltage $eV=2\Delta$. We also describe the general numerical method to solve the problem self-consistently
and apply it for quantitative description of the $0$--$\pi$ transition and $I_c$ enhancement in SFIFS junctions.

\subsection{Model}

We consider the structure of SFIFS type, where I is an insulating barrier of arbitrary strength (see
Fig.~\ref{fig:sketch}a). We assume that the S layers are bulk and that the dirty limit conditions are fulfilled in the S
and F metals. Although our method is applicable in the general situation of different ferromagnets and superconductors,
for simplicity below we illustrate our results in the case when equivalent S and F materials are used on both sides of
the structure (although the directions of the exchange field in the two F layers may be different), both F layers have
the thickness $d_F$, and the two SF interfaces have the same transparency. At the same time, we do not put any
limitations on $d_F$ and the transparency.

The Usadel functions $G$, $F$, $\bar F$ obey the normalization condition $G^2 + F \bar F =1$, which allows the following
parameterizations in terms of the new function $\Phi$:
\begin{gather}
G(\omega_n)= \frac{\widetilde\omega_n} {\sqrt{\widetilde\omega_n^2+\Phi(\omega_n) \Phi^*(-\omega_n)}},\qquad F(\omega_n)
=\frac{\Phi(\omega_n)} {\sqrt{\widetilde\omega_n^2+\Phi(\omega_n)
\Phi^*(-\omega_n)}}, \label{def_f} \\
\bar F(\omega_n) = F^* (-\omega_n). \notag
\end{gather}
The quantity $\widetilde\omega_n=\omega_n+ih$ corresponds to the general case when the exchange field $h$ is present.
However, in the S layers $h=0$ and we have simply $\widetilde\omega_n=\omega_n$.

We choose the $x$ axis perpendicular to the plane of the interfaces with the origin at the barrier I. The Usadel
equations \cite{3:Usadel} in the S and F layers have the form
\begin{gather}
\xi_S^2 \frac{\pi T_c}{\omega_n G_S} \frac\partial{\partial x} \left[ G_S^2 \frac\partial
{\partial x} \Phi_S \right] -\Phi_S =-\Delta,  \label{EqUS}\\
\xi_F^2 \frac{\pi T_c}{\widetilde\omega_n G_F} \frac\partial{\partial x} \left[ G_F^2 \frac\partial {\partial x} \Phi_F
\right] -\Phi_F =0, \label{EqUSF}
\end{gather}
where $T_c$ is the critical temperature of the superconductors, $\Delta$ is the order parameter (which is nonzero only
in the S layers), $\omega_n$ is the Matsubara frequency, and the coherence lengths $\xi$ are related to the diffusion
constants $D$ as $\xi_{S(F)}=\sqrt{D_{S(F)}/2\pi T_c}$. The order parameter satisfies the self-consistency equations
\begin{equation}
\Delta \ln \frac T{T_c} +\pi T\sum_{\omega_n} \frac{\Delta -G_S \Phi_S \sgn\omega_n}{|\omega_n|} =0. \label{EqDEL}
\end{equation}
We restrict ourselves to the cases of parallel and antiparallel orientations of the exchange fields $h$ in the
ferromagnets.

The boundary conditions at the SF interfaces ($x=\mp d_F$) have the form \cite{3:KL} (see Ref.~\cite{3:Koshina2} for
details)
\begin{gather}
\frac{\xi_S G_S^2}{\omega_n} \frac\partial{\partial x} \Phi_S =\gamma \frac{\xi_F G_F^2}
{\widetilde\omega_n} \frac\partial{\partial x} \Phi_F, \label{BC_Fi2} \\
\pm\gamma_B \frac{\xi_F G_F}{\widetilde\omega_n} \frac\partial{\partial x} \Phi_F = G_S \left(
\frac{\Phi_F}{\widetilde\omega_n} -\frac{\Phi_S}{\omega_n} \right), \label{BC_Fi1} \\
\text{with}\quad \gamma_B= \frac{R_B \mathcal{A}}{\rho_F \xi_F},\quad \gamma= \frac{\rho_S \xi_S}{\rho_F \xi_F}, \notag
\end{gather}
where $R_B$ and $\mathcal{A}$ are the resistance and the area of the SF interfaces; $\rho_{S(F)}$ is the resistivity of
the S (F) layer. At the I interface ($x=0$) the boundary conditions read
\begin{gather}
\frac{G_{F1}^2}{\widetilde\omega_{n1}} \frac\partial{\partial x} \Phi_{F1} =
\frac{G_{F2}^2}{\widetilde\omega_{n2}} \frac\partial{\partial x} \Phi_{F2}, \label{BI_1} \\
\gamma_{B,I} \frac{\xi_F G_{F1}}{\widetilde\omega_{n1}} \frac\partial{\partial x} \Phi_{F1} = G_{F2} \left(
\frac{\Phi_{F2}}{\widetilde\omega_{n2}} -\frac{\Phi_{F1}}{\widetilde\omega_{n1}} \right),
\label{BI_2} \\
\text{with}\quad \gamma_{B,I}= \frac{R_{B,I} \mathcal{A}}{\rho_F \xi_F}, \notag
\end{gather}
where the indices $1,2$ refer to the left and right hand side of the I interface, respectively.

In the bulk of the S electrodes we assume a uniform current-carrying superconducting state
\begin{equation}
\Phi(x=\mp \infty) =\frac{\Delta_0 \exp\left(i[\mp \varphi /2+ 2m v_s x]\right)}{1+2D_S m^2 v_s^2
/\sqrt{\omega_n^2+|\Phi|^2}}, \label{BC_bulk}
\end{equation}
where $m$ is the electron's mass, $v_s$ is the superfluid velocity, and $\varphi$ is the phase difference across the
junction.

The supercurrent density is constant across the system. In the F part it is given by the expression
\begin{equation}
J=\frac{i\pi T}{2e\rho}\sum_{\omega_n} \frac{G^2(\omega_n)}{\widetilde\omega_n^2}\left[ \Phi_{\omega_n}
\frac\partial{\partial x} \Phi_{-\omega_n}^* -\Phi_{-\omega_n}^* \frac\partial{\partial x} \Phi_{\omega_n} \right],
\label{cur}
\end{equation}
while analogous formula for the S part is obtained if we substitute $\widetilde\omega_n\to \omega_n$. This expression,
together with the boundary condition (\ref{BI_2}) and the symmetry relation $F(-\omega_n,h)=F(\omega_n,-h)$, yields the
formula for the supercurrent across the I interface:
\begin{equation} \label{cur1}
I = \frac{\pi T}{e R_{B,I}} \sum_{\omega_n} \Im \left[ F_{F1}^*(-h_1) F_{F2} (h_2) \right]
\end{equation}
[the functions $F$ are related to $\Phi$ via Eq. (\ref{def_f})].

\subsection{Limit of thin F layers}

Let us consider the limit of thin F layers: $d_F \ll \min(\xi_F,\sqrt{D_F/2h})$. Under the condition $\gamma_B/\gamma\gg
1$ we can neglect the suppression of superconductivity in the superconductors. We assume further that the transparency
of the I barrier is small, $\gamma_{B,I}\gg\max(1,\gamma_B)$, hence the SF bilayers are decoupled. In this case we can
set $v_s=0$ and expand the solution of Eq. (\ref{EqUSF}) in the F layers up to the second order in small spatial
gradients. Applying the boundary condition (\ref{BC_Fi1}), we obtain the solution in the form similar to that in SN
bilayer \cite{3:Gol1,3:Koshina2}:
\begin{gather}
\Phi_{F1,F2} = \frac{\widetilde\omega_{n 1,2} /\omega_n}{1+\gamma_{BM} \widetilde\omega_{n 1,2} / \pi
T_c G_S} \Delta_0 \exp(\mp i\varphi/2),  \label{Sol2}\\
\text{with}\quad \gamma_{BM}=\gamma_{B} \frac{d_F}{\xi_F},\quad G_S=\frac{\omega_n}{\sqrt{\omega_n^2+\Delta_0^2}}.\notag
\end{gather}
This solution corresponds to the $\gamma_M=0$ limit of Refs.~\cite{3:Koshina2,3:KK_p,3:KK_a}, therefore the analytical
results (\ref{CurPF})--(\ref{nuF0}) presented below are explicitly or implicitly contained in those works. We employ
them to discover the physical mechanisms leading to the effects of the $0$--$\pi$ transition and the enhancement of the
critical current by the exchange field.

Substituting Eq. (\ref{Sol2}) into the expression for the supercurrent (\ref{cur1}) we obtain $I(\varphi)=I_c
\sin\varphi$. For the parallel orientation of the exchange fields, $h_1=h_2=h$, the critical current is
\begin{equation} \label{CurPF}
I_c^{(p)}=\frac{2\pi T}{eR_{B,I}} \sum_{\omega_n >0} \frac{\Delta_0^2}\Omega \cdot \frac A{A^2+B^2},
\end{equation}
while for the antiparallel orientation, $h_1=-h_2=h$, we obtain
\begin{equation} \label{CurAF}
I_c^{(a)}=\frac{2\pi T}{eR_{B,I}} \sum_{\omega_n >0} \frac{\Delta_0^2}\Omega \cdot \frac 1{\sqrt{A^2+B^2}}.
\end{equation}
Here we have introduced the following notations:
\begin{gather}
A = (\pi T_c)^2 \Omega - \left( \gamma_{BM} h \right)^2 \Omega + \gamma_{BM} \omega_n^2
\left( 2 + \gamma_{BM} \Omega \right), \\
B= 2 \omega_n h \gamma_{BM} \Omega \left( 1+\gamma_{BM} \Omega \right), \label{B} \\
\Omega = \frac{\sqrt{\omega_n^2 +\Delta_0^2}}{\pi T_c}. \notag
\end{gather}

At $h / \pi T_c =1/\gamma_{BM}$ and small $\omega_n$, the expression under the sum in Eq. (\ref{CurAF}) behaves as
$1/\omega_n$, thus at low $T$ the critical current diverges logarithmically: $I_c^{(a)} \propto \ln(T_c/T)$. This effect
was pointed out in Refs.~\cite{3:BVE,3:KK_a}.

The above results become physically transparent in the real energy $E$ representation. Performing the analytical
continuation in Eqs. (\ref{def_f}), (\ref{Sol2}) by replacement $\omega_n \rightarrow -iE$, we obtain the expression for
the DoS per one spin projection (spin ``up'') $\nu_F(E)=\Re G_F(E)$ in the F layers:
\begin{gather}
\nu_F(E)= \left| \Re \frac{\widetilde E}{\sqrt{\widetilde E^2-\Delta_0^2}} \right|, \label{DENS}\\
\widetilde E =E +\gamma_{BM}(E -h)\frac{\sqrt{\Delta_0^2-E^2}}{\pi T_c}, \notag
\end{gather}
which demonstrates the energy renormalization due to the exchange field. Equation (\ref{DENS}) yields
\begin{equation} \label{nuF0}
\nu_F(0)=\Re \frac{\gamma_{BM} h/ \pi T_c}{\sqrt{(\gamma_{BM} h/ \pi T_c)^2-1}},
\end{equation}
which shows that at $h / \pi T_c =1/\gamma_{BM}$ the singularity in the DoS is shifted to the Fermi level. Exactly at
this value of $h / \pi T_c$ the maximum of $I_c^{(a)}$ is achieved due to the overlap of two $E^{-1/2}$ singularities.
This leads to logarithmic divergency of the critical current (\ref{CurAF}) in the limit $T\to 0$, similarly to the well
known Riedel singularity of nonstationary supercurrent in SIS tunnel junctions at voltage $eV=2\Delta_0$, where the
energy shift is due to the electric potential. At the same value of the exchange field $h / \pi T_c =1/\gamma_{BM}$ the
critical current changes its sign (i.e., the crossover from $0$ to $\pi$ contact occurs) for parallel magnetizations of
the F layers [see Eq. (\ref{CurPF})].

We emphasize that the scenario of the $0$--$\pi$ transition in our case differs from those studied before where the
$\pi$-shift of the phase was either due to spatial oscillations of the order parameter in F layers (see, e.g.,
Ref.~\cite{3:Ryazanov} and references therein) or due to the proximity-induced phase rotation in S layers \cite{3:KK_p}.
In our case the phase does not change in either layer; instead, it jumps at the SF interfaces. This scenario is most
clearly illustrated in the limit of large $h$ where Eqs. (\ref{def_f}), (\ref{Sol2}) yield $F_F \propto -i\Delta\sgn h $
whereas $F_S \propto \Delta$; thus the phase jumps by $\pi/2$ at each of the SF interfaces, providing the total
$\pi$-shift between $F_{F1}(-h)$ and $F_{F2}(h)$ [it is the phase difference between these two functions that determines
the supercurrent according to Eq. (\ref{cur1})]. Physically, these jumps are not a property of the SF interfaces as
such. Indeed the condition for the $0$--$\pi$ transition is determined by the $\gamma_{BM}$ parameter, which depends on
\textit{both} the interface transparency and the thickness of the F layer. Probably, the $\pi/2$ phase is acquired due
to multiple passage of quasiparticles, reflected at the SF interface, through the F layer (while only a very small phase
is acquired at a single passage).

The considered effects take place only for sufficiently low I-barrier transparency. Indeed, it follows from Eq.
(\ref{Sol2}) that $G_F (\omega_n)\propto 1/\sqrt{\omega_n}$ for small $\omega_n$ under condition $h / \pi T_c
=1/\gamma_{BM}$. As a result, the boundary condition (\ref{BI_2}) yields that at
\begin{equation} \label{condT}
\frac{\omega_n}{\pi T_c} \leqslant \min \left( \frac {\xi_F} {d_F\gamma_{B,I}},\; \frac{\gamma_B}{\gamma_{B,I}} \right)
\end{equation}
the solutions (\ref{Sol2}) are not valid, since in this frequency range the effective transparency of the I interface
(the parameter $G_{F1} G_{F2} / \gamma_{B,I}$) increases and the spatial gradients in the F layers become large,
similarly to the Kulik--Omelyanchuk case \cite{3:KO1,3:Likharev}. At these frequencies, the Green functions $G$, $F$
(and hence the contribution to the critical current from these frequencies) are $h$-independent. As a result, the
barrier transparency parameter $\gamma_{B,I}$ provides the cutoff of the low-temperature logarithmic singularity of
$I_c^{(a)}$ at $h / \pi T_c =1/\gamma_{BM}$ [see Eq. (\ref{CurAF})]. According to Eq. (\ref{condT}), the critical
current saturates at low temperature $T^*=T_c\min ( \xi_F/d_F\gamma_{B,I},\; \gamma_B/\gamma_{B,I} )$. We note that any
asymmetry in the SFIFS junction will also lead to the cutoff of $I_c^{(a)}$ divergency. The above estimates are done for
the case of low barrier transparency, $\xi_F/d_F\gamma_{B,I}\ll1$ and $\gamma_B/\gamma_{B,I}\ll1$. The opposite regime
of high transparency requires separate study.

\subsection{General case}

For arbitrary F-layer thicknesses and the interface parameters, we solved the boundary problem
(\ref{def_f})--(\ref{BC_bulk}) numerically using an iterative procedure. Starting from trial values of the complex order
parameters $\Delta$ and the Green functions $G_{S,F}$ we solve the resulting linear equations and the boundary
conditions for the functions $\Phi_{S,F}$. After that we recalculate $G_{S,F}$ and $\Delta$. Then we repeat the
iterations until convergency is reached. The self-consistency of calculations is checked by the condition of
conservation of the supercurrent (\ref{cur}) across the junction. We emphasize that our method is \textit{fully}
self-consistent: in particular, it includes the self-consistency over the superfluid velocity $v_s$, which is essential
(contrary to the constriction case) in the quasi-one-dimensional geometry.

\begin{figure}
 \centerline{\includegraphics[width=90mm]{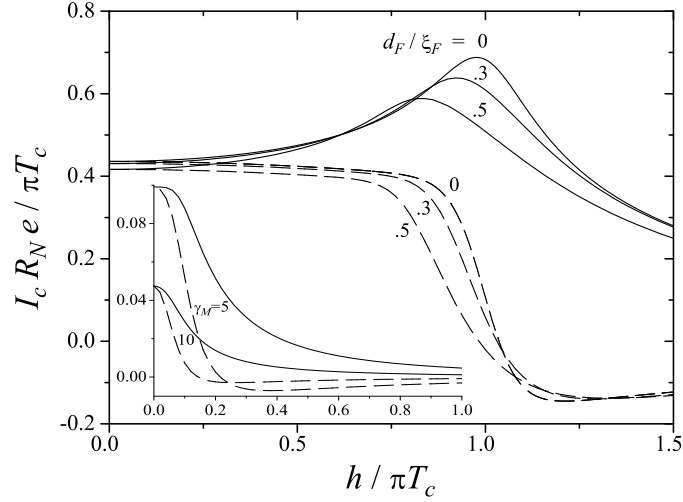}}
\caption{Enhancement of the critical current (antiparallel magnetizations, solid lines) and the $0$--$\pi$ transition at
which $I_c$ changes its sign (parallel magnetizations, dashed lines) in the SFIFS junction at $T/T_c = 0.05$,
$\gamma_{BM}=1$, and $\gamma_M=0$. Inset: the same for large values of $\gamma_M$. Note that the limit $d_F/\xi_F=0$ is
taken at fixed $\gamma_{BM}=\gamma_B d_F/\xi_F$. Since it is $\gamma_{BM}$ that determines the strength of the proximity
effect in the limit of thin F layers, $I_c$ demonstrates dependence on $h$.}
 \label{fig:fig1}
\end{figure}

Figure~\ref{fig:fig1} shows $I_c(h)$ dependencies calculated at $T=0.05\, T_c$ from the numerical solution of the
boundary problem (\ref{def_f})--(\ref{BC_bulk}) for the fixed value of $\gamma_{BM}=1$ and a set of different F-layers
thicknesses and the SF interface parameters $\gamma$. The normal junction resistance is $R_N=R_{B,I}+2R_B+ 2\rho_F d_F/
\mathcal{A}$. The curves $d_F/\xi_F=0$ are the limits of vanishing $d_F/\xi_F$ ratio at fixed $\gamma_{BM}$ and are
calculated from Eqs. (\ref{CurPF}), (\ref{CurAF}). For thin F layers the results depend only on the combination
$\gamma_M=\gamma d_F/\xi_F$. The enhancement of $I_c$ and the crossover to the $\pi$-state are clearly seen for the
antiparallel and parallel orientations, respectively. In accordance with the estimates given above, these effects take
place for the values of the exchange field $h$ close to $\pi T_c$. The enhancement disappears with increasing gradients
in the F layers since the solution Eq. (\ref{Sol2}) loses its validity. This is illustrated in Fig.~\ref{fig:fig1} by
increasing the thickness $d_F$ or $\gamma_M$.

Thus in the case of large $\gamma_M$ the enhancement effect is absent, in contrast to the statement of
Ref.~\cite{3:KK_a}. This contradiction is due to the fact that the calculation in the large $\gamma_M$ limit in
Ref.~\cite{3:KK_a} is valid only at $T\gg T_c/\gamma_M^2$, whereas the effect of $I_c^{(a)}$ enhancement exists only at
small $T$. Therefore the enhancement effect at large $\gamma_M$ and small $T$ in Ref.~\cite{3:KK_a} was obtained beyond
the applicability range of the approximate method. At the same time, at small $\gamma_M$, the $I_c(h)$ dependencies in
Refs.~\cite{3:KK_p,3:KK_a} are qualitatively correct but quantitatively inaccurate because in the approximate analytical
calculation the correction over small $\gamma_M$ was taken into account only in the Green function but not in the order
parameter $\Delta$. The accurate calculation requires to consider this correction self-consistently (similarly to
Ref.~\cite{3:Lukichev}).

\begin{figure}
 \centerline{\includegraphics[width=90mm]{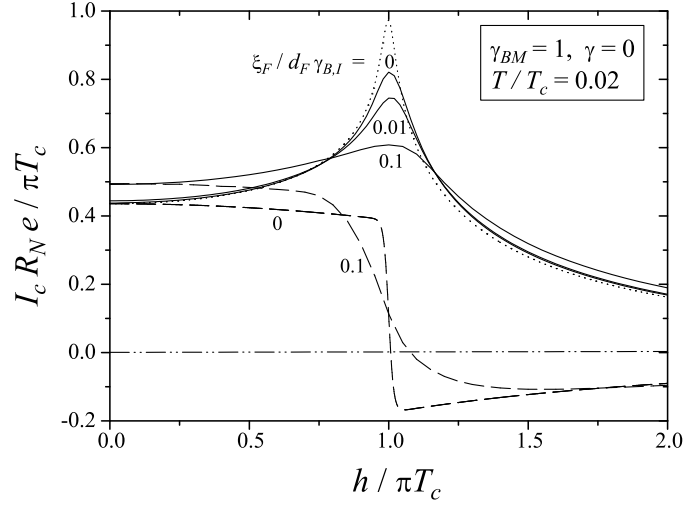}}
\caption{Enhancement of the critical current (antiparallel magnetizations, solid lines) and the $0$--$\pi$ transition at
which $I_c$ changes its sign (parallel magnetizations, dashed lines) in the SFIFS junction: influence of temperature and
barrier transparency. The dotted line corresponds to $T/T_c=0.01$ and $\xi_F / d_F \gamma_{B,I}=0$; the parameters for
other curves are given in the figure.}
 \label{fig:fig2}
\end{figure}

Influence of temperature and barrier transparency on the critical current anomaly is shown in Fig.~\ref{fig:fig2}. One
can see that, in accordance with the above estimate, the cutoff of $I_c^{(a)}$ singularity is provided by finite
temperature or barrier transparency. Namely, with the decrease of the barrier strength parameter $\gamma_{B,I}$ the peak
magnitude starts to drop when the ratio $d_F\gamma_{B,I}/\xi_F$ becomes comparable to $T/T_c$. With further decrease of
$d_F\gamma_{B,I}/\xi_F$ the singularity disappears, while the transition to the $\pi$-state shifts to large values of
$h$.

\begin{figure}
 \centerline{\includegraphics[width=90mm]{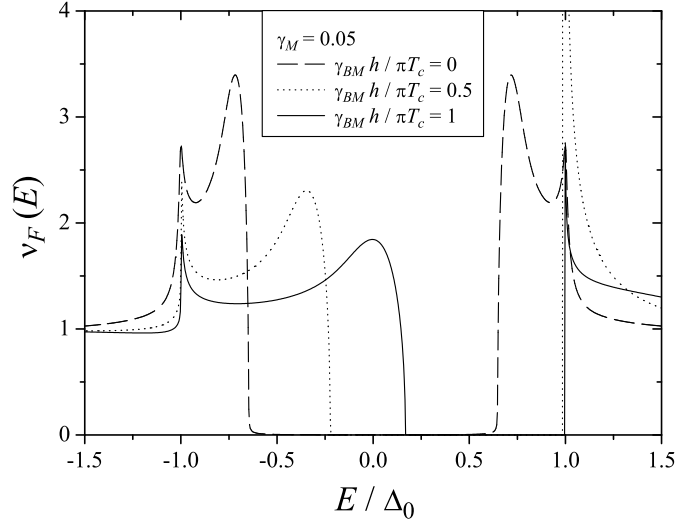}}
\caption{Normalized density of states for spin ``up'' in the F layer for various exchange fields.}
 \label{fig:fig3}
\end{figure}

Figure~\ref{fig:fig3} demonstrates the DoS in the F layers for one spin projection, calculated numerically in the limit
of small I-barrier transparency. At $h=0$ we reproduce the well-known minigap existing in SN bilayer. At finite $h$ the
gap shifts in energy (asymmetrically) and the peak in the DoS reaches zero energy at $h / \pi T_c =1/\gamma_{BM}$. One
can see that even for a small value $\gamma_M = 0.05$ the peaks are rather broad, therefore the singularity in
$I_c^{(a)}$ is suppressed by $\gamma_M$ very rapidly.

\begin{figure}
 \centerline{\includegraphics[width=90mm]{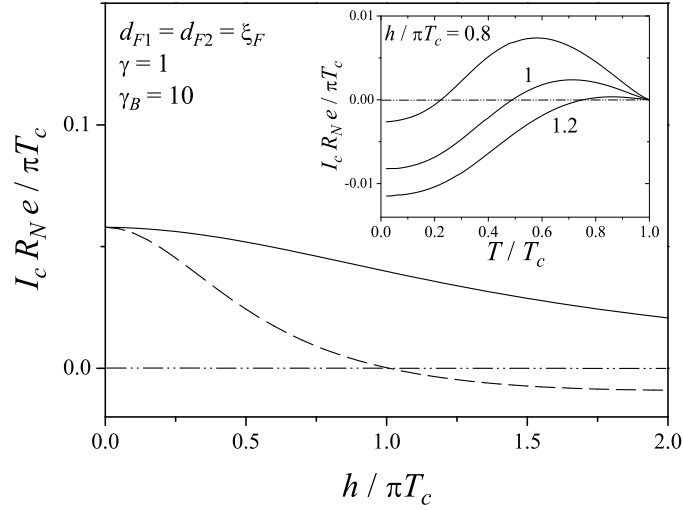}}
\caption{Critical current in the general case: switching effect. $T/T_c=0.5$, the solid and dashed lines correspond to
the antiparallel and parallel orientations of magnetizations, respectively. Inset: thermally induced $0$--$\pi$
crossover in the parallel case.}
 \label{fig:fig4}
\end{figure}

In the practically interesting limit of finite F-layer thickness (see Fig.~\ref{fig:fig4}) the numerical calculations
demonstrate monotonic suppression of $I_c$ with increase of the exchange field $h$ for antiparallel magnetizations of
the F layers and the $0$--$\pi$ crossover for the parallel case. One can see from Fig.~\ref{fig:fig4} that for given
temperature and thickness of the F layers it is possible to find the value of the exchange field at which switching
between parallel and antiparallel orientations will lead to switching of $I_c$ from nearly zero to a finite value (or to
switching between $0$ and $\pi$ states).

The case of parallel F-layers magnetizations in the absence of the I barrier corresponds to the standard SFS junction
where the $0$--$\pi$ transition is possible due to spatial oscillations of induced superconducting ordering in the F
layer. The thermally induced $0$--$\pi$ crossover in SFS junction was observed in Ref.~\cite{3:Ryazanov}, where simple
theory based on the linearized Usadel equations was also presented. Here we obtain such a crossover (see the inset in
Fig.~\ref{fig:fig4}) from the fully self-consistent solution in the range of the exchange fields corresponding to that
of Ref.~\cite{3:Ryazanov}.

\subsection{Conclusions}

In this section, we have developed a general method to solve the Usadel equations in SFIFS junctions self-consistently.
Using our method, we have investigated theoretically the Josephson current in SFIFS and SFS junctions. To clarify the
physical mechanisms behind the observed effects, we analytically considered the limiting case of thin F layers.

If the magnetizations of the two F layers in the SFIFS junction are antiparallel, then the critical current can be
enhanced by the exchange field, as it was predicted before in similar structures. We demonstrate that this effect is
similar to the Riedel singularity that takes place in SIS junction due to shifting of the density of states by voltage;
in the SFIFS junction, the shift is due to the exchange field. The logarithmic divergence of the maximal critical
current is cut off by finite temperature or barrier transparency.

If the magnetizations are parallel, then the junction can undergo the transition to the $\pi$-state even in the limit of
thin F layers. We demonstrate that this effect can be due to the effective $\pi/2$ shifts at the two SF interfaces.

\clearpage

\section{Nonsinusoidal current--phase relation in SFcFS and SIFIS Josephson junctions}

\subsection{Introduction}

The relation between the supercurrent $I$ across a Josephson junction and the difference $\varphi$ between the phases of
the order parameters in the superconducting banks is an important characteristic of the structure. The form of
$I(\varphi)$ dependence is essentially used for analyzing the dynamics of systems containing Josephson junctions
\cite{3:LikharevK}. Studying $I(\varphi)$ also provides information on pairing symmetry in superconductors
\cite{3:Il'ichev}.

In structures with tunnel-type conductivity of a weak link (SIS) the current--phase relation is sinusoidal, $I
(\varphi)=I_c \sin\varphi$ with $I_c>0$, in the whole temperature range below the critical temperature. At the same
time, in point contacts (ScS) and junctions with metallic type of conductivity (SNS) strong deviations from the
sinusoidal form take place at low temperatures $T$ \cite{3:Likharev} with the maximum of $I (\varphi)$ achieved at
$\pi/2 < \varphi_\mathrm{max} < \pi$.

The situation drastically changes if there is magnetoactive material in the region of weak link. The transition from the
$0$-state ($I_c>0$) to the $\pi$-state ($I_c<0$) has been theoretically predicted in a variety of Josephson structures
containing ferromagnets
\cite{3:Bulaevski,3:Buzdin_pi,3:Buzdin1,3:Buzdin2,3:Tanaka,3:KK_p,3:Dobro,3:Yip,3:Wilhelm,3:Fogelstrom,3:Radovic01,
3:Chtch,3:Barash_Bobkova} and experimentally observed in SFS and SIFS junctions
\cite{3:Ryazanov,3:Ryazanov_new,3:Kontos}. In the general case modifications of $I(\varphi)$ do not reduce to the change
of the critical current sign in the sinusoidal Josephson relation. It was shown that the presence of a ferromagnet may
result in a nonsinusoidal shape of $I(\varphi)$. In the diffusive regime, this situation occurs in long SFS junctions
with ideally transparent interfaces \cite{3:Yip,3:Wilhelm}. However, in this case the effects take place only in a
narrow interval of very low temperatures (due to smallness of the Thouless energy), while in the present section we
shall consider short-length structures where the effects are more pronounced and exist practically in the whole
temperature range.

In this section we investigate anomalies of the $I(\varphi)$ relation in several types of SFS structures which allow
analytical solution: the SFcFS point contact with clean or diffusive constriction as a weak link, and the double-barrier
SIFIS junction; the ferromagnetic layers are assumed to be thin, and the magnetization is homogeneous throughout the F
part of the system. In particular, we show that the maximum of $I(\varphi)$ can shift from $\pi/2 \leqslant
\varphi_\mathrm{max} < \pi$ to $0 < \varphi_\mathrm{max} < \pi /2$ as a function of the exchange field in the
ferromagnet. Previously, current--phase relation of this type was theoretically predicted either if superconductivity in
the S electrodes was suppressed by the supercurrent in the SNS structure \cite{3:Meriakri,3:Zubkov,3:Kupr} or in the
vicinity of $T=0$ in long SFS junctions \cite{3:Yip,3:Wilhelm}.

The outline of the section is as follows. We start with studying the SFcFS structure composed of two SF sandwiches
linked by a clean Sharvin constriction with arbitrary transparency $D$. We show that the energy--phase relation of this
junction can have \textit{two} minima: at $\varphi=0$ and $\varphi=\pi$ (while the energy of the junction in the pure
$0$- or $\pi$-state has a \textit{single} minimum --- at $\varphi=0$ or $\varphi=\pi$, respectively); such situation was
discussed in Ref.~\cite{3:Radovic01} for a clean SFS junction. As a result, the $I(\varphi)$ dependence can intersect
zero not only at $\varphi=0$ and $\varphi =\pi$ but also at an arbitrary value $\varphi_0$ from the interval $0<
\varphi_0< \pi$. The salient effects which occur in junctions with clean constriction survive averaging over the
distribution of transmission eigenvalues and thus occur also in diffusive point contacts. Physically, the properties of
SFS structures are explained by splitting of Andreev levels due to the exchange field; to demonstrate this, we study the
spectral supercurrent (this physical mechanism was formulated for long SFS junctions at low temperatures in
Refs.~\cite{3:Yip,3:Wilhelm}). Finally, we show that the same mechanism provides shifting of the $I(\varphi)$ maximum to
$\varphi < \pi /2$ in the double-barrier SIFIS junctions which can be more easily realized in experiment.

\subsection{SFcFS with clean constriction}

We start with a model structure (see Fig.~\ref{fig:sketch}b) composed of two superconducting SF bilayers connected by a
clean constriction with transparency $D$ (the size of the constriction $a$ is much smaller than the mean free path $l$:
$a\ll l$). We assume that the S layers are bulk and that the dirty limit conditions are fulfilled in the S  and F
metals. For simplicity we also assume that the parameters of the SF interfaces $\gamma$ and $\gamma_B$ obey the
condition
\begin{gather} \label{cond1}
\gamma \ll \max (1,\gamma _B).
\end{gather}
We shall consider the symmetric structure and restrict ourselves to the limit of thin F layers:
\begin{equation} \label{cond2}
d_F \ll \min \left( \xi_F, \sqrt{D_F/2h}\right).
\end{equation}

Under condition (\ref{cond1}), we can neglect the suppression of superconductivity in the S electrodes by the
supercurrent and the proximity effect, and reduce the problem to solving the Usadel equation in the F layers:
\begin{equation}
\xi_F^2 \frac \partial {\partial x}\ \left[ G_F^2 \frac \partial {\partial x} \Phi_F \right]
-\frac{\widetilde\omega_n}{\pi T_c} G_F \Phi_F =0, \label{UsadelFi}
\end{equation}
with the boundary conditions at the SF interfaces ($x=\mp d_F$):
\begin{gather}
\pm \gamma_B \frac{\xi_F G_F}{\widetilde\omega_n} \frac \partial {\partial x} \Phi_F =G_S \left(
\frac{\Phi_F}{\widetilde\omega_n} -\frac{\Phi_S}{\omega_n} \right) , \label{bound_cond} \\
G_S = \frac{\omega_n}{\sqrt{\omega_n^2+\Delta_0^2}},\qquad \Phi_S (\mp d_F)=\Delta_0 \exp \left( \mp i \varphi/2
\right). \notag
\end{gather}
Here $\Delta_0$ is the absolute value of the order parameter in the superconductors.

Under condition (\ref{cond2}), the spatial gradients in the F layers arising due to the proximity effect and current are
small. Then we can employ the solution (\ref{Sol2}), obtained in Sec.~\ref{sec:SFIFS}:
\begin{equation} \label{Phi_0}
\Phi_{F1,F2} = \Phi_0 \exp(\mp i\varphi/2),\qquad \Phi_0 = \frac{\widetilde\omega_n}W \Delta_0,
\end{equation}
where
\begin{equation}
W=\omega_n + \widetilde\omega_n \gamma_{BM} \Omega,\qquad \Omega =\frac{\sqrt{\omega_n^2 +\Delta_0^2}}{\pi T_c}, \qquad
\gamma_{BM} =\gamma_B \frac{d_F}{\xi_F},
\end{equation}
and the indices $1$ and $2$ refer to the left- and right-hand side of the constriction, respectively.

The supercurrent in the constriction geometry is given by the general expression \cite{3:Zaitsev}
\begin{equation}
I=\frac{4\pi T}{e R_N} \Im \sum_{\omega_n>0} \frac{(\bar F_1 F_2 -F_1 \bar F_2)/2}{2 - D \left[ 1 - G_1 G_2 - (\bar F_1
F_2 +F_1 \bar F_2)/2 \right] },
\end{equation}
where $R_N$ is the normal-state resistance of the junction. Inserting Eq. (\ref{Phi_0}) into this expression, we obtain
\begin{equation} \label{Constr_D}
I=\frac{2\pi T}{e R_N} \Re \sum_{\omega_n >0} \frac{\Delta_0^2 \sin\varphi}{W^2+\Delta_0^2 \left[ 1 -D \sin^2(\varphi
/2) \right]}.
\end{equation}
Finally, the current--phase relation takes the form
\begin{gather}
I(\varphi)=\frac{2\pi T}{e R_N}\sum_{\omega_n >0} \frac{C \Delta_0^2 \sin\varphi}{C^2+B^2}, \\
C = \Delta_0^2 \left[ 1-D\sin^2 \left( \varphi/2 \right) \right] - h^2 \left( \gamma_{BM} \Omega \right)^2 +\omega_n^2
\left( 1+\gamma_{BM} \Omega \right)^2 , \notag
\end{gather}

\begin{figure}
 \centerline{\includegraphics[width=90mm]{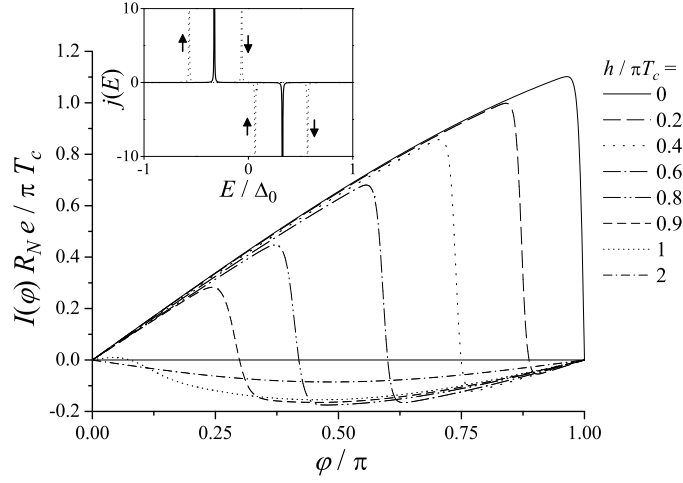}}
\caption{Current--phase relation in clean SFcFS junction with ideally transparent constriction ($D=1$) at $T/T_c=0.01$,
$\gamma_{BM}=1$ for different values of the exchange field $h$. Inset: spectral supercurrent density at $\varphi=2 \pi
/3$ for $h / \pi T_c =0$ (solid line) and $h / \pi T_c =0.4$ (dotted line, the arrows near the peaks denote the
corresponding spin direction).}
 \label{fig:fig1_curph}
\end{figure}

\begin{figure}
 \centerline{\includegraphics[width=90mm]{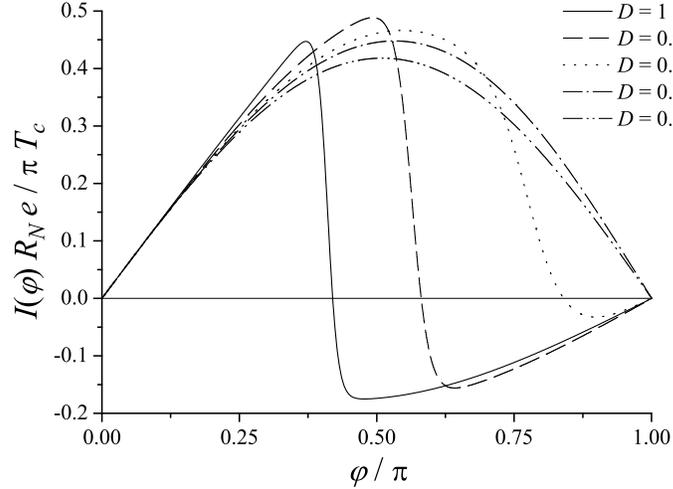}}
\caption{Current--phase relation in clean SFcFS junction at $T/T_c=0.01$, $\gamma_{BM}=1$, $h / \pi T_c =0.8$ for
different values of the barrier transparency $D$.}
 \label{fig:fig2_curph}
\end{figure}

At small $\omega_n$, the function $C$ [and hence $I(\varphi)$] changes its sign at finite phase difference
\begin{equation}
\varphi_c =2\arcsin \sqrt{\frac{1-(\gamma_{BM} h / \pi T_c )^2}D}
\end{equation}
if the exchange field is in the range $1-D < (\gamma_{BM} h / \pi T_c )^2 <1$. The results for $I(\varphi)$ are shown in
Figs.~\ref{fig:fig1_curph}, \ref{fig:fig2_curph} and can be understood (similarly to Refs.~\cite{3:Yip,3:Wilhelm}) by
considering the spectral supercurrent density $j(E)$ that is defined as the quantity which determines the supercurrent
according to the formula
\begin{equation} \label{j}
I=\frac 1{2 e R_N} \int_{-\infty}^\infty dE\, f(E) \left[ j_\uparrow (E) +j_\downarrow (E) \right],
\end{equation}
where $f(E) = 1/(e^{E/T}+1)$ is the Fermi distribution function. The physical meaning of $j_\uparrow (E)$ and
$j_\downarrow (E)$ is the current carried by the quasiparticles with energy $E$ and spin up or down (denoted as
$\uparrow$ or $\downarrow$) if the corresponding states are occupied, while $f(E)$ describes their
occupation.\footnote{In the paper A.\,A.~Golubov, M.\,Yu.~Kupriyanov, Ya.\,V.~Fominov, Pis'ma Zh. Eksp. Teor. Fiz.
\textbf{75}, 709 (2002) [JETP Lett. \textbf{75}, 588 (2002)] we used a different form of expression for the total
current, which can be written in our present notations as
\begin{equation*}
I=\frac 1{4 e R_N} \int_{-\infty}^\infty dE\, \tanh\! \left( \frac E {2T} \right) \left[ -j_\uparrow (E) - j_\downarrow
(E) \right].
\end{equation*}
However, here we prefer the expression in the form (\ref{j}) because its physical meaning is clearer.} To calculate the
spectral supercurrent density, we perform the analytical continuation in Eq. (\ref{Constr_D}); the result is
\begin{equation}
j_\sigma (E) = - \Im \frac{\Delta_0^2 \sin\varphi}{\Delta_0^2 \left[ 1-D \sin^2 (\varphi/2) \right] - \left[ E+
\gamma_{BM} (E- \sigma h) \sqrt{\Delta_0^2 -E^2}/\pi T_c \right]^2},
\end{equation}
where $\sigma = \pm 1$ corresponds to spin up or down. Now we should recall that this formula originates from the
retarded Green functions, hence we should add the infinitesimal imaginary part $i0$ to the energy $E$. As a result,
$j(E)$ is given by a sum of delta-functions $\delta (E-E_A)$ where $E_A$ are the energies of the Andreev bound states.
At $\gamma_{BM}=0$, the well-known result $E_A =\pm \Delta_0 \sqrt{1-D \sin^2 (\varphi /2)}$ is reproduced, while at
finite $\gamma_{BM}$ the exchange field splits each bound state into two (see inset in Fig.~\ref{fig:fig1_curph}).

At small temperature, the integration in Eq. (\ref{j}) is effectively restricted to the negative energy domain. The
reversing of the supercurrent sign is then explained by the fact that at $\varphi=\varphi_c$ the positive peak for spin
down crosses zero leaving the domain $E<0$, and simultaneously the negative peak for spin up moves from the region $E>0$
into the region $E<0$.

The sign-reversal of the supercurrent (the $0$--$\pi$ transition) can also be achieved at a \textit{fixed} $h$ due to
nonequilibrium population of levels. This phenomenon has been studied in long diffusive SNS
\cite{3:Volkov,3:Wilhelm1,3:Basel} and SFS junctions \cite{3:Yip,3:Wilhelm}.

\subsection{SFcFS with diffusive constriction}

To get the $I(\varphi)$ relation for the diffusive point contact [$l \ll a\ll \min( \xi_F, \sqrt{D_F/2h} )$] we
integrate $\int_0^1 \rho(D) I(D) dD$, where $I(D)$ is given by Eq. (\ref{Constr_D}) for the clean-constriction case
(note that $R_N \propto D^{-1}$ in this equation) and $\rho(D)$ is Dorokhov's density function $\rho(D)= 1/ 2 D
\sqrt{1-D}$ \cite{3:Dorokhov}. Finally, we arrive at the result
\begin{equation} \label{I_varphi_SFcFS}
I(\varphi) = \frac{4\pi T}{e R_N} \Re \sum_{\omega_n >0} \frac{\Delta_0 \cos (\varphi /2)}{\sqrt{W^2+\Delta_0^2 \cos^2
(\varphi /2)}} \arctan \left( \frac{\Delta_0 \sin (\varphi /2)}{\sqrt{W^2+\Delta_0^2 \cos^2 (\varphi /2)}} \right).
\end{equation}
This expression coincides with the direct solution of the Usadel equations (solving the Usadel equations, we should take
into account that the gradients in the constriction region are large). In the $\gamma_{BM}=0$ limit, the current does
not depend on the characteristics of the F layers, and Eq. (\ref{I_varphi_SFcFS}) reproduces the Kulik--Omelyanchuk
formula for the diffusive ScS junction \cite{3:KO1,3:Likharev}.

\begin{figure}
 \centerline{\includegraphics[width=90mm]{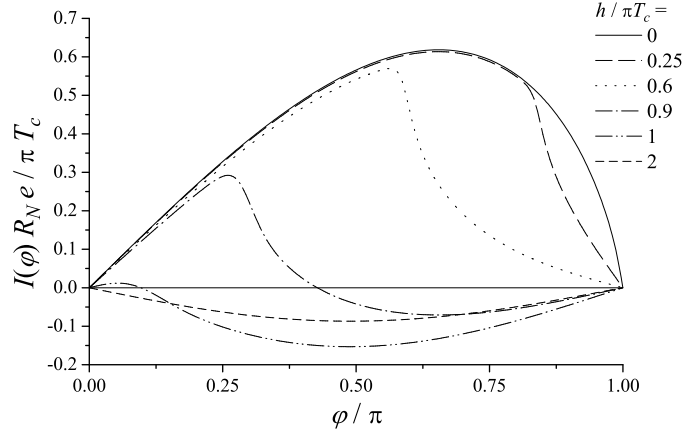}}
\caption{Current--phase relation in diffusive SFcFS point contact at $T/T_c=0.01$, $\gamma_{BM}=1$ for different values
of the exchange field $h$.}
 \label{fig:fig3_curph}
\end{figure}

Calculation of $I(\varphi)$ using the above expression yields results similar to those for the clean point contact,
however the transition from $0$- to $\pi$-state becomes less sharp (see Fig.~\ref{fig:fig3_curph}).

Temperature dependence of the critical current in this case shows thermally-induced $0$--$\pi$ crossover with nonzero
critical current at the transition point, in agreement with results of Refs.~\cite{3:Wilhelm,3:Radovic01,3:Chtch}. This
is a natural result since the barrier transparency is high and the current--phase relation is strongly nonsinusoidal.

\subsection{SIFIS}

Now we turn to the double-barrier SIFIS junction (I denotes an insulating barrier)
--- see Fig.~\ref{fig:sketch}c. This structure is easier for experimental implementation than an SFcFS junction.

We assume that condition (\ref{cond1}) is satisfied; then we can neglect the suppression of superconductivity in the S
electrodes by the supercurrent and the proximity effect. In this case the system is described by Eqs.
(\ref{UsadelFi})--(\ref{bound_cond}), although now instead of two F layers connected by a constriction we have a
continuous F layer (at $-d_F <x< d_F$).

We also assume that the F layer is thin [condition (\ref{cond2})] and that $\gamma_B \gg d_F/\xi_F$, hence the spatial
gradients in the F layer are small. Then we can expand the solution of Eqs. (\ref{UsadelFi})--(\ref{bound_cond}) up to
the second order in small gradients, arriving at
\begin{gather}
\Phi_F =\Phi_0 \cos( \varphi/ 2 ) + i\frac{\widetilde\omega_n G_S}{\omega_n G_F} \frac{\Delta_0 \sin
(\varphi /2)}{\gamma_B} \frac x{\xi_F}, \label{Sol1_} \\
G_F=\frac{\widetilde\omega_n}{\sqrt{\widetilde\omega_n^2+\Phi_0^2 \cos^2 ( \varphi/ 2 )}}, \label{Sol1a}
\end{gather}
with $\Phi_0$ defined in Eq. (\ref{Phi_0}) [in the final result (\ref{Sol1_}) we retained only the first order in
gradients --- this accuracy is sufficient for calculating the current].

\begin{figure}
 \centerline{\includegraphics[width=90mm]{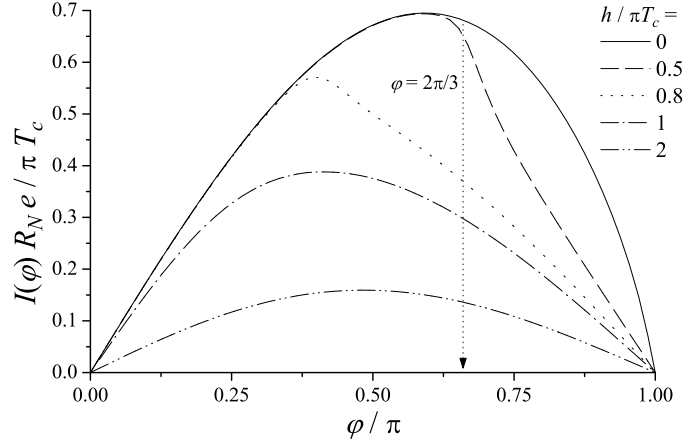}}
\caption{Current--phase relation in double-barrier SIFIS junction at $T/T_c=0.02$, $\gamma_{BM}=1$ for different values
of the exchange field $h$. The value $\varphi=2\pi/3$ will be used in Fig.~\ref{fig:fig5_curph}.}
 \label{fig:fig4_curph}
\end{figure}

Inserting the solution (\ref{Sol1_}), (\ref{Sol1a}) into the general expression (\ref{cur}) for the supercurrent, we
obtain
\begin{equation}
I(\varphi) =\frac{2\pi T}{e R_N} \Re \sum_{\omega_n >0} \frac{\Delta_0^2 \sin\varphi}{\sqrt{\omega_n^2+\Delta_0^2}
\sqrt{W^2+\Delta_0^2 \cos^2(\varphi /2)}} \label{IotFi}
\end{equation}
(our assumptions imply that $R_N\approx 2 R_B$). This result demonstrates that the SIFIS junction with thin F layer is
always in the $0$-state.\footnote{In the case under discussion, when the F layers are thin and the interface parameters
obey condition (\ref{cond1}), the phase of the order parameter is constant in the S part and almost constant in the F
part, however it jumps at the two SF interfaces (see. Sec.~\ref{sec:SFIFS}). The two jumps compensate each other in the
SIFIS junction with a single F layer, whereas in the SFcFS junctions they add up at the weak link thus opening a
possibility for the $\pi$-state.} Nevertheless, $I(\varphi)$ is strongly modified by finite $h$ (see
Fig.~\ref{fig:fig4_curph}), especially at small temperatures. Figure~\ref{fig:fig4_curph} clearly demonstrates that an
increase of $h$ results not only in suppression of the critical current, but also in the shift of the $I(\varphi)$
maximum from $\varphi_\mathrm{max} \approx 1.86$ at $h=0$ to the values smaller than $\pi /2$. In the limit of large
exchange fields, $h / \pi T_c \gg 1/ \gamma_{BM}$, the $I(\varphi)$ dependence returns to the sinusoidal form.

\begin{figure}
 \centerline{\includegraphics[width=80mm]{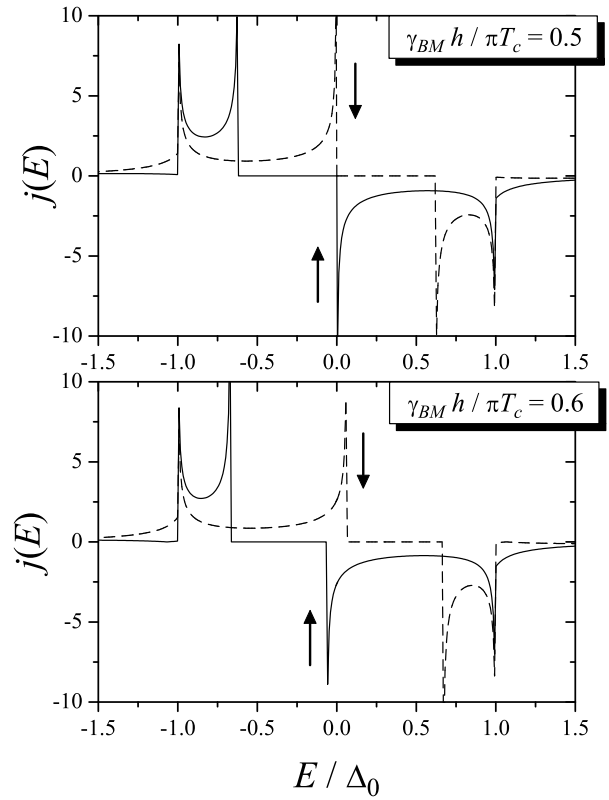}}
\caption{Spectral supercurrent density in diffusive double-barrier SIFIS junction with thin ferromagnetic interlayer at
$\gamma_{BM}=1$, $\varphi=2\pi/3$ for two values of the exchange field $h$. The chosen value of $\varphi$ corresponds to
$\varphi_c$ at $\gamma_{BM} h / \pi T_c =0.5$, and the figure demonstrates that the positive peak for the spin down
projection disappears from while the negative peak for the spin up projection appears in the $E<0$ domain.}
 \label{fig:fig5_curph}
\end{figure}

The physical origin of these results becomes clear in the real energy $E$ representation. Performing the analytical
continuation in Eq. (\ref{IotFi}) by replacement $\omega_n \rightarrow -iE$, we obtain the spectral supercurrent
density:
\begin{gather}
j_\sigma (E)= -\Im \frac{\Delta_0^2 \sin\varphi}{\sqrt{\Delta_0^2 -E^2}
\sqrt{\Delta_0^2 \cos^2 (\varphi /2)-\widetilde E^2}}, \label{SpectrJ} \\
\widetilde E =E +\gamma_{BM} (E -\sigma h) \frac{\sqrt{\Delta_0^2-E^2}}{\pi T_c}. \notag
\end{gather}
Equation (\ref{SpectrJ}) implies that at
\begin{equation}
\varphi_c =2 \arccos \left( \frac{\gamma_{BM} h}{\pi T_c} \right),
\end{equation}
singularities in $j(E)$ are shifted to the Fermi level. At $\varphi > \varphi_c$, the negative singularity in
$j_\uparrow (E)$ crosses the Fermi level and appears in the $E<0$ domain (recall that at low temperatures only $j(E)$ at
$E<0$ contribute to the current), whereas the positive peak in $j_\downarrow (E)$ leaves the $E<0$ domain (this process
is illustrated in Fig.~\ref{fig:fig5_curph}). As a result, the contribution to the supercurrent from low energies
changes its sign, and the supercurrent $I(\varphi)$ becomes suppressed at $\varphi
> \varphi_c$ (see Fig.~\ref{fig:fig4_curph}). However, at energies with larger absolute values $E \sim -\Delta_0$,
modifications in $j(E)$ are weak, and the resulting $I(\varphi)$ does not change its sign. This mechanism is similar to
the one that takes place in long SFS junctions at low temperatures \cite{3:Yip,3:Wilhelm}.

\subsection{Conclusions}

In this section, we have studied the current--phase relation in several types of Josephson junctions with thin
ferromagnetic interlayers: SFcFS junction with ballistic or diffusive constriction and SIFIS planar junction. The
current--phase relation can be highly nonsinusoidal in these junctions due to splitting of Andreev bound states by the
exchange field.

In particular, the maximum of $I(\varphi)$ can be achieved at $\varphi< \pi/2$. In SFcFS junctions, the supercurrent can
change its sign at arbitrary value of the phase between $0$ and $\pi$. When the maximum absolute value corresponds to
negative current, the junction is in the $\pi$-state.

Another consequence of this sign change is that the energy--phase relation for the junction has two minima: at
$\varphi=0$ and $\varphi=\pi$. The phenomenon can be used for engineering qubits.

\clearpage
\renewcommand{\@evenhead}{}

\chapter[Decoherence due to nodal quasiparticles in \textit{d}-wave Josephson junctions]
{\huge Decoherence due to nodal quasiparticles in \textit{d}-wave Josephson junctions}

\renewcommand{\@evenhead}
    {\raisebox{0pt}[\headheight][0pt]
     {\vbox{\hbox to\textwidth{\thepage \hfil \strut \textit{Chapter \thechapter}}\hrule}}
    }

\renewcommand{\@oddhead}
    {\raisebox{0pt}[\headheight][0pt]
     {\vbox{\hbox to\textwidth{\textit{\leftmark} \strut \hfil \thepage}\hrule}}
    }

\section{Introduction}

In addition to the fundamental importance, Josephson junctions between \textit{d}-wave superconductors are of great
interest for the rapidly developing field of quantum computations. In this field, the mathematical aspects (so to speak,
writing programs) are rather well elaborated, while the quantum computer itself has not been realized so far, and there
are only few successful realizations of a qubit (quantum bit). It is not clear at present, which of the proposed qubit
implementation will prove most successful from the practical point of view.

\begin{figure}
 \centerline{\includegraphics[width=70mm]{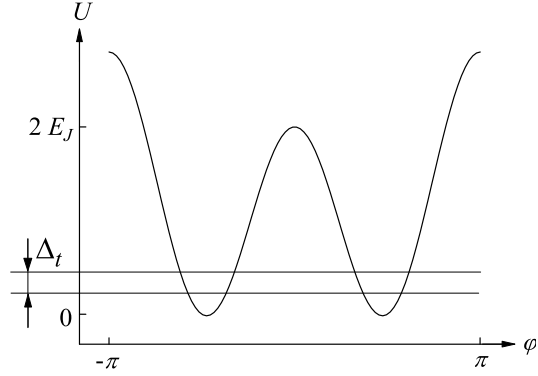}}
\caption{Schematic dependence of the Josephson energy $U$ on the phase difference $\varphi$. The barrier of the height
$2 E_J$ separates two nontrivial minima. The splitting of the lowest energy level due to the tunneling across the
barrier is denoted $\Delta_t$.}
 \label{fig:2well}
\end{figure}

Solid-state proposals, and in particular superconducting devices, have a number of advantages, e.g., scalability and
variability \cite{4:MSS,4:Blatter}. Particularly interesting are the so-called quiet qubits, which are
\textit{intrinsically} degenerate, i.e., do not require any external source for maintaining the degeneracy. Such qubits
can be realized in systems involving \textit{d}-wave superconductors \cite{4:Ioffe,4:Zagoskin}, which are of the
so-called ``phase qubit'' type (the information is encoded by the phase difference $\varphi$ across the junction).
Recently, it was experimentally demonstrated that a double-well dependence of the energy versus the phase difference
(inside a single period) is indeed realized in the Josephson junctions between \textit{d}-wave superconductors
\cite{4:Il'ichev} (see Fig.~\ref{fig:2well}).

For any qubit implementation, there are processes that hamper its successful operating --- the so-called decoherence
processes, which destroy a quantum state of the qubit. Although the quiet phase qubits are rather well isolated from the
environment, there are intrinsic mechanisms of decoherence even at low temperatures. The quantum tunneling of the phase
between the two minima leads to fluctuating voltage across the junction, which excites quasiparticles. The dissipative
current across the interface arises, leading to a finite decoherence time $\tau_\varphi$.

The knowledge of $\tau_\varphi$ is essential for estimating the efficiency of the qubit: short decoherence time makes
the qubit senseless from the practical point of view, while a long enough decoherence time opens the way for quantum
correction algorithms that in principle allow to perform an infinitely long computation \cite{4:corr}.

The relevance of the quasiparticle processes at low temperatures is specific for \textit{d}-wave superconductors. In the
conventional \textit{s}-wave case, the quasiparticle transport below the gap is suppressed. At the same time, in gapless
anisotropic superconductors the gap vanishes in certain directions (the nodal directions), hence the low-energy
quasiparticle appear. In the present chapter, we consider a DID Josephson junction (D = \textit{d}-wave superconductor,
I = insulator), and study the decoherence due to nodal quasiparticles (quasiparticles moving along the nodal
directions).

\section{Decoherence time (general strategy)}

Theoretical description of the quantum dynamics of a tunnel junction between two \textit{s}-wave superconductors was
developed in Ref.~\cite{4:ESA} (see Ref.~\cite{4:SZ} for a review); in particular, the effective action for the phase
difference $\varphi$ was obtained. Later this description was generalized to the case of \textit{d}-wave superconductors
in Refs.~\cite{4:Bruder,4:BGZ}. The effective action for $\varphi$ is similar to the general case considered by
Caldeira, Leggett et al. \cite{4:CL,4:Leggett_review}, who studied influence of dissipation on quantum tunneling in
macroscopic systems. The dissipation was described as being due to the interaction with a bath of oscillators (the
environment). The ``strength'' of the environment, depending on the frequency $\omega$, is characterized by the spectral
function $J(\omega)$. In the Josephson junction, the environment is represented by the quasiparticles, and the spectral
function is given by $\hbar I(\hbar\omega/e) /e$, where $I$ is the dissipative quasiparticle current taken at
``voltage'' $\hbar \omega /e$ \cite{4:ESA}.

A system living in a double-well potential and described by an extended coordinate can be ``truncated'' to the two-state
system (spin $1/2$) with the two states ($\sigma_z =\pm 1$) corresponding to the minima of the potential (see
Fig.~\ref{fig:2well}). The theory of dissipative two-state systems is thoroughly elaborated \cite{4:Leggett_review} for
the cases when the spectral function behaves as $J(\omega) \propto \omega^s$ up to some high-frequency cutoff. The
situations when $s=1$, $s>1$, and $0<s<1$ are called ohmic, superohmic, and subohmic, respectively. In this language,
the dissipation due to nodal quasiparticles in the Josephson junction is superohmic, as we demonstrate below.

What is the decoherence in such a system? Assume that during the time $t<0$ the system is held in the right well (i.e.,
at $\sigma_z=1$). At $t=0$ the constraint is released, and we consider the expectation value of the system coordinate:
$P(t) = \left< \sigma_z(t) \right>$. Below we shall encounter the superohmic case at zero temperature. Then
\cite{4:Leggett_review}
\begin{equation}
P(t) = \cos(\Delta_t t/\hbar) \exp(- t / \tau_\varphi)
\end{equation}
--- the cosine describes coherent oscillations between the two wells ($\Delta_t$ is the tunnel splitting of levels,
see Fig.~\ref{fig:2well}) while the exponential leads to their incoherent damping.

The decoherence time $\tau_\varphi$ is expressed in terms of the spectral function \cite{4:Leggett_review}. Returning
from the general theory to the particular case of the Josephson junction, we write the corresponding result as
\begin{equation}
\tau_\varphi = \frac{4e}{\delta\varphi^2 I(\Delta_t/e)} =\frac{4\pi\hbar}{\delta\varphi^2 e R_q I(\Delta_t/e)},
\end{equation}
where $\delta\varphi$ is the distance between the potential minima and $R_q = \pi \hbar / e^2 \approx
13\,\mathrm{k}\Omega$ is the quantum resistance. Comparing the decoherence time with the characteristic time of
oscillations between the wells, $\hbar/ \Delta_t$, we obtain the quality factor
\begin{equation} \label{Q}
Q = \frac{\tau_\varphi \Delta_t}{2\hbar} = \frac{2\pi \Delta_t} {\delta\varphi^2 e R_q I(\Delta_t/e)},
\end{equation}
which must be large for successful operating of the qubit.

In the DID junction, the tunnel splitting $\Delta_t$ is much smaller than the order parameter $\Delta$, hence
$\tau_\varphi$ is determined by the quasiparticle current at low ``voltage''.

\section{Quasiparticle current}

\begin{figure}
 \centerline{\includegraphics[width=80mm]{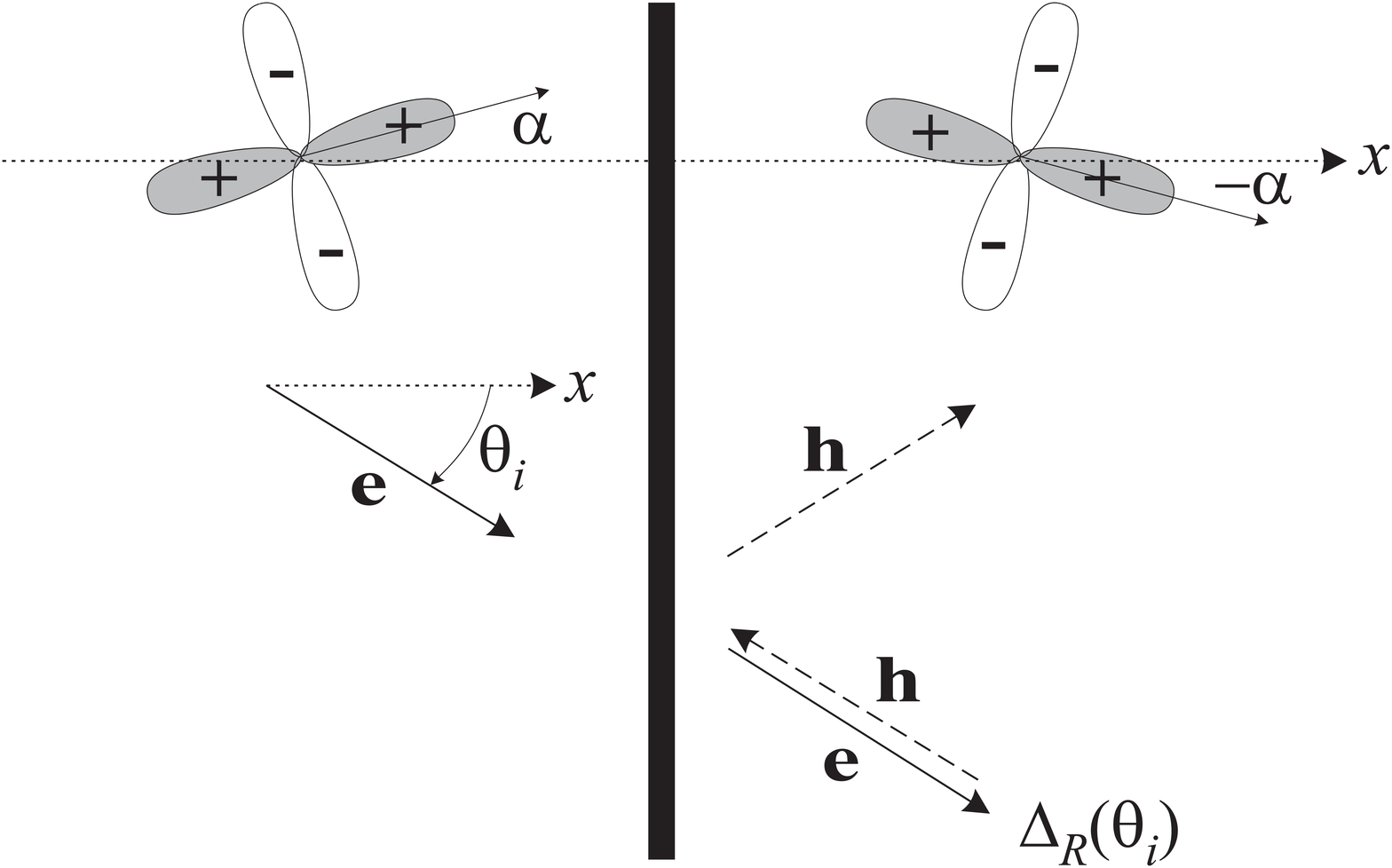}}
\caption{DID junction of mirror orientation $\alpha/-\alpha$. The positive lobes of the order parameter are shaded. An
electron \textbf{e} moving along a truly nodal direction $\theta_i$ of the left superconductor, tunnels into an induced
nodal direction of the right superconductor. $\Delta_R (\theta_i) \ne 0$, therefore the electron experiences the Andreev
reflection; the hole \textbf{h} returns to the interface, and after reflection at the interface escapes into the bulk
along the truly nodal direction $-\theta_i$. In this process, the total current into the bulk of the right
superconductor is composed of the Cooper pair along $\theta_i$ and the hole along $-\theta_i$.}
 \label{fig:system}
\end{figure}

We consider the grain-boundary Josephson junction between two quasi-two-dimensional $d_{x^2-y^2}$-wave superconductors
with cylindrical Fermi surfaces. The orientations of the superconductors are characterized by the angles between the
\textit{a}-axes and the normal to the interface (the $x$ axis) --- see Fig.~\ref{fig:system}. According to
Ref.~\cite{4:Il'ichev}, we consider the mirror junction, in which the misorientation angles on both sides are equal in
magnitude but opposite in sign, $\alpha/-\alpha$ (we take $-45^\circ \leqslant \alpha \leqslant 45^\circ$ because all
physically different situations in the mirror junction are realized in this interval). The order parameter depends on
the direction (parametrized by the angle $\theta$) and the distance to the interface:
\begin{equation} \label{Delta_dwave}
\Delta_{L,R} (x,\theta) = \widetilde\Delta_{L,R} (x) e^{i\varphi_{L,R}} \cos\left( 2 \theta \mp 2 \alpha \right),
\end{equation}
where the indices $L$ and $R$ refer to the left- and right-hand side of the junction, respectively.

The two generate states of the DID junction correspond to two minima of the junction energy and carry no current across
the interface (the Josephson current is proportional to the derivative of the Josephson energy with respect to the
phase; the derivative is zero at the minima). At the same time, physically, the two states differ by the direction of
spontaneous currents flowing along the interface \cite{4:Huck,4:Amin}. These currents produce fluxes that can interact
with the environment and make the \textit{d}-wave qubit not completely quiet \cite{4:Blatter}. From this point of view,
the mirror orientation of the junction is the most preferable one, because in the mirror junction there is no
\textit{total} current along the interface (the currents are nonzero but compensated). In principal, the presence of the
spontaneous currents implies the phase change along the interface. However, the currents appear as a result of the
Andreev reflections, which are of the second order over transparency. In the tunneling limit, the currents and the
corresponding phase change along the interface are small, and $\Delta$ has the form (\ref{Delta_dwave}) with constant
$\varphi_L$ and $\varphi_R$.

The quasiparticle current in the tunneling limit at low temperatures, $k_B T\ll \hbar\omega$, is given by
\begin{gather}
I(\hbar\omega/e) = \frac 1{e R_N} \int_{-\pi /2}^{\pi /2} d\theta \frac{D(\theta) \cos\theta}{\widetilde
D} \int_0^{\hbar\omega} dE\;  \nu \left( E- \hbar\omega,\theta \right) \nu \left(E,\theta \right), \label{I_V_general}\\
\widetilde D = \int_{-\pi/2}^{\pi/2} d\theta D(\theta)\cos\theta. \notag
\end{gather}
Here $R_N$ is the normal-state resistance of the interface, $\nu(E,\theta)$ is the density of states (DoS) at the
interface, normalized to the normal-metal value, and $D(\theta)$ is the angle-dependent transparency of the interface.
We have not labelled the DoS by the indices $L$ and $R$ because $\nu_L(E,\theta) =\nu_R(E,\theta)$ in the mirror
junction.

Below we calculate the nodal contribution to the current (\ref{I_V_general}) at $\hbar\omega \ll \widetilde\Delta_0$,
where $\widetilde\Delta_0 = \widetilde\Delta(\pm\infty)$ is the bulk amplitude of the order parameter. The angle
integration contributing to the current is then limited to narrow angles around the nodal directions, where the
low-energy DoS is nonzero (as we shall see below, the width of the angles is $\delta\theta = \hbar\omega
/\widetilde\Delta_0$).

To calculate the DoS, we employ the quasiclassical approach. The quasiclassical matrix Green function
\begin{equation}
\widehat G = \begin{pmatrix} g & f \\ \Bar f & -g \end{pmatrix}
\end{equation}
obeys the Eilenberger equation \cite{4:Eilenberger,4:LO_qc} and satisfies the normalization condition $\widehat
G^2=\widehat 1$. It can be parametrized as
\begin{equation} \label{parametrization}
g=\frac{1-ab}{1+ab},\qquad f=\frac{2a}{1+ab},\qquad \Bar f =\frac{2b}{1+ab},
\end{equation}
then the normalization condition is automatically satisfied. The equations for the new functions $a(x,\theta)$ and
$b(x,\theta)$ take the form of the Riccati equations \cite{4:Schopohl}:
\begin{gather}
\hbar v_\mathrm{F} \cos\theta\, \frac{da}{dx} - 2 i E a +\Delta^* a^2 -\Delta =0, \notag \\
\hbar v_\mathrm{F} \cos\theta\, \frac{db}{dx} + 2 i E b -\Delta b^2 +\Delta^* =0, \label{b}
\end{gather}
where $v_\mathrm{F}$ is the absolute value of the Fermi velocity $\mathbf{v}_\mathrm{F}$, and $\theta$ denotes the angle
between $\mathbf{v}_\mathrm{F}$ and the $x$ axis.

In the tunneling limit, the DoS is calculated at an impenetrable interface. Let us consider, e.g., the right
superconductor (the right half-space). We need to find the low-energy DoS in two cases: 1)~in the vicinity of a nodal
direction, so that $E, \Delta(\theta) \ll \widetilde\Delta_0$, 2)~at a gapped direction, so that $E \ll \Delta(\theta)$.
In the first case, the spatial scale $\xi_E = \hbar v_\mathrm{F} \cos\theta / |\mathcal{E}_+|$ on which the
quasiclassical Green functions vary (we denote $\mathcal{E}_\pm = \sqrt{E^2- \left| \Delta (\infty,\pm \theta)
\right|^2}$), is much larger than the coherence length $\xi = \hbar v_\mathrm{F} / 2\pi k_B T_c$ on which variations of
$\Delta$ occur. This allows us to regard $\Delta$ as constant when integrating Eqs. (\ref{b}) over $x$. In other words,
the functions $a$ and $b$ at low energies do not feel the suppression of $\Delta$ near the interface, because it takes
place on a small scale. In the second case, the spatially dependent parts of $a$ and $b$ are proportional to
$E/\Delta(\theta) \ll 1$ and hence small. Thus $a$ and $b$ at the interface are equal to their bulk values, as if
$\Delta$ was constant.

Thus we can regard $\Delta(x,\theta)$ as equal to the bulk value $\Delta_0(\theta)=\Delta(\infty,\theta)$. The
integration of the functions $a$ and $b$ over $x$ in Eqs. (\ref{b}) is stable only in the directions determined by the
sign of $\cos\theta$. At $\cos\theta >0$, the function $b(x,\theta)$ is stably integrated from $x=\infty$ to the
interface ($x=0$), hence
\begin{equation} \label{b_surface}
b(0,\theta) = b(\infty,\theta) = i\, \frac{E-\mathcal{E}_+ \sgn E}{\Delta_0(\theta)}.
\end{equation}
At the same time at $\cos\theta >0$, the function $a$ is stably integrated from the interface to $x=\infty$. Therefore
to find $a(0,\theta)$, we consider the trajectory directed along $\pi-\theta$. Since $\cos(\pi-\theta) <0$, the function
$a$ is stably integrated from $x=\infty$ to the interface. Finally, the direction $\pi-\theta$ is converted to $\theta$
upon reflection at the specular interface:
\begin{equation}
a(0,\theta) = a(0,\pi-\theta) =a(\infty,\pi-\theta) = i\, \frac{E-\mathcal{E}_- \sgn E}{\Delta_0^*(-\theta)}.
\end{equation}
As a result, the DoS $\nu = \Re g$ at the interface is
\begin{equation} \label{g_interface}
\nu (E,\theta) = \Re \frac{|E| \left( \mathcal{E}_+ +\mathcal{E}_- \right)} {E^2 - \Delta_0(\theta) \Delta_0^*(-\theta)
+ \mathcal{E}_+ \mathcal{E}_-}.
\end{equation}
The gap in the spectrum is $E_g(\theta)=\min\left( \left| \Delta_0(\theta) \right|, \left| \Delta_0(-\theta) \right|
\right)$.

The DoS is symmetric, $\nu(\theta) =\nu(-\theta)$, because the Green functions are continuous upon reflection. Thus in
each superconductor there are two ``truly'' nodal directions $\theta_i$ ($i=1,2$) in the interval $-\pi/2 <\theta <
\pi/2$, and also two ``induced'' nodal directions $-\theta_i$. Near a nodal direction $E_g(\theta)= 2 \widetilde\Delta_0
|\theta-\theta_i|$. Along a truly nodal direction, the gap vanishes and the DoS is the same as in the normal metal,
$\nu(E)=1$. For an ``induced'' nodal direction this is so only near the interface.

In the left superconductor, the truly nodal directions are $\theta_{1,2} = \alpha \pm 45^\circ$. Due to the mirror
symmetry, the truly nodal directions of the right superconductor coincide with the induced nodal directions of the left
one, and vice versa. In total, there are four nodal directions in the junction, which are symmetric with respect to the
interface normal.

In this situation, the transport is due to the processes of the following type. An electron moving along a truly nodal
direction $\theta_i$ of the left superconductor, tunnels into an induced nodal direction of the right superconductor
(see Fig.~\ref{fig:system}). However, the electron cannot escape into the bulk of the right superconductor because
$\Delta_R (\theta_i) \ne 0$. Therefore the electron experiences the Andreev reflection \cite{4:Andreev}; the hole
returns to the interface, and after reflection at the interface escapes into the bulk along the truly nodal direction
$-\theta_i$. In this process, the total current into the bulk of the right superconductor is composed of the Cooper pair
along $\theta_i$ and the hole along $-\theta_i$, which is overall equivalent to the transfer of one electron.

The nodal contribution to the current (\ref{I_V_general}) appears only due to integrating in the vicinity of the nodal
directions where $E_g < \hbar \omega$. The DoS near the nodal directions at small energies can be found from Eq.
(\ref{g_interface}). Below we distinguish the general case when $\Delta_0 (\theta) \ne \pm \Delta_0 (-\theta)$, and two
special cases: $\Delta_0(\theta) = \Delta_0(-\theta)$ (at $\alpha = 0^\circ$) and $\Delta_0 (\theta) =-\Delta_0
(-\theta)$ (at $\alpha = 45^\circ$).

At $\alpha=0^\circ$, the truly nodal and induced nodal directions coincide in each superconductor, and Eq.
(\ref{g_interface}) yields the BCS-like DoS:
\begin{equation} \label{nu_0}
\nu_{0^\circ} (E,\theta) = \Re \frac{ |E| } { \sqrt{E^2- \left| \Delta_0(\theta) \right|^2} } .
\end{equation}

At $\alpha=45^\circ$, the truly nodal and induced nodal directions again coincide, and Eq. (\ref{g_interface}) yields
the DoS of the inverse BCS type:
\begin{equation} \label{nu_45}
\nu_{45^\circ} (E,\theta) = \Re \frac{ \sqrt{E^2- \left| \Delta_0(\theta) \right|^2} } { |E| } .
\end{equation}

Finally, if $|\alpha| \gg \hbar\omega / \widetilde\Delta_0$ and $45^\circ - |\alpha| \gg \hbar\omega /
\widetilde\Delta_0$ (i.e., $\alpha$ is not too close to $0^\circ$ and $\pm 45^\circ$), then $\Delta_0(\theta)$ in the
essential angle of the width $\delta\theta =\hbar \omega / \widetilde\Delta_0$ around a nodal direction is much smaller
than $\Delta_0(-\theta)$. Then in the region of energies that contribute to the quasiparticle current, $\left|
\Delta_0(\theta) \right| < |E| < \hbar\omega \ll \left| \Delta_0(-\theta) \right|$, the DoS is again given by the
inverse BCS formula:
\begin{equation} \label{nu_i}
\nu_g (E, \theta\approx \theta_i) =\Re \frac{ \sqrt{E^2-\left| \Delta_0(\theta) \right|^2} } { |E| } .
\end{equation}

\begin{figure}
 \centerline{\includegraphics[width=80mm]{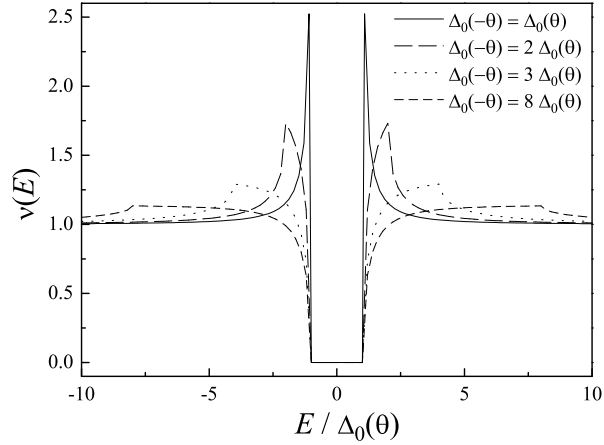}}
\caption{Density of states following from Eq. (\ref{g_interface}). The energy is normalized to $\Delta_0(\theta)$, while
$\Delta_0(-\theta)$ is varied.}
 \label{fig:dos_DID}
\end{figure}

Figure~\ref{fig:dos_DID} demonstrates the DoS at different angles $\theta$, which are parametrized by different ratios
$\Delta_0(-\theta) / \Delta_0(\theta)$. At $\Delta_0(-\theta) = \Delta_0(\theta)$, the DoS has the BCS-like square-root
singularity near $E_g$ [see Eq. (\ref{nu_0})]. At $\Delta_0(-\theta) \ne \Delta_0(\theta)$, the DoS has the inverse-BCS
behavior near $E_g$ [see Eq. (\ref{nu_i})].

Inserting Eqs. (\ref{nu_0})--(\ref{nu_i}) into Eq. (\ref{I_V_general}), we obtain:
\begin{equation} \label{I_V_J}
I(\hbar\omega/e) = \frac{A(\alpha)}{e R_N} \frac{(\hbar\omega)^2}{\widetilde\Delta_0} \sum_{i=1,2} \frac{D(\theta_i)
\cos\theta_i}{\widetilde D},
\end{equation}
where $\theta_{1,2} =\alpha \pm 45^\circ$ and $A$ is a number, which depends on the orientation of crystals: $A(0^\circ)
\approx 0.46$, $A(45^\circ) \approx 0.19$, and $A(\alpha) = 2 A(45^\circ) \approx 0.37$ when $\alpha$ is not too close
to $0^\circ$ or $\pm 45^\circ$.

In Refs.~\cite{4:Bruder,4:BGZ}, the quadratic current--voltage characteristic, $I\propto \omega^2$, was obtained for the
case of aligned nodal directions (i.e., for the $\alpha/\alpha$ orientation).

\section{Estimate}

Equations (\ref{Q}), (\ref{I_V_J}) yield:
\begin{equation} \label{Q_1}
Q = \Biggl( \frac{2\pi} {\delta\varphi^2 A(\alpha) \sum\limits_{i=1,2} D(\theta_i) \cos\theta_i / \widetilde D} \Biggr)
\frac{R_N}{R_q} \frac{\widetilde\Delta_0}{\Delta_t}.
\end{equation}

To proceed further, we need to estimate the tunnel splitting $\Delta_t$ (see Fig.~\ref{fig:2well}). This is easily done
due to the formal equivalence between quantum mechanics of a Josephson junction described by the phase $\varphi$ and a
particle described by the coordinate $x$ \cite{4:Tinkham}. To employ this analogy, we should substitute
\begin{equation}
m \leftrightarrow \frac{\hbar^2}{8 E_C},
\end{equation}
where $m$ is the particle mass and $E_C = e^2 /2 C$ is the charging energy of the Josephson junction ($C$ is its
capacitance). For the estimate, we assume that the second harmonic dominates in the energy--phase relation, $U(\varphi)=
E_J (1+\cos 2\varphi)$, and the energy of the levels is small compared to $E_J$. Then the attempt frequency is $\omega_0
= 4 \sqrt{2 E_J E_C}/ \hbar$, the tunneling action is calculated between the points $\varphi=-\pi/2$ and $\pi/2$, and we
obtain:
\begin{equation} \label{Delta_t}
\Delta_t = \frac{ 4 \sqrt{2 E_J E_C} }\pi \exp\left( -\sqrt{ \frac{2 E_J}{E_C} } \right) .
\end{equation}

To obtain a numerical estimate, we take the characteristics of the junction as in the experiment of Il'ichev et al.
\cite{4:Il'ichev}. The capacitance of the junction is $C \sim 10^{-14}$\,F \cite{4:Grajcar}, hence $E_C / k_B \sim
0.1$\,K. The characteristic Josephson energy is of the order of several Kelvin. For estimate, we take $2 E_J / k_B =
7$\,K. The resistance of the interface is $R_N \sim 50\,\Omega$ \cite{4:Grajcar}.

As a result, $\Delta_t / k_B \sim 2.5\cdot 10^{-4}$\,K. Finally, we estimate $\delta\varphi\sim\pi$, $\widetilde\Delta_0
/ k_B \sim 200$\,K, and assume a thin $\delta$-functional barrier with transparency $D(\theta) = D_0 \cos^2 \theta$,
then the quality factor is $Q \sim 10^3\div 10^4$. Here we have retained only the order of magnitude for $Q$, because we
cannot expect a higher accuracy in the case when important characteristics of the junction (e.g., $C$ and $E_J$) are
known only by the order of magnitude. We also made an essential assumption that the second Josephson harmonic dominates,
which is the most favorable case, hence the obtained estimate is ``optimistic''.

The latter assumption can be realized under special conditions, while in a more common situation the first and the
second harmonics are of the same order. Estimates for this case were done in a recent work \cite{4:Tzalenchuk}, where
the characteristics of mesoscopic junctions between high-$T_c$ superconductors were experimentally studied and
theoretically analyzed. A characteristic value of $\Delta_t \sim 0.1$\,K was reported under the conditions that
correspond to $R_N \sim 100\,\Omega$. Assuming such parameters for the mirror junction, we obtain $Q \sim 10 \div 10^2$,
which is a more realistic estimate than that for the experiment of Ref.~\cite{4:Il'ichev}.

The above estimates for $Q$ are very different. At the same time, a general consequence of Eq. (\ref{Q_1}) is that the
quality factor grows as the splitting $\Delta_t$ becomes smaller. We note in this respect, that the values of the
critical current (and hence the Josephson energy) measured in Refs.~\cite{4:Il'ichev,4:Tzalenchuk}, are much smaller
than expected. If the critical current is enhanced to the expected value, then $\Delta_t$ decreases, which finally leads
to an increase of $Q$.

If $\alpha\ne 0^\circ$, the low-energy quasiparticles are presented not only by the nodal quasiparticles, but also by
the midgap states (MGS) with zero energy \cite{4:Hu}. In the case of specular interface and clean superconductors,
considered in this chapter, the DoS corresponding to the MGS is proportional to $\delta(E)$, hence the MGS on the two
sides of the interface do not overlap and do not contribute to the current at a finite voltage.

In the asymmetric case, when $\alpha_L \ne \pm \alpha_R$ (precisely speaking, when $\bigl| |\alpha_L| - |\alpha_R|
\bigr| > \hbar \omega / \widetilde\Delta_0$), the nodal directions of the left and right superconductors do not match
each other. At first sight, this solves the problem of decoherence since the transport from nodal to nodal direction is
suppressed. However, a more important transport ``channel'' arises: between the nodal directions and the MGS. This leads
to a stronger decoherence than in the symmetric case.

In the mirror junction, the MGS contribute to the quasiparticle current if they are split and/or broadened
\cite{4:splitting}. To take into account the contribution of the MGS into decoherence, the present approach should be
considerably modified. This issue requires a separate study.

The low-energy quasiparticles (and dissipation due to them) can be suppressed due to finite size of the \textit{d}-wave
superconductors. This issue is schematically discussed in Sec.~\ref{sec:finite_size} below.

\section{Finite size effects}
\label{sec:finite_size}

\begin{figure}
 \centerline{\includegraphics[width=80mm]{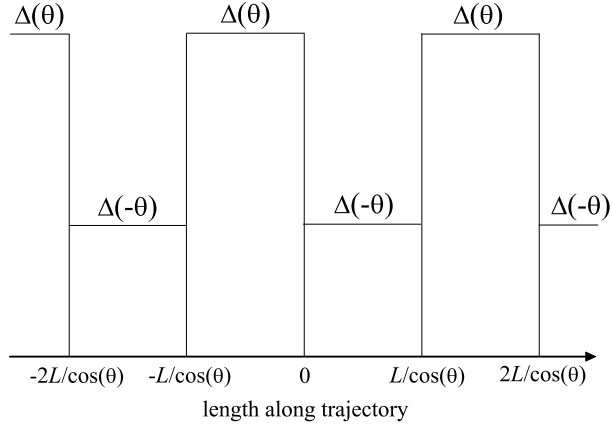}}
\caption{Schematic illustration of a periodic order parameter profile along a trajectory in a \textit{d}-wave layer of
width $L$. We have chosen the trajectory with the incidence angle $\theta$, hence the part of the trajectory between two
successive reflections has the length $L/\cos\theta$. Actually, at $\alpha\ne 0^\circ$ the order parameter is suppressed
near the interfaces, however, to capture the qualitative effects we neglect this suppression and consider the model of
piecewise constant $\Delta$.}
 \label{fig:AKP}
\end{figure}

The effect of finite size on the low-energy quasiparticles in DID junctions can be most easily studied in the
quasi-one-dimensional geometry assuming finite thickness $L$ (the dimension in the $x$ direction) of the \textit{d}-wave
superconductors. Then each trajectory is successively reflected from the I interface and from the outer surface of the
layer. This leads to periodicity of the order parameter profile along a trajectory (see Fig.~\ref{fig:AKP}).

The finite size leads to two main effects:

\begin{enumerate}

\item[1.] For nodal quasiparticles (this corresponds to $\Delta(-\theta)=0$ in Fig.~\ref{fig:AKP}), the Andreev
quantization arises, similarly to the SNS junction \cite{4:Andreev,4:Kulik}. The system of equidistant energy levels
appears, and the level spacing is $\pi v_\mathrm{F} \cos\theta / L$ \cite{4:Andreev}, while the lowest level is $\pi
v_\mathrm{F} \cos\theta / 2 L$ \cite{4:Kulik}. Thus, the energy gap is formed.

\item[2.] The MGS, which exist if $\Delta(\theta)$ and $\Delta(-\theta)$ have different signs, split. This effect,
similar to the level splitting in a double-well quantum potential, is due to the overlap of the localized states across
the lowest of the barriers ($\Delta(-\theta)$ in Fig.~\ref{fig:AKP}). To estimate the corresponding splitting, below we
consider a simplified model shown in Fig.~\ref{fig:AKP_two}.

\end{enumerate}

\subsection{Tunnel splitting of the zero Andreev level}

\begin{figure}
 \centerline{\includegraphics[width=70mm]{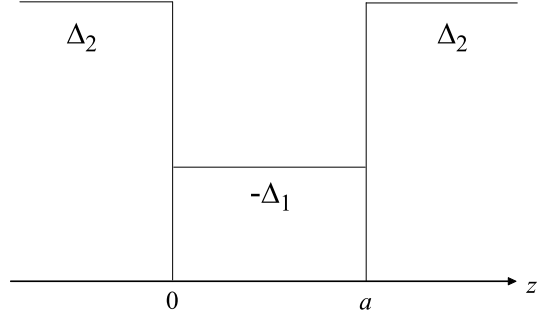}}
\caption{The profile of the order parameter, corresponding to two point contacts.}
 \label{fig:AKP_two}
\end{figure}

A classical quantum-mechanical problem is splitting of the level in a symmetric double-well potential. A similar effect
takes place in the case of two coupled superconducting point contacts. Let us consider the splitting of the zero-energy
levels in two coupled point contacts with the phase difference of $\pi$ across each contact (see
Fig.~\ref{fig:AKP_two}).

We shall use the technique of the Bogolyubov -- de Gennes equation:
\begin{equation}
\begin{pmatrix} -\frac 1{2m} \frac{d^2}{dx^2} -\mu & \Delta \\ \Delta^* & \frac 1{2m} \frac{d^2}{dx^2} +\mu
\end{pmatrix}
\begin{pmatrix} u \\ v \end{pmatrix} = E \begin{pmatrix} u \\ v \end{pmatrix}.
\end{equation}
In the Andreev approximation \cite{4:Andreev}, we separate rapid oscillations of the solution:
\begin{equation} \label{Aa_sep}
\begin{pmatrix} u \\ v \end{pmatrix} = \begin{pmatrix} u_0 \\ v_0 \end{pmatrix} \exp(ik_F x),
\end{equation}
and retain only the first derivatives of the slow functions $u_0$ and $v_0$. The Bogolyubov -- de Gennes equation then
takes the form
\begin{equation} \label{Aa}
\begin{pmatrix} -i v_\mathrm{F} \frac d{dx} & \Delta \\ \Delta^* & i v_\mathrm{F} \frac d{dx} \end{pmatrix}
\begin{pmatrix} u_0 \\ v_0 \end{pmatrix} = E \begin{pmatrix} u_0 \\ v_0 \end{pmatrix}.
\end{equation}

The boundary conditions are the continuity of $u_0$ and $v_0$ at the interfaces. The sign of the momentum is a good
quantum number, and we can first consider the anzatz~(\ref{Aa_sep}), which corresponds to the momentum directed to the
right. In the general case of arbitrary phase difference, the energy levels for the momentum directed to the left can
then be obtained after inverting the sign of the phase. At the same time, in the case of real $\Delta$ (when the phase
difference is either $0$ or $\pi$), the energy levels are degenerate with respect to the momentum direction.

The solutions in the three spatial regions are
\begin{align}
\begin{pmatrix} u_0 \\ v_0 \end{pmatrix} &= A \begin{pmatrix} \Delta_2 \\ E+i\sqrt{\Delta_2^2-E^2}\end{pmatrix}
e^{q_2 z},\qquad z<0, \notag \\
\begin{pmatrix} u_0 \\ v_0 \end{pmatrix} &= B \begin{pmatrix} -\Delta_1 \\ E+i\sqrt{\Delta_1^2-E^2}
\end{pmatrix}
e^{q_1 z} + C \begin{pmatrix} -\Delta_1 \\ E-i\sqrt{\Delta_1^2-E^2} \end{pmatrix} e^{-q_1 z},\quad 0<z<a, \notag \\
\begin{pmatrix} u_0 \\ v_0 \end{pmatrix} &= D \begin{pmatrix} \Delta_2 \\ E-i\sqrt{\Delta_2^2-E^2}
\end{pmatrix}
e^{-q_2 z},\qquad a<z,
\end{align}
where
\begin{equation}
q_{1,2}=\frac{\sqrt{\Delta_{1,2}^2-E^2}}{v_\mathrm{F}}.
\end{equation}
Substituting these solutions into the boundary conditions at the two interfaces ($x=0$ and $x=a$), we obtain four linear
equations for the four coefficients $A$, $B$, $C$, and $D$. The spectrum is determined from the condition that the
corresponding determinant is zero. The straightforward algebra leads to the equation
\begin{equation}
\left( E^2 +\Delta_1 \Delta_2 \right) \tanh (q_1 a)- \sqrt{ (\Delta_1^2-E^2) (\Delta_2^2-E^2) } =0.
\end{equation}
In the limit $a \Delta_1/v_\mathrm{F} \gg 1$, the splitting is small, and the levels are
\begin{equation} \label{A_split}
E=\pm \frac{2 \Delta_1 \Delta_2}{\Delta_1+\Delta_2} \exp\left( -\frac{a \Delta_1}{v_\mathrm{F}} \right).
\end{equation}

\section{Conclusions}

In this chapter, we have proposed an approach that allows to calculate the decoherence time due to nodal quasiparticles
in the DID junctions, which can be used as phase qubits. The dissipation in the junctions of mirror orientation is
weaker than in the asymmetric ones. We find the superohmic dissipation with $s=2$ in the mirror junction, which becomes
weak at small splitting of the ground state. The superohmic case is most favorable (compared to ohmic and subohmic) for
possible qubit applications. For available experimental data, we estimate the quality factor. Finally, we illustrate the
possibility to suppress quasiparticle decoherence due to finite size of the \textit{d}-wave superconductors.

\clearpage
\renewcommand{\@evenhead}{}

\appendix

\chapter[Interpretation of $\tau$ as escape times]{\huge Interpretation of $\tau$ as escape times}
\label{app:tau}

\renewcommand{\@evenhead}
    {\raisebox{0pt}[\headheight][0pt]
     {\vbox{\hbox to\textwidth{\thepage \hfil \strut \textit{Appendix \thechapter}}\hrule}}
    }

\renewcommand{\@oddhead}
    {\raisebox{0pt}[\headheight][0pt]
     {\vbox{\hbox to\textwidth{\textit{\leftmark} \strut \hfil \thepage}\hrule}}
    }

The quantities $\tau_S$ and $\tau_N$ introduced in Eq. (\ref{tau_original}) may be interpreted as escape times from the
corresponding layers. The arguments go as follows.

If the layers are thin, then the diffusion inside the layers is ``fast'' and the escape time from a layer is determined
by the interface resistance. The time of penetration through a layer or the interface is determined by the corresponding
resistance: $R_{S(N)}$ or $R_B$, hence the diffusion is ``fast'' if $R_{S(N)} \ll R_B$.

Let us use the detailed balance approach, and consider an interval of energy $dE$. In the S layer, the charge in this
interval is $Q_S= e \nu_S\, dE\, \mathcal{A} d_S$. Let us define the escape time from the S layer $t_S$, so that the
current from S to N is equal to $Q_S/t_S$. On the other hand, this current can be written as $dE/eR_B$, hence
\begin{equation}
\frac{Q_S}{t_S}=\frac{dE}{eR_B},
\end{equation}
and we immediately obtain
\begin{equation}
t_S = \frac{\sigma_S d_S R_B \mathcal{A}}{D_S}.
\end{equation}
Similarly, we obtain the expression for the escape time from the F layer $t_N$. As a result, the relations between the
quantities $\tau$ defined in Eq. (\ref{tau_original}) and the escape times $t$ are simply
\begin{equation}
\tau_S = 2 t_S,\qquad \tau_N = 2 t_N.
\end{equation}

\clearpage
\renewcommand{\@evenhead}{}

\chapter[Analytical results for SF bilayers]{\huge Analytical results for SF bilayers} \label{ap:sec:analytics}

\renewcommand{\@evenhead}
    {\raisebox{0pt}[\headheight][0pt]
     {\vbox{\hbox to\textwidth{\thepage \hfil \strut \textit{Appendix \thechapter}}\hrule}}
    }

\renewcommand{\@oddhead}
    {\raisebox{0pt}[\headheight][0pt]
     {\vbox{\hbox to\textwidth{\textit{\leftmark} \strut \hfil \thepage}\hrule}}
    }

\section{Limit of thin S layer}

(i) When $d_S \ll \xi_S$ and $h \gg \pi T_{cS}$, problem (\ref{self_cons})--(\ref{usadel_b}) can be solved analytically.
The first of the above conditions implies that $\Delta$ can be considered constant, and $F^+$ weakly depends on the
spatial coordinate; so $F^+(x,\omega_n)=2\Delta/\omega_n + A(\omega_n)\cosh(k_S[x-d_S])$. The boundary conditions
determine the coefficient $A$; as a result
\begin{equation} \label{thin_S_F^+}
F^+(\omega_n)\equiv F^+(x=0,\omega_n)=\frac{2\Delta}{\omega_n} \left[
\frac{A_S(\omega_n)}{A_S(\omega_n)+W(\omega_n)}\right],
\end{equation}
where $k_S$, $A_S$, and $W$ are defined in Eq. (\ref{W_def}). Finally, the self-consistency equation for $T_c$ takes the
form
\begin{equation} \label{thin_S}
\ln\frac{T_{cS}}{T_c} = \Re \psi\left( \frac 12 +\frac\gamma 2 \frac{\xi_S}{d_S} \frac 1{\gamma_B+B_h}
\frac{T_{cS}}{T_c} \right) -\psi\left( \frac 12 \right),
\end{equation}
where $B_h$ does not depend on $\omega_n$ due to the condition $h \gg \pi T_{cS}$:
\begin{equation}
B_h=\left[ \widetilde k_h \xi_F \tanh (\widetilde k_h d_F) \right]^{-1},\qquad \widetilde k_h \approx \frac 1
{\xi_F}\sqrt{\frac{ih}{\pi T_{cS}}}.
\end{equation}

In the limiting case of perfect interface ($\gamma_B=0$) and thick F layer ($d_F\to \infty$), Eq. (\ref{thin_S})
reproduces the result of Ref.~\cite{app:Buzdin_EL}.

(ii) If the F layer is also thin, $d_F\ll \sqrt{D_F/2 h}$, Eq. (\ref{thin_S}) is further simplified:
\begin{equation}
\ln\frac{T_{cS}}{T_c} = \Re \psi\left( \frac 12 + \frac{\tau_F}{\tau_S} \left[ \frac 1{-i+\tau_F h} \right]
\frac{h}{2\pi T_c} \right) -\psi\left( \frac 12 \right),
\end{equation}
where $\tau_S$, $\tau_F$ are defined similarly to Eq. (\ref{tau_original}):
\begin{equation} \label{tau}
\tau_S = \frac{2 d_S R_B \mathcal{A}}{\rho_S D_S},\qquad \tau_F = \frac{2 d_F R_B \mathcal{A}}{\rho_F D_F},
\end{equation}
and have the physical meaning of the escape time from the corresponding layer. They are related to the quantities
$\gamma$, $\gamma_B$ used in the body of the paper as
\begin{equation} \label{tau_gamma}
\tau_S =\frac{\gamma_B}\gamma \frac 1{\pi T_{cS}} \frac{d_S}{\xi_S},\qquad \tau_F =\gamma_B \frac 1{\pi T_{cS}}
\frac{d_F}{\xi_F}.
\end{equation}

(iii) If the S layer is thin, $d_S\ll \xi_S$, and the SF interface is opaque, $\gamma_B\to\infty$, the critical
temperature of the bilayer only slightly deviates from $T_{cS}$. In this limit Eq. (\ref{thin_S_F^+}) applies with
$W=\gamma/\gamma_B \ll 1$, and we finally obtain:
\begin{equation} \label{Tc_opaque}
T_c = T_{cS}-\frac\pi{4\tau_S}.
\end{equation}
Interestingly, characteristics of the F layer ($d_F$, $h$, etc.) do not enter the formula. In particular, this formula
is valid for an SN bilayer \cite{app:McMillan} (where N is a nonmagnetic normal material, $h=0$) because Eq.
(\ref{Tc_opaque}) was obtained without any assumptions about the value of the exchange energy.

\section{Cooper limit}

In the Cooper limit, when both layers are very thin [$d_S\ll \sqrt{D_S / 2 \omega_D}$, $d_F \ll \min(\sqrt{D_F /
2\omega_D}, \sqrt{D_F/2 h})$, with $\omega_D$ the Debye energy of the S metal] and the interface is transparent, the
bilayer is equivalent to a homogeneous superconducting layer with internal exchange field. This layer is described by
effective parameters: the order parameter $\Delta^{(\mathrm{eff})}$, the exchange field $h^{(\mathrm{eff})}$, and the
pairing constant $\lambda^{(\mathrm{eff})}$. In this subsection we develop the ideas of Ref.~\cite{app:Bergeret-Efetov},
demonstrate a simple derivation of this description, and find the limits of its applicability.

The Usadel equations (\ref{U_1}), (\ref{U_2}) for the two layers can be written as a single equation:
\begin{equation}
\label{U_thin} \frac{D_F\theta(-x)+D_S\theta(x)}{2}\frac{d^2 F}{dx^2} -|\omega_n|F-ih\sgn(\omega_n)
\theta(-x)F+\Delta\theta(x)=0,
\end{equation}
where $\theta$ is the Heaviside function [$\theta(x>0)=1$, $\theta(x<0)=0$]. The self-consistency equation (\ref{U_3})
can be rewritten as
\begin{equation}
\Delta(x)=\lambda \theta(x) \pi T\sum_{\omega_n} F(x,\omega_n),
\end{equation}
where $\lambda$ is the pairing constant.

First, we consider the ideal SF interface: $\gamma_B=0$ [see Eq. (\ref{bound_2})], then $F(x)$ is continuous at the
interface and nearly constant across the \textit{whole} bilayer, i.e., $F_S(x)\approx F_F(x)=F$. Applying the integral
operator to Eq. (\ref{U_thin}):
\begin{equation}
\frac{\nu_F}{\nu_Sd_S+\nu_F d_F}\int_{-d_F}^0 dx + \frac{\nu_S}{\nu_Sd_S+\nu_F d_F}\int_0^{d_S}dx
\end{equation}
(here $\nu$ is the normal-metal density of states), and cancelling gradient terms due to the boundary condition
(\ref{bound_1}), we obtain the equations describing a homogeneous layer:
\begin{gather}
-|\omega_n| F(\omega_n)-i h^{(\mathrm{eff})} \sgn(\omega_n) F(\omega_n)+
\Delta^{(\mathrm{eff})}=0, \label{homo} \\
\Delta^{(\mathrm{eff})}=\lambda^{(\mathrm{eff})} \pi T\sum_{\omega_n} F(\omega_n),
\end{gather}
with the effective parameters (see also Ref.~\cite{app:Bergeret-Efetov}):
\begin{gather}
\label{U_eff} h^{(\mathrm{eff})} = \frac{\tau_F}{\tau_S+\tau_F} h,\qquad
\Delta^{(\mathrm{eff})} = \frac{\tau_S}{\tau_S+\tau_F} \Delta,\\
\lambda^{(\mathrm{eff})} = \frac{\tau_S}{\tau_S+\tau_F} \lambda,\qquad T_{cS}^{(\mathrm{eff})} = \frac{\exp(C)}\pi
2\omega_D \exp\left( -\frac 1{\lambda^{(\mathrm{eff})}} \right),\notag
\end{gather}
where $C$ is Euler's constant and $T_{cS}^{(\mathrm{eff})}$ is the critical temperature of the layer in the absence of
ferromagnetism (i.e., at $h^{(\mathrm{eff})} = 0$). The critical temperature is determined by the equation
\begin{equation}
\ln\frac{T_{cS}^{(\mathrm{eff})}}{T_c} = \Re \psi\left( \frac 12 + i\frac{h^{(\mathrm{eff})}}{2\pi T_c} \right)
-\psi\left( \frac 12 \right).
\end{equation}

Actually, the description in terms of effective parameters (\ref{U_eff}) is applicable at an arbitrary temperature
(i.e., when the Usadel equations are nonlinear) and has a clear physical interpretation: the superconducting ($\Delta$,
$\lambda$) and ferromagnetic ($h$) parameters are renormalized according to the part of time spent by quasiparticles in
the corresponding layer. This physical picture is based on interpretation of $\tau$ as escape times, which we present in
the next subsection.

Now we discuss the applicability of the above description for a nonideal interface ($\gamma_B\neq 0$). In this case $F$
is nearly constant in each layer, but these constants are different: $F_S(x)\approx F_S + C_S (x-d_S)^2$, $F_F(x)\approx
F_F+C_F(x+d_F)^2$, where $|F_S|\gg |C_S| d_S^2$ and $|F_F|\gg |C_F| d_F^2$. Using the Usadel equation (\ref{U_thin}) and
the boundary conditions (\ref{bound_1}), (\ref{bound_2}), we find the difference $\delta F\equiv F_S-F_F$:
\begin{equation}
\delta F =\frac\Delta{\displaystyle \frac 1{\tau_S} +|\omega_n| \left[ 1+\frac 1{\tau_F \left( \left| \omega_n \right|+i
h \sgn\omega_n \right) } \right] }.
\end{equation}
Finally, the homogeneous description is valid when $|\delta F/F|\ll 1$ [with $F$ determined by Eq. (\ref{homo})], which
yields:
\begin{equation}
\max(h, \omega_D) \max(\tau_S,\tau_F)\ll 1
\end{equation}
(here $\omega_n \sim \omega_D$ has been taken as the largest characteristic energy scale in the quasi-homogeneous
bilayer).

\clearpage
\renewcommand{\@evenhead}{}

\chapter[Applicability of the single-mode approximation for calculating $T_c$ in SF bilayers]
{\huge Applicability of the single-mode approximation for calculating $T_c$ in SF bilayers} \label{ap:sma}

\renewcommand{\@evenhead}
    {\raisebox{0pt}[\headheight][0pt]
     {\vbox{\hbox to\textwidth{\thepage \hfil \strut \textit{Appendix \thechapter}}\hrule}}
    }

\renewcommand{\@oddhead}
    {\raisebox{0pt}[\headheight][0pt]
     {\vbox{\hbox to\textwidth{\textit{\leftmark} \strut \hfil \thepage}\hrule}}
    }

\section{Case study}

As pointed out in Sec.~\ref{sec:multi-mode}, the single-mode approximation (SMA) is applicable only if the parameters of
the bilayer are such that $W$ [see Eq. (\ref{W_def})] can be considered $\omega_n$-independent. An example is the case
when $\gamma_B \gg |B_h|$, hence $W= \gamma/ \gamma_B$.

The condition $\gamma_B \gg |B_h|$ can be written in a simpler form; to this end we should estimate $|B_h|$. We
introduce the real and imaginary parts of $\widetilde k_h$: $\widetilde k_h= k_h' +ik_h''$, and note that $k_h'>k_h''$.
Then using the properties of the trigonometric functions and the estimate $\tanh x \sim \min(1,x)$ we obtain
\begin{equation} \label{est1}
|B_h| \sim \left[ k_h' \xi_F \tanh (k_h' d_F) \right]^{-1},
\end{equation}
and finally cast the condition $\gamma_B \gg |B_h|$ into the form
\begin{equation} \label{condition_SMA}
\frac 1{\gamma_B} \ll \min\left\{ \sqrt{\max\left( \frac{T_c}{T_{cS}}, \frac{h}{\pi T_{cS}} \right)}; \frac{d_F}{\xi_F}
\max\left( \frac{T_c}{T_{cS}}, \frac{h}{\pi T_{cS}} \right) \right\},
\end{equation}
where the ratio $T_c /T_{cS}$ originates from $\omega_n / \pi T_{cS}$ with $\omega_n \sim \pi T_c$ as the characteristic
energy scale in the bilayer.

If condition (\ref{condition_SMA}) is satisfied, then the SMA is valid and $T_c$ is determined by the equations
\begin{gather}
\ln\frac{T_{cS}}{T_c}=\psi\left(\frac 12+ \frac{\Omega^2}2 \frac{T_{cS}}{T_c}\right)- \psi\left(\frac
12\right),  \label{sc}\\
\Omega \tan\left(\Omega \frac{d_S}{\xi_S} \right) = \frac\gamma{\gamma_B}. \label{bc}
\end{gather}

These equations can be further simplified in two limiting cases which we consider below.

(1) $\displaystyle \frac\gamma{\gamma_B} \frac{d_S}{\xi_S} \ll 1$:

in this case Eq. (\ref{bc}) yields $\Omega^2 =\frac\gamma{\gamma_B} \frac{\xi_S}{d_S}$, and Eq. (\ref{sc}) takes the
form
\begin{equation} \label{sma_res1}
\ln\frac{T_{cS}}{T_c}=\psi\left(\frac 12+ \frac 12 \frac\gamma{\gamma_B} \frac{\xi_S}{d_S} \frac{T_{cS}}{T_c} \right)-
\psi\left(\frac 12\right),
\end{equation}
which reproduces the $\gamma_B \gg |B_h|$ limit of Eq. (\ref{thin_S}).

(2) $\displaystyle \frac\gamma{\gamma_B} \frac{d_S}{\xi_S} \gg 1$:

in this case Eq. (\ref{bc}) yields $\Omega \frac{d_S}{\xi_S} =\frac\pi 2$, and Eq. (\ref{sc}) takes the form
\begin{equation} \label{sma_res2}
\ln\frac{T_{cS}}{T_c}=\psi\left(\frac 12+ \frac{\pi^2}8 \left[ \frac{\xi_S}{d_S} \right]^2 \frac{T_{cS}}{T_c} \right)-
\psi\left(\frac 12\right).
\end{equation}

Equations (\ref{sc})--(\ref{sma_res2}) can be used for calculating the critical temperature $T_c$ and the critical
thickness of the S layer $d_{cS}$ below which the superconductivity in the SF bilayer vanishes (i.e., $T_c=0$).

\section{Results for the critical temperature}

In the limit when $T_c$ is close to $T_{cS}$, Eqs. (\ref{sma_res1}), (\ref{sma_res2}) yield
\begin{equation} \label{T_c_weak_dev}
T_c =T_{cS} \left( 1-\frac{\pi^2}4 \frac\gamma{\gamma_B} \frac{\xi_S}{d_S}\right) \qquad\text{if~~}
\frac\gamma{\gamma_B} \ll \min\left( \frac{d_S}{\xi_S},\frac{\xi_S}{d_S} \right),
\end{equation}
and
\begin{equation}
T_c=T_{cS} \left[1 - \left( \frac{\pi^2}4 \frac{\xi_S}{d_S} \right)^2 \right]\qquad\text{if~~} \frac{d_S}{\xi_S} \gg
\max\left( 1, \frac{\gamma_B}\gamma \right).
\end{equation}
Using relations (\ref{tau_gamma}) one can check that result (\ref{T_c_weak_dev}) is equivalent to Eq. (\ref{Tc_opaque}).

\section{Results for the critical thickness}

The critical thickness of the S layer $d_{cS}$ is defined as the thickness below which there is no superconductivity in
the SF bilayer: $T_c ( d_{cS} )=0$.\footnote{When calculating $d_{cS}$, we assume that the phase transition is of the
second order. However, this issue may require a separate study. In principle, under some circumstances, the order of the
phase transition in SF systems can change from second to first one
--- see Sec.~\ref{sec:num_res}.} When $T_c \to 0$, Eq. (\ref{sc}) yields $\Omega =1/\sqrt{2\exp(C)}$ (where $C \approx
0.577$ is Euler's constant), and Eq. (\ref{bc}) takes the form
\begin{equation}
\frac 1{\sqrt{2\exp(C)}} \tan\left(\frac 1{\sqrt{2\exp(C)}} \frac{d_{cS}}{\xi_S} \right) = \frac\gamma{\gamma_B}.
\end{equation}
Explicit results for $d_{cS}$ can be obtained in limiting cases:
\begin{equation}
\frac{d_{cS}}{\xi_S} =2\exp(C) \frac\gamma{\gamma_B} \qquad\text{if~~} \frac\gamma{\gamma_B} \frac{d_S}{\xi_S} \ll 1,
\end{equation}
and
\begin{equation}
\frac{d_{cS}}{\xi_S} =\pi \sqrt{\frac{\exp(C)}2} \qquad\text{if~~} \frac\gamma{\gamma_B} \frac{d_S}{\xi_S} \gg 1.
\end{equation}

\clearpage
\renewcommand{\@evenhead}{}

 \chapter[Spatial dependence of the order parameter in SF bilayers]
 {\huge Spatial dependence of the order parameter in SF bilayers}
 \label{ap:sec:space dep expl}

\renewcommand{\@evenhead}
    {\raisebox{0pt}[\headheight][0pt]
     {\vbox{\hbox to\textwidth{\thepage \hfil \strut \textit{Appendix \thechapter}}\hrule}}
    }

\renewcommand{\@oddhead}
    {\raisebox{0pt}[\headheight][0pt]
     {\vbox{\hbox to\textwidth{\textit{\leftmark} \strut \hfil \thepage}\hrule}}
    }

According to the self-consistency equation, in the S layer the proximity order parameter $F(x,\tau=0)$ is proportional
to $\Delta(x)$:
\begin{equation}
F_S(x,\tau=0) = \frac{\Delta(x)}{\pi\lambda},
\end{equation}
where $\lambda$ is the pairing constant which can be expressed via the Debye energy:
\begin{equation}
\lambda^{-1} =\ln\left( \frac{2\exp(C) \omega_D}{\pi T_{cS}} \right).
\end{equation}
$\Delta(x)$ can be found as the eigenvector of the matrix $\hat L- \hat 1 \ln(T_{cS}/T_c)$ [see Eq. (\ref{det})],
corresponding to the zero eigenvalue.

After that we can express $F(x,\tau=0)$ in the F layer via $\Delta(x)$ in the superconductor. The Green function
$F_F(x,\omega_n)$ in the F layer is given by Eq. (\ref{F_F}), with $C(\omega_n)$ found from the boundary conditions:
\begin{equation}
C(\omega_n) = \left( \frac{B_h}{\gamma_B+B_h} \right) \frac{F_S(0,\omega_n)}{\cosh(\widetilde k_h d_F)}.
\end{equation}
The Green function at the S side of the SF interface is
\begin{equation}
F_S(0,\omega_n)=\frac{F_S^+(0,\omega_n)+F_S^-(0,\omega_n)}2.
\end{equation}
The symmetric part $F_S^+$ is given by Eq. (\ref{fsplus}). The antisymmetric part is
\begin{equation}
F_S^- = C^-(\omega_n) \cosh\left( k_S [x-d_S] \right),
\end{equation}
with $C^-(\omega_n)$ found from the boundary conditions:
\begin{equation}
C^-(\omega_n) = \left[ \frac{i \gamma\Im B_h}{A_S |\gamma_B+B_h|^2 +\gamma (\gamma_B+\Re B_h)} \right]
\frac{F_S^+(0,\omega_n)}{\cosh(k_S d_S)}.
\end{equation}
Finally, the proximity order parameter in the F layer is the Fourier transform [see Eq. (\ref{order_parameter})] of
\begin{gather}
F_F(x,\omega_n) = \left[ 1+ \frac{i \gamma\Im B_h}{A_S |\gamma_B+B_h|^2 +\gamma
(\gamma_B+\Re B_h)} \right] \notag\\
\times \left( \frac{B_h}{\gamma_B+B_h} \right) \frac{\cosh\left( \widetilde k_h [x+d_F] \right)}{\cosh(\widetilde k_h
d_F)} \int_0^{d_S} G(0,y;\omega_n) \Delta(y) dy.
\end{gather}

\clearpage
\renewcommand{\@evenhead}{}

\chapter*{\huge Summary}
\addcontentsline{toc}{chapter}{Summary}

\renewcommand{\@evenhead}
    {\raisebox{0pt}[\headheight][0pt]
     {\vbox{\hbox to\textwidth{\thepage \hfil \strut \textit{Summary}}\hrule}}
    }

\renewcommand{\@oddhead}
    {\raisebox{0pt}[\headheight][0pt]
     {\vbox{\hbox to\textwidth{\textit{Summary} \strut \hfil \thepage}\hrule}}
    }

The research described in this thesis deals with thermodynamic and transport properties of various superconducting
junctions. The systems under consideration are composed of conventional \textit{s}-wave superconductors in contact with
normal metal or ferromagnet, and of anisotropic \textit{d}-wave superconductors.\footnote{The notations: S
--- \textit{s}-wave superconductor, D
--- \textit{d}-wave superconductor, N --- normal metal, F --- ferromagnetic metal, I --- insulator, c
--- constriction.} Various aspects of the proximity and Josephson effects in such systems are studied theoretically.

When a superconductor contacts a normal metal, a number of phenomena known as the proximity effect takes place. The two
materials influence each other on a spatial scale of the order of the coherence length in the vicinity of the interface.
In particular, the superconducting correlations between quasiparticles are induced into the normal metal, because the
Cooper pairs penetrating into the normal metal have a finite lifetime there. Until they decay into two independent
electrons, they preserve the superconducting properties. Alternatively, the proximity effect can be viewed as resulting
from the fundamental process known as the Andreev reflection, at which an electron impinging from the normal metal onto
the interface with the superconductor is reflected back as a hole, while a Cooper pair is transferred into the
superconductor.

The Josephson effect is a macroscopic quantum effect that occurs when two superconductors are connected via a weak link,
which can be either a nonsuperconducting material (insulator or metal) or a geometrical constriction. The
superconducting condensate in each of the two weakly coupled superconductors is described by its wave function and the
corresponding phase. The essence of the Josephson effect is the appearance of the supercurrent (nondissipative current)
between the superconductors at finite phase difference $\varphi$ between them. The supercurrent can flow in the absence
of voltage, and in the simplest case is given by the sinusoidal Josephson relation $I = I_c \sin\varphi$. The quantity
$I_c$ (or, more generally, the maximal supercurrent) is called the critical current.

\section*{Superconductivity in thin SN bilayers}

In Chapter~1, the proximity effect in thin SN bilayers is studied.

Although the investigation of the proximity effect in SN systems was started about forty years ago, the technology
allowing to produce and measure experimental samples of mesoscopic dimensions was achieved relatively recently. In
particular, it became possible to study SN structures consisting of thin layers (having thickness smaller than the
coherence length). Such structures behave as a single superconductor with nontrivial properties. From practical point of
view, the proximity structures can be used as superconductors with relatively easily adjustable parameters, in
particular, the energy gap and the critical temperature. The parameters of the proximity structures can be tuned, e.g.,
by varying the thicknesses of the layers. This method has already found its application in superconducting transition
edge bolometers and photon detectors for astrophysics.

While most of theoretical works on SN proximity structures have focused on the limit of ideally transparent interfaces,
the experimental progress requires the advances in theory that take into account arbitrary interface transparency. This
crucial parameter determines the strength of the proximity effect and at the same time is not directly measurable.

In Chapter~1, the theory of superconductivity in thin SN sandwiches (bilayers) in the diffusive limit is developed, with
particular emphasis on the case of very thin superconductive layers, $d_S \ll d_N$. The proximity effect in the system
is governed by the SN interface resistance (per channel) $\rho_B$. The energy gap, the order parameter, the critical
temperature, the density of superconducting electrons, the parallel and perpendicular critical magnetic fields of the
bilayer are investigated as functions of $\rho_B$.

The case of relatively low resistance (which can still have large absolute values) can be completely studied
analytically. This case corresponds to the situation when the diffusion of quasiparticles between the layers is fast
compared to the characteristic time $\Delta^{-1}$. The theory describing the bilayer in this limit is of
Bardeen--Cooper--Schrieffer (BCS) type but with the minigap (in the single-particle density of states) $E_g \ll \Delta$
substituting the order parameter $\Delta$ in the standard BCS relations; the original relations are thus severely
violated. In the opposite limit of an opaque interface, the behavior of the system is in many respects close to the BCS
predictions. Over the entire range of $\rho_B$, the properties of the bilayer are found numerically. It is demonstrated
that the gap is a nonmonotonic function of the interface resistance, reaching its maximum in the region of moderate
resistances. Also in the region of moderate resistances, a jump of the parallel critical field (due to a redistribution
of supercurrents in the bilayer) is discovered. Finally, it is shown that, due to geometrical symmetry, the results
obtained for the bilayer also apply to more complicated structures such as SNS and NSN trilayers, SNINS and NSISN
systems, and SN superlattices.

\section*{Proximity effect in SF systems}

In Chapter 2, the critical temperature in SF bilayers and FSF trilayers is studied.

Compared to SN structures, the physics of SF systems is even richer. In contrast to the SN case, the superconducting
order parameter does not simply decay into the nonsuperconducting metal but also oscillates. This behavior is due to the
exchange field in the ferromagnet that acts as a potential of different signs for two electrons in a Cooper pair and
leads to a finite momentum of the pair (similarly to the Larkin--Ovchinnikov--Fulde--Ferrell state in bulk materials).
This oscillations reveal itself in nonmonotonic dependence of the critical temperature $T_c$ of SF systems as a function
of the F layers thickness. At the same time, in most of the works investigating this effect, the methods to calculate
$T_c$ were approximate. An exact method to calculate $T_c$ at arbitrary parameters of the system was lacking. The need
for such a method was also motivated by a recent experiment that did not correspond to the previously considered
approximations and limiting cases.

In Chapter~2, numerical methods are developed to exactly calculate the critical temperature of a dirty SF bilayer at
arbitrary parameters of the structure (thicknesses of the layers $d_S$ and $d_F$, interface transparency). The methods
are applied to study the nonmonotonic behavior of the critical temperature versus thickness of the F layer. Good
agreement with experimental data is demonstrated. In limiting cases, analytical results for the critical temperature and
the critical thickness of the S layer are obtained.

Another interesting effect in SF systems takes place if the magnetization of the ferromagnet is inhomogeneous. Then the
triplet superconducting component can arise in the system. The triplet component corresponds to pairing between
electrons with the same spin projection (while in the conventional case the Cooper pairs are formed by electrons with
opposite spin projections). Recently, it was demonstrated that the triplet component also arises in the case of several
homogeneous but noncollinearly oriented ferromagnets. However, the conditions at which the superconductivity is not
destroyed in this system were not found. The simplest system of the above type is an FSF trilayer. The answer to the
question about the conditions for the superconductivity to exist can be obtained when studying the critical temperature
of the system. At the same time, a method to calculate $T_c$ in a situation when the triplet component is generated, was
lacking.

Therefore, the methods developed for SF bilayers in the first part of Chapter~2, are then applied to study the critical
temperature of the dirty FSF trilayer at arbitrary parameters of the system (mutual orientation of the magnetizations
$\alpha$, thicknesses of the layers, interface transparency). Changing of characteristic types of the $T_c(d_F)$
dependence while varying the mutual orientation is demonstrated. In the general case, it is analytically proven that
$T_c(\alpha)$ is a monotonic function. In interesting limiting cases, analytical results for the critical temperature of
the system and the critical thickness of the S layer are obtained. The necessary conditions for the existence of the odd
(in energy) triplet superconductivity in multilayered SF structures are formulated.

A possible practical application of FSF structures is a spin valve, a system that switches between superconducting and
nonsuperconducting states when the relative orientation of the magnetizations is varied. Although the superconductive
spin valve is not yet experimentally realized, the work in this direction has already started.

\section*{Josephson effect in SFS junctions}

The Josephson effect in the systems containing ferromagnets (e.g., structures of the SFIFS, SIFIS or SFcFS type) also
has a number of peculiarities. Among them are the transition from the ordinary ($0$-state) to the so-called $\pi$-state
(in other words, the inversion of the critical current sign or the additional $\pi$ phase shift in the Josephson
relation) when the two ferromagnets are aligned in parallel, the enhancement of the critical current by the exchange
field in the case of antiparallel orientation, and nonsinusoidal current--phase relation. The interest to SFS junctions
with nontrivial current--phase relation is in particular due to their possible employment for engineering logic circuits
of novel types (both classical and quantum bits).

In Chapter~3, the Josephson effect in dirty SFIFS junctions is studied by a self-consistent method at arbitrary system
parameters (F layers thickness, interface transparencies). When the magnetizations of the ferromagnets are antiparallel
the effect of the critical current enhancement by the exchange field is observed, while in the case of parallel
magnetizations the junction exhibits the transition to the $\pi$-state. In the limit of thin F layers, these
peculiarities of the critical current are studied analytically and explained qualitatively. The mechanism of the
$0$--$\pi$ transition, at which the phase jumps by $\pi/2$ at the two SF interfaces, is discovered. It is demonstrated
that the effect of the critical current enhancement by the exchange field (at the antiparallel magnetizations alignment)
is similar to the Riedel singularity in SIS junctions. The logarithmic divergence of the maximal critical current is cut
off by finite temperature or finite transparency of the tunnel interface. The effect of switching between $0$ and $\pi$
states by changing the mutual orientation of the F layers is demonstrated.

Nonsinusoidal dependence of the Josephson current on the phase difference in SFcFS point contacts and planar
double-barrier SIFIS junctions is studied in the limit of thin F layers. It is demonstrated that in SFcFS junctions, the
current can cross zero not only at $\varphi=0$ and $\varphi=\pi$ but also at an intermediate value. This implies that
the energy of the junction has two minima --- at $\varphi=0$ and $\varphi=\pi$, hence the $0$ and $\pi$ states of the
junction coexist. If the minima are of the same depth, the system can be used as a quantum bit (qubit). The physical
mechanisms leading to highly nontrivial $I(\varphi)$ dependence are identified by studying the spectral supercurrent
density.

\section*{Decoherence due to nodal quasiparticles in \textit{d}-wave Josephson junctions}

Another interesting type of nonuniform superconducting systems is a junction between superconductors of nontrivial
symmetry. The anisotropic superconductors with the \textit{d}-wave symmetry of the order parameter are widely discussed
because this symmetry is realized in the high-temperature superconductors. A possibility to implement a so-called qubit
(quantum bit) based on \textit{d}-wave junctions was proposed theoretically. Quantum bit is, simply speaking, a quantum
mechanical system with two states (which can be imagined as spin $1/2$). While a classical bit can be either in one
state or in the other, a qubit can also be in a superposition of the two states. If a quantum computer is built of such
qubits, it would have the advantage of natural computational parallelism that can enormously speed up certain types of
computational tasks. A possibility to implement a qubit based on DID junction stems from the fact that the energy of
such a junction as a function of the phase difference can have a double-well form with two minima. This degeneracy of
the ground state arises due to the nontrivial symmetry of the superconductors. Due to tunneling between the wells, the
ground state splits, and the two resulting levels effectively form the quantum-mechanical two-state system. At the same
time, the gapless nature of the \textit{d}-wave superconductors leads to appearance of low-energy quasiparticles which
can destroy the quantum coherence of the qubit and hence hamper its successful functioning. Calculation of the
corresponding decoherence time is necessary for estimating the efficiency of the proposed qubits.

In Chapter~4, the decoherence time due to the low-energy nodal quasiparticles in the Josephson junction between two
\textit{d}-wave superconductors is calculated. The decoherence is due to an intrinsic dissipative process: quantum
tunneling between the two minima of the double-well potential excites nodal quasiparticles which lead to incoherent
damping of quantum oscillations. In DID junctions of the mirror orientation ($\alpha/-\alpha$), the contribution to the
dissipation from the nodal quasiparticles is found to be superohmic, hence the quality factor (the number of coherent
oscillations that occur before the decoherence overcomes) grows as the tunnel splitting of the ground state decreases.
The quality factor is estimated for available experimental data on DID junction. The suppression of low-energy
quasiparticles due to finite size of the \textit{d}-wave superconductors is discussed.

\clearpage
\renewcommand{\@evenhead}{}

\chapter*{\huge Samenvatting (Summary in Dutch)}
\addcontentsline{toc}{chapter}{Samenvatting (Summary in Dutch)}

\renewcommand{\@evenhead}
    {\raisebox{0pt}[\headheight][0pt]
     {\vbox{\hbox to\textwidth{\thepage \hfil \strut \textit{Samenvatting (Summary in Dutch)}}\hrule}}
    }

\renewcommand{\@oddhead}
    {\raisebox{0pt}[\headheight][0pt]
     {\vbox{\hbox to\textwidth{\textit{Samenvatting (Summary in Dutch)} \strut \hfil \thepage}\hrule}}
    }

{\hyphenpenalty=10000

De onderzoeksthema's van dit proefschrift zijn de thermodynamische eigenschappen en transporteigenschappen van
verscheidene supergeleidende contacten. De onderzochte systemen zijn samengesteld uit conventionele \textit{s}-type
supergeleiders in contact met een normaal metaal of ferromagneet en uit anisotrope \textit{d}-type
supergeleiders.\footnote{De notaties zijn: S --- \textit{s}-type supergeleider, D --- \textit{d}-type supergeleider, N
--- normaal metaal, F --- ferromagnetisch metaal, I --- isolator, c --- constrictie.} Een aantal aspecten van het
Josephson effect en het effect van de nabijheid van een supergeleider in dergelijke systemen is theoretisch onderzocht.

Als een supergeleider in contact wordt gebracht met een normaal metaal vindt een aantal verschijnselen plaats dat onder
de noemer nabijheidseffect valt. De twee materialen be\"{\i}nvloeden elkaar op het grensvlak over een lengteschaal van
de ordegrootte van de coherentielengte. De supergeleidende correlaties tussen quasideeltjes worden in het normale metaal
ge\"{\i}nduceerd vanwege de eindige verblijftijd van de binnendringende Cooperparen. Tot het moment van uiteenvallen in
twee onafhankelijke elektronen blijven de supergeleidende eigenschappen bewaard. Aan de andere kant kan het
nabijheidseffect worden gezien als een manifestatie van het fundamentele proces dat Andreev reflectie wordt genoemd,
waarbij een elektron vanuit het normale metaal op het grensvlak met de supergeleider wordt gereflecteerd als een gat
terwijl een Cooperpaar wordt doorgegeven aan de supergeleider.

Het Josephson effect is een macroscopisch kwantumeffect dat optreedt wanneer twee supergeleiders verbonden zijn via een
zwakke koppeling, hetgeen een niet-supergeleidend materiaal (isolerend of metallisch) kan zijn of een geometrische
constrictie. Het supergeleidende condensaat in elk van de twee supergeleiders wordt door een golffunctie beschreven met
elke een fase. De essentie van het Josephson effect is het optreden van een superstroom (niet-dissiperende stroom) bij
een eindig faseverschil $\varphi$ tussen de supergeleiders. Superstroom kan zonder aangelegd potentiaalverschil stromen
en is in het meest eenvoudige geval gegeven door de sinuso\"{\i}dale Josephson relatie $I = I_c \sin\varphi$. De
grootheid $I_c$ (in het algemeen de maximale superstroom) wordt de kritieke stroom genoemd.

\section*{Supergeleiding in dunne SN-bilagen}

In hoofdstuk~1 is het nabijheidseffect in dunne SN-bilagen bestudeerd.

Hoewel het onderzoek naar het nabijheidseffect in SN-structuren reeds veertig jaar geleden is begonnen, is de
technologie om experimentele structuren met mesoscopische afmetingen te prepareren en te meten pas recent bereikt. In
het bijzonder werd het mogelijk SN-structuren te bestuderen bestaande uit dunne lagen (met een dikte die kleiner is dan
de coherentielengte). Dergelijke structuren gedragen zich als een enkele supergeleider met niet-triviale eigenschappen.
De SN-structuren kunnen, praktisch gezien, worden gebruikt als supergeleiders met relatief makkelijk aan te passen
eigenschappen zoals de `gap' en de kritieke temperatuur. De eigenschappen kunnen bijvoorbeeld worden aangepast door de
laagdiktes te varieren. Deze methode wordt reeds gebruikt in bolometers die zijn gebaseerd op de overgang van de
supergeleidende naar normale toestand en fotondetectoren voor astrofysische doeleinden.

Hoewel de meeste theoretische studies naar SN-structuren zich hebben geconcentreerd op de limiet van volledig
transparante grensvlakken, vereist de experimentele ontwikkeling een vooruitgang in de theorie voor een willekeurige
transparantie van het grensvlak. Deze cruciale parameter bepaalt de sterkte van het nabijheidseffect maar is niet direct
meetbaar.

De theorie van de supergeleiding in dunne SN-structuren (bilagen) in de limiet van diffuus transport is ontwikkeld in
hoofdstuk~1, met een specifieke nadruk op structuren met een hele dunne supergeleidende laag, $d_S \ll d_N$. Het
nabijheidseffect in het systeem wordt bepaald door de weerstand van het SN-grensvlak (per kanaal) $\rho_B$. De `gap', de
ordeparameter, de kritieke temperatuur, de dichtheid van supergeleidende elektronen en de parallelle en loodrechte
kritieke magneetvelden van de bilaag zijn onderzocht als functie van $\rho_B$.

Het geval van een relatief lage weerstand (hetgeen nog steeds absoluut hoge waardes toelaat)  kan geheel analytisch
worden bestudeerd. Dit geval komt overeen met de situatie waarin de diffusie tussen quasideeltjes snel is vergeleken met
de karakteristieke tijd $\Delta^{-1}$. De theorie die de bilaag in deze limiet beschrijft is de
Bardeen--Cooper--Schrieffer (BCS) theorie waarin nu de ordeparameter $\Delta$ vervangen wordt door de `minigap' (in de
toestandsdichtheid van de quasideeltjes) $E_g \ll \Delta$, hetgeen dus een ernstige schending van de oorspronkelijke
vergelijkingen veroorzaakt. Het systeem in de tegenovergestelde limiet van een beperkt transparant grensvlak gedraagt
zich echter in veel opzichten ongeveer naar de voorspellingen uit de BCS theorie. De eigenschappen voor een bilaag zijn
numeriek bepaald voor alle $\rho_B$. Het is afgeleid dat de `gap' een niet-monotone functie van de weerstand van het
grensvlak is, waarbij het maximum wordt bereikt in het gebied met middelmatige weerstandswaarden. In dit gebied met
middelmatige weerstandswaarden is een sprong ontdekt in het parallelle kritieke veld (vanwege een herverdeling van
superstromen in de bilaag). Tenslotte is aangetoond dat, vanwege geometrische symmetrie, de resultaten voor een bilaag
ook gelden voor ingewikkeldere structuren zoals SNS- en NSN-trilagen, SNINS- en NSISN-structuren en SN-superstructuren.

\section*{Het nabijheidseffect in SF-structuren}

In hoofdstuk~2 is de kritieke temperatuur van SF bilagen en FSF-trilagen bestudeerd.

Vergeleken met de SN-structuren is de fysica van SF-structuren zelfs nog gevarieerder. In tegenstelling tot in
SN-structuren, dringt de supergeleidende ordeparameter niet slechts eenvoudigweg het normale metaal binnen, maar kan
daarbij ook oscilleren. Dit gedrag wordt veroorzaakt door het `exchange' veld in de ferromagneet, hetgeen zich gedraagt
alsof er voor elk van de twee elektronen in een Cooperpaar een potentiaal met tegengestelde polariteit bestaat, waardoor
het paar een impuls krijgt (vergelijkbaar met de Larkin--Ovchinnikov--Fulde--Ferrell toestand in materialen). Deze
oscillaties manifesteren zichzelf in het niet-monotone verband tussen de kritieke temperatuur $T_c$ van SF-systemen en
de laagdikte van de F-laag. In de meeste studies naar dit effect zijn de methoden om $T_c$ uit te rekenen slechts
benaderingen en een precieze methode om $T_c$ uit te rekenen als functie van de parameters van het systeem ontbrak. De
behoefte aan een dergelijke methode kwam ook voort uit recente experimentele bevindingen die niet overeenkomen met de
tot nu toe beschouwde limietgevallen en benaderingen.

In hoofdstuk~2 zijn numerieke methodes ontwikkeld om de kritieke temperatuur van een SF-bilaag in de diffuse limiet
exact uit te rekenen als functie van de parameters van het systeem (laagdiktes $d_S$ en $d_F$, grensvlak transparantie).
De methoden zijn toegepast in een studie naar het niet-monotone verband tussen de kritieke temperatuur en de laagdikte
van de F-laag. Een goede overeenstemming met de experimentele gegevens is hierbij bereikt. In speciale limietgevallen
zijn analytische uitdrukkingen gevonden voor de kritieke temperatuur en de kritieke dikte van de S-laag.

Een ander interessant effect vindt plaats in SN-structuren wanneer de magnetisatie van de ferromagneet niet homogeen is.
In dit geval kan een triplet component in de supergeleiding ontstaan. De triplet component komt overeen met de paring
tussen elektronen met dezelfde projectie van de spin (terwijl in het conventionele geval de Cooperparen worden gevormd
door elektronen met tegengestelde spin projecties). Het is recent aangetoond dat de triplet component ook kan ontstaan
in systemen met homogene, maar niet-colineair geori\"{e}nteerde ferromagneten. De voorwaarden waaronder de
supergeleiding niet wordt tenietgedaan in zulke structuren waren echter nog niet gevonden. Het meest eenvoudige
voorbeeld van een dergelijke structuur is de FSF-trilaag. Het antwoord op de vraag naar de voorwaarden voor het bestaan
van supergeleiding kan worden beantwoord door de kritieke temperatuur van het systeem te onderzoeken. Een methode om
$T_c$ te berekenen in structuren waarbij een triplet component wordt gegenereerd, ontbrak echter nog.

Vanwege dit feit, zijn de methodes voor de SF-bilaag uit hoofdstuk~2 gebruikt om de kritieke temperatuur van een diffuse
FSF-trilaag te bepalen als functie van de parameters van het systeem (wederzijdse ori\"{e}ntatie van de magnetisaties
$\alpha$, laagdiktes en grensvlak transparantie). Het is aangetoond hoe de functies $T_c(d_F)$ veranderen wanneer de
ori\"{e}ntaties worden gewijzigd. In het meest algemene geval is het analytisch bewezen dat $T_c(\alpha)$ een
niet-monotone functie is. In de interessante limietgevallen zijn analytische uitdrukkingen voor de kritieke temperatuur
en de kritieke dikte van laag S verkregen. De noodzakelijke voorwaarden voor het bestaan van oneven (in energie) triplet
supergeleiding in multilaags SF-structuren is geformuleerd.

Een mogelijke praktische toepassing van FSF-structuren is de spin-transistor, een systeem dat schakelt tussen de
supergeleidende en de niet-supergeleidende toestand wanneer de relatieve ori\"{e}ntatie van de magnetisaties wordt
veranderd. Hoewel de supergeleidende spin transistor experimenteel nog niet is gerealiseerd, is het onderzoek in deze
richting wel begonnen.

\section*{Het Josephson effect in SFS contacten}

{\exhyphenpenalty=10000

Het Josephson effect in structuren die ferromagneten bevatten (bijvoorbeeld SFIFS-, SIFIS-, SFcFS-structuren) heeft een
aantal bijzondere eigenschappen, waaronder de overgang van de gewone toestand ($0$-toestand) naar de zogenaamde
$\pi$-toestand (met andere woorden: het teken van de kritieke stroom wordt ge\"{\i}nverteerd, of: een additionele
faseverschuiving van $\pi$ vindt plaats) wanneer de ferromagneten parallel geori\"{e}nteerd zijn, de toename van de
kritieke stroom door het `exchange' veld wanneer de ferromagneten anti-parallel geori\"{e}nteerd zijn en een
niet-sinuso\"{\i}dale stroom--fase relatie. De interesse in SFS-contacten met een niet-triviale stroom--fase relatie
komt met name voort uit de mogelijke toepasbaarheid in het realiseren van nieuwe elektronische circuits (met klassieke
of kwantum bits).

}

In hoofdstuk~3 is het Josephson effect bestudeerd in diffuse SFIFS-structuren met willekeurig gekozen parameters
(laagdikte van F, grensvlak transparantie) door middel van een consistente methode. Wanneer de magnetisaties van de
ferromagneten anti-parallel zijn geori\"{e}nteerd, wordt een toename van de kritieke stroom door het `exchange' veld
gevonden, terwijl in het geval van een parallelle ori\"{e}ntatie het contact een overgang naar de $\pi$-toestand kan
hebben. Deze bijzonderheden van de kritieke stroom zijn analytisch bestudeerd en kwalitatief verklaard in de limiet
waarin de laagdikte van F klein is. Het mechanisme van de overgang van $0$ naar $\pi$, waarbij de fase met $\pi/2$
verspringt op de twee SF-grensvlakken, is ontdekt. Het is aangetoond dat de toename van de kritieke stroom door het
`exchange' veld (bij een anti-parallelle ori\"{e}ntatie) gelijkenis vertoont met de Riedel singulariteit in
SIS-contacten. De logaritmische divergentie van de maximale kritieke stroom wordt afgezwakt door de eindige temperatuur
of de niet-ideale transparantie van het grensvlak. Het verschijnsel van schakelen tussen de $0$- en de $\pi$-toestand
door het veranderen van de wederzijdse ori\"{e}ntatie van de ferromagnetische lagen is aangetoond.

Het niet-sinuso\"{\i}dale verband tussen de Josephson stroom en het faseverschil in SFcFS-puntcontacten en planaire
dubbele-barri\`{e}re SIFIS-contacten is bestudeerd in het limietgeval waarbij de F laag dun is. Het is aangetoond dat de
stroom in SFcFS-contacten niet alleen de nulwaarde kan passeren bij $\varphi=0$ en $\varphi=\pi$, maar ook bij een
tussenliggende waarde. Dit betekent dat de energie van het contact twee minima heeft (op $\varphi=0$ en $\varphi=\pi$)
en de $0$- en de $\pi$-toestand tegelijk bestaan. Wanneer de minima een gelijke diepte hebben, kan het systeem worden
gebruikt als een kwantumbit (qubit). Het fysische mechanisme dat voor de niet-triviale $I(\varphi)$ relatie zorgt, is
ontrafeld door het bestuderen van de spectrale superstroomdichtheid.

\section*{Decoherentie door quasideeltjes in de nodale richting van \textit{d}-type Josephson contacten}

Een ander interessant type niet-uniform supergeleidend systeem is een contact tussen supergeleiders met een
niet-triviale symmetrie. Veelal worden anisotrope supergeleiders bestudeerd met de symmetrie van de ordeparameter van
het \textit{d}-type omdat deze symmetrie wordt gerealiseerd in hoge-temperatuur supergeleiders. De mogelijkheid om een
zogenaamd qubit (kwantumbit) gebaseerd op \textit{d}-type contacten te implementeren is reeds theoretisch voorgesteld.
Een qubit is in feite een kwantummechanisch systeem met twee toestanden (dit kan worden voorgesteld als spin $1/2$).
Terwijl een klassieke bit in de ene of in de andere toestand is, kan een qubit zich ook in een superpositie van de twee
toestanden bevinden. Een kwantumcomputer gebaseerd op qubits zou dan het voordeel hebben dat een natuurlijke vorm van
parallel rekenen kan worden gebruikt, wat bepaalde soorten numerieke taken enorm zou kunnen versnellen. De mogelijkheid
een qubit te implementeren met behulp van DID-contacten komt voort uit het feit dat de energie van dit contact een
dubbele potentiaalput met twee minima vormt als functie van het faseverschil. Deze gedegenereerde grondtoestand ontstaat
vanwege de niet-triviale symmetrie van de supergeleiders. De grondtoestand splitst vanwege tunnelprocessen tussen de
energieminima en de twee resulterende energietoestanden vormen een kwantumechanisch twee-toestandensysteem. Het
ontbreken van de `gap' in \textit{d}-type supergeleiders leidt tot de aanwezigheid van laag-energetische quasideeltjes
die de kwantumcoherentie van de qubit kunnen onderdrukken en daarmee een succesvol gebruik bemoeilijken.

In hoofdstuk~4 is de tijd van de decoherentie door de laag-energetische quasideeltjes uitgerekend voor Josephson
contacten met twee \textit{d}-type supergeleiders. De decoherentie wordt veroorzaakt door een intrinsiek dissipatief
proces: een kwantumtunnelproces tussen de twee minima van de dubbele potentiaalput slaat nodale quasideeltjes aan, wat
aanleiding geeft tot een incoherente demping van de kwantumoscillaties. De bijdrage aan de dissipatie van de nodale
quasideeltjes in DID contacten met een gespiegelde orientatie ($\alpha/-\alpha$) is super-ohmisch. Hierdoor stijgt de
kwaliteitsfactor (het aantal coherente oscillatie dat plaatsvindt voordat decoherentie optreedt) bij een afnemend
verschil tussen de grondtoestanden. De kwaliteitsfactor is geschat voor de beschikbare experimentele gegevens. De
onderdrukking van laag-energetische quasideeltjes door een beperkte laagdikte van de \textit{d}-type supergeleiders is
toegelicht.

}

\clearpage
\renewcommand{\@evenhead}{}

\chapter*{\huge Acknowledgments}
\addcontentsline{toc}{chapter}{Acknowledgments}

\renewcommand{\@evenhead}
    {\raisebox{0pt}[\headheight][0pt]
     {\vbox{\hbox to\textwidth{\thepage \hfil \strut \textit{Acknowledgments}}\hrule}}
    }

\renewcommand{\@oddhead}
    {\raisebox{0pt}[\headheight][0pt]
     {\vbox{\hbox to\textwidth{\textit{Acknowledgments} \strut \hfil \thepage}\hrule}}
    }

First of all, I would like to thank my promotor, Horst Rogalla, for providing me the opportunity to carry out this
research as a joint project between the Landau Institute and the University of Twente.

I want to express my deep gratitude to my scientific advisors, Alexander Golubov and Mikhail V. Feigel'man, for
permanent attention and help in my work. I sincerely appreciate your advices and criticism, and I have learned a lot
from you with respect to both the style of scientific research and understanding specific topics or calculations.

Also I have gained a lot from joint work with my other co-authors, Nikolai Chtchelkatchev and Mikhail Yu. Kupriyanov.
Our fruitful collaboration has always been very motivating for me.

I want to thank one of my first teachers, Yaroslav M. Blanter. I highly appreciate your advices which have always been
timely when I had to take serious decisions concerning my scientific life.

As this research has been carried out both at the Landau Institute in Russia and at the University of Twente in the
Netherlands, I wish to thank both groups for creating the stimulating environment.

I am also grateful to all my colleagues with whom I discussed the results presented in this thesis and whose opinion was
very important for me. I would specially like to thank Jan Aarts, Mohammad Amin, Marco Aprili, Yuri S. Barash, Igor
Burmistrov, Alexandre Buzdin, Dima Geshkenbein, Dmitri Ivanov, Alebker Yu. Kasumov, Anatoly I. Larkin, Yuri Makhlin,
Yuli V. Nazarov, Pavel Ostrovsky, Zoran Radovi\'c, Alexander Rusanov, Valery Ryazanov, Alexander Shnirman, Mikhail
Skvortsov, Lenar Tagirov, Yukio Tanaka, and Alexandre Zagoskin.

Special thanks go to Alexander Brinkman for his hospitable help during my stay in the Netherlands and, in particular,
for helping me in preparation of this thesis (Summary in Dutch, etc.).

Большое спасибо родителям, родственникам и друзьям за помощь и поддержку. Особый привет Пуху и Тимону.

\clearpage
\renewcommand{\@evenhead}{}

 \chapter*{\vspace*{-20mm}\huge List of publications}
 \addcontentsline{toc}{chapter}{List of publications}

\renewcommand{\@evenhead}
    {\raisebox{0pt}[\headheight][0pt]
     {\vbox{\hbox to\textwidth{\thepage \hfil \strut \textit{List of publications}}\hrule}}
    }

\renewcommand{\@oddhead}
    {\raisebox{0pt}[\headheight][0pt]
     {\vbox{\hbox to\textwidth{\textit{List of publications} \strut \hfil \thepage}\hrule}}
    }

\hangindent=1em \noindent
1.~Ya.\,V.~Fominov, M.\,V.~Feigel'man,\\
\textit{Superconductive properties of thin dirty superconductor--normal-metal bilayers},\\
Phys. Rev.~B \textbf{63}, 094518 (2001).\\
(cond-mat/0008038)
\bigskip

\hangindent=1em \noindent
2.~Ya.\,V.~Fominov, N.\,M.~Chtchelkatchev, A.\,A.~Golubov,\\
\textit{Critical temperature of superconductor/ferromagnet bilayers},\\
Pis'ma Zh. Eksp. Teor. Fiz. \textbf{74}, 101 (2001) [JETP Lett. \textbf{74}, 96 (2001)].\\
(cond-mat/0106185)
\bigskip

\hangindent=1em \noindent
3.~Ya.\,V.~Fominov, N.\,M.~Chtchelkatchev, A.\,A.~Golubov,\\
\textit{Nonmonotonic critical temperature in superconductor/ferromagnet bilayers},\\
Phys. Rev.~B \textbf{66}, 014507 (2002).\\
(cond-mat/0202280)
\bigskip

\hangindent=1em \noindent
4.~A.\,A.~Golubov, M.\,Yu.~Kupriyanov, Ya.\,V.~Fominov,\\
\textit{Critical current in SFIFS junctions},\\
Pis'ma Zh. Eksp. Teor. Fiz. \textbf{75}, 223 (2002) [JETP Lett. \textbf{75}, 190 (2002)].\\
(cond-mat/0201249)
\bigskip

\hangindent=1em \noindent
5.~A.\,A.~Golubov, M.\,Yu.~Kupriyanov, Ya.\,V.~Fominov,\\
\textit{Nonsinusoidal current--phase relation in SFS Josephson junctions},\\
Pis'ma Zh. Eksp. Teor. Fiz. \textbf{75}, 709 (2002) [JETP Lett. \textbf{75}, 588 (2002)].\\
Erratum: Pis'ma Zh. Eksp. Teor. Fiz. \textbf{76}, 268 (2002) [JETP Lett. \textbf{76}, 231 (2002)].\\
(cond-mat/0204568)
\bigskip

\hangindent=1em \noindent
6.~A.\,A.~Golubov, M.\,Yu.~Kupriyanov, Ya.\,V.~Fominov,\\
\textit{Mechanisms of $0$--$\pi$ transition and current--phase relations in SFS Josephson junctions},\\
in Proceedings of MS+S2002, p.~137. Title of the book: TOWARDS THE CONTROLLABLE QUANTUM STATES (Mesoscopic
Superconductivity and Spintronics, Atsugi, Japan, 4--6 March 2002), edited by H.~Takayanagi and J.~Nitta. World
Scientific Publishing.
\bigskip

\hangindent=1em \noindent
7.~Ya.\,V.~Fominov, M.\,Yu.~Kupriyanov, M.\,V.~Feigel'man,\\
\textit{A comment on the paper ``Competition between superconductivity and magnetism in ferromagnet/superconductor
heterostructures'' by Yu.\,A.~Izyumov, Yu.\,N.~Proshin, and M.\,G.~Khusainov},\\
Usp. Fiz. Nauk \textbf{173}, 113 (2003) [Phys.--Usp. \textbf{46}, 105 (2003)]. \clearpage

\hangindent=1em \noindent
8.~Ya.\,V.~Fominov, A.\,A.~Golubov, M.\,Yu.~Kupriyanov,\\
\textit{Triplet proximity effect in FSF trilayers},\\
Pis'ma Zh. Eksp. Teor. Fiz. \textbf{77}, 609 (2003) [JETP Lett. \textbf{77}, 510 (2003)].\\
(cond-mat/0303534)
\bigskip

\hangindent=1em \noindent
9.~Ya.\,V.~Fominov, A.\,A.~Golubov, M.\,Yu.~Kupriyanov,\\
\textit{Decoherence due to nodal quasiparticles in d-wave qubits},\\
Pis'ma Zh. Eksp. Teor. Fiz. \textbf{77}, 691 (2003) [JETP Lett. \textbf{77}, 587 (2003)].\\
(cond-mat/0304383)


\begin{thebibliography}{99}
\addcontentsline{toc}{section}{References}

\bibitem{Kamerlingh}
H.~Kamerlingh Onnes, Leiden Commun. \textbf{119b}, \textbf{120b}, \textbf{122b}, \textbf{124c} (1911).

\bibitem{Londons}
F.~London, H.~London, Proc. Roy. Soc.~A \textbf{149}, 71 (1935).

\bibitem{GL}
V.\,L.~Ginzburg, L.\,D.~Landau, Zh. Eksp. Teor. Fiz. \textbf{20}, 1064 (1950).

\bibitem{Kapitza}
P.\,L.~Kapitza, Dokl. Akad. Nauk SSSR \textbf{18}, 29 (1938); Nature \textbf{141}, 74 (1938); Zh. Eksp. Teor. Fiz.
\textbf{11}, 1 (1941); \textbf{11}, 581 (1941).

\bibitem{Landau}
L.\,D.~Landau, Zh. Eksp. Teor. Fiz. \textbf{11}, 592 (1941).

\bibitem{Cooper_pairing}
L.\,N.~Cooper, Phys. Rev. \textbf{104}, 1189 (1956).

\bibitem{BCS}
J.~Bardeen, L.\,N.~Cooper, J.\,R.~Schrieffer, Phys. Rev. \textbf{108}, 1175 (1957).

\bibitem{Bogolyubov}
N.\,N.~Bogolyubov, Zh. Eksp. Teor. Fiz. \textbf{34}, 58 (1958); \textbf{34}, 73 (1958) [Sov. Phys. JETP \textbf{7}, 41
(1958); \textbf{7}, 51 (1958)].

\bibitem{Gor'kov}
L.\,P.~Gor'kov, Zh. Eksp. Teor. Fiz. \textbf{34}, 735 (1958) [Sov. Phys. JETP \textbf{7}, 505 (1958)].

\bibitem{Van_Harlingen}
D.\,J.~Van~Harlingen, Rev. Mod. Phys. \textbf{67}, 515 (1995).

\bibitem{Tsuei}
C.\,C.~Tsuei, J.\,R.~Kirtley, Rev. Mod. Phys. \textbf{72}, 969 (2000).

\bibitem{Andreev}
A.\,F.~Andreev, Zh. Eksp. Teor. Fiz. \textbf{46}, 1823 (1964); \textbf{49}, 655 (1965) [Sov. Phys. JETP \textbf{19},
1228 (1964); \textbf{22}, 455 (1966)].

\bibitem{Josephson}
B.\,D.~Josephson, Phys. Lett. \textbf{1}, 251 (1962).

\bibitem{Likharev}
K.\,K.~Likharev, Rev. Mod. Phys. \textbf{51}, 101 (1979).

\bibitem{Cooper}
L.\,N.~Cooper, Phys. Rev. Lett. \textbf{6}, 689 (1961).

\bibitem{Werthamer}
N.\,R.~Werthamer, Phys. Rev. \textbf{132}, 2440 (1963).

\bibitem{deGennes_review}
P.\,G.~de~Gennes, Rev. Mod. Phys. \textbf{36}, 225 (1964).

\bibitem{Poelaert}
A.~Poelaert, Ph.D. thesis, University of Twente, The Netherlands (1999).

\bibitem{Verhoeve}
P.~Verhoeve, N.~Rando, A.~Peacock, D.~Martin, R.~den~Hartog, Opt. Eng. \textbf{41}, 1170 (2002).

\bibitem{LOFF}
P.~Fulde, R.\,A.~Ferrell, Phys. Rev. \textbf{135}, A550 (1964);\\
A.\,I.~Larkin, Yu.\,N.~Ovchinnikov, Zh. Eksp. Teor. Fiz. \textbf{47}, 1136 (1964) [Sov. Phys. JETP \textbf{20}, 762
(1965)].

\bibitem{BK}
A.\,I.~Buzdin, M.~Yu.~Kupriyanov, Pis'ma Zh. Eksp. Teor. Fiz. \textbf{52}, 1089 (1990) [JETP Lett. \textbf{52}, 487
(1990)].

\bibitem{Radovic}
Z.~Radovi\'c, M.~Ledvij, Lj.~Dobrosavljevi\'c--Gruji\'c, A.\,I.~Buzdin, J.\,R.~Clem, Phys. Rev.~B \textbf{44}, 759
(1991).

\bibitem{PKhI}
Yu.\,N.~Proshin, M.\,G.~Khusainov, Zh. Eksp. Teor. Fiz. \textbf{113}, 1708 (1998);
\textbf{116}, 1887 (1999) [JETP \textbf{86}, 930 (1998); \textbf{89}, 1021 (1999)];\\
M.\,G.~Khusainov, Yu.\,N.~Proshin, Phys. Rev.~B \textbf{56}, R14283 (1997); \textbf{62}, 6832 (2000).

\bibitem{Tagirov}
L.\,R.~Tagirov, Physica~C \textbf{307}, 145 (1998).

\bibitem{ROP}
V.\,V.~Ryazanov, V.\,A.~Oboznov, A.\,S.~Prokofiev, S.\,V.~Dubonos, Pis'ma Zh. Eksp. Teor. Fiz. \textbf{77}, 43 (2003)
[JETP Lett. \textbf{77}, 39 (2003)].

\bibitem{Bergeret}
F.\,S.~Bergeret, K.\,B.~Efetov, A.\,I.~Larkin, Phys. Rev.~B \textbf{62}, 11872 (2000);\\
F.\,S.~Bergeret, A.\,F.~Volkov, K.\,B.~Efetov, Phys. Rev. Lett. \textbf{86}, 4096 (2001).

\bibitem{Kadigrobov}
A.~Kadigrobov, R.\,I.~Shekhter, M.~Jonson, Europhys. Lett. \textbf{54}, 394 (2001); Fiz. Nizk. Temp. \textbf{27}, 1030
(2001) [Low Temp. Phys. \textbf{27}, 760 (2001)].

\bibitem{JLTP}
L.\,N.~Bulaevskii, A.\,I.~Rusinov, M.~Kuli\'c, J. Low Temp. Phys. \textbf{39}, 255 (1980).

\bibitem{VBE}
A.\,F.~Volkov, F.\,S.~Bergeret, K.\,B.~Efetov, Phys. Rev. Lett. \textbf{90}, 117006 (2003).

\bibitem{Tagirov_PRL}
L.\,R.~Tagirov, Phys. Rev. Lett. \textbf{83}, 2058 (1999).

\bibitem{Buzdin_EL}
A.\,I.~Buzdin, A.\,V.~Vedyayev, N.\,V.~Ryzhanova, Europhys. Lett. \textbf{48}, 686 (1999).

\bibitem{ANL}
J.\,Y.~Gu, C.-Y.~You, J.\,S.~Jiang, J.~Pearson, Ya.\,B.~Bazaliy, S.\,D.~Bader, Phys. Rev. Lett. \textbf{89}, 267001
(2002).

\bibitem{Bulaevski}
L.\,N.~Bulaevskii, V.\,V.~Kuzii, A.\,A.~Sobyanin, Pis'ma Zh. Eksp. Teor. Fiz. \textbf{25}, 314 (1977) [JETP Lett.
\textbf{25}, 290 (1977)].

\bibitem{Buzdin_pi}
A.\,I.~Buzdin, L.\,N.~Bulaevskii, S.\,V.~Panyukov, Pis'ma Zh. Eksp. Teor. Fiz. \textbf{35}, 147 (1982) [JETP Lett.
\textbf{35}, 178 (1982)].

\bibitem{Ryazanov}
V.\,V.~Ryazanov, V.\,A.~Oboznov, A.\,Yu.~Rusanov, A.\,V.~Veretennikov, A.\,A.~Golubov, J.~Aarts, Phys. Rev. Lett.
\textbf{86}, 2427 (2001).

\bibitem{BVE}
F.\,S.~Bergeret, A.\,F.~Volkov, K.\,B.~Efetov, Phys. Rev. Lett. \textbf{86}, 3140 (2001).

\bibitem{Beasley}
E.\,Terzioglu, M.\,R.~Beasley, IEEE Trans. Appl. Superconduct. \textbf{8}, 48 (1998).

\bibitem{Blatter}
G.~Blatter, V.\,B.~Geshkenbein, L.\,B.~Ioffe, Phys. Rev.~B \textbf{63}, 174511 (2001).

\bibitem{Ioffe}
L.\,B.~Ioffe, V.\,B.~Geshkenbein, M.\,V.~Feigel'man, A.\,L.~Fauch\`{e}re, G.~Blatter, Nature (London) \textbf{398}, 679
(1999).

\bibitem{Zagoskin}
A.\,M.~Zagoskin, cond-mat/9903170 (1999);\\
A.~Blais, A.\,M.~Zagoskin, Phys. Rev.~A \textbf{61}, 042308 (2000).

\bibitem{Il'ichev}
E.~Il'ichev, M.~Grajcar, R.~Hlubina, R.\,P.\,J.~IJsselsteijn, H.\,E.~Hoenig, H.-G.~Meyer, A.~Golubov, M.\,H.\,S.~Amin,
A.\,M.~Zagoskin, A.\,N.~Omelyanchouk, M.\,Yu.~Kupriyanov, Phys. Rev. Lett. \textbf{86}, 5369 (2001).

\bibitem{AGD}
A.\,A.~Abrikosov, L.\,P.~Gor'kov, I.\,E.~Dzyaloshinski, \textit{Methods of Quantum Field Theory in Statistical Physics}
(Dover, New York, 1977).

\bibitem{Eilenberger}
G.~Eilenberger, Z.~Phys. \textbf{214}, 195 (1968).

\bibitem{LO_qc}
A.\,I.~Larkin, Yu.\,N.~Ovchinnikov, Zh. Eksp. Teor. Fiz. \textbf{55}, 2262 (1968) [Sov. Phys. JETP \textbf{28}, 1200
(1969)].

\bibitem{Usadel}
K.\,D.~Usadel, Phys. Rev. Lett. \textbf{25}, 507 (1970).

\bibitem{Zaitsev}
A.\,V.~Zaitsev, Zh. Eksp. Teor. Fiz. \textbf{86}, 1742 (1984) [Sov. Phys. JETP \textbf{59}, 1015 (1984)].

\bibitem{KL}
M.\,Yu.~Kupriyanov, V.\,F.~Lukichev, Zh. Eksp. Teor. Fiz. \textbf{94}, 139 (1988) [Sov. Phys. JETP \textbf{67}, 1163
(1988)].

\end{thebibliography}

\begin{thebibliography}{99}
\addcontentsline{toc}{section}{References}

\bibitem{1:BCS}
J.~Bardeen, L.\,N.~Cooper, J.\,R.~Schrieffer, Phys. Rev. \textbf{108}, 1175 (1957).

\bibitem{1:Kasumov}
A.\,Yu.~Kasumov, R.~Deblock, M.~Kociak, B.~Reulet, H.~Bouchiat, I.\,I.~Khodos, Yu.\,B.~Gorbatov, V.\,T.~Volkov,
C.~Journet, M.~Burghard, Science \textbf{284}, 1508 (1999).

\bibitem{1:AB}
V.~Ambegaokar, A.~Baratoff, Phys. Rev. Lett. \textbf{10}, 486 (1963).

\bibitem{1:Kasumov_new}
M.~Kociak, A.\,Yu.~Kasumov, S.~Gu\'eron, B.~Reulet, L.~Vaccarini, I.\,I.~Khodos, Yu.\,B.~Gorbatov, V.\,T.~Volkov,
H.~Bouchiat, Phys. Rev. Lett. \textbf{86}, 2416 (2001).

\bibitem{1:Cooper}
L.\,N.~Cooper, Phys. Rev. Lett. \textbf{6}, 689 (1961).

\bibitem{1:deGennes_review}
P.\,G.~de~Gennes, Rev. Mod. Phys. \textbf{36}, 225 (1964).

\bibitem{1:LO}
A.\,I.~Larkin, Yu.\,N.~Ovchinnikov, in \textit{Nonequilibrium Superconductivity}, edited by D.~N.~Langenberg and
A.~I.~Larkin (Elsevier, New York, 1986), p.~530, and references therein.

\bibitem{1:Usadel}
K.\,D.~Usadel, Phys. Rev. Lett. \textbf{25}, 507 (1970).

\bibitem{1:KL}
M.\,Yu.~Kupriyanov, V.\,F.~Lukichev, Zh. Eksp. Teor. Fiz. \textbf{94}, 139 (1988) [Sov. Phys. JETP \textbf{67}, 1163
(1988)].

\bibitem{1:Kasumov_priv}
A.\,Yu.~Kasumov, private communication.

\bibitem{1:Khusainov1}
M.\,G.~Khusainov, Pis'ma Zh. Eksp. Teor. Fiz. \textbf{53}, 554 (1991) [JETP Lett. \textbf{53}, 579 (1991)].

\bibitem{1:McMillan}
W.\,L.~McMillan, Phys. Rev. \textbf{175}, 537 (1968).\\ The quantities $\Gamma_S$ and $\Gamma_N$ used by McMillan are
analogous to our quantities $1/\tau_S$ and $1/\tau_N$, respectively. The equations obtained by McMillan in the framework
of the tunneling Hamiltonian method were later derived microscopically (from the Usadel equations) by A.~A.~Golubov and
M.~Yu.~Kupriyanov, Physica~C \textbf{259}, 27 (1996).\\ To avoid confusion, we note that Eq. (39) from McMillan's paper
determining the critical temperature of the SN bilayer is misprinted. The correct equation, following from Eqs. (37),
(38), (40) and leading to Eq. (41) of McMillan, reads $\ln (T_{CS}/T_C) = (\Gamma_S/\Gamma)\, [ \psi ( \frac 12
+\Gamma/2\pi T_C ) -\psi (\frac 12 )]$.

\bibitem{1:Golubov}
A.\,A.~Golubov, in \textit{Superconducting Superlattices and Multilayers}, edited by I.~Bozovic, SPIE proceedings
Vol.~2157 (SPIE, Bellingham, WA, 1994), p.~353.

\bibitem{1:Gol1}
A.\,A.~Golubov, M.\,Yu.~Kupriyanov, Zh. Eksp. Teor. Fiz. \textbf{96}, 1420 (1989) [Sov. Phys. JETP \textbf{69}, 805
(1989)].

\bibitem{1:Maki_parallel}
K.~Maki, Progr. Theor. Phys. (Kyoto) \textbf{29}, 603 (1963).

\bibitem{1:Abrikosov}
A.\,A.~Abrikosov, Zh. Eksp. Teor. Fiz. \textbf{32}, 1442 (1957) [Sov. Phys. JETP \textbf{5}, 1174 (1957)].

\bibitem{1:HW}
E.~Helfand, N.\,R.~Werthamer, Phys. Rev. Lett. \textbf{13}, 686 (1964).

\bibitem{1:Maki}
K.~Maki, Physics \textbf{1}, 21 (1964). The mistake made in this paper was corrected in C.~Caroli, M.~Cyrot,
P.\,G.~de~Gennes, Solid State Commun. \textbf{4}, 17 (1966).

\bibitem{1:deGennes}
P.\,G.~de~Gennes, Phys. Konden. Mater. \textbf{3}, 79 (1964).

\bibitem{1:deGennes_book}
P.\,G.~de~Gennes, \textit{Superconductivity of Metals and Alloys} (Benjamin, New York, 1966).

\bibitem{1:Radovic_Hc2}
Z.~Radovi\'{c}, M.~Ledvij, L.~Dobrosavljevi\'{c}-Gruji\'{c}, Phys. Rev.~B \textbf{43}, 8613 (1991).

\bibitem{1:Luders}
G.~L\"{u}ders, K.\,D.~Usadel, \textit{The Method of the Correlation Function in Superconductivity Theory}
(Springer-Verlag, Berlin, 1971), p.~213.

\end{thebibliography}

\begin{thebibliography}{99}
\addcontentsline{toc}{section}{References}

\bibitem{2:Ryazanov}
V.\,V.~Ryazanov, V.\,A.~Oboznov, A.\,Yu.~Rusanov, A.\,V.~Veretennikov, A.\,A.~Golubov, J.~Aarts, Phys. Rev. Lett.
\textbf{86}, 2427 (2001);\\
V.\,V.~Ryazanov, V.\,A.~Oboznov, A.\,V.~Veretennikov, A.\,Yu.~Rusanov, Phys. Rev.~B \textbf{65}, 020501(R) (2001).

\bibitem{2:Kontos}
T.~Kontos, M.~Aprili, J.~Lesueur, X.~Grison, Phys. Rev. Lett. \textbf{86}, 304 (2001); T.~Kontos, M.~Aprili, J.~Lesueur,
F.~Gen\^{e}t, B.~Stephanidis, R.~Boursier, Phys. Rev. Lett. \textbf{89}, 137007 (2002).

\bibitem{2:Radovic}
Z.~Radovi\'c, M.~Ledvij, Lj.~Dobrosavljevi\'c--Gruji\'c, A.\,I.~Buzdin, J.\,R.~Clem, Phys. Rev.~B \textbf{44}, 759
(1991).

\bibitem{2:Tagirov_PRL}
L.\,R.~Tagirov, Phys. Rev. Lett. \textbf{83}, 2058 (1999).

\bibitem{2:Buzdin_DOS}
A.~Buzdin, Phys. Rev.~B \textbf{62}, 11377 (2000).

\bibitem{2:Nazarov_DOS}
M.~Zareyan, W.~Belzig, Yu.\,V.~Nazarov, Phys. Rev. Lett. \textbf{86}, 308 (2001).

\bibitem{2:Jiang}
J.\,S.~Jiang, D.~Davidovi\'c, D.\,H.~Reich, C.\,L.~Chien, Phys. Rev. Lett. \textbf{74}, 314 (1995).

\bibitem{2:Muhge}
Th.~M\"{u}hge, N.\,N.~Garif'yanov, Yu.\,V.~Goryunov, G.\,G.~Khaliullin, L.\,R.~Tagirov, K.~Westerholt,
I.\,A.~Garifullin, H.~Zabel, Phys. Rev. Lett. \textbf{77}, 1857 (1996).

\bibitem{2:Aarts}
J.~Aarts, J.\,M.\,E.~Geers, E.~Br\"uck, A.\,A.~Golubov, R.~Coehoorn, Phys. Rev.~B \textbf{56}, 2779 (1997).

\bibitem{2:Lazar}
L.~Lazar, K.~Westerholt, H.~Zabel, L.\,R.~Tagirov, Yu.\,V.~Goryunov, N.\,N.~Garif'yanov, I.\,A.~Garifullin, Phys. Rev.~B
\textbf{61}, 3711 (2000).

\bibitem{2:Ryazanov_JETPL}
V.\,V.~Ryazanov, V.\,A.~Oboznov, A.\,S.~Prokofiev, S.\,V.~Dubonos, Pis'ma Zh. Eksp. Teor. Fiz. \textbf{77}, 43 (2003)
[JETP Lett. \textbf{77}, 39 (2003)].

\bibitem{2:Aarts_recent}
A.~Rusanov, R.~Boogaard, M.~Hesselberth, H.~Sellier, J.~Aarts, Physica~C \textbf{369}, 300 (2002).

\bibitem{2:BVK}
A.\,I.~Buzdin, B.~Vuji\v{c}i\'c, M.~Yu.~Kupriyanov, Zh. Eksp. Teor. Fiz. \textbf{101}, 231 (1992) [Sov. Phys. JETP
\textbf{74}, 124 (1992)].

\bibitem{2:Demler}
E.\,A.~Demler, G.\,B.~Arnold, M.\,R.~Beasley, Phys. Rev.~B \textbf{55}, 15174 (1997).

\bibitem{2:PKhI}
Yu.\,N.~Proshin, M.\,G.~Khusainov, Zh. Eksp. Teor. Fiz. \textbf{113}, 1708 (1998);
\textbf{116}, 1887 (1999) [JETP \textbf{86}, 930 (1998); \textbf{89}, 1021 (1999)];\\
M.\,G.~Khusainov, Yu.\,N.~Proshin, Phys. Rev.~B \textbf{56}, R14283 (1997); \textbf{62}, 6832 (2000).

\bibitem{2:Tagirov}
L.\,R.~Tagirov, Physica~C \textbf{307}, 145 (1998).

\bibitem{2:Usadel}
K.\,D.~Usadel, Phys. Rev. Lett. \textbf{25}, 507 (1970).

\bibitem{2:LO}
A.\,I.~Larkin, Yu.\,N.~Ovchinnikov, in \textit{Nonequilibrium Superconductivity}, edited by D.\,N.~Langenberg and
A.\,I.~Larkin (Elsevier, New York, 1986), p.~530, and references therein.

\bibitem{2:RS}
J.~Rammer, H.~Smith, Rev. Mod. Phys. \textbf{58}, 323 (1986).

\bibitem{2:Golubov1}
A.\,A.~Golubov, M.\,Yu.~Kupriyanov, V.\,F.~Lukichev, A.\,A.~Orlikovskii, Mikroelektronika \textbf{12}, 355 (1983) [Sov.
J. Microelectronics \textbf{12}, 191 (1984)].

\bibitem{2:KL}
M.\,Yu.~Kupriyanov, V.\,F.~Lukichev, Zh. Eksp. Teor. Fiz. \textbf{94}, 139 (1988) [Sov. Phys. JETP \textbf{67}, 1163
(1988)].

\bibitem{2:digamma}
M.~Abramowitz, I.\,A.~Stegun, \textit{Handbook of Mathematical Functions} (Dover, New York, 1974).

\bibitem{2:Morse}
P.\,M.~Morse, H.~Feshbach, \textit{Methods of Theoretical Physics} (McGraw-Hill, New York, 1953), Vol.~1.

\bibitem{2:private_com}
V.V. Ryazanov, private communication.

\bibitem{2:Sarma}
G.~Sarma, J.~Phys. Chem. Solids \textbf{24}, 1029 (1963); see also D.~Saint-James, G.~Sarma, E.\,J.~Thomas, \textit{Type
II Superconductivity} (Pergamon, Oxford, 1969), p.~159.

\bibitem{2:Feynman}
R.\,P.~Feynman, A.\,R.~Hibbs, \textit{Quantum Mechanics and Path Integrals} (McGraw-Hill, New York, 1965).









\bibitem{2:Bergeret}
F.\,S.~Bergeret, K.\,B.~Efetov, A.\,I.~Larkin, Phys. Rev.~B \textbf{62}, 11872 (2000);\\
F.\,S.~Bergeret, A.\,F.~Volkov, K.\,B.~Efetov, Phys. Rev. Lett. \textbf{86}, 4096 (2001).

\bibitem{2:Kadigrobov}
A.~Kadigrobov, R.\,I.~Shekhter, M.~Jonson, Europhys. Lett. \textbf{54}, 394 (2001); Fiz. Nizk. Temp. \textbf{27}, 1030
(2001) [Low Temp. Phys. \textbf{27}, 760 (2001)].

\bibitem{2:VBE}
A.\,F.~Volkov, F.\,S.~Bergeret, K.\,B.~Efetov, Phys. Rev. Lett. \textbf{90}, 117006 (2003).

\bibitem{2:JLTP}
L.\,N.~Bulaevskii, A.\,I.~Rusinov, M.~Kuli\'c, J. Low Temp. Phys. \textbf{39}, 255 (1980).



\bibitem{2:Buzdin_EL}
A.\,I.~Buzdin, A.\,V.~Vedyayev, N.\,V.~Ryzhanova, Europhys. Lett. \textbf{48}, 686 (1999).

\bibitem{2:Khusainov}
M.\,G.~Khusainov, Yu.\,A.~Izyumov, Yu.\,N.~Proshin, Pis'ma Zh. Eksp. Teor. Fiz. \textbf{73}, 386 (2001) [JETP Lett.
\textbf{73}, 344 (2001)].

\bibitem{2:Buzdin_cm}
I.~Baladi\'e, A.~Buzdin, Phys. Rev.~B \textbf{67}, 014523 (2003).

\bibitem{2:Deutscher}
G.~Deutscher, F.~Meunier, Phys. Rev. Lett. \textbf{22}, 395 (1969).

\bibitem{2:ANL}
J.\,Y.~Gu, C.-Y.~You, J.\,S.~Jiang, J.~Pearson, Ya.\,B.~Bazaliy, S.\,D.~Bader, Phys. Rev. Lett. \textbf{89}, 267001
(2002).

\bibitem{2:Buzdin}
I.~Baladi\'e, A.~Buzdin, N.~Ryzhanova, A.~Vedyayev, Phys. Rev.~B \textbf{63}, 054518 (2001).

\bibitem{2:Berezinskii}
V.\,L.~Berezinskii, Pis'ma Zh. Eksp. Teor. Fiz. \textbf{20}, 628 (1974) [JETP Lett. \textbf{20}, 287 (1974)].

\bibitem{2:Larkin}
A.\,I.~Larkin, Pis'ma Zh. Eksp. Teor. Fiz. \textbf{2}, 205 (1965) [Sov. Phys. JETP Lett. \textbf{2}, 130 (1965)].

\bibitem{2:reentrant}
I.\,A.~Garifullin, D.\,A.~Tikhonov, N.\,N.~Garif'yanov, L.~Lazar, Yu.\,V.~Goryunov, S.\,Ya.~Khlebnikov, L.\,R.~Tagirov,
K.~Westerholt, H.~Zabel, Phys. Rev.~B \textbf{66}, 020505(R) (2002).

\end{thebibliography}

\begin{thebibliography}{99}
\addcontentsline{toc}{section}{References}

\bibitem{3:Bulaevski}
L.\,N.~Bulaevskii, V.\,V.~Kuzii, A.\,A.~Sobyanin, Pis'ma Zh. Eksp. Teor. Fiz. \textbf{25}, 314 (1977) [JETP Lett.
\textbf{25}, 290 (1977)].

\bibitem{3:Buzdin_pi}
A.\,I.~Buzdin, L.\,N.~Bulaevskii, S.\,V.~Panyukov, Pis'ma Zh. Eksp. Teor. Fiz. \textbf{35}, 147 (1982) [JETP Lett.
\textbf{35}, 178 (1982)].








\bibitem{3:Buzdin1}
A.\,I.~Buzdin, M.\,Yu.~Kupriyanov, Pis'ma Zh. Eksp. Teor. Fiz. \textbf{53}, 308 (1991) [JETP Lett. \textbf{53}, 321
(1991)].

\bibitem{3:Buzdin2}
A.\,I.~Buzdin, B.~Vuji\v{c}i\'{c}, M.\,Yu.~Kupriyanov, Zh. Eksp. Teor. Fiz. \textbf{101}, 231 (1992) [Sov. Phys. JETP
\textbf{74}, 124 (1992)].

\bibitem{3:Tanaka}
Y.~Tanaka, S.~Kashiwaya, Physica~C \textbf{274}, 357 (1997).

\bibitem{3:KK_p}
E.\,A.~Koshina, V.\,N.~Krivoruchko, Pis'ma Zh. Eksp. Teor. Fiz. \textbf{71}, 182 (2000) [JETP Lett. \textbf{71}, 123
(2000)];\\
E.\,A.~Koshina, V.\,N.~Krivoruchko, Phys. Rev.~B \textbf{63}, 224515 (2001).

\bibitem{3:Dobro}
L.~Dobrosavljevi\'{c}--Gruji\'{c}, R.~Ziki\'{c}, Z.~Radovi\'{c}, Physica~C \textbf{331}, 254 (2000).

\bibitem{3:Yip}
S.-K.~Yip, Phys. Rev.~B \textbf{62}, R6127 (2000).

\bibitem{3:Wilhelm}
T.\,T.~Heikkil\"{a}, F.\,K.~Wilhelm, G.~Sch\"{o}n, Europhys. Lett. \textbf{51}, 434 (2000).

\bibitem{3:Fogelstrom}
M.~Fogelstr\"{o}m, Phys. Rev.~B \textbf{62}, 11812 (2000).

\bibitem{3:Radovic01}
Z.~Radovi\'{c}, L.~Dobrosavljevi\'{c}--Gruji\'{c}, B.~Viji\'{c}i\'{c}, Phys. Rev.~B \textbf{63}, 214512 (2001).

\bibitem{3:Chtch}
N.\,M.~Chtchelkatchev, W.~Belzig, Yu.\,V.~Nazarov, C.~Bruder, Pis'ma Zh. Eksp. Teor. Fiz. \textbf{74}, 357 (2001) [JETP
Lett. \textbf{74}, 323 (2001)].

\bibitem{3:Barash_Bobkova}
Yu.\,S.~Barash, I.\,V.~Bobkova, Phys. Rev.~B \textbf{65}, 144502 (2002).

\bibitem{3:Ryazanov}
V.\,V.~Ryazanov, V.\,A.~Oboznov, A.\,Yu.~Rusanov, A.\,V.~Veretennikov, A.\,A.~Golubov, J.~Aarts, Phys. Rev. Lett.
\textbf{86}, 2427 (2001).

\bibitem{3:BVE}
F.\,S.~Bergeret, A.\,F.~Volkov, K.\,B.~Efetov, Phys. Rev. Lett. \textbf{86}, 3140 (2001).

\bibitem{3:KK_a}
V.\,N.~Krivoruchko, E.\,A.~Koshina, Phys. Rev.~B \textbf{64}, 172511 (2001).

\bibitem{3:Usadel}
K.\,D.~Usadel, Phys. Rev. Lett. \textbf{25}, 507 (1970).

\bibitem{3:KL}
M.\,Yu.~Kupriyanov, V.\,F.~Lukichev, Zh. Eksp. Teor. Fiz. \textbf{94}, 139 (1988) [Sov. Phys. JETP \textbf{67}, 1163
(1988)].



\bibitem{3:Koshina2}
E.\,A.~Koshina, V.\,N.~Krivoruchko, Fiz. Nizk. Temp. \textbf{26}, 157 (2000) [Low Temp. Phys. \textbf{26}, 115 (2000)].

\bibitem{3:Gol1}
A.\,A.~Golubov, M.\,Yu.~Kupriyanov, Zh. Eksp. Teor. Fiz. \textbf{96}, 1420 (1989) [Sov. Phys. JETP \textbf{69}, 805
(1989)].

\bibitem{3:KO1}
I.\,O.~Kulik and A.\,N.~Omelyanchuk, Pis'ma Zh. Eksp. Teor. Fiz. \textbf{21}, 216 (1975) [JETP Lett. \textbf{21}, 96
(1975)].

\bibitem{3:Likharev}
K.\,K.~Likharev, Rev. Mod. Phys. \textbf{51}, 101 (1979).

\bibitem{3:Lukichev}
V.\,F.~Lukichev, Fiz. Nizk. Temp. \textbf{10}, 1219 (1984) [Sov. J. Low Temp. Phys. \textbf{10}, 639 (1984)].








\bibitem{3:LikharevK}
K.\,K.~Likharev, \textit{Dynamics of Josephson Junctions and Circuits} (Gordon and Breach, Amsterdam, 1991).

\bibitem{3:Il'ichev}
E.~Il'ichev, M.~Grajcar, R.~Hlubina, R.\,P.\,J.~IJsselsteijn, H.\,E.~Hoenig, H.-G.~Meyer, A.~Golubov, M.\,H.\,S.~Amin,
A.\,M.~Zagoskin, A.\,N.~Omelyanchouk, M.\,Yu.~Kupriyanov, Phys. Rev. Lett. \textbf{86}, 5369 (2001).

\bibitem{3:Ryazanov_new}
V.\,V.~Ryazanov, V.\,A.~Oboznov, A.\,V.~Veretennikov, A.\,Yu.~Rusanov, Phys. Rev.~B \textbf{65}, 020501(R) (2001);\\
V.\,V.~Ryazanov, V.\,A.~Oboznov, A.\,V.~Veretennikov, A.\,Yu.~Rusanov, A.\,A.~Golubov, J.~Aarts, Usp. Fiz. Nauk (Suppl.)
\textbf{171}, 81 (2001).

\bibitem{3:Kontos}
T.~Kontos, M.~Aprili, J.~Lesueur, X.~Grison, Phys. Rev. Lett. \textbf{86}, 304 (2001);\\
T.~Kontos, M.~Aprili, J.~Lesueur, F.~Gen\^{e}t, B.~Stephanidis, R.~Boursier, Phys. Rev. Lett. \textbf{89}, 137007
(2002).



\bibitem{3:Meriakri}
Z.\,G.~Ivanov, M.\,Yu.~Kupriyanov, K.\,K.~Likharev, S.\,V.~Meriakri, O.\,V.~Snigirev, Fiz. Nizk. Temp. \textbf{7}, 560
(1981) [Sov. J. Low Temp. Phys. \textbf{7}, 274 (1981)].

\bibitem{3:Zubkov}
A.\,A.~Zubkov, M.\,Yu.~Kupriyanov, Fiz. Nizk. Temp. \textbf{9}, 548 (1983) [Sov. J. Low Temp. Phys. \textbf{9}, 279
(1983)].

\bibitem{3:Kupr}
M.\,Yu.~Kupriyanov, Pis'ma Zh. Eksp. Teor. Fiz. \textbf{56}, 414 (1992) [JETP Lett. \textbf{56}, 399 (1992)].

\bibitem{3:Zaitsev}
A.\,V.~Zaitsev, Zh. Eksp. Teor. Fiz. \textbf{86}, 1742 (1984) [Sov. Phys. JETP \textbf{59}, 1015 (1984)].

\bibitem{3:Volkov}
A.\,F.~Volkov, Phys. Rev. Lett. \textbf{74}, 4730 (1995).

\bibitem{3:Wilhelm1}
F.\,K.~Wilhelm, G.~Sch\"{o}n, A.\,D.~Zaikin, Phys. Rev. Lett. \textbf{81}, 1682 (1998).

\bibitem{3:Basel}
J.\,J.\,A.~Baselmans, A.\,F.~Morpurgo, B.\,J.~van Wees, T.\,M.~Klapwijk, Nature \textbf{397}, 43 (1999).

\bibitem{3:Dorokhov}
O.\,N.~Dorokhov, Solid State Commun. \textbf{51}, 381 (1984).

\end{thebibliography}

\begin{thebibliography}{99}
\addcontentsline{toc}{section}{References}

\bibitem{4:MSS}
Y.~Makhlin, G.~Sch\"{o}n, A.~Shnirman, Rev. Mod. Phys. \textbf{73}, 357 (2001).

\bibitem{4:Blatter}
G.~Blatter, V.\,B.~Geshkenbein, L.\,B.~Ioffe, Phys. Rev.~B \textbf{63}, 174511 (2001).

\bibitem{4:Ioffe}
L.\,B.~Ioffe, V.\,B.~Geshkenbein, M.\,V.~Feigel'man, A.\,L.~Fauch\`{e}re, G.~Blatter, Nature (London) \textbf{398}, 679
(1999).

\bibitem{4:Zagoskin}
A.\,M.~Zagoskin, cond-mat/9903170 (1999);\\
A.~Blais, A.\,M.~Zagoskin, Phys. Rev.~A \textbf{61}, 042308 (2000).

\bibitem{4:Il'ichev}
E.~Il'ichev, M.~Grajcar, R.~Hlubina, R.\,P.\,J.~IJsselsteijn, H.\,E.~Hoenig, H.-G.~Meyer, A.~Golubov, M.\,H.\,S.~Amin,
A.\,M.~Zagoskin, A.\,N.~Omelyanchouk, M.\,Yu.~Kupriyanov, Phys. Rev. Lett. \textbf{86}, 5369 (2001).





\bibitem{4:corr}
M.~Nielsen, I.~Chuang, \textit{Quantum Computation and Quantum Information} (Cambridge University Press, Cambridge,
2000).

\bibitem{4:ESA}
U.~Eckern, G.~Sch\"{o}n, V.~Ambegaokar, Phys. Rev.~B \textbf{30}, 6419 (1984).

\bibitem{4:SZ}
G.~Sch\"{o}n, A.\,D.~Zaikin, Phys. Rep. \textbf{198}, 237 (1990).

\bibitem{4:Bruder}
C.~Bruder, A.~van Otterlo, G.\,T.~Zimanyi, Phys. Rev.~B \textbf{51}, 12904 (1995).

\bibitem{4:BGZ}
Yu.\,S.~Barash, A.\,V.~Galaktionov, A.\,D.~Zaikin, Phys. Rev.~B \textbf{52}, 665 (1995).

\bibitem{4:CL}
A.\,O.~Caldeira, A.\,J.~Leggett, Phys. Rev. Lett. \textbf{46}, 211 (1981).

\bibitem{4:Leggett_review}
A.\,J.~Leggett, S.~Chakravarty, A.\,T.~Dorsey, M.\,P.\,A.~Fisher, A.~Garg, W.~Zwerger, Rev. Mod. Phys. \textbf{59}, 1
(1987).

\bibitem{4:Huck}
A.~Huck, A.~van~Otterlo, M.~Sigrist, Phys. Rev.~B \textbf{56}, 14163 (1997).

\bibitem{4:Amin}
M.\,H.\,S.~Amin, A.\,N.~Omelyanchouk, S.\,N.~Rashkeev, M.~Coury, A.\,M.~Zagoskin, Physica~B \textbf{318}, 162 (2002).

\bibitem{4:Eilenberger}
G.~Eilenberger, Z.~Phys. \textbf{214}, 195 (1968).

\bibitem{4:LO_qc}
A.\,I.~Larkin, Yu.\,N.~Ovchinnikov, Zh. Eksp. Teor. Fiz. \textbf{55}, 2262 (1968) [Sov. Phys. JETP \textbf{28}, 1200
(1969)].

\bibitem{4:Schopohl}
N.~Schopohl, K.~Maki, Phys. Rev.~B \textbf{52}, 490 (1995);\\
N.~Schopohl, cond-mat/9804064 (1998).

\bibitem{4:Andreev}
A.\,F.~Andreev, Zh. Eksp. Teor. Fiz. \textbf{46}, 1823 (1964); \textbf{49}, 655 (1965) [Sov. Phys. JETP \textbf{19},
1228 (1964); \textbf{22}, 455 (1966)].

\bibitem{4:Tinkham}
M.~Tinkham, \textit{Introduction to Superconductivity} (McGraw-Hill, Singapore, 1996), Ch.~6.4 and 7.3.

\bibitem{4:Grajcar}
M.~Grajcar, private communication.

\bibitem{4:Tzalenchuk}
A.\,Ya.~Tzalenchuk, T.~Lindstr\"om, S.\,A.~Charlebois, E.\,A.~Stepantsov, Z.~Ivanov, A.\,M.~Zagoskin, Phys. Rev.~B
\textbf{68}, 100501(R) (2003).

\bibitem{4:Hu}
C.-R.~Hu, Phys. Rev. Lett. \textbf{72}, 1526 (1994).

\bibitem{4:splitting}
Y.~Tanaka, S.~Kashiwaya, Phys. Rev.~B \textbf{53}, 9371 (1996);\\
R.\,A.~Riedel, P.\,F.~Bagwell, Phys. Rev.~B \textbf{57}, 6084 (1998);\\
Yu.\,S.~Barash, Phys. Rev.~B \textbf{61}, 678 (2000);\\
A.\,A.~Golubov, M.\,Yu.~Kupriyanov, Pis'ma Zh. Eksp. Teor. Fiz. \textbf{69}, 242 (1999) [JETP Lett. \textbf{69}, 262
(1999)];\\
A.~Poenicke, Yu.\,S.~Barash, C.~Bruder, V.~Istyukov, Phys. Rev.~B \textbf{59}, 7102 (1999).

\bibitem{4:Kulik}
I.\,O.~Kulik, Zh. Eksp. Teor. Fiz. \textbf{57}, 1745 (1969) [Sov. Phys. JETP \textbf{30}, 944 (1970)].

\end{thebibliography}

\begin{thebibliography}{99}
\addcontentsline{toc}{section}{References}

\bibitem{app:Buzdin_EL}
A.\,I.~Buzdin, A.\,V.~Vedyayev, N.\,V.~Ryzhanova, Europhys. Lett. \textbf{48}, 686 (1999).

\bibitem{app:McMillan}
W.\,L.~McMillan, Phys. Rev. \textbf{175}, 537 (1968).\\
The equations obtained by McMillan in the framework of the tunneling Hamiltonian method were later derived
microscopically (from the Usadel equations) in the work A.\,A.~Golubov, M.\,Yu.~Kupriyanov, Physica~C \textbf{259}, 27
(1996).\\
To avoid confusion, we note that Eq. (39) from McMillan's paper determining the critical temperature of a SN bilayer is
incorrect. The correct equation, following from Eqs. (37), (38), (40) and leading to Eq. (41) of McMillan's paper, reads
$\ln (T_{CS}/T_C) = (\Gamma_S/\Gamma)\, [ \psi ( \frac 12 +\Gamma/2\pi T_C ) -\psi (\frac 12 )]$.

\bibitem{app:Bergeret-Efetov}
F.\,S.~Bergeret, A.\,F.~Volkov, K.\,B.~Efetov, Phys. Rev. Lett. \textbf{86}, 3140 (2001); Phys. Rev.~B \textbf{64},
134506 (2001).

\end{thebibliography}
\end{document}